\documentclass[aip,jcp,reprint,longbibliography,floatfix]{revtex4-2}
\usepackage{amsmath,amssymb,mathtools,bm}
\usepackage{booktabs,graphicx,xcolor}
\usepackage{tikz}
\usetikzlibrary{arrows.meta,positioning,fit}
\usepackage{hyperref}
\graphicspath{{figures/}}
\hypersetup{colorlinks=true,linkcolor=blue!55!black,citecolor=blue!55!black,urlcolor=blue!55!black}
\newcommand{\vect}[1]{\bm{#1}}
\begin{document}
\title{Effective Harmonic and Anharmonic Models for Vibronic Spectra}
\author{Vincenzo Barone}
\email{vincebarone52@gmail.com}
\affiliation{INSTM, National Interuniversity Consortium for Materials Science and Technology, via G. Giusti 9, 50121 Firenze, Italy}
\begin{abstract}
Vibrationally resolved spectra of medium-sized molecules contain dense progressions, weak borrowed-intensity bands and large-amplitude structural motions. Their interpretation requires state-to-state assignments together with a complete spectral envelope, while a fully anharmonic treatment of every normal coordinate is usually impractical. We introduce a general workflow that builds both outputs from automatically generated nonredundant internal coordinates and a common transition-operator representation. The harmonic level combines time-independent (TI) assignments with time-dependent (TD) thermal correlations and an explicit omitted-strength bound for the positive harmonic dipole sum. A discrete-variable representation (DVR) promotes selected flexible coordinates while the remaining modes are contracted analytically. Perturbative and variational nuclear states extend the same operator treatment to separate-state modes with Duschinsky mixing or to shared modes with residual quadratic couplings. The workflow recovers approximately 99\% of the harmonic intensity for strongly reorganized medium-sized examples, reproduces measured absorption and fluorescence trends for anthracene and p-terphenyl, and gives signed electronic circular dichroism (ECD) profiles for camphor. Water photoionization, a diffuse-basis two-photon response tensor for the pyrazine $n\to3s$ Rydberg band, and nonplanar 4-pyrone calculations test continuum and anharmonic extensions. The result is a unified route from molecular structure to resolved assignments and numerically converged finite-model spectra.
\end{abstract}
\maketitle
\raggedbottom

\section{Introduction}
Vibrationally resolved spectroscopy is most informative when a molecular spectrum can be interpreted through assigned transitions and their collective envelope. This is also the regime in which the calculation becomes difficult. Medium-sized molecules combine dozens of vibrational modes, strong changes in equilibrium geometry, weak Herzberg--Teller intensity borrowing and flexible torsions or inversions. A useful calculation must preserve those chemical distinctions while accounting for essentially all of the intensity in the selected spectral window. It must also provide a practical route from a resolved line list to a thermal spectrum.

The central problem is therefore a coupled representation problem. Franck--Condon (FC) amplitudes describe nuclear overlap, Herzberg--Teller (HT) terms describe the first coordinate dependence of the electronic transition operator, and the nuclear Hamiltonian determines how both contributions are distributed among vibrational states. The time-independent (TI) formulation resolves individual transitions; the time-dependent (TD) formulation contracts the same operator over the full thermal manifold. Agreement between the two is the most direct way to connect assignments with the observed spectral envelope.

Our solution starts from automatically generated extended symmetry-adapted local coordinates (ESALC): ESALC-I retains nonlinear internal-coordinate functions, while ESALC-C supplies their Cartesian tangent representation. The molecular perception and coordinate construction are provided by ORACLE, with the ESALC definitions described in Refs.~\cite{Barone2026ORACLE,Barone2026SONIC,Barone2026ESALC}. Bond stretches, bends, torsions and out-of-plane motions retain their family identity and local support. This makes the coordinates suitable for medium-sized molecules with localized bending, torsional and out-of-plane displacements.

The workflow follows a controlled hierarchy. A complete harmonic calculation supplies the baseline: every positive-curvature mode is retained, TI selects resolved lines until the requested intensity and profile criteria are met, and TD provides the independent thermal correlation. Flexible coordinates can then be promoted to a coupled discrete-variable representation (DVR) while the remaining harmonic bath is integrated analytically. Small-amplitude anharmonicity is treated perturbatively when the corrected states are stable; resonant or large-amplitude subspaces are treated variationally with vibrational configuration interaction (VCI). The two electronic surfaces can use their own normal modes with Duschinsky mixing, or a shared nuclear basis whose residual quadratic couplings are retained explicitly. In the latter representation the two force matrices are jointly reduced by an orthogonal common-mode rotation, and the residual couplings are retained in the finite Hamiltonians.

The resulting method preserves the chemical identity of internal coordinates, returns state-to-state assignments and complete profiles from the same Hamiltonian and operator, exposes an absolute omitted-strength bound for the positive harmonic dipole sum, and increases the nuclear model only where the spectrum requires it. The general position--momentum operator also covers velocity, magnetic and continuum couplings; the applications include dipole, chiroptical, continuum and two-photon vertices.

We assess the hierarchy on controlled fields, anthracene, p-terphenyl, camphor, water, pyrazine and 4-pyrone. The tests include measured absorption, fluorescence and chiroptical profiles, continuum photoionization, common-mode anharmonic states and coupled torsions. Comparisons with established vibronic programs use identical physical inputs and are used to verify the harmonic reference. The remainder of the paper introduces the molecular representation, derives the TI and TD contractions, and then follows the progression from complete harmonic spectra to selected anharmonic models.

The contributions developed and assessed here are:
\begin{enumerate}
\item A general linear position--momentum operator contracted through shared Gaussian amplitudes, with a positive harmonic strength sum rule and a computable global omission bound. The molecular operators use $p_{aj}=0$; the momentum term supplies a formal interface for velocity and nonadiabatic operators.
\item A window-directed TI construction that returns state-to-state FC/HT assignments and checks positive-strength completeness and each signed observable against an independent TD profile in the requested interval.
\item A coupled-DVR/harmonic-bath contraction that combines adiabatic bath displacement, its Born--Huang contribution and torsion-dependent displaced Gaussian optical overlaps, retaining linear FC/HT interference on both subspaces.
\item A common-mode contraction of two anharmonic nuclear Hamiltonians, retaining residual position and kinetic quadratic couplings and transporting the transition operator consistently. The resulting spectral amplitudes are direct contractions of the two nuclear eigenvector sets, providing a compact shared representation whose reuse and cost are evaluated on the matched examples.
\end{enumerate}
The first two constructions supply a controlled harmonic reference and assigned spectra; the latter two extend the nuclear model while preserving the operator treatment.

Harmonic FC and HT theory provides an established starting point. The normal-coordinate mixing and displacement underlying the two-surface harmonic construction go back to the Duschinsky and Sharp--Rosenstock formulations, with the multidimensional coherent-state form given by Doktorov and co-workers.\cite{Duschinsky1937,SharpRosenstock1964,Doktorov1977} The linear property expansion is the Herzberg--Teller construction.\cite{HerzbergTeller1933} TI sums resolve individual transitions; TD correlations include the oscillator manifold without explicit final-state enumeration.\cite{Baiardi2013,FCclasses2023,Heller1975} Recurrence and generating-function implementations span Hermite, Gaussian and class-based formulations.\cite{Gruner1987,Ruhoff1994,Santoro2007,Dodonov1994} Internal coordinates can improve the representation of bending and torsional displacements, although a quadratic potential in those coordinates remains a harmonic approximation.\cite{Baiardi2016} Anharmonic wavefunctions and the continuation of local force data to finite displacements introduce further, distinct modeling choices.\cite{Luis2004,Rodriguez2006,Bowman1986,Christiansen2004VCC}

Integrated FC/HT optical workflows and anharmonic frequency corrections already form part of the \href{https://gaussian.com/}{Gaussian}-based vibronic literature.\cite{Barone2009Integrated,Bloino2010OnePhoton,Baiardi2013} Time-dependent wavepacket and Raman formulations provide the complementary propagation viewpoint used by TD spectra.\cite{Heller1975,LeeHeller1979} A direct precursor combines a numerical DVR treatment of one large-amplitude coordinate with harmonic transverse modes in topology-derived internal coordinates; it also permits the transverse normal modes to vary along that coordinate.\cite{Baiardi2017HybridDVR} This establishes the starting partition for the coupled nuclear and optical constructions developed here.

The applications instantiate this hierarchy in sequence: complete harmonic FC/HT spectra establish the reference, a two-torsion DVR resolves a flexible subspace in p-terphenyl, and perturbative or variational nuclear states extend the same TI/TD operators in separate-state and common-mode bases. Matched harmonic timings and coverage curves are given in the Supporting Information (SI).

The implementation is publicly distributed as \href{https://github.com/yogibubu/SATI-standalone}{SATI} (Symmetry-Adapted Transition Intensities), a self-contained spectroscopy package.\cite{SATISoftware2026} It includes the coordinate builder, force-field services, harmonic and anharmonic solvers, and input adapters required by the reported workflows. The package includes its required scientific dependencies. The SI specifies installation checks and electronic-program interfaces.

The electronic quantities needed by the different spectroscopies are supplied by distinct electronic-structure programs: ground- and excited-state energies and force fields, transition moments and their derivatives, continuum amplitudes, and chiroptical or nonlinear-response tensors are expressed in a common property representation. Automatic interfaces connect these programs to the same nuclear and vibronic engine, so that changing the electronic source does not change the coordinate construction, TI/TD contractions or anharmonic solvers.

Controlled oscillator and coordinate-covariance tests establish numerical consistency. Water photoionization connects continuum-coupled harmonic amplitudes to an internal-coordinate zero-asymptote fitted force-field (ZAFF)/DVR reference, with a separate matched-field perturbation control and comparison to established water photoelectron measurements and calculations.\cite{BotterRosenstock1969,Reutt1986Water,Engin2019Water,EzSpectra2022} The molecular assessment then follows coupled torsions, common-mode nuclear states and signed spectral assignments. The SI contains the coherent Raman contraction,\cite{RamanTI2014,RamanTD2014} the diatomic solver check, detailed timing measurements and continuum-operator validation.

\section{Molecular representation and potential models}
\subsection{Automatic ESALC-I and ESALC-C representations}
\label{sec:esalc-representation}
Let $\vect{s}$ denote a frozen common ESALC representation for the two electronic states. Around a reference point $\vect{s}_r$, the potential is
\begin{equation}
 V(\vect{s})=E_r+\vect{g}^{\mathsf T}\Delta\vect{s}
 +\tfrac12\Delta\vect{s}^{\mathsf T}\vect{F}\Delta\vect{s},
 \qquad \Delta\vect{s}=\vect{s}-\vect{s}_r .
\end{equation}
ESALC-I construction preserves the identity of primitive families (bond stretches, bends, torsions and out-of-plane motions) and their local molecular supports. Symmetry adaptation and redundancy removal are applied within the prescribed family and locality structure. A global delocalization can mix distant primitives and different families while spanning the same vibrational tangent space. The distinction matters for following a chemically meaningful displacement between electronic states and for selecting subsequent anharmonic subspaces.

Both realizations start from the same molecular perception, primitive inventory and symmetry adaptation. ESALC-I retains the nonlinear internal-coordinate functions. ESALC-C freezes their Cartesian tangent representation at a named reference, supplying a local Cartesian realization without changing the automatically assigned chemical families. Finite torsional displacements use ESALC-I; the local tangent representation supplies Cartesian reference and covariance controls.\cite{Barone2026SONIC,Barone2026ESALC}

\begin{figure*}[t]
\centering
\includegraphics[width=.88\textwidth]{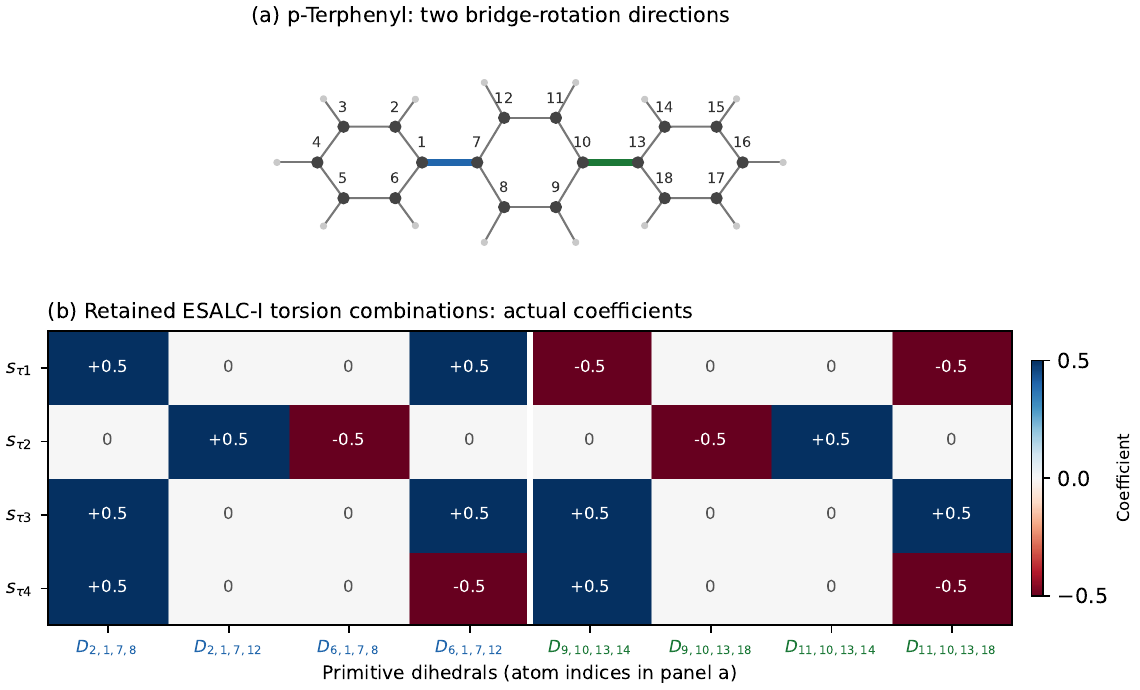}
\caption{ESALC-I torsional block of p-terphenyl. The two inter-ring bonds and the four retained combinations of their eight proper-dihedral primitives are shown with the original coefficients. The two independent bridge-rotation directions retain the torsion family identity.}
\label{fig:esalc-torsions}
\end{figure*}

Concretely, molecular connectivity supplies primitive bond lengths, valence angles and dihedrals $f_p(\vect{x})$. Their increments form the coordinate representation through fixed coefficients,
\begin{equation}
 \begin{aligned}
 \Delta s_a(\vect{x})&=\sum_p C_{ap}\,\Delta f_p(\vect{x}),\\
 \vect{B}_s&=\vect{C}\vect{B}_{\rm prim},\qquad
 \vect{G}=\vect{B}_s\vect{M}^{-1}\vect{B}_s^{\mathsf T}.
 \end{aligned}
 \label{eq:esalc-construction}
\end{equation}
Here $\vect{M}$ is the Cartesian atomic-mass matrix and each Wilson row is a coordinate derivative. Dihedral increments are continued about their reference using $\operatorname{atan2}(\sin\Delta\phi,\cos\Delta\phi)$. Symmetry adaptation groups chemically equivalent primitives within their family and molecular support; redundancy removal retains the nonlinear molecule's $3N-6$ tangent directions. Explicit bridge-rotation responses are checked in the torsional block as well as in the full Wilson matrix. The same coefficients and primitive references define both electronic surfaces. ESALC-C replaces the primitive functions by their Cartesian tangent at the chosen reference. Figure~\ref{fig:esalc-torsions} shows actual torsional coefficients for p-terphenyl; symmetry-related bridges can enter one combination while its primitive family remains identifiable.

A positive local Wilson matrix $\vect{G}$ specifies the kinetic metric. Wilson G/F analysis supplies frequencies and a transformation $\vect{T}$ from mass-normalized harmonic coordinates to ESALC displacements. Both the force field and the metric belong to a named coordinate representation; fields with different coordinate representation identities must be transported before they can be combined.

For a nonzero gradient, the minimum of the local quadratic potential and its electronic relaxation energy are
\begin{equation}
 \vect{s}_*=\vect{s}_r-\vect{F}^{-1}\vect{g},\qquad
 \Delta E_{\mathrm{rel}}=-\tfrac12\vect{g}^{\mathsf T}\vect{F}^{-1}\vect{g}.
 \label{eq:relax}
\end{equation}
This operation evaluates the supplied quadratic model and its relaxation energy. The electronic gap supplied at the two expansion points is corrected by the difference between their relaxation energies. The zero-point contribution is then included to obtain the vibronic origin.

Nonstationary Cartesian derivatives require the coordinate-curvature term. In terms of the back transformation $\vect{x}(\vect{s})$, with $A_{ai}=\partial x_a/\partial s_i$,
\begin{equation}
 (F_s)_{ij}=(\vect{A}^{\mathsf T}\vect{H}_x\vect{A})_{ij}
 +\sum_a (g_x)_a\frac{\partial^2 x_a}{\partial s_i\partial s_j}.
 \label{eq:curvature}
\end{equation}
The Gaussian ingestion route evaluates this correction using the existing analytic Wilson rows and central differences of those rows. The finite-difference step is refined explicitly; no analytic second-derivative implementation is implied. A linear projection of a Cartesian Hessian is insufficient for a vertical calculation at a nonstationary point. Native ESALC fields are assumed to contain the appropriate curvilinear derivatives.

For finite displacements, the common coordinate representation uses nonlinear bond lengths, valence angles and reference-centered dihedrals. Explicit torsional coordinates are retained and their response rank is checked independently of the full Wilson-matrix rank. This identifies the coordinate model before either spectral solver is used; acquisition and coordinate representation checks are documented in the SI.

\subsection{Adiabatic and vertical approximations}
The potential model and the transition-property expansion are independent choices. The potential models follow the established adiabatic and vertical families.\cite{FCclasses2023} Adiabatic Hessian (AH) uses the two equilibrium geometries and their harmonic force fields. Adiabatic shift (AS) retains the reference-state force field and metric and uses the other equilibrium geometry only for the displacement. The adiabatic shift-frequency (ASF) model updates the diagonal curvatures in the reference normal coordinates while suppressing mixing between distinct reference modes.

Vertical Hessian (VH) expands the other surface at the reference-state minimum and obtains its quadratic relaxation from Eq.~\eqref{eq:relax}. Vertical gradient (VG) uses the reference Hessian with the other-state gradient; vertical gradient-frequency (VGF) additionally updates the diagonal projected curvatures. Within exactly degenerate reference-frequency subspaces, the projected other-state curvature is diagonalized before constructing ASF or VGF, avoiding dependence on an arbitrary degenerate-mode gauge. General local fields, including two nonstationary expansions, are accepted by the generalized quadratic formulation.

The reference may be either supplied electronic surface. This permits a vertical model at the ground-state minimum for absorption or at the emitting-state minimum for emission. The input electronic gap always refers to the supplied expansion points. Explicit adiabatic models reject nonstationary minimum data, and explicit vertical models require a gradient at the reference minimum. The present harmonic reference uses positive curvatures. Branch-cut-free formulations also treat unbound and imaginary-frequency vibrations.\cite{Changala2022Imaginary} Separating a genuinely flexible coordinate into a different vibrational treatment is a physical model decision; the established FCclasses3 program provides iterative projectors for defining the corresponding reduced harmonic space.\cite{FCclasses2023}

\section{Transition amplitudes and complementary spectral formulations}
\subsection{Gaussian generator and operator insertions}
The two local normal-coordinate systems are related by
\begin{equation}
 \vect{Q}_f=\vect{J}\vect{Q}_i+\vect{k},\quad
 \vect{J}=\vect{T}_f^{-1}\vect{T}_i,\quad
 \vect{k}=\vect{T}_f^{-1}(\vect{s}_{i,*}-\vect{s}_{f,*}).
\end{equation}
State-specific local metrics can produce a nonorthogonal map. The overlap integral retains the coordinate Jacobian $|\det\vect{J}|^{1/2}$, normalizing the local oscillator representations in a common integration measure.

The normalized overlap generating function has the Gaussian form
\begin{equation}
 \mathcal{G}(\vect{z})=C_0\exp\!\left(
 \tfrac12\vect{z}^{\mathsf T}\vect{R}\vect{z}
 +\vect{r}^{\mathsf T}\vect{z}\right),
\end{equation}
where $\vect{z}$ contains initial and final creation variables. Gaussian integration determines $C_0$, $\vect{R}$ and $\vect{r}$ from the two frequencies, coordinate map and displacement. The normalized Hermite coefficients obey
\begin{equation}
 I_{\vect{n}}=\frac{r_j I_{\vect{n}-\vect{e}_j}
 +\sum_{k:\,n_k>\delta_{jk}} R_{jk}\sqrt{\bigl(n_k-\delta_{jk}\bigr)}\,
 I_{\vect{n}-\vect{e}_j-\vect{e}_k}}{\sqrt{n_j}},
 \label{eq:recurrence}
\end{equation}
for $n_j>0$, with $I_{\vect{0}}=C_0$. The square-root factor follows from normalized oscillator states; the stated summation range keeps every referenced occupation nonnegative. The same coefficients provide FC amplitudes and linear HT insertions. Transition-dipole derivatives supplied in ESALC coordinates are transformed to the initial dimensionless normal coordinates and transported to the relaxed property reference.

More generally, a channel-resolved coupling can be written
\begin{equation}
 O_a=c_a+\sum_j(m_{aj}x_j+p_{aj}\pi_j),\qquad
 \pi_j=-i\frac{\partial}{\partial x_j}.
 \label{eq:operator}
\end{equation}
The momentum term gives the general linear operator insertion; electric-dipole FC/HT calculations retain the position terms. Acting on the initial harmonic generator converts it to another linear Gaussian insertion. At thermal equilibrium, with $\bar n_j$ the Bose occupation, its total strength is
\begin{align}
 S_O={}&\sum_a|c_a|^2\nonumber\\
 &+\tfrac12\sum_{aj}\left[
 |m_{aj}-ip_{aj}|^2\bar n_j+
 |m_{aj}+ip_{aj}|^2(\bar n_j+1)\right].
 \label{eq:sumrule}
\end{align}
This positive expression retains the position--momentum commutator and complex interference. Electric-dipole FC/HT spectra use $p_{aj}=0$.

\subsection{A few coupled DVR coordinates and an analytic harmonic bath}
\label{sec:hybrid-dvr-hierarchy}
Anharmonicity is introduced where the harmonic description changes the
observable being interpreted. A large-amplitude torsion, inversion or bond
stretch is represented by its finite-range coordinate function and its full
one- or few-dimensional kinetic operator. The remaining modes retain the
analytic harmonic contraction. This separation preserves the dense bath of a
medium-sized molecule while resolving the coordinate that controls the band
positions, wavefunction localization or intensity borrowing.

Active-space selection ranks potential, kinetic and FC/HT contributions separately;
then test nested one- and few-coordinate blocks against assigned lines and the
corresponding TD profile while retaining the harmonic bath analytically. A
coordinate with a small direct Franck--Condon amplitude can still be essential
when it modulates the transition moment or couples two bright configurations.
In the p-terphenyl application the two torsions were specified a priori and
then checked by nested contractions; the general ranking protocol is reported
for reuse on other molecules. The same transition-operator representation serves the enlarged active Hamiltonian.

The three terms that control the intermediate model are the finite-coordinate
potential, the coordinate-dependent kinetic metric and the response of the
transition operator. Keeping only the first term can reproduce low levels while
missing intensity redistribution. Keeping the metric but freezing the operator
can reproduce energies while missing FC/HT interference. The implementation
transports all three terms together and checks them against a nested variational
calculation or an independent TD profile.

Let $a$ denote a small set of explicit anharmonic coordinates and $b$ the
remaining harmonic coordinates. A useful intermediate level is
\begin{equation}
 H_s=H_{a,s}+\tfrac12 p_b^T G_{b,s}p_b
             +\tfrac12 b^T F_{b,s}b .
 \label{eq:hybrid-dvr-hierarchy}
\end{equation}
The linear optical operator is $\mu(a,b)=\mu_a(a)+D_b b$.
The active Hamiltonian retains its coupled potential, inverse kinetic metric
and volume connection. The bath retains its two-state harmonic mixing.
The active transition matrices are evaluated by numerical contraction, while
Gaussian correlations sum the harmonic bath analytically. Signed mixed
operator traces retain interference between HT insertions in the active
space and in the bath. Thus adding a second DVR coordinate enlarges the
active nuclear problem, without adding an explicit occupation cutoff to
all remaining modes. The harmonic bath and active subspace are defined
before either TI or TD evaluation, preventing double counting of a flexible
motion.

The direct precursor treats one large-amplitude coordinate with DVR in a
modified Gaussian implementation.\cite{Baiardi2017HybridDVR} FCclasses3
provides iterative projectors that can remove several flexible coordinates
from its harmonic subset; the published package does not itself compute
anharmonic DVR spectra.\cite{FCclasses2023} These are distinct capabilities.
Here the molecular example treats the two inter-ring torsions of
p-terphenyl jointly and retains an 88-mode harmonic bath.

Leading potential coupling to the bath can be examined from a periodic
linear force $g_s(a)^Tb$. Completing the harmonic square gives
\begin{equation}
 \Delta V_s(a)=-\tfrac12 g_s(a)^T F_{b,s}^{-1}g_s(a).
 \label{eq:adiabatic-bath-response}
\end{equation}
In dimensionless bath normal coordinates this is
$-\sum_j g_{sj}(a)^2/(2\omega_{sj})$. The implemented adiabatic response
also includes the diagonal Born--Huang displacement contribution. Forces
are ranked separately on the two surfaces by their population-weighted
contribution to Eq.~\eqref{eq:adiabatic-bath-response}; reduced selections
are checked against the complete response on the emission profile.
The reference-fiber calculation retains the original Gaussian optical bath
while this response enters the active Hamiltonians. A further step translates
the optical fibers themselves by $d_s(a)=-g_s(a)/\omega_s$. In dimensionless
normal coordinates, the Gaussian displacement then becomes
\begin{equation}
 k(a)=k+A d_i(a)-d_f(a),\qquad
 A=\omega_f^{1/2}J\omega_i^{-1/2}.
 \label{eq:translated-bath-shift}
\end{equation}
Here the diagonal frequency matrices supply the indicated square roots.
The reference curvatures and mode map are retained. Interpolating the bath
transition amplitudes in the active coordinates gives a finite bank of
cross-source Gaussian kernels. Their contraction with signed active operator
traces retains bath Duschinsky mixing, temperature and linear FC/HT
interference, without a bath occupation cutoff. One factorization per time
serves all interpolation nodes. The SI validates this optical extension and
checks the underlying adiabatic response against explicit oscillator
coupling in a contracted variational space.

\subsection{Perturbative anharmonic states on both surfaces}
Small-amplitude anharmonicity can be included by correcting the nuclear wavefunctions on both electronic surfaces while retaining the harmonic integrals as reusable building blocks. Perturbative anharmonic FC calculations with Duschinsky mixing have a substantial methodological history.\cite{Luis2004,Luis2005,Bonness2006,Bloino2012HDC,Bloino2012Moments,Bloino2015Intensities,Yang2021} Integrated optical frameworks also use second-order vibrational perturbation theory (VPT2) frequencies to renormalize harmonic overlaps.\cite{Barone2009Integrated,Bloino2010OnePhoton,Baiardi2013} The implementation contracts explicit sparse wavefunction expansions, retaining the non-Gaussian corrections required for state-to-state intensities.

For each surface $s$, write $H_s=H_{0s}+\lambda V_{3s}+\lambda^2V_{4s}$. The operators include the supplied cubic and quartic potential terms and, for a curvilinear field, the corresponding kinetic-metric contributions. Rayleigh--Schr\"odinger perturbation theory gives
\begin{equation}
 |\Psi_{sn}\rangle=|n;s\rangle+
 \lambda|n;s\rangle^{(1)}+
 \lambda^2|n;s\rangle^{(2)}.
 \label{eq:samstate}
\end{equation}
The second-order correction contains the direct quartic term, two successive cubic interactions, and the normalization contribution $-\tfrac12\bigl\||n;s\rangle^{(1)}\bigr\|^2|n;s\rangle$. Full wavefunction corrections generally require quartic elements beyond the semidiagonal field sufficient for selected energy corrections. All tensors must share the same mode order, signs and coordinate gauge as their harmonic surface.

Perturbative reliability is assessed for each target state through its energy and wavefunction corrections. For every intermediate state $m$, the coupling
$\langle m|V_{3s}+V_{4s}|n\rangle$ is compared with the harmonic gap
$E_{n}^{(0)}-E_m^{(0)}$. We use the dimensionless indicator
$\eta_{mn}=|\langle m|V_{3s}+V_{4s}|n\rangle|/|E_n^{(0)}-E_m^{(0)}|$; values approaching unity, or clusters with several comparable denominators, are a recommendation to define a variational block. The user-selected variational blocks are diagonalized together to retain intensity borrowing. The resulting path connects isolated VPT2 states to selected VCI states without assigning a resonance to a single unperturbed oscillator; the polyad-transport construction provides the corresponding perturb-then-diagonalize context.\cite{Barone2026ESALC}

Duschinsky mixing continues to connect the underlying harmonic bases. For a supplied FC/HT operator, contraction of Eq.~\eqref{eq:samstate} on both surfaces gives $M=M^{(0)}+\lambda M^{(1)}+\lambda^2M^{(2)}$. Thus $M^{(2)}$ includes both second-order wavefunction corrections and the product of the two first-order corrections. The consistently truncated strength is
\begin{align}
 I^{[2]}={}&|M^{(0)}|^2+2\operatorname{Re}[M^{(0)*}M^{(1)}]\nonumber\\
 &+|M^{(1)}|^2+2\operatorname{Re}[M^{(0)*}M^{(2)}],
 \label{eq:samstrength}
\end{align}
with Cartesian components summed for a vector operator. Squaring the corrected total moment instead retains additional higher-order terms; that positive completion is recorded separately. Neither quantity is silently clipped or renormalized. This mechanical perturbation order is distinct from the Taylor order of the electronic property; the present extension retains constant and linear FC/HT operators.

The perturbative route is used for weakly coupled states, with first- and
second-order correction norms reported for every target. These norms diagnose
series stability but are not rigorous bounds on the omitted anharmonic space.
Resonant or strongly mixed states are selected for a variational subspace, where
the target states are represented explicitly, including states whose harmonic
intensity is weak or zero. TI then assigns the resulting eigenstates, while TD
contracts the same finite Hamiltonian when a complete profile is required. The analytic
Gaussian TD reference remains the harmonic limit of this hierarchy.

\subsection{Common modes for anharmonic TI and TD contractions}
A common nuclear basis provides an alternative to separate-state normal modes. Starting from the two normal-mode descriptions, an affine transformation $s=s_0+Lq$ is chosen to reduce the off-diagonal quadratic potential terms on both surfaces. In a reference mass coordinate representation, a weighted Jacobi optimization minimizes the sum of squared off-diagonal elements of the transformed force matrices. The two state-specific kinetic metrics, displacement vectors and scalar energy origins are transported consistently.

The resulting Hamiltonians retain their residual quadratic position and momentum couplings. Their variational or perturbative treatment can differ between surfaces and must be checked in the occupied and virtual nuclear spaces relevant to the spectrum. With a common set of modal functions, an FC amplitude becomes a direct contraction of nuclear eigenvectors; the mode mixing is carried by the Hamiltonians. Native anharmonic functions must be transported in full rather than truncated again after the coordinate change.

The validation target is equality of harmonic and anharmonic spectra against the separate-state Duschinsky formulation on the same Hamiltonians and transition operators. For the harmonic water model, the complete common-Hamiltonian FC correlation agrees with the standard result within $5.2\times10^{-12}$ on all 4097 time points. A shared variational basis with 1330 configurations reproduces the first 40 line positions within $4.45\times10^{-4}$~cm$^{-1}$. Their FC/HT strengths, including signed interference at five electron energies, agree within $1.7\times10^{-14}$ of the FC+HT sum rule. These lines retain more than 99.995\% of the total intensity. Thus the common-coordinate representation reproduces both assigned transitions and the harmonic envelope.

The anharmonic extension evaluates the complete accepted native potentials in the common coordinate representation. A full-product sine-Galerkin reference transports the coordinate-dependent kinetic metric and physical-volume connection exactly. Nuclear-domain, basis and quadrature convergence are checked separately before selecting or perturbatively dressing VCI configurations. Weak-band intensities require their own domain and basis controls. The water anharmonic check uses 2600 selected products and reproduces five assigned bands of the independently converged bond--angle reference within 0.0103~cm$^{-1}$ and $8.5\times10^{-6}$ relative FC+HT strength. Separate complete-product controls distinguish selection error from coordinate-domain effects; the SI reports the retained spectral profiles and their limits. Optical spectra of 4-pyrone provide a further target; its decay-rate application belongs to a separate photophysics study.

The numerical benefit depends on the task. A matched checkpoint-based harmonic
benchmark for 4-pyrone uses the same 27-mode fields, FC/HT operator and 4097 time
samples. Median correlation times over five warmed, alternating-order runs are
0.832~s in common modes and 0.959~s in the standard representation for
absorption; emission gives 0.832 and 0.958~s. Common-frame preparation adds
0.65--0.83~s and is reusable. The preparation cost exceeds the contraction saving for a single profile and can be shared across repeated operator calculations. For the matched anharmonic water
operator, selecting 2600 products instead of the complete 13824 reduces the
basis dimension by a factor of 5.32 while preserving the five assigned FC+HT
strengths within $6.3\times10^{-8}$ relative to that product reference.
This demonstrates useful compression of one fixed Hamiltonian. The comparison measures compression of this matched Hamiltonian. The SI separates preparation, solution and contraction costs.

\subsection{TI amplitudes and missing strength}
TI evaluates each selected transition with Eq.~\eqref{eq:recurrence}, including its initial thermal population. Candidates are ranked by one- and two-mode Gaussian occupation marginals assembled into a conditional-tree proposal. Linear-operator neighbors supplement the guide for HT transitions. The guide controls which transitions are visited; their physical intensities always use the full coupled recurrence. Excitations may involve every mode, subject to explicit occupation and memory budgets. The selection algorithm and overlap-cache implementation are described in the SI.

Convergence is assessed from the physical, positive dipole-strength sum,
\begin{equation}
 S_{\mathrm{miss}}=S_O-\sum_{\mathrm{reported}\;i,f}p_i
 |\langle f|O|i\rangle|^2 .
\end{equation}
For the complete harmonic model this gives an a posteriori omission bound and, at every time, a bound on the absolute difference between the selected TI correlation and the complete harmonic correlation. A weak or zero parent amplitude never discards its descendants. Explicit occupation, transition and memory limits accompany each retained-strength fraction, without renormalization. This harmonic bound complements the local batch-prescreening bounds of Jankowiak \textit{et al.}\cite{Jankowiak2007} The bound applies to positive harmonic strengths. Anharmonic convergence is established by nested nuclear spaces, and signed chiroptical signals require window-resolved comparison against TD. The ECD and circularly polarized luminescence (CPL) profiles are therefore checked against TD in the requested window, while the positive harmonic dipole sum supplies the rigorous omitted-strength certificate.

\subsection{TD thermal correlation and Fourier transform}
The TD formulation evaluates the thermal trace directly,
\begin{equation}
 C_O(t)=\operatorname{Tr}\!\left[
 \rho_i O^\dagger e^{-iH_f t/\hbar}Oe^{iH_i t/\hbar}\right],
 \end{equation}
 \begin{equation}
 S_O(\widetilde\nu)=\frac{1}{2\pi}
 \int C_O(\tau)e^{i\widetilde\nu\tau}\,d\tau,
 \label{eq:correlation}
\end{equation}
where $\tau=2\pi ct$ when $\widetilde\nu$ is a wavenumber. The analytic Gaussian contraction includes the full thermal oscillator sum and is independent of TI truncation. For stable positive-frequency modes, its determinant is evaluated on the analytic unit-disk branch. The thermal executor uses complex symmetric resolvents with positive-real-part pivot blocks to obtain this branch without repeated eigenvalue calculations, retaining the spectral evaluator as a numerical fallback. This prescription is tested here within the stable oscillator domain. Changala, Genossar and Baraban provide a broader branch-cut-free correlation formulation that includes imaginary and unbound vibrations;\cite{Changala2022Imaginary} the oscillator-domain restriction used here is summarized with the other model assumptions below. The zero-time trace is checked against Eq.~\eqref{eq:sumrule}.

\begin{figure}[t]
\centering
\begin{tikzpicture}[node distance=3mm and 3mm,box/.style={draw=blue!55!black,rounded corners,align=center,font=\scriptsize,inner sep=3pt,text width=24mm},arr/.style={-{Latex[length=2mm]},thick,blue!55!black}]
\node[box] (qc) {Electronic-structure data\\geometries, energies, derivatives\\dipoles, spin--orbit (SOC) or nonadiabatic coupling (NAC) operators};
\node[box,right=of qc] (esalc) {Common internal-coordinate representation\\curvilinear derivative transport\\Wilson G/F metrics};
\node[box,right=of esalc] (model) {AH / AS / ASF\\VH / VG / VGF / local fields\\consistent property references};
\node[box,below=of esalc] (ti) {Compiled TI\\state selection and shared overlaps\\resolved vibronic lines};
\node[box,right=of ti] (td) {Parallel TD\\analytic Gaussian propagators\\Fourier transform / quadrature};
\node[box,left=of ti] (obs) {Observables\\FC/HT and resonance Raman\\spectra and assigned transitions};
\draw[arr] (qc)--(esalc);\draw[arr] (esalc)--(model);
\draw[arr] (model.south)--(td.north);\draw[arr] (model.south west)--(ti.north east);
\draw[<->,dashed,blue!55!black] (td)--(ti);\draw[arr] (ti)--(obs);
\draw[arr] (td.south) to[out=-90,in=-90] (obs.south);
\end{tikzpicture}
\caption{Information flow for the harmonic calculation. TI assigns individual transitions and TD recovers the complete thermal profile from the same molecular model. NAC denotes a nonadiabatic coupling operator.}
\label{fig:workflow}
\end{figure}
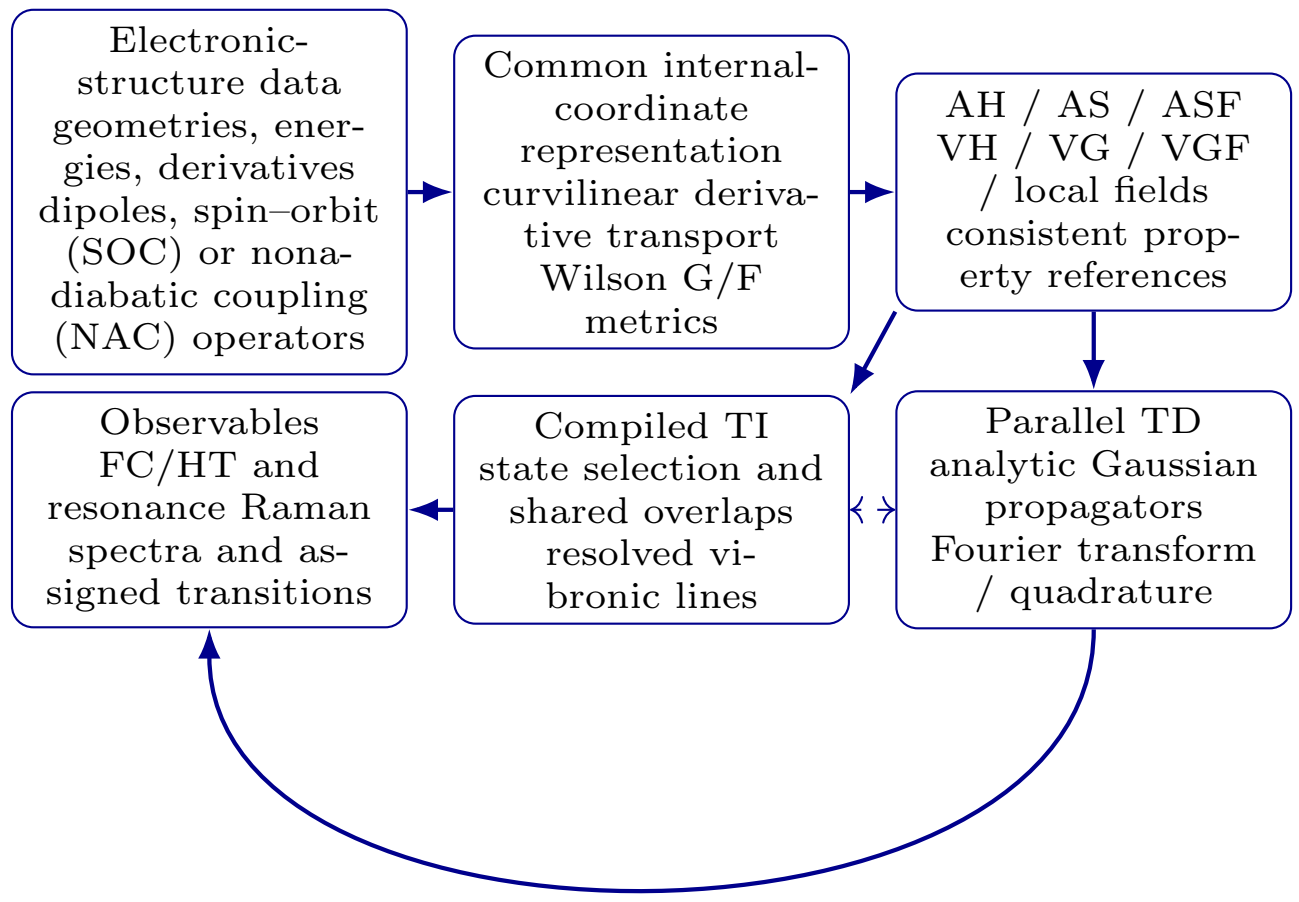

FC, HT and momentum insertions share each Gaussian factorization. At zero temperature, exactly vanishing thermal factors reduce the matrix dimension algebraically. The Fourier transform retains absolute strengths and signed interference, with the electronic carrier restored on the output frequency axis. Independent time-step and time-window controls test aliasing and the omitted tail, respectively. Figure~\ref{fig:workflow} summarizes the information flow from electronic structure to assigned lines and complete spectra.

\section{Results and numerical assessment}
\subsection{Controlled harmonic examples}
A displaced oscillator reproduces the Poisson FC progression, whereas a zero-displacement, purely HT transition reproduces its single-quantum selection rule. Complex position/momentum insertions reproduce the thermal commutator sum rule and agree between TI and TD for absorption and emission. These checks isolate the amplitude algebra and its thermal consistency.

\begin{figure*}[t]
\includegraphics[width=.68\textwidth]{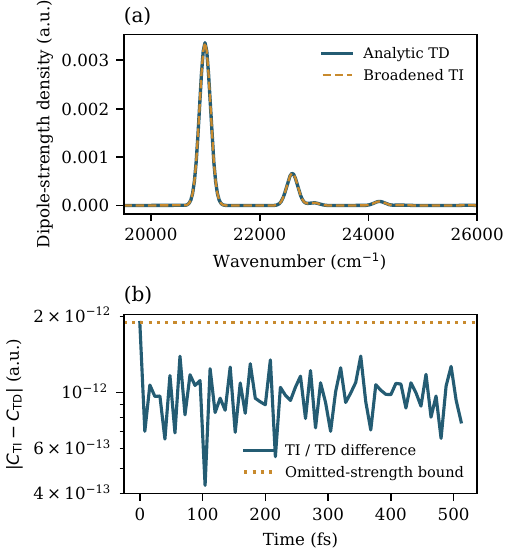}
\caption{Two-state ESALC example. The analytic TD spectrum, Gaussian-broadened TI progression and correlation difference are shown together with the omitted TI strength.}
\label{fig:consistency}
\end{figure*}

The common-coordinate example in Fig.~\ref{fig:consistency} uses two native two-dimensional ESALC fields with different positive force constants and Wilson metrics. Their frequencies are approximately $(1475.84,2138.43)$ and $(1604.55,1997.82)$~cm$^{-1}$. At 300~K, a 660-line TI progression retains all but about $1.9\times10^{-12}$ atomic units of the total strength. Its correlation differs from the independent analytic TD result by $1.9\times10^{-12}$ atomic units, consistent with the omitted-strength bound to floating-point accuracy. An arbitrary invertible change of the common coordinate representation preserves the correlation, isolating the contraction from molecular-specific details.

\subsection{Molecular harmonic spectra and coupled-DVR extensions}
The archived input set contains Gaussian force fields for anthracene (66 modes) and p-terphenyl (90 modes), including ground-state minima, excited-state minima, and excited-state derivatives at the ground-state geometry. The fields use the Perdew--Burke--Ernzerhof hybrid (PBE0) functional with 6-31G(d) and 6-31G(d,p) bases, respectively; the same fields are supplied to every vibronic calculation in this comparison. Transition properties are taken at the Franck--Condon geometry. The original anthracene derivative record is used at FC level in the completeness comparison below. Two analytic property calculations supply its full FC/HT operator for the experimental comparison. The complete p-terphenyl record supports FC/HT. Related torsional studies of para-phenylene oligomers motivate the coupled-coordinate treatment used here.\cite{Heimel2005Torsion,Lukes2008Torsion}

Canonical curvilinear ESALCs are built automatically from the molecular-topology primitive contract. For p-terphenyl, four proper dihedrals are retained for each inter-ring bond, and the explicit torsional ESALC block retains rank two at both minima. Numerical derivatives of coordinate values agree with their Wilson rows to approximately $10^{-10}$~\AA$^{-1}$ in the checks performed. An orthogonal recombination of the same valence ESALCs changes the sampled correlation by less than $7\times10^{-12}$ atomic units. A matched comparison uses the same compiled adaptive selector, FC operator, 0~K temperature and 100000-transition budget for three coordinate models built from the same minimum fields. ESALC-I and full-pool delocalized valence coordinates recover 72.945\% and 73.447\%, respectively. Their broadened TD profiles differ by 0.789\% of the ESALC-I peak. A state-specific mass-weighted Eckart Cartesian construction recovers 0.02072\% at this finite budget and gives a substantially different local harmonic profile. These are different finite-displacement approximations to the molecular potential. Their similar spectral recovery separates the structural role of locality from the selector performance: ESALC-I coordinates involve a mean of 9.27 atoms and one primitive family, whereas the delocalized coordinates involve all 32 atoms and three families. The SI gives the construction, exact selector settings and results.

\begin{figure*}[t]
\centering
\includegraphics[width=.96\textwidth]{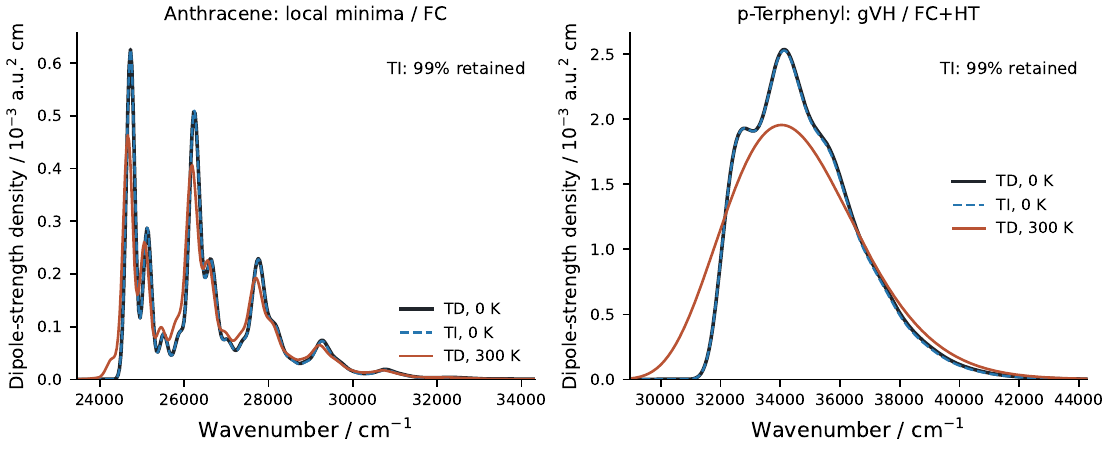}
\caption{Harmonic ESALC spectra for anthracene and p-terphenyl. Complete TD profiles at 0 and 300~K are compared with the selected 0~K TI expansion using the same Gaussian broadening.}
\label{fig:molecular}
\end{figure*}

Figure~\ref{fig:molecular} and Table~\ref{tab:molecular} summarize two harmonic examples: a locally relaxed anthracene model and a generalized vertical p-terphenyl model at the supplied ground-state geometry. Both calculations retain every mode of their positive-curvature fields; the subsequent DVR calculation treats torsional anharmonicity explicitly.

\begin{table}[t]
\caption{Molecular TI intensity recovery at 0~K. Retained strengths are reported as integer percentages; exact fractions and omission bounds are archived. The rounded 99\% criterion accepts an unrounded recovery of at least 98.6\%.}
\label{tab:molecular}
\begin{ruledtabular}
\begin{tabular}{lrrr}
Model & Modes & Transitions & Retained\\
\hline
Anthracene, local minima & 66 & $8\,192$ & 99\%\\
p-Terphenyl, generalized V & 90 & $10\,182\,656$ & 99\%\\
\end{tabular}
\end{ruledtabular}
\end{table}

The TI calculation recovers approximately 99\% of the absolute intensity even for the strongly reorganized 90-mode p-terphenyl example, while retaining explicit transition assignments. The selector visits classes up to seven and eleven, respectively. Its exact retained fraction is preserved in the numerical evidence on the absolute scale. Independent complete TD correlations at nonzero times remain within the positive omitted-strength bound for both examples. This verifies the selected harmonic expansion of the specified electronic fields.

TD spectra use a 512~fs window and Gaussian broadening with $\sigma=100$~cm$^{-1}$. Comparing nested 0.25 and 0.125~fs time grids gives integrated absolute differences below $10^{-12}$ of the total strength for these final ESALC models at 300~K. Independent time-step refinement checks aliasing in addition to damping and integrated strength.

\begin{figure}[t]
\centering
\includegraphics[width=.96\linewidth]{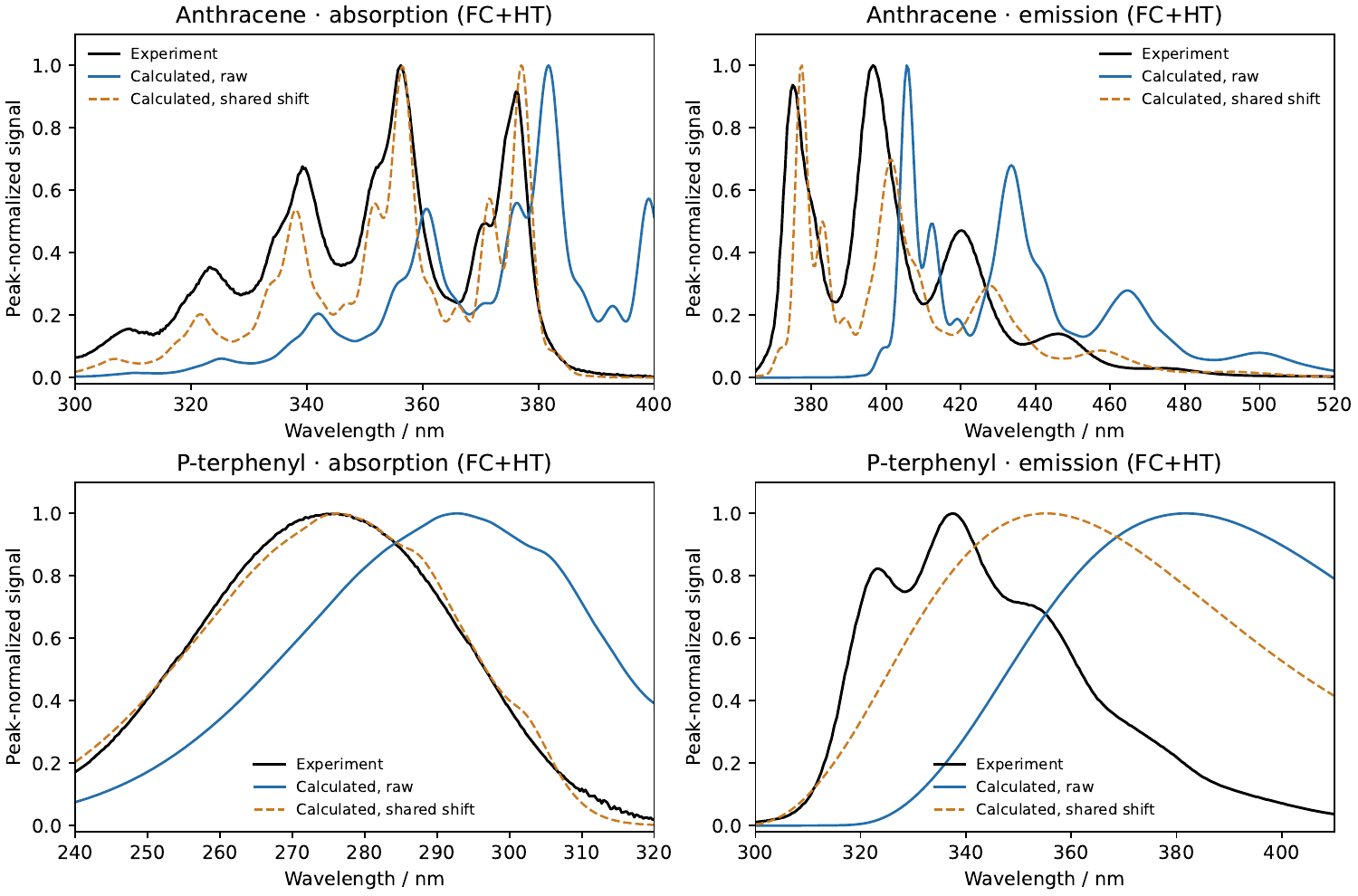}
\caption{Calculated and measured solution spectra for anthracene and p-terphenyl. Curves are peak-normalized; orange curves use the absorption-derived shift for both absorption and emission.}
\label{fig:organic-experiment}
\end{figure}

\paragraph{Experimental absorption and fluorescence.}
Calculations connect these molecular fields to measured solution spectra using AH-like local relaxation of the two supplied minimum fields. The comparison assesses band patterns and their sensitivity to the optical operator; its energy translation is fitted to absorption data.
Figure~\ref{fig:organic-experiment} compares absorption and fluorescence of
anthracene and p-terphenyl with the numerical PhotochemCAD records in
cyclohexane.\cite{PhotochemCAD2018} The spectra use both relaxed minima,
all 66 and 90 modes, respectively, the common curvilinear coordinate representation and
state-specific local metrics. Absorption uses properties at the ground-state
geometry, while emission uses those at the excited-state minimum. The
independently evaluated thermal TD profiles use 300~K and a fixed Gaussian
standard deviation of 100~cm$^{-1}$.

The comparison retains the calculated photon-frequency factors. Absorption
is proportional to $\widetilde\nu S(\widetilde\nu)$; the photon-emission
profile includes $\widetilde\nu^3$ and the wavelength-density Jacobian.
Each displayed curve is normalized to its own peak in the specified band
window. Solid calculated curves retain their electronic energy origins.
The dashed comparison uses an empirical energy calibration: one rigid wavenumber translation determined
from the lowest-energy resolved absorption band and calculated origin for anthracene, and from the absorption-envelope maxima for p-terphenyl and transfers it unchanged to
fluorescence. The nuclear amplitudes and vibrational spacings are unchanged by this translation; photon-frequency factors are evaluated on the displayed energy scale.

After the stated absorption calibration, the peak-normalized root-mean-square (RMS) differences are 0.127 and 0.183 for anthracene absorption and emission, and 0.030 and 0.388 for p-terphenyl. In the harmonic model, the p-terphenyl fluorescence maximum remains at 355.0~nm compared with 337.5~nm measured, a discrepancy of 1461~cm$^{-1}$; its resolved shoulder and peak are not reproduced by the broad calculated envelope. The fluorescence discrepancy motivates the explicit torsional treatment below. The anthracene band positions improve after calibration, while their relative intensities, including the ordering of the first two fluorescence bands, remain different.

On identical anthracene nuclear Hamiltonians, adding the calculated HT derivatives changes the absorption RMS from 0.147 to 0.127 and the emission RMS from 0.193 to 0.183. A finite sensitivity analysis varies the window edges by up to 10~nm, the shared translation by $\pm100$~cm$^{-1}$ and peak versus area normalization. HT decreases the discrepancy in all 18 variants per spectrum; for peak normalization the RMS decrease spans 0.0166--0.0228 in absorption and 0.0046--0.0189 in emission. These ranges measure sensitivity to the analysis convention. The SI records every variant. Both fluorescence profiles retain the absorption-derived calibration; cyclohexane is the experimental medium, and no solvent-dependent spectral correction is fitted in this comparison.

Area-based shape measures distinguish the effects hidden by a single RMS.
For anthracene absorption, HT increases the overlap coefficient from 0.855
to 0.875 and decreases the wavelength Wasserstein distance from 5.50 to
4.10~nm. For emission the overlap is nearly unchanged (0.745 to 0.744),
while the transport distance increases from 2.16 to 4.30~nm despite the
smaller RMS. The HT redistribution therefore improves some aspects of
the envelope while worsening others. All metrics retain the stated
window and the absorption-derived energy translation.

\begin{figure}[t]
\centering
\includegraphics[width=.96\linewidth]{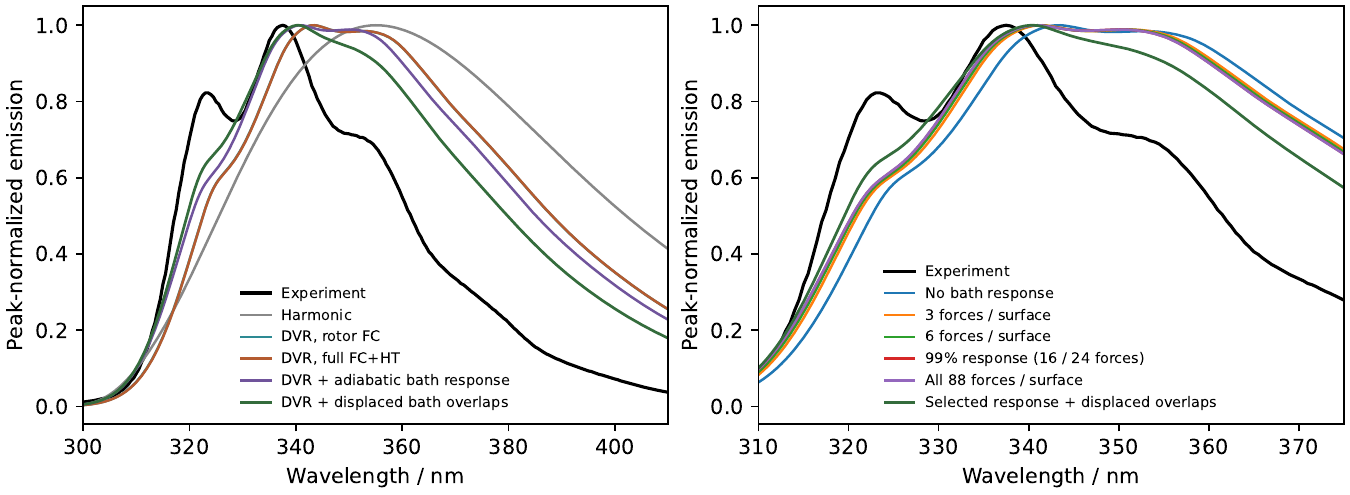}
\caption{p-Terphenyl fluorescence at 298.15~K with harmonic, one- and two-torsion DVR models and the selected bath response. All curves use the same absorption translation and 100~cm$^{-1}$ Gaussian broadening.}
\label{fig:torsional-emission}
\end{figure}

The two inter-ring torsions provide the intermediate level of the hierarchy
for p-terphenyl fluorescence (Fig.~\ref{fig:torsional-emission}). Coupled
curvilinear DVR states replace the two harmonic rotor directions; the
remaining 88 modes retain their two-state harmonic correlation. The torsional energies and gradients use gas-phase PBE0/6-31G(d,p) for $S_0$ and its time-dependent density-functional theory (TDDFT) form for $S_1$. Anharmonic torsions were already used to explain the absorption--emission asymmetry of para-phenylene oligomers,\cite{Heimel2005Torsion} and full-dimensional self-consistent-charge density-functional tight-binding (SCC-DFTB) simulations provide another treatment of their torsional dynamics.\cite{Lukes2008Torsion} The DVR--harmonic treatment of 9-phenylanthracene is a direct architectural precedent.\cite{Baiardi2017HybridDVR} All models
in this refined comparison use 298.15~K, a Gaussian standard deviation of
100~cm$^{-1}$ and the same absorption-derived translation. The harmonic
maximum is 355.0~nm. Two coupled DVR coordinates move it to 343.5~nm,
compared with 337.5~nm measured, and reduce the peak-normalized RMS
from 0.387 to 0.290 with full FC+HT.

\begin{figure}[t]
\centering
\includegraphics[width=.79\linewidth]{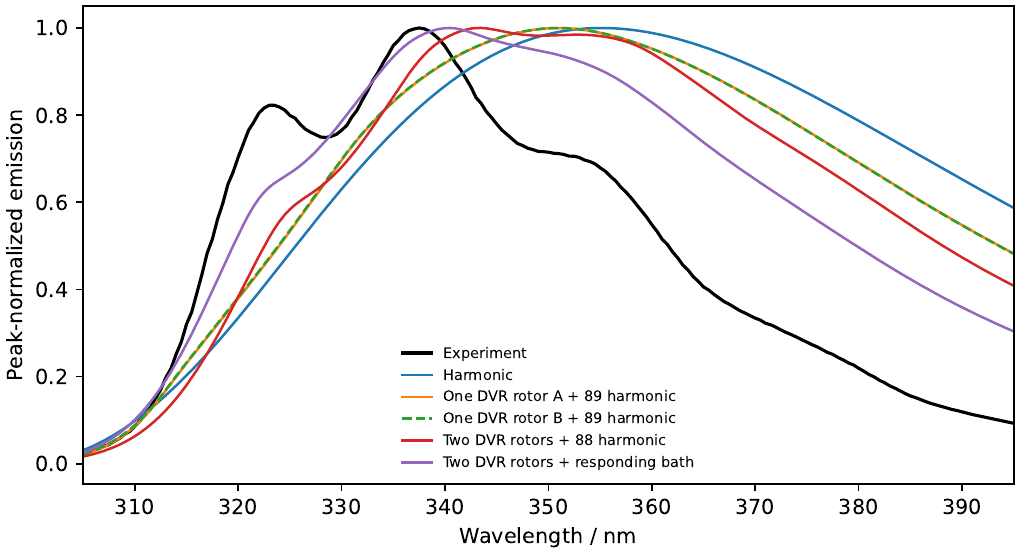}
\caption{p-Terphenyl fluorescence with one or two explicit DVR torsions and 89 or 88 harmonic modes. Both one-rotor choices retain full FC+HT. Curves are peak-normalized at 298.15~K with $\sigma=100$~cm$^{-1}$ and the same absorption-derived translation.}
\label{fig:one-two-rotor}
\end{figure}

A matched one-coordinate ablation restores the other torsion to the
harmonic bath, retaining all 90 vibrations. Either choice of the explicit
DVR rotor gives a 351.0~nm maximum and RMS 0.326. Treating both rotors
explicitly gives 343.5~nm and RMS 0.290 before adding bath response
(Fig.~\ref{fig:one-two-rotor}). Both flexible coordinates therefore
contribute to the resolved spectral improvement.
The one-rotor profiles change by less than $10^{-5}$ of their peak on
the final 81-to-101-function refinement.

The optical contraction retains HT derivatives on both the torsions and
the harmonic bath, including their signed interference. Removing the rotor
derivatives recovers the factorized correlation within
$3\times10^{-16}$ relative error. Torsional HT has a small effect here:
the RMS changes from 0.28921 to 0.28967. The nuclear model supplies the dominant change in this example.

The existing Cartesian gradients also determine the leading linear
bath forces. Their adiabatic displacement response, including the
diagonal Born--Huang contribution, enters the active Hamiltonians while
the optical contraction retains the reference Gaussian bath. Ranking
forces independently on each surface selects 16 for $S_0$ and 24 for
$S_1$; the resulting profile differs from the complete 88-force response
by 0.083\% of its peak and 2.22~cm$^{-1}$ in centroid. The maximum is
341.0~nm and the RMS is 0.256. The integrated shoulder-to-main-band ratio
increases from 0.393 harmonically to 0.454 with the selected response,
compared with 0.632 measured. No fluorescence origin or intensity ratio
is fitted.

Including the torsion-dependent displaced bath overlaps then moves the
maximum to 340.5~nm, lowers the RMS to 0.201 and increases the shoulder/main
ratio to 0.488. The residual maximum displacement is 3.0~nm (about 261~cm$^{-1}$) relative to experiment; the experimental shoulder/main ratio is 0.632. These profiles retain the gas-phase electronic surfaces while the measured spectrum is in cyclohexane. Both electric-dipole HT insertions and their active--bath
interference are retained. These changes use the same force selection,
electronic data and absorption calibration. Increasing the optical
interpolation from 13 to 17 nodes per rotor changes the profile by 0.0029\%
of its peak; increasing the sine basis from 51 to 61 changes it by 0.0081\%.
An explicit variational check of the slow and dominant bath couplings gives
differences below 0.30~cm$^{-1}$ in the first 24 low excitation energies on
each surface at zero bath temperature. This separately checks the leading
adiabatic response used in the fluorescence model. Independently, the
translated optical kernel agrees with explicit coordinate quadrature and
oscillator sums to $4.34\times10^{-14}$ absolute error in a one-mode
bath test with unequal frequencies, three translation pairs and linear
FC/HT operators; the full 88-mode zero-translation limit recovers the
reference correlation within $1.75\times10^{-15}$ relative error.

The spectral comparison also uses area-normalized profiles on the declared
300--410~nm wavelength window. Their overlap coefficient
$\mathcal O=\int\min(P,Q)\,d\lambda$ measures shared area, while
$W_1=\int|F_P-F_Q|\,d\lambda$ measures the intensity-transport distance
in nm; $F$ is the cumulative area. These metrics use the same
absorption-derived translation and introduce no additional alignment.
Table~\ref{tab:profile-shapes} shows the successive improvements.
The SI gives the piecewise-linear integration and all molecular values.

\begin{table}[t]
\caption{p-Terphenyl fluorescence shape on 300--410~nm. Profiles are
normalized to unit area on this window; $\mathcal O=1$ and $W_1=0$
mean identical shapes. Every model retains the same absorption-derived
energy translation.}
\label{tab:profile-shapes}
\begin{ruledtabular}
\begin{tabular}{lrr}
Nuclear model & $\mathcal O$ & $W_1$ (nm)\\
\hline
Harmonic & 0.706 & 16.35\\
Coupled DVR & 0.773 & 12.62\\
DVR + bath response & 0.806 & 10.68\\
DVR + displaced overlaps & 0.843 & 8.64\\
\end{tabular}
\end{ruledtabular}
\end{table}

A product-state ablation separates explicit inter-rotor correlation from
the anharmonicity of each rotor. Two self-consistent one-dimensional
anharmonic modal problems replace the coupled two-dimensional eigenproblem,
with the same physical Hamiltonians, both torsions and the complete optical
bath retained. At the common 51-function basis and 13-node interpolation,
the product model gives 341.5~nm, RMS 0.220 and shoulder/main ratio 0.465,
compared with 340.5~nm, 0.201 and 0.488 for coupled DVR. The coupling is
therefore a smaller, resolved correction to the main torsional improvement.
This comparison removes inter-rotor correlation; it does not make one of
the two torsions harmonic.

The full FC+HT DVR profile changes by 0.0011\% of its peak between 61 and
71 sine functions per rotor. For the complete bath response, increasing
the basis from 51 to 61 changes it by 0.0079\%. The finite-domain
landscape retains the same four $S_0$ and two $S_1$ minima. The SI gives
the force selection, independent gradient checks, operator contractions
and profile controls.

Finite Coulomb-attenuating-method Becke three-parameter Lee--Yang--Parr (CAM-B3LYP) force controls assess electronic-surface sensitivity at
the same geometries. For $S_0$, a periodic correction fitted to six
energy/gradient points predicts two reserved controls within
10.0~cm$^{-1}$ and 42.6~cm$^{-1}$/rad. The eight $S_1$ controls retain
the same bright electronic state: atomic-orbital (AO) metric transition-density
fingerprint overlaps with PBE0 are 0.989--1.000, and the sampled
$S_2-S_1$ separations are 0.31--0.84~eV. The SI separates this state
assignment from the more demanding interpolation of its force field,
including direct CAM-B3LYP and difference-field fits. A directly fitted periodic
CAM-B3LYP field predicts the two control energies within 0.72 and 5.60~cm$^{-1}$
and has two symmetry-related wells. In a matched fixed-bath sensitivity
calculation, replacing the torsional fields changes the fluorescence
maximum from 343.5 to 344.0~nm at fixed electronic gap. The additional
CAM-B3LYP gap change of +1924.68~cm$^{-1}$ moves it to 323.0~nm with the
unchanged absorption calibration. Thus the gap change dominates the
torsional-shape change in this diagnostic. Both profiles converge within
0.049\% of their peak from 41 to 51 functions per rotor. The SI reports
the force residuals, fit selection and separate profiles. Three additional
representations selected from training energies and gradients and two-well landscape checks
retain a 344.0~nm maximum and have a 1.85\% peak-normalized profile spread;
their 41-to-51-function basis changes remain below 0.053\%. Their external
force checks exclude adopting them as improved production fields. The responding-bath
fluorescence profiles above retain the PBE0 emission model.

\paragraph{Signed chiroptical spectra.}
Harmonic vibronic ECD and circularly polarized luminescence (CPL) with FC/HT operators and Duschinsky mixing
are established approaches, including applications to camphor derivatives.\cite{Baiardi2013,Barone2014Chiroptical,DelGaldo2020Camphor}
Here the implementation uses the same TI/TD set of transition operators for signed
rotatory strengths and positive electric strengths, with window-directed
selection connecting state-to-state assignments to a complete TD profile.

The $S_0\!\rightarrow S_1$ absorption of (1$S$,4$S$)-camphor adds a
75-mode chiroptical example. Both minima and their analytic Hessians are
calculated with the Becke three-parameter Lee--Yang--Parr (B3LYP) functional, D3 dispersion and Becke--Johnson (BJ) damping, using the def2-TZVP basis; the excited state uses TDDFT.
The final $S_1$ structure has 75 positive frequencies, the lowest being
118.22~cm$^{-1}$. A common automatic curvilinear ESALC-I coordinate representation connects the
two local fields. Electric and magnetic transition moments and their
Cartesian derivatives, calculated at the ground-state minimum, supply the
same FC/HT set of transition operators used for ordinary absorption. This gives the
electric strength, magnetic strength and signed electric--magnetic
interference in one TD calculation.

\begin{figure}[!t]
\centering
\includegraphics[width=.98\linewidth]{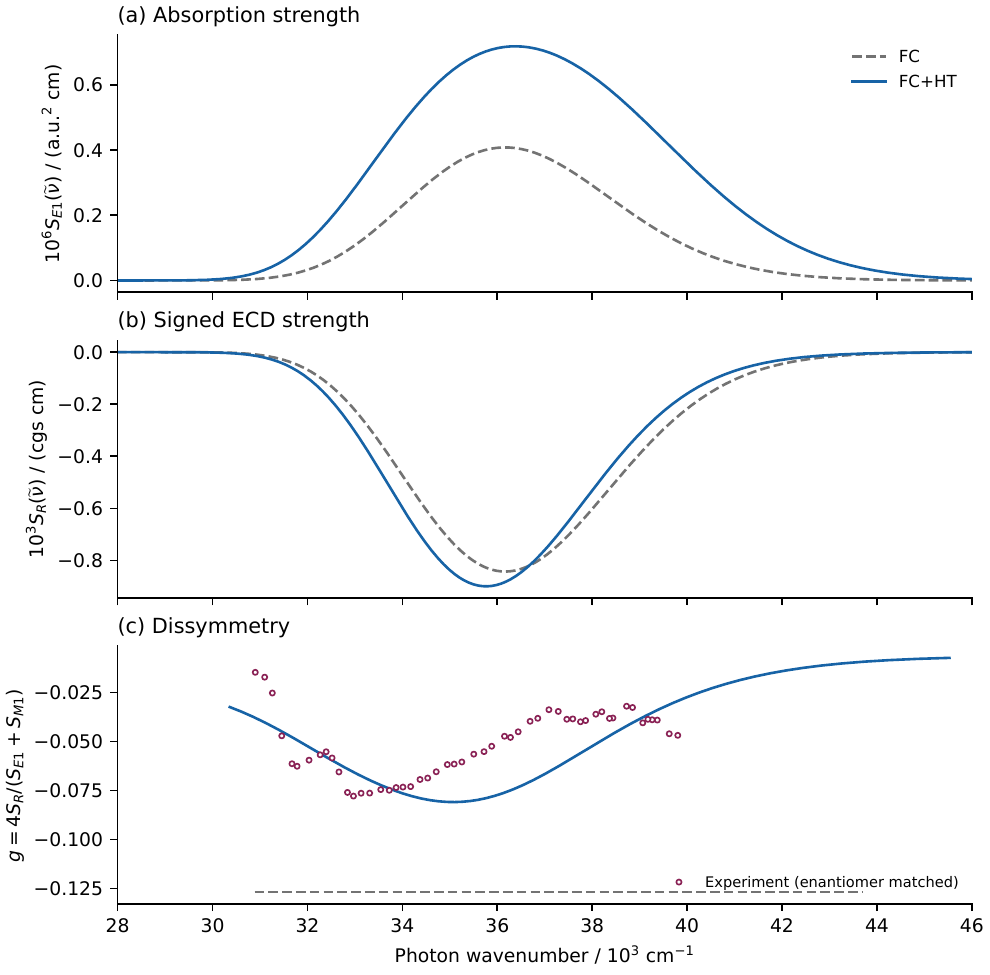}
\caption{Camphor spectra at 298.15~K, with all 75 modes and $\sigma=100$~cm$^{-1}$. Dashed curves: FC; solid curves: FC+HT. (a) Electric strength. (b) Signed rotatory strength in $10^{-40}$~erg~esu~cm/G. (c) Dissymmetry and enantiomer-matched experimental points.\cite{Pulm1997Camphor,Kerber2025ECD} Calculated curves retain absolute strengths; dissymmetry is shown above 1\% of the electric-profile maximum.}
\label{fig:camphor-ecd}
\end{figure}

Figure~\ref{fig:camphor-ecd} shows the resulting absorption strength, ECD
and dissymmetry at 298.15~K. HT increases the integrated electric strength
from 0.002182 to 0.004930 atomic units, whereas the signed rotatory strength
changes by only 2.8\%. Consequently, the band dissymmetry changes from
$-0.1269$ to $-0.0580$. At 0~K the FC+HT value is $-0.0652$: thermal
population and the coordinate-dependent operator contribute separately.
The frequency dependence of the dissymmetry makes this distinction visible
within the band. Earlier calculations and gas-phase measurements provide
the context for this weak electric-dipole-allowed carbonyl transition.\cite{Pulm1997Camphor,Kerber2025ECD,Kerber2026Correction}
The experimental band values of Pulm \textit{et al.}, as re-expressed by
Kerber \textit{et al.}, give $D=0.00515$ atomic units and $|g|=0.054$.
FC+HT brings the electric strength within 4.3\%, the rotatory strength
within 2.6\% and the conventional band ratio $4|R|/D=0.0582$ within
7.8\% of these observations. The RMS difference from the measured
frequency-dependent $g$ decreases from 0.0777 with FC to 0.0171 with
FC+HT over 251.25--324.25~nm. The calculated $|g|$ maximum is at 285~nm,
whereas the experimental trace peaks near 303~nm. Figure~\ref{fig:camphor-ecd}
retains this unshifted comparison. Experimental points are extracted from
the published vector trace, with the enantiomer sign matched once; no
energy translation or intensity scale is fitted. The SI gives the source,
band-strength table, magnetic convention and temporal controls.

Additional TD-PBE0 and TD-CAM-B3LYP properties, evaluated at the same
ground-state geometry with def2-TZVP and D3(BJ), separate electronic
operator sensitivity from the nuclear model. Keeping the B3LYP nuclear
fields and relaxation, their vertical-gap changes are +150 and
+680~cm$^{-1}$, respectively. The resulting integrated $D$ values are
0.004685 and 0.003765 atomic units, compared with 0.004930 for B3LYP;
the corresponding $4|R|/D$ ratios are 0.05645 and 0.05130.
Thus the ratio varies less than its individual electric and rotatory
strengths. The unshifted $|g|$ maxima are 285.25 and 279.0~nm, compared
with 285.0~nm for B3LYP and approximately 303~nm measured. These fixed-field
controls, plotted and tabulated in the SI, distinguish the favorable
B3LYP integrated strengths from the frequency-dependent electronic error.

A targeted 0~K TI control assigns the low-energy region
30000--33000~cm$^{-1}$. The calculation stops at 4,417,792 selected
transitions, recovering 99.28\% of its positive electric-dipole plus magnetic-dipole (E1+M1) strength. All three
observable profiles satisfy 2\% relative peak and 1\% integrated absolute
error tolerances against TD. For signed ECD these errors are 1.56\% and
0.62\%, respectively. The window-directed calculation retains 25.04\% of the global
positive norm. Its state-to-state assignments are checked within the
specified window; Fig.~\ref{fig:camphor-ecd} gives the full TD spectrum.

\begin{figure}[!t]
\centering
\includegraphics[width=.98\linewidth]{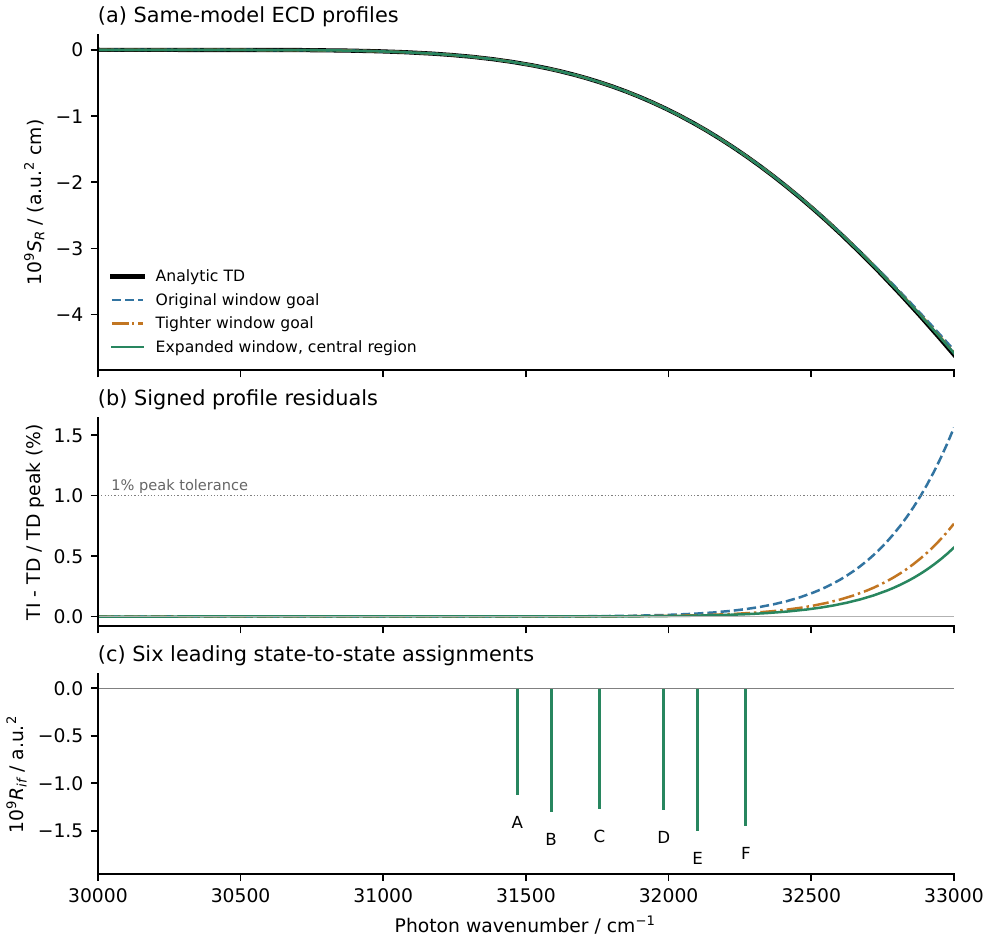}
\caption{Assigned camphor ECD at 0~K with $\sigma=100$~cm$^{-1}$. (a) TD and selected TI profiles. (b) Signed residuals relative to the TD peak; dotted line: 1\% tolerance. (c) Six leading transitions and their rotatory strengths; occupations A--F are listed in the SI. The green curve uses a selection window enlarged by 250~cm$^{-1}$ on each side.}
\label{fig:camphor-ti-assignments}
\end{figure}

The assignments also survive tighter profile goals and a wider selection
window (Fig.~\ref{fig:camphor-ti-assignments}). Raising the positive-strength
target to 99.5\% and halving the peak/integrated tolerances to 1\%/0.5\%
gives 8,689,408 transitions, 99.66\% window recovery and 0.29\%
integrated ECD error. Extending the window to 29750--33250~cm$^{-1}$
then gives 11,393,024 transitions. The 20 leading positive-strength assignments
and the 20 leading ECD assignments retain their identities and FC/HT
components in both controls. In the original interval the final ECD
peak and integrated errors are 0.57\% and 0.21\%; the window
expansion changes the central profile by at most 0.197\% of the
TD peak. These controls connect the assigned TI lines to a TD
reference envelope within the chosen spectral region.

The six listed transitions share one quantum of $v_9$ and combine three
or four quanta of $v_1$, two or three of $v_2$, and zero or one of $v_{15}$.
Their physical content follows directly from the final-state ESALC modes:
$v_1$ (118.11~cm$^{-1}$) is a collective carbonyl/cage wag with methyl
torsion; $v_2$ (169.45~cm$^{-1}$) is dominated by coupled gem-dimethyl
torsion; $v_9$ (297.99~cm$^{-1}$) mixes cage/methylene deformation with
methyl motion; and $v_{15}$ (513.25~cm$^{-1}$) combines carbonyl
bending/pyramidalization and cage-angle motion. The SI relates these labels
to Cartesian displacements and geometric primitive projections using the
same normal-mode ordering as the assigned lines.

\subsection{Water photoionization: harmonic and internal-coordinate anharmonic models}
\label{sec:waterexample}
Water ionization provides a small molecular test in which the nuclear Hamiltonian and the transition operator can both be examined explicitly. Neutral singlet and cationic doublet geometries and harmonic Hessians were calculated with Gaussian at the revised dispersion-corrected spin-component-scaled double-hybrid (rDSD) level with jun-cc-pVTZ. The anharmonic molecular reference uses analytic zero-asymptote fitted force-field (ZAFF) potentials in the two O--H distances and the H--O--H angle, together with the complete curvilinear $J=0$ kinetic operator. Occupation labels below identify bend, symmetric stretch and antisymmetric stretch character; anharmonic states are assigned by harmonic projection rather than by energy ordering alone.

The photoionization operator is supplied by signed non-diagonal renormalized second-order (NR2) electron-propagator Dyson orbitals with jun-cc-pVTZ and a van Leeuwen--Baerends (LB94) local continuum field constructed from the current rDSD density. At five outgoing-electron energies between 0.05 and 1.00~hartree, its complex amplitudes are differentiated along all three initial normal modes. Rigid AO-tensor alignment and overlap-based Dyson phase tracking keep the displaced calculations in a common electronic frame. The resulting constant and linear terms supply FC and HT amplitudes, including their interference. These calculations give conditional dipole-strength distributions at fixed electron kinetic energy; photon, spin and orientation prefactors are not inserted.

For the harmonic ESALC reference based on Wilson force and inverse-mass matrices ($F$ and $G$), independent TI and TD calculations agree over all 4097 correlation points within $3.4\times10^{-12}$ of the combined FC/HT (FCHT) sum rule. Gaussian-broadened TI lines and the TD spectrum agree within $6.7\times10^{-14}$ of the peak on the complete spectral grid. The TI expansion contains 680 final states and recovers the sum rule to better than $10^{-14}$ relatively. This agreement connects resolved transition assignments to the independently calculated envelope. In particular, the antisymmetric fundamental is FC-dark but HT-active, although its first pole-strength derivative vanishes by symmetry: pole strengths alone do not determine transition-amplitude derivatives.

The ZAFF construction retains the reference gradient and Hessian and fits finite-displacement energies, gradients and Hessians, without reconstructing higher electronic derivatives. Each surface uses 30 electronic geometries, including four checks excluded from coefficient selection. Radial functions are Simmons--Parr--Finlan (SPF) expansions in $1-R_0/R$; angular terms use cosine coordinates, and native support switches control connected products. Hydrogen-exchange symmetry ties equivalent coefficients. The lowest mixed degree passing calibration is five, with all 29 coefficient groups identified. Subsequent radial regularization starts at order 14 and preserves reference derivatives through fourth order. Maximum errors at the excluded points are 5.02 and 3.49~cm$^{-1}$ for neutral and cation, respectively. These errors measure the accuracy over the fitted molecular domain. The full construction and shape checks are given in the SI.

Both nuclear states are solved on the same physical internal-coordinate representation. All off-diagonal kinetic terms and the volume density $\rho\propto R_1^2R_2^2\sin\theta$ are retained. The electronic linear operator is tangent-transported from the initial normal coordinates to that coordinate representation. Nuclear integrals then contract the two wavefunctions directly on a common quadrature, without a harmonic projection cutoff or a separate Duschinsky transformation of the integration grid. The reference continuum amplitudes and their first derivatives are unchanged; their finite-displacement continuation is the declared linear internal-coordinate model.

\begin{figure}[!t]
\centering
\includegraphics[width=.98\linewidth]{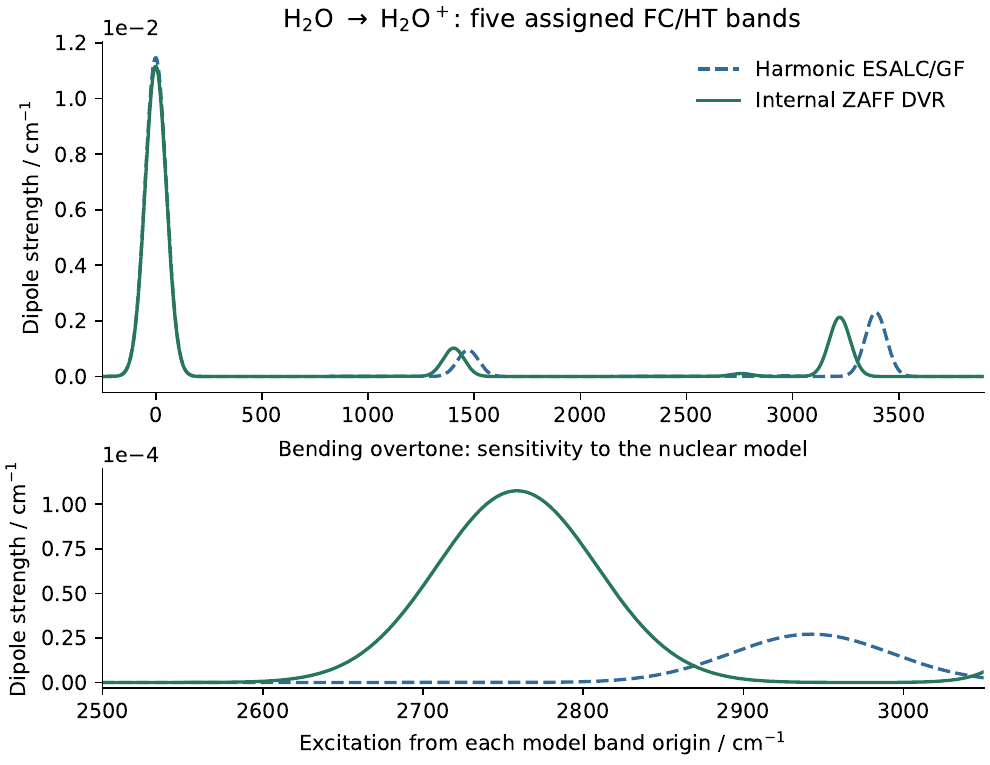}
\caption{Harmonic and ZAFF DVR profiles for the five water bands of Table~\ref{tab:water}, with $\sigma=50$~cm$^{-1}$ and each model's origin at zero. Absolute conditional strengths are retained. The lower panel resolves the bending overtone.}
\label{fig:water}
\end{figure}

Table~\ref{tab:water} reports the five assigned bands; Figure~\ref{fig:water} compares their broadened profiles.

\begin{table*}[t]
\caption{Five assigned water-ionization bands at electron energy 0.20~hartree. Excitations refer to each model's origin. Conditional FC+HT strengths compare the harmonic ESALC model with the full curvilinear ZAFF DVR model at $J=0$.}
\label{tab:water}
\begin{ruledtabular}
\begin{tabular}{lrrrr}
Band character & Harmonic excitation (cm$^{-1}$) & ZAFF excitation (cm$^{-1}$) & Harmonic strength & ZAFF strength\\
\hline
000 & 0.00 & 0.00 & 1.438011 & 1.399631\\
100 & 1471.20 & 1403.42 & 0.119192 & 0.128041\\
010 & 3391.25 & 3221.34 & 0.287793 & 0.266080\\
001 & 3450.22 & 3268.89 & 0.001963 & 0.001784\\
200 & 2942.39 & 2758.59 & 0.003405 & 0.013486\\
\end{tabular}
\end{ruledtabular}
\end{table*}

The final DVR uses 28 sine functions per coordinate and 60 quadrature points per coordinate. For the five assigned bands in Table~\ref{tab:water}, increasing the basis from 24 to 28 changes spacings by less than $1.3\times10^{-6}$~cm$^{-1}$. An independent quadrature refinement changes them by less than $6\times10^{-11}$~cm$^{-1}$. The final domain expansion changes their spacings by at most 0.0151~cm$^{-1}$ and their FC/HT strengths by less than 0.0020\% over all five electron energies. The first twelve final states retain 99.07--99.19\% of the corresponding anharmonic sum rule. Higher saved levels have larger domain changes, up to 3.94~cm$^{-1}$; the tight five-band controls are not assigned to the entire anharmonic spectrum.

The fundamentals shift downward by 68--181~cm$^{-1}$, while the weak bending overtone is particularly sensitive to the nuclear model. A separate matched-Hamiltonian test isolates perturbative truncation.
For each original rectilinear rDSD quartic surface, VPT2 and the local
variational calculation use identical potentials and constant kinetic
metrics. Table~\ref{tab:watermatched} shows that strict second-order
VPT2 overestimates the weak bending-overtone strength by a factor of 5.1,
while the bending fundamental changes much less. This factor applies to the explicit strict second-order wavefunction expansion used here. The separately evaluated frequency-renormalized Gaussian model gives 0.000767 for this strength, compared with 0.000634 variationally; its ratio is 1.21 rather than 5.1. For the fundamental, however, the relative errors are 22\% harmonically, 20\% with frequency-renormalized Gaussian functions and 3.5\% with strict VPT2; for the overtone they are 54\%, 21\% and 414\%, respectively. Strict VPT2 is therefore better for the fundamental, while frequency renormalization is better for the overtone. Neither perturbative approximation is uniformly more accurate for these two bands. The overtone's local
variational strength changes by 1.21\% in the last domain expansion at
this electron energy; harmonic-projection refinement changes all five
assigned strengths by less than 0.00040\%. Its first- and second-order
cationic correction norms are 0.279 and 0.120. Thus accurate amplitude
contraction and moderate wavefunction-correction norms do not guarantee
relative intensity accuracy for a weak band. This is an approximation
test on the stated local quartic Hamiltonian, separate from the
harmonic--ZAFF model comparison above.

\begin{table}[t]
\caption{Matched-field FC/HT test at 0.20~hartree. VPT2, variational and harmonic columns use the same rectilinear rDSD quartic Hamiltonian and continuum operator; the renormalized column is a separate Gaussian-width model.}
\label{tab:watermatched}
\begin{ruledtabular}
\setlength{\tabcolsep}{3pt}
\small
\begin{tabular}{lrrrr}
Band & Harmonic & Renorm. & Strict VPT2 & Variational\\
\hline
100 & 0.117384 & 0.114952 & 0.099378 & 0.095998\\
200 & 0.000974 & 0.000767 & 0.003258 & 0.000634\\
\end{tabular}
\end{ruledtabular}
\end{table}

\begin{figure}[t]
\centering
\includegraphics[width=.90\linewidth]{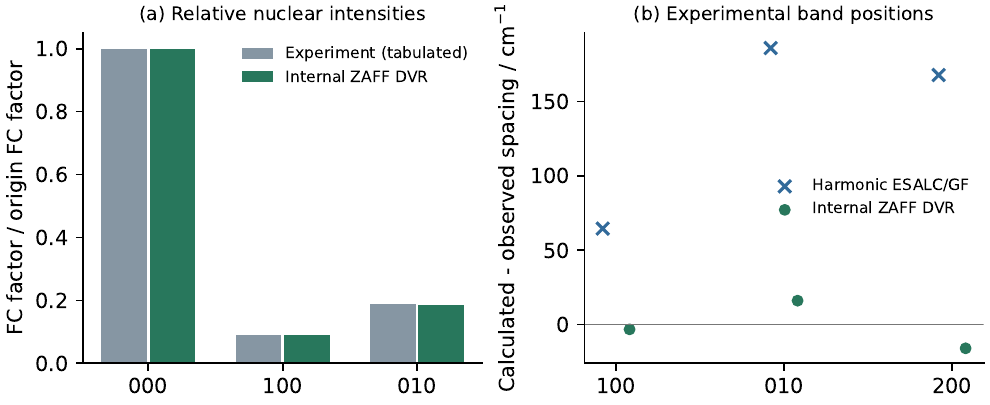}
\caption{External water check using tabulated measurements: (a) nuclear FC
factors relative to the origin band; (b) calculated-minus-observed spacings.
Origin normalization defines relative band intensities, without fitting the
three-band area. These FC data test nuclear overlaps, rather than the full
photon-energy-dependent continuum FC/HT cross section.}
\label{fig:waterexternal}
\end{figure}

\paragraph{External check against photoelectron measurements.}
Experimental spacings measured by Reutt \emph{et al.} and tabulated by
Engin \emph{et al.} provide an external test of the nuclear energies;
Brundle--Turner FC factors tabulated by Botter and Rosenstock test the
Condon intensity sector.\cite{Reutt1986Water,Engin2019Water,BotterRosenstock1969}
No potential parameter or intensity scale is fitted to these data.
Table~\ref{tab:waterexternal} and Figure~\ref{fig:waterexternal} compare the assigned bands using our bend-first convention. The
ZAFF ground-band FC factor is 0.7499, compared with 0.757 experimentally;
the bending and symmetric-stretch factors are 0.0668 and 0.1389, compared
with 0.069 and 0.143. The calculated spacing errors decrease from roughly
65--186~cm$^{-1}$ harmonically to 3--16~cm$^{-1}$ with ZAFF.
The bending-overtone spacing is tested, but its experimental FC factor
is not tabulated in these sources.

\begin{table}[t]
\caption{Water photoelectron comparison. $\delta\nu$ is the ZAFF-minus-measured
spacing (cm$^{-1}$). Experimental FC error limits are the original probable-error
estimates, not standard deviations. A dash denotes unavailable intensity.}
\label{tab:waterexternal}
\begin{ruledtabular}
\begin{tabular}{lrrr}
Band & $\delta\nu$ & ZAFF FC & Experimental FC\\
\hline
000 & 0.0 & 0.7499 & $0.757\pm0.005$\\
100 & $-3.2$ & 0.0668 & $0.069\pm0.005$\\
010 & $+16.1$ & 0.1389 & $0.143\pm0.005$\\
200 & $-16.0$ & 0.0068 & ---\\
\end{tabular}
\end{ruledtabular}
\end{table}

Electronic-model sensitivities remain distinct from these nuclear controls. Increasing the continuum angular cutoff from $l_{\max}=6$ to 7 changes the total harmonic FCHT strength by less than 0.10\%, while the weak antisymmetric HT band changes by up to 2.4\%. Changing the explicit exchange--correlation (XC) tail switching interval changes integrated harmonic strengths by up to 4.1\%. Thus harmonic TI/TD equivalence, convergence of the selected DVR bands, and sensitivity to the chosen molecular and continuum models answer separate questions.

The same native water Hamiltonians also validate the common-modal anharmonic
representation. Selected orthonormal modal products retain the complete
coordinate-dependent metric and potential through projection of the coordinate-dependent Hamiltonian. With 2600 configurations, the first 12 excitation energies
agree with the 13824-product reference within 0.0202~cm$^{-1}$; relative FC+HT
intensity errors for five assigned bands at five electron energies are below
$6.3\times10^{-8}$. The common full-product reference separately agrees with
the physical bond--angle calculation within 0.0103~cm$^{-1}$ on those bands.
These controls support this low-energy molecular model; the SI separates
contraction, domain and quadrature errors.

The retained-state comparison also compares FC, HT and signed interference
separately. With Gaussian $\sigma=50$~cm$^{-1}$, the integrated absolute
FC+HT profile difference is below $1.15\times10^{-7}$ of the retained
strength against the complete common-product operator, and below
$6.65\times10^{-5}$ against the independent bond--angle reference,
at all five electron energies. These profile tests use 12 retained levels;
the five assigned bands define the independently checked low-energy window.

\subsection{4-pyrone: harmonic spectra and anharmonic interference sensitivity}
\label{sec:pyrone-optical}

The optical spectra of 4-pyrone extend the harmonic assessment to two excited
surfaces with different structural responses. Geometries and Hessians are
computed at B3LYP/6-31+G(d) for $S_0$, $S_1$ and $S_2$. Both excited-state
minima have 27 real vibrational frequencies. The $S_1$ structure remains
essentially planar, whereas following the out-of-plane instability of the
initial $S_2$ stationary point gives a nonplanar minimum. Its heavy-atom
root-mean-square distance from the best-fit plane is 0.0662~\AA.

\begin{figure}[t]
\includegraphics[width=0.88\linewidth]{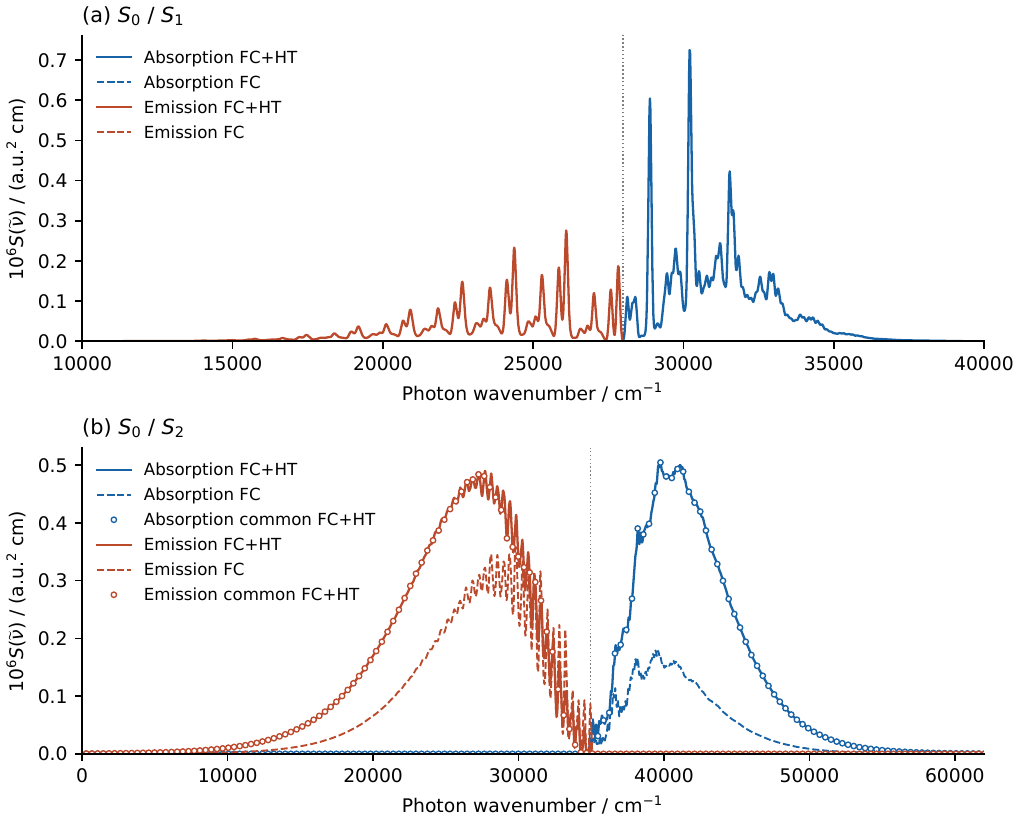}
\caption{Calculated 4-pyrone absorption and emission profiles at 0~K. Solid curves include FC+HT, dashed curves FC only; open circles show the common-mode $S_2$ result.}
\label{fig:pyrone-optical}
\end{figure}

Figure~\ref{fig:pyrone-optical} shows adiabatic harmonic FC+HT spectra in a
shared curvilinear ESALC-I coordinate representation, with state-specific local Wilson metrics
and the full intersurface mode transformation. The dipole and its Cartesian
derivatives are evaluated at the $S_0$ equilibrium geometry for absorption
and at the corresponding excited-state minimum for emission, then
transported to the internal-coordinate representation. The calculated
zero-point-corrected origins are 27991.6 and 34934.8~cm$^{-1}$ for $S_1$
and $S_2$, respectively. No empirical energy shift is applied.

The two surfaces illustrate different roles of the transition operator.
For $S_1$, the equilibrium FC contribution is negligible on the scale of
the figure: linear HT terms provide the intensity in both directions.
For $S_2$, FC intensity is substantial, and HT terms strongly reshape and
increase the absorption envelope. The complete integrated dipole strengths
are $9.571\times10^{-4}$ and $3.931\times10^{-3}$~a.u.$^2$ for absorption
to $S_1$ and $S_2$, and $4.951\times10^{-4}$ and
$5.041\times10^{-3}$~a.u.$^2$ for the respective emission models.
Absorption and emission therefore need separate property expansions;
an intensity-reflected absorption profile does not supply the emission
spectrum of these models.

The profiles are evaluated by independent, complete harmonic TD
correlations. Halving the time step from 0.25 to 0.125~fs at fixed
1024~fs window changes the density by less than $3.1\times10^{-8}$ of
its peak for all four spectra. TI recovers 99.002\% and 99.001\% of
the exact integrated strength for $S_1$ absorption and emission.
The TI implementation explicitly evaluates millions of transitions while
retaining individual assignments for the nonzero lines. This capability
is demonstrated by the more than 10 million transitions used for
p-terphenyl and by the 50 million absorption transitions evaluated here
for $S_2$. The broader $S_2$ progression is more demanding: these
50 million transitions recover 91.370\%, so its TI assignment remains incomplete.
The SI records this distinction between a converged TD envelope and
a finite TI assignment. It also defines the comparison with common
intermediate modes: the same two quadratic Hamiltonians, the same
FC+HT operator and an independently measured intensity omission.

The SI compares three alternative transition-selection strategies with the adopted selector on the same $S_2$ model.

The complete harmonic limit of the common-mode Hamiltonians provides
an independent operator-level comparison. For both $S_2$ absorption
and emission, the separate FC, HT and signed interference correlations
agree with the standard formulation within $1.7\times10^{-13}$ of
the total strength over all 4097 time points. The corresponding density
differences are below $5.4\times10^{-14}$ of the full spectral peak.
Thus eliminating the intersurface overlap transformation preserves the
full linear-property spectrum when all residual quadratic terms are
retained. This exact quadratic check precedes the separate convergence
assessment of a finite dressed-VCI expansion.

\paragraph{Anharmonic interference sensitivity.}
The nonplanar $S_2$ surface tests the sensitivity of weak optical lines
to nuclear correlation and FC/HT cancellation in a shared representation.
This calculation uses the Tamm--Dancoff approximation (TDA) with B3LYP/6-31+G(d) nuclear fields and
transition properties. It describes the inversion motion by a periodic
potential and retains all 26 transverse coordinates in a coupled quadratic
field whose forces and Hessian depend on inversion. All transverse mixed
quadratic terms are included variationally. This conditional model captures
inversion-dependent coupling without requiring a complete molecular cubic
and quartic force field. The DVR--harmonic partition and the variation of transverse modes along a large-amplitude path have a direct precedent.\cite{Baiardi2017HybridDVR} Here the inversion and retained transverse states enter a joint selected variational Hamiltonian; transition operators are contracted in its shared coordinate representation.

\begin{figure}[t]
\includegraphics[width=0.94\linewidth]{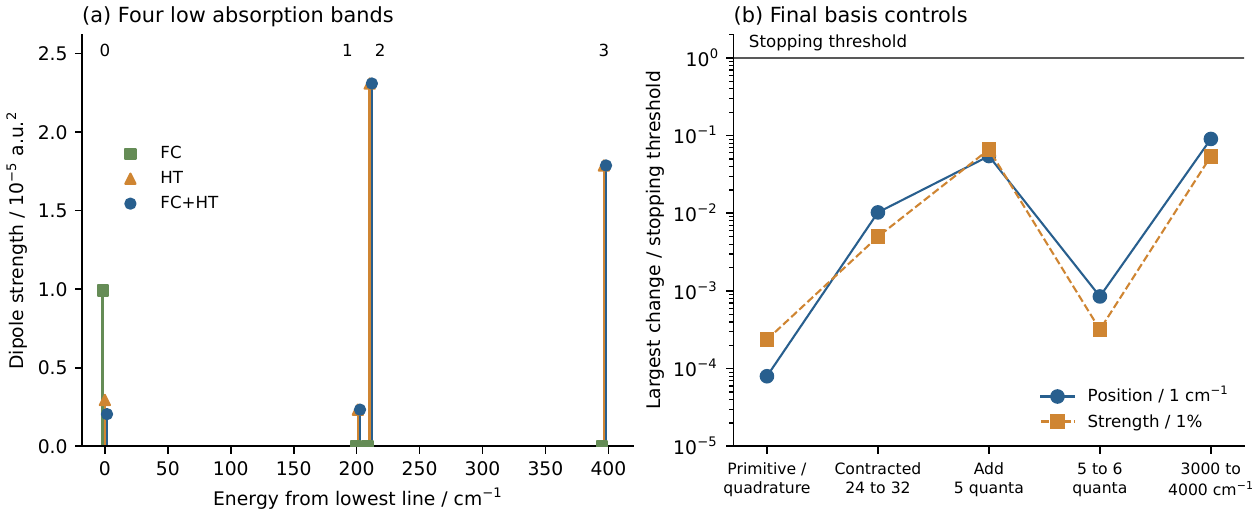}
\caption{Conditional anharmonic $S_2$ model of 4-pyrone. (a) FC, HT and coherent FC+HT strengths for four accepting states; integer labels give state order. (b) Maximum position and strength changes divided by the 1~cm$^{-1}$ and 1\% stopping thresholds. Values below unity satisfy the corresponding criterion; basis definitions are in the SI.}
\label{fig:pyrone-s2-anharmonic}
\end{figure}

Figure~\ref{fig:pyrone-s2-anharmonic} and
Table~\ref{tab:pyrone-s2-low-bands} report the four lowest accepting states
from the harmonic $S_0$ vacuum at 0~K. Their excitation intervals are
200.995, 210.405 and 396.433~cm$^{-1}$ above the lowest line. Linear HT
terms dominate the three upper bands. For the lowest band, the separate
FC and HT strengths are $9.912\times10^{-6}$ and
$2.948\times10^{-6}$~a.u.$^2$, but their signed interference is
$-1.081\times10^{-5}$~a.u.$^2$. The resulting FC+HT strength is only
$2.049\times10^{-6}$~a.u.$^2$. Retaining amplitudes and their relative
signs is therefore essential even in this narrow energy window.

Nuclear-state populations give a physical interpretation of this window.
The 200.995~cm$^{-1}$ line has 94.86\% weight in the first excited contracted
inversion function with the transverse vacuum. The 210.405 and
396.433~cm$^{-1}$ lines have 95.44\% and 95.37\% weight in single excitations
of the two lowest planar-reference transverse modes, both outside the
molecular plane, with the lowest inversion function. Their reference
frequencies are 204.216 and 381.113~cm$^{-1}$. These are dominant-character
assignments in the declared product basis; inversion averaging and coupling
are retained in the reported energies. The lowest accepting state has
95.42\% weight in the corresponding product vacuum. Reduced inversion-state
purities of 0.951--0.982 quantify the residual inversion/transverse
correlation, despite the simple leading characters. The lowest optical line
retains only 15.93\% of its incoherent FC-plus-HT strength because of destructive
interference. Nuclear-state character and optical intensity therefore provide
complementary diagnostics.

AO-transported electronic transition fingerprints retain root-2 identity at
all five acquired geometries, with adjacent overlaps of 0.930--0.986.
The lowest optical line is invariant to a common FC/HT phase; a 1\% change
in HT amplitude changes its strength by about 2.4\%. These controls support
state and phase consistency while exposing property-model sensitivity.
The smallest sampled electronic gap is 0.096~eV at the outer point;
Section~\ref{sec:model-scope} discusses the corresponding electronic-model
assumption. The SI records state tracking and optical-response controls.

A finite-path property control adds a bounded periodic curvature response
fitted to the same five dipoles and analytic derivatives, while preserving
the planar value and first derivative of the added response at zero.
The three upper strengths change by less than 0.27\%, whereas the strongly
cancelled lowest line increases by a factor of 2.06--2.14 for the two tested
response orders. This separates the optical-response sensitivity from nuclear-basis
convergence. Table~\ref{tab:pyrone-s2-low-bands}
therefore identifies the linear-operator model explicitly.

The final tested basis contains 24 contracted inversion functions and
$43\,298$ transverse configurations. Independent changes of the primitive
basis/quadrature, inversion contraction and transverse occupation space
satisfy the chosen 1~cm$^{-1}$ position and 1\% strength criteria.
The last enlargement changes the four positions by at most
0.091~cm$^{-1}$ and the FC+HT strengths by 0.055\%.
Separately expanding the inversion walls from $\pm3.5$ to $\pm4.0$ at the
final occupation space changes the positions by at most
$6.56\times10^{-4}$~cm$^{-1}$ and strengths by 0.00064\%.
The physical model retains fixed tangent optical transport and a
conditional quadratic bath. These four lines
carry 1.678\% of the full initial-state dipole-strength sum rule and
define the low-energy sensitivity window. The electronic models and interpretation of their comparison are
specified in Section~\ref{sec:model-scope}.
The SI specifies the model, source contraction and finite basis controls.

\begin{table}[t]
\caption{Four low $S_0\to S_2$ absorption lines of the conditional
TDA-B3LYP model with the linear FC/HT operator at 0~K. $n$ labels accepting states in energy order;
$\Delta\widetilde\nu$ is measured from the lowest line.
Strengths retain their computed scale without renormalization.}
\label{tab:pyrone-s2-low-bands}
\begin{ruledtabular}
\begin{tabular}{rrrr}
$n$ & $\widetilde\nu$ (cm$^{-1}$) &
$\Delta\widetilde\nu$ (cm$^{-1}$) & FC+HT ($10^{-5}$~a.u.$^2$)\\
\hline
0 & 34810.98 & 0.00 & 0.20487\\
1 & 35011.97 & 200.99 & 0.23337\\
2 & 35021.38 & 210.41 & 2.30858\\
3 & 35207.41 & 396.43 & 1.78699\\
\end{tabular}
\end{ruledtabular}
\end{table}

\subsection{Two-photon response and vibronic spectrum of pyrazine}
\label{sec:pyrazine-tpa}
The two-photon example uses pyrazine in the $D_{2h}$ geometry from the 2025 library of chemical building blocks (LCB25), canonicalized by ORACLE before the electronic calculations.\cite{Lazzari2025LCB25} The target band is the
cold-molecule $n\!\to\!3s$, $^1A_g\leftarrow{}^1A_g$ Rydberg transition. We use
CAM-B3LYP with the diffuse aug-cc-pVTZ basis for the electronic response. The
state is identified from the TDDFT root dominated by the $n\to3s$ orbital
transition; the Gaussian optimization and analytic Hessian provide an independent state-specific
nuclear reference, while \href{https://www.turbomole.org/}{TURBOMOLE} supplies the direct quadratic response tensor
at the same electronic model. The profile shown below uses the full Gaussian CAM-B3LYP/aug-cc-pVTZ field, built from the Gaussian ground- and excited-state Hessians and the mass-weighted displacement. Both surfaces use the adiabatic harmonic model at their Gaussian CAM-B3LYP/aug-cc-pVTZ minima, with all 24 modes and their Duschinsky mixing.

For a degenerate two-photon transition the electronic property is a symmetric
rank-two tensor $S_{ab}(\omega,\omega)$, obtained from the quadratic response
function. Its rotational invariants are
\begin{align*}
 d_F&=\frac{1}{30}\sum_{ab}S_{aa}S_{bb},\\
 d_G&=\frac{1}{30}\sum_{ab}S_{ab}S_{ab},\\
 d_H&=\frac{1}{30}\sum_{ab}S_{ab}S_{ba}.
\end{align*}
with linearly parallel, linearly perpendicular and circularly polarized strengths
\begin{align*}
\delta_{\parallel}&=2d_F+2d_G+2d_H,\\
\delta_{\perp}&=-d_F+4d_G-d_H,\\
\delta_{\mathrm{circ}}&=-2d_F+3d_G+3d_H.
\end{align*}
The
vibronic contraction uses the tensor as a property vertex,
$A_{ab}(\mathbf n)=S_{ab}\langle\mathbf n_f|0_i\rangle$; derivatives
$\partial S_{ab}/\partial Q_j$ can be added through the same HT interface. In the
present benchmark, the direct response at the planar reference geometry gives,
for the equal-frequency process, $S_{xx}=-2.33005$, $S_{yy}=-21.65277$ and
$S_{zz}=10.57599$ a.u., and the CAM-B3LYP vertical root is 54\,110.4~cm$^{-1}$. For this tensor, $d_F=5.9914$ and $d_G=d_H=19.5374$, giving $\delta_{\parallel}=90.1326$, $\delta_{\perp}=52.6209$, and $\delta_{\mathrm{circ}}=105.2417$ in the electronic response convention. The propagated circular profiles are the two helicity channels; $\delta_{\perp}$ is reported as an invariant and is not a third propagated profile.

\begin{figure}[t]
\centering
\includegraphics[width=0.92\linewidth]{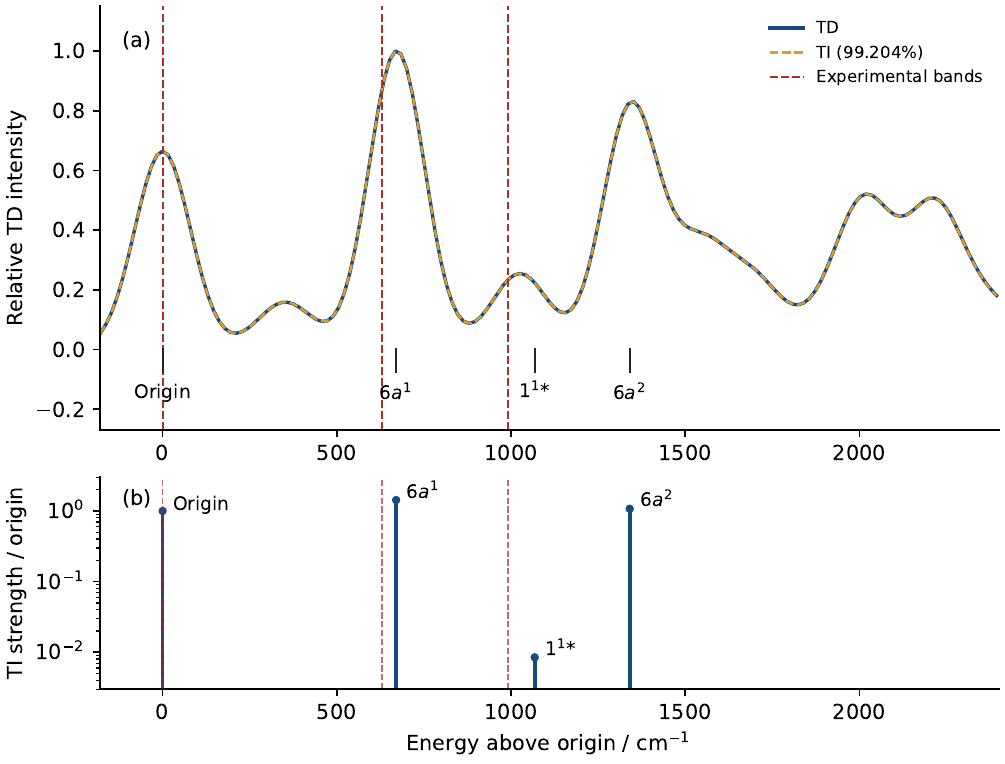}
\caption{Pyrazine two-photon absorption. (a) Calculated TD profile aligned with the experimental origin; dashed lines mark measured bands. (b) Wilson assignments of TI fundamentals and the $6a^2$ overtone, with strengths relative to the origin; the asterisk marks the proposed $1^1$ assignment. All calculations use the complete Gaussian CAM-B3LYP/aug-cc-pVTZ field.}
\label{fig:pyrazine-tpa}
\end{figure}

The TI and TD contractions use the same nuclear Hamiltonian and tensor vertex.
TI assigns individual vibronic transitions, whereas TD forms the continuous
finite-window profile; their agreement is the numerical contraction check. Because the tensor is a constant FC vertex, its polarization factors multiply the same nuclear overlap probabilities. Adaptive TI retains 99.204\% of this probability in 1024 assigned transitions, with an omitted fraction of 0.796\% in every polarization. The Gaussian-broadened TI profile differs from TD by 0.222\% of the TD peak; its integrated absolute difference equals the omitted fraction, without intensity renormalization. The TD tail is negligible. The integrated strength equals the electronic invariant by completeness for this constant FC vertex. The assigned TI lines and their broadened profile supply the nuclear test. The full Gaussian field gives a vibronic origin of 53\,347.3~cm$^{-1}$. Relative to the experimental origin at
50\,835~cm$^{-1}$, the calculated origin is higher by 2\,512.3~cm$^{-1}$
(0.312 eV). TI resolves the $6a^1$ fundamental ($a_g$, program index 6) at 670.34~cm$^{-1}$, compared with the experimental $6a^1$ spacing of 629~cm$^{-1}$ (+6.6\%). The $6a^2$ overtone lies at 1340.68~cm$^{-1}$; the broadened feature near 1352~cm$^{-1}$ contains this overtone and overlapping lines. The proposed $1^1$ ring-breathing fundamental ($a_g$, program index 12) at 1068.04~cm$^{-1}$ corresponds to the experimental band at 992~cm$^{-1}$ (+7.7\%). Its FC strength is only 0.59\% of $6a^1$, so it appears as a weak feature on the overlapping combination-band envelope. The TI strengths relative to the origin are 1.4280, 0.00848 and 1.0741 for $6a^1$, the proposed $1^1$, and $6a^2$, respectively. This weak prediction for the observed ring-breathing band motivates extending the constant-FC vertex. Including $\partial S_{ab}/\partial Q_j$ through the implemented HT interface supplies coordinate-dependent two-photon intensity borrowing and provides the direct route to testing this intensity discrepancy. Figure~\ref{fig:pyrazine-tpa} shows the calculated
profile after shifting its origin to the experimental value. The polarization
ratio $\Omega=\delta_{\mathrm{circ}}/\delta_{\parallel}=1.1676$ is consistent
with the computed totally symmetric transition; no experimental $\Omega$ value
is used here. The bands and assignments are taken from Dion and Bernstein.\cite{Dion1995Rydberg}
This example demonstrates that the same ESALC coordinate and TI/TD machinery
accepts a diffuse-basis quadratic electronic response, in addition to
 electric-dipole and chiroptical operators.

\subsection{External harmonic reference}
Matched physical inputs connect the harmonic calculation to established vibronic programs. Independent harmonic implementations reproduce the water TI profile and the complete 27-mode 4-pyrone FC+HT models; the long-grid maximum profile residual is $5.22\times10^{-4}$ of the external peak without an energy or intensity fit. A compact recurrence implementation provides an additional reference for displacement and mode mixing.\cite{FCclasses2023,MolFCSoftware} These checks establish the harmonic reference for the subsequent nuclear-model hierarchy.

An established harmonic spectral solver evaluates absorption and emission through an analytic path-integral correlation, with HT terms and Duschinsky rotation. The corresponding module is termed Excited State Dynamics (ESD).\cite{DeSouza2018ESD,ORCA61ESD} Executed FC and FC+HT comparisons reuse the same Cartesian force fields and operators. For the thermal 27-mode FC+HT profile, the maximum total-profile residual is 1.81\% on the unmodified reference output and 0.074\% after the separately identified analytic zero-endpoint audit. Absolute units and the original outputs are retained. The SI supplies each program's grids, endpoint conventions, spectral coverage and independent discretization controls.

The C++ implementation gives shorter execution times in the tested resource-matched harmonic cases, including finite-temperature FC+HT. Complete-process measurements include physical-input preparation and spectral export; TI cost is reported against recovered strength, while TD cost is compared at a stated time step and window. The SI contains the timing tables, cost--coverage curves, compiler/library records and unrounded repetitions. Anharmonic assessment instead uses matched-field perturbative/variational comparisons and the experimental molecular profiles. Trajectory-based extended thawed Gaussian and Hagedorn approaches provide complementary nuclear approximations.\cite{Patoz2018ETGA,Begusic2020Thermal,Zhang2026Hagedorn}

\subsection{Computational implementation}
General C++ kernels evaluate harmonic FC/HT models with dynamic mode
dimensions and explicit resource budgets.
The kernels reconstruct the Gaussian from physical inputs and share exact
recurrences and reusable factorization workspaces across complex position
and momentum operators. The SI describes the shared Hermitian operator representation and reports
both complete-grid and converged-window execution costs.
Large TI expansions use grouped transition evaluations and persistent overlap workspaces. A degree-level contraction path can distribute independent
integrals across threads while preserving the ordered sum for each integral.
The transition list supports restarting large calculations while preserving state identities and absolute strengths. The SI documents data reuse and its measured cost.

The window-directed TI option provides state-to-state assignments in a
chosen spectral interval, with TD as the independent completeness and
profile reference. Each TI line retains its initial and final vibrational
state labels, transition energy and physical strength. A same-model TD calculation supplies the
positive operator strength in each requested interval. Exponentially
biased one- and two-mode occupation probabilities rank candidate occupations, while the exact
recurrence evaluates their unchanged physical amplitudes. Selection stops when every requested window reaches both its specified
positive-strength fraction and the peak and integrated absolute profile
error tolerances for every requested observable. Distinct occupation
rankings serve the unfinished windows, sharing computed overlaps
and a single physical transition list. An automatic assignment
report resolves FC, HT and interference contributions and separates
line centers inside each interval from external line tails. Gaussian or Lorentz tails are included across window boundaries;
signed FC/HT interference and chiroptical strengths remain signed. The result
reports window recovery and profile errors alongside the global omitted-strength
bound. The resulting assigned transition list satisfies the stated tests in the requested
spectral region; the TD profile supplies the corresponding aggregate
reference without resolving individual transitions. The SI gives the general protocol and
checks for absorption, emission, finite temperature and restart consistency.

\subsection{Model selection and scope}
\label{sec:model-scope}
A practical calculation starts from a named electronic model, a common
coordinate definition and a consistent property expansion. The harmonic TI/TD
comparison supplies the reference and its intensity omission bound. Stable
weakly coupled blocks use second-order vibrational perturbation theory (VPT2),
with wavefunction corrections and target intensities monitored together.
Small denominators identify resonant blocks; those blocks are included in a
variational vibrational configuration-interaction (VCI) space. Large-amplitude
coordinates are represented by coupled DVR functions, while the remaining
modes stay in the analytic harmonic bath. Common-mode contractions transport
the potential, kinetic metric and transition operator in one shared basis.

Numerical sensitivity is reported alongside the physical model used for each
observable. For the four low $S_2$ bands, the final basis change is below
0.091~cm$^{-1}$ and 0.055\% in FC+HT strength. The SI separates potential-fit,
transverse-field, optical-embedding and electronic-surface controls so that
each reported spectrum has an explicit numerical context.

The local harmonic reference freezes each state metric; finite-path metric derivatives
are retained in the specified internal-coordinate nuclear models. The nuclear
spectra use separate adiabatic electronic surfaces, and linear HT terms provide
coordinate-dependent intensity borrowing and weak electronic-state mixing in
the transition operator.\cite{Patoz2018ETGA} The reported $S_2$ lines are
adiabatic vibronic observables for the sampled surfaces; internal-conversion
branching and fluorescence yields require an additional nonadiabatic coupling
model.

\section{Discussion and conclusions}
The results support a three-level spectral hierarchy. Automatic ESALC-I/ESALC-C molecular representations feed a complete harmonic FC/HT engine with complementary TI assignments and TD thermal profiles. Matched-input tests establish agreement; the resource-matched benchmarks give an 18.11-fold TI advantage at a common $\geq54\%$ recovery threshold with at most seven simultaneously excited modes, and 2.6--4.4-fold TD advantages at 0~K on the stated finite grids, with a factor of 1.15 in the 298.15~K FC+HT control (full conditions are in the SI). Coupled DVR coordinates then add selected large-amplitude anharmonicity without enumerating the harmonic bath. Finally, perturbative and variational states extend TI/TD spectra beyond the harmonic model, using either Duschinsky mixing between state-specific modes or a common nuclear basis. The harmonic result measures completeness of the selected transition manifold, whereas the anharmonic result concerns convergence of a finite nuclear basis; these checks are reported separately.
The pyrazine application adds a diffuse-basis quadratic response tensor for the $n\to3s$ Rydberg band and shows how TI assignments distinguish a fundamental from its overtone in the broadened TD profile.

The larger harmonic examples exercise complete curvilinear coordinate representations. Approximately 99\% intensity recovery in selected TI examples measures the completeness achieved by the transition selector. Primitive-family identity and locality describe the coordinate construction and keep structural assignments interpretable; the nearly equal intensity recoveries connect the spectral equivalence of the two internal-coordinate representations to their different structural interpretability. TD supplies the complementary complete profile and an efficient treatment of thermal populations. Their agreement connects individual assignments to the spectral envelope.
Common modes provide a second representation of the same two-surface problem. Residual quadratic position and momentum couplings remain in the Hamiltonians, while FC amplitudes become direct eigenvector contractions without a separate Duschinsky transformation. Harmonic water checks reproduce both the spectral envelope and assigned FC/HT lines. For the anharmonic water model, the selected common basis reproduces five bands of the independent bond--angle reference within 0.0103~cm$^{-1}$ and $8.5\times10^{-6}$ relative FC+HT strength. These matched-Hamiltonian controls connect spectral equivalence to a compact shared nuclear representation.

Measured solution absorption and fluorescence profiles provide external comparisons for the 66-mode anthracene and 90-mode p-terphenyl models. Full FC/HT operators are used in both directions, with the anthracene derivatives also testing the effect of HT terms on the measured progression.
For p-terphenyl, two coupled torsions, a selected adiabatic bath response
and torsion-dependent displaced Gaussian overlaps reduce the fluorescence
RMS from 0.387 harmonically to 0.201 with the same absorption calibration.
The low-energy explicit-oscillator check supports this response model,
while independent interpolation and nuclear-basis enlargements support
the spectral contraction.

The 75-mode camphor example extends the same harmonic operator bank to
signed ECD. Electric and magnetic FC/HT moments share the nuclear
contraction. The thermal band dissymmetry changes from $-0.1269$ at FC
level to $-0.0580$ with HT, demonstrating why absorption and chiroptical
interference must be treated together. The comparison follows the established
chiroptical and camphor literature.\cite{Stephens1970,Tinoco1975,Pulm1997Camphor,Barone2014Chiroptical,DelGaldo2020Camphor,Kerber2025ECD,Kerber2026Correction}

The water photoionization calculation connects signed continuum FC/HT amplitudes to harmonic and anharmonic nuclear states, with comparisons to measured band positions and nuclear FC factors. Its internal-coordinate ZAFF reference retains the complete curvilinear kinetic operator and demonstrates stable selected-band changes relative to the harmonic model. A separate same-field quartic control shows that a weak overtone can have a large intensity error with the strict second-order wavefunction expansion used here, even when the amplitude contraction is accurate. The frequency-renormalized Gaussian baseline instead gives errors of about 20\% for the bending fundamental and 21\% for its overtone, whereas strict wavefunction VPT2 gives 3.5\% and 414\%, respectively. Neither approximation is uniformly more accurate across these two bands. Correcting Gaussian widths and expanding non-Gaussian nuclear functions are distinct approximations. Supplementary coherent-optical and diatomic tests separately check the two-vertex contraction and nuclear level solver; the comparison is placed alongside the selected-mode Fermi golden rule treatment of Ito and Kurashige.\cite{ItoKurashige2026}

The nonplanar $S_2$ example extends the spectral hierarchy to a periodic inversion coordinate coupled to all 26 transverse coordinates. Four low absorption lines are numerically stable under the final basis controls, with strength changes below 0.055\%. A finite-path optical-property response increases the weakest line at 34810.98~cm$^{-1}$ by a factor of 2.06--2.14 relative to the linear operator. This sensitivity reflects strong FC/HT cancellation and limits the predictive reliability of that intensity despite its numerical convergence. The conditional transverse field and fixed tangent optical embedding specify the approximation; the result establishes the low-energy interference-sensitivity window. Related variational and selected-configuration approaches provide the broader anharmonic context.\cite{Luis2006CoupledFC,ViBra2026,Madsen2019Oligothiophene,Phosphorescence2019,Shuai2010,Begusic2022DoubleWell}

Together these results connect a harmonic reference with explicit intensity accounting to flexible-coordinate and fully anharmonic nuclear contractions. State assignments and complete profiles remain complementary outputs of one specified Hamiltonian and transition operator.

\section*{Data availability}
The SATI source distribution, version 0.1.0rc8, is publicly available at \url{https://github.com/yogibubu/SATI-standalone}; the associated release provides source and offline Linux installation archives.\cite{SATISoftware2026} It includes C++ sources, the required dependency packages for the stated platforms, licenses, examples and installation checks. A self-contained finite-temperature example reproduces TI/TD profiles checked in specified windows, state-to-state FC/HT assignments and calculation restart using the installed package. Electronic-structure executables are external prerequisites for generating new inputs. Electronic-structure inputs, force-field and continuum-operator conventions, spectral validation summaries, source checksums and figure scripts accompany the manuscript repository. The SI identifies the evidence and numerical controls for each retained molecular example, including the separate published-constant iodine model. No new experimental measurements were generated. The water comparison uses unfitted experimental observations. The organic absorption and fluorescence comparisons use the archived numerical PhotochemCAD solution spectra.\cite{PhotochemCAD2018} The iodine constants and solution Raman bands are documented in the SI.

\bibliography{references,validation_references}
\end{document}


\maketitle
\begingroup\small
\tableofcontents
\endgroup
\clearpage
This document records spectral state selection, additional optical observables, harmonic molecular benchmarks and the electronic-continuum validation supporting the water spectra. Potential, nuclear-domain and electronic-model controls remain distinct.
Execution benchmarks use a single stated AMD EPYC server and the same physical inputs for both implementations. ORACLE is the molecular-perception framework.\cite{Barone2026ORACLE} SATI denotes Symmetry-Adapted Transition Intensities, the spectroscopy package described in the main text. Throughout this document, FC and HT
denote Franck--Condon and Herzberg--Teller terms; TI and TD denote
time-independent and time-dependent formulations; DVR and VCI denote discrete
variable representation and vibrational configuration interaction. ESALC denotes
extended symmetry-adapted local coordinates, GF the Wilson GF vibrational
construction, FCHT the combined FC/HT operator, ECD electronic circular
dichroism, CPL circularly polarized luminescence, MCD magnetic circular
dichroism, TPA two-photon absorption, TPCD two-photon circular dichroism,
ZAFF the zero-asymptote fitted force field, and SOC/NAC
spin--orbit and nonadiabatic coupling operators. Adiabatic Hessian (AH), adiabatic shift (AS), adiabatic shift-frequency (ASF), vertical Hessian (VH), vertical gradient (VG) and vertical gradient-frequency (VGF) denote the harmonic potential models. Second-order vibrational perturbation theory is abbreviated VPT2; vibrational self-consistent field is VSCF. Electronic-structure abbreviations include atomic orbital (AO), molecular orbital (MO), density-functional theory (DFT), time-dependent DFT (TDDFT), Tamm--Dancoff approximation (TDA), self-consistent field (SCF), and exchange--correlation (XC). RMS denotes root-mean-square; HWHM denotes half width at half maximum. E1 and M1 denote electric-dipole and magnetic-dipole operators. SPF denotes Simmons--Parr--Finlan radial coordinates. The electronic models rDSD, NR2 and LB94 denote the revised dispersion-corrected spin-component-scaled double hybrid, non-diagonal renormalized second-order electron propagator and van Leeuwen--Baerends potential, respectively. Internal conversion and intersystem crossing are abbreviated IC and ISC. Single-vibronic-level fluorescence is abbreviated SVL.

The numerical implementation uses a fast Fourier transform (FFT), orthogonal--triangular (QR) factorization, singular-value decomposition (SVD) and diagonal-pivot symmetric ($LDL^{\mathsf T}$) or lower--upper (LU) factorization. BLAS denotes Basic Linear Algebra Subprograms, LAPACK the Linear Algebra Package and FFTW Fastest Fourier Transform in the West. Computational resources are reported as processor threads, elapsed seconds and memory. Central processing unit is abbreviated CPU; the operating-system interface is the Portable Operating System Interface (POSIX), and parallel loops use Open Multi-Processing (OpenMP). The data formats include JavaScript Object Notation (JSON), tab-separated values (TSV), Portable Document Format (PDF) and Gaussian formatted checkpoints (FCHK). File checksums use the 256-bit Secure Hash Algorithm (SHA-256). Molecular-orbital labels HOMO and LUMO denote the highest occupied and lowest unoccupied orbitals. The one- and two-excited-mode classes are C1 and C2; higher classes count simultaneously excited modes.

Electronic fields are read from \href{https://gaussian.com/}{Gaussian} and \href{https://www.faccts.de/orca/}{ORCA}. ESD denotes the latter program's Excited State Dynamics spectral module. PBE0 is the Perdew--Burke--Ernzerhof hybrid functional; B3LYP is Becke three-parameter Lee--Yang--Parr, and CAM-B3LYP adds Coulomb attenuation. D3(BJ) denotes D3 dispersion with Becke--Johnson damping. Laboratory-unit SI denotes the International System of Units, while document references to SI denote this Supporting Information.

Table~\ref{tab:literature-map} groups the literature supporting the spectral formulations and molecular comparisons.

\begin{table}[!htbp]
\centering
\caption{Literature grouped by the physical and numerical aspects of the present work.}
\label{tab:literature-map}
\begin{tabular}{p{0.24\linewidth}p{0.66\linewidth}}
\toprule
Role & Representative references\\
\midrule
FC/HT and Duschinsky foundations & Duschinsky, Sharp--Rosenstock, Doktorov, Herzberg--Teller\\
Harmonic recurrences and state selection & Gruner--Brumer, Ruhoff, Santoro, Jankowiak, FCclasses3\cite{FCclasses2023}\\
Anharmonic nuclear models & Luis--Bishop--Kirtman, Rodriguez, Bonness, Bowman, Christiansen, ViBra\cite{ViBra2026}\\
Time-dependent and Raman formulations & Heller, Lee--Heller, Baiardi, Egidi\\
Electronic and chiroptical applications & Stephens, Tinoco, Pulm, Kerber, de Souza\\
Photoionization and molecular benchmarks & Botter--Rosenstock, Reutt, Engin, PhotochemCAD\\
Software and alternative implementations & ORCA Excited State Dynamics,\cite{DeSouza2018ESD,ORCA61ESD} ezSpectra,\cite{EzSpectra2022} MolFC,\cite{MolFCSoftware} Gaussian boson methods\\
\bottomrule
\end{tabular}
\end{table}

\section{Spectral state selection and amplitude reuse}
\subsection{Conditional TI selection and overlap implementation}
The TI intensity is the squared operator matrix element multiplied by the initial thermal population. A finite class/occupation enumeration is retained as a small-system reference. With $N$ modes, total occupation $Q$, and at most $C$ excited modes, its state count is
\begin{equation}
 N_{\mathrm{states}}=1+\sum_{k=1}^{\min(N,Q,C)}{N\choose k}{Q\choose k}.
\end{equation}
Production selection instead ranks candidates using the Gaussian occupation distribution. Multiplication of the initial creation variables by $\sqrt{q_j}$, where $q_j=\exp(-\hbar\omega_j/k_BT)$, represents the thermally weighted FC amplitudes as a purified Gaussian state. Its reduced one- and two-mode occupation distributions retain squeezing, parity structure and mode correlations. Such Gaussian number distributions admit multidimensional Hermite representations.\cite{Dodonov1994} Large pair tables are obtained by Cauchy coefficient extraction of their number-generating function, with radial damping and padding; this transform concerns the proposal, not the physical spectrum.

A maximum-correlation spanning tree connects the occupation variables. Its positive proposal has the form
\begin{equation}
 P_T(\vect{n})=P_r(n_r)\prod_{j\ne r}P_j(n_j\mid n_{p(j)}).
\end{equation}
Linear-operator creation neighbors supplement the FC guide for HT selection. The proposal is a ranking approximation: it is not substituted for the coupled FC/HT intensity. Max-product subtree messages identify optimal completions of a fixed occupation prefix. A compiled priority queue then visits nonnegative deviations from these completions, using lazy append iterators and disjoint prefixes to avoid duplicate configurations. The number of simultaneously excited modes can reach $N$; there is no two-mode excitation cap.

Every selected transition is evaluated with the full recurrence in Eq.~\eqref{M-eq:recurrence}. The cache uses packed multiword occupation keys, with a complete key comparison after hashing; hash collisions cannot identify distinct states. A compiled iterative dependency traversal removes a least-occupied mode first, so high overtones can be reused across weak combinations. Dependency edges are contracted immediately and discarded, while their amplitudes remain cached. This separates state prioritization from the established reuse of multidimensional overlap recurrences.\cite{Gruner1987,Ruhoff1994} Independent transition rows run in parallel; compact column arrays avoid millions of individual \href{https://www.python.org/}{Python} output records.

Convergence is assessed from the physical, positive dipole-strength sum,
\begin{equation}
 S_{\mathrm{miss}}=S_O-\sum_{\mathrm{reported}\;i,f}p_i
 |\langle f|O|i\rangle|^2 .
\end{equation}
This gives an a posteriori omission bound and, at every time, a bound on the absolute difference between the selected TI correlation and the complete harmonic correlation. A weak or zero parent amplitude never discards its descendants. Exhausted occupation, transition or cache budgets remain explicitly incomplete, and retained strengths are never renormalized. This global certificate is distinct from the rigorous local batch-prescreening bounds of Jankowiak \textit{et al.}\cite{Jankowiak2007} The present conditional selector does not implement those bounds. Its stopping test applies to positive FC/HT dipole strengths, not to a possibly cancelling signed chiral signal.

\section{Additional observables, molecular tests and implementation benchmarks}
Table~\ref{tab:inputs} collects the molecular data and numerical controls required by each spectral model.

\begin{table*}[!htbp]
\caption{Electronic-structure and molecular inputs required by the harmonic calculations. All nuclear fields must refer to the same identified ESALC representation. SAM and LAM denote small- and large-amplitude motions, respectively.}
\label{tab:inputs}
\begin{ruledtabular}
\begin{tabular}{p{0.17\textwidth}p{0.41\textwidth}p{0.33\textwidth}}
Calculation & Required molecular/electronic data & Numerical or physical controls\\
\hline
AH / AS / ASF & Reference minimum, Hessian and metric; other minimum and electronic gap. Other Hessian/metric for AH, projected other Hessian for ASF. & Common coordinate representation; stable harmonic space.\\
VH / VG / VGF & Reference minimum, Hessian and metric; other-state gradient and vertical gap at that minimum. Other Hessian for VH/VGF. & Quadratic relaxation; full curvilinear Hessian where required.\\
FC / HT spectra & Transition dipole and, for HT, its ESALC derivatives at an identified property reference. & Temperature and absorption/emission orientation; TI budgets or TD time step, window and broadening.\\
Resonance Raman & Common lower surface, one or more intermediate electronic surfaces, and complex transition dipoles with optional derivatives. & Incident frequencies, intermediate-state amplitude HWHM, final-state selection and polarization geometry.\\
\end{tabular}
\end{ruledtabular}
\end{table*}

\subsection{Chiral, magnetic and two-photon observables}
An operator bank extends the same Gaussian machinery to several supplied electronic response tensors. For complex amplitudes $A_a$, each observable is expressed as a Hermitian form,
\begin{equation}
 S_\alpha=\sum_{ab}A_a W^{(\alpha)}_{ab}A_b^*.
\end{equation}
TI shares the overlap graph among all tensor components. TD shares each Gaussian factorization and uses an exact QR reduction of the projected HT right-hand sides. Signed spectra and interference terms are retained without clipping. A positive bank norm, rather than cancellation in a signed observable, supplies the finite-space omission bound $\|W^{(\alpha)}\|_2 S_{\mathrm{miss,bank}}$. Adaptive selection uses the positive total strength of the supplied operator set. For a constant two-photon FC tensor, every polarization is the corresponding electronic invariant multiplied by a shared nuclear FC probability; the pyrazine calculation uses this factorization to retain 99.204\% of the strength.

ECD and CPL use supplied electric and magnetic moments in a common phase, frame, origin and unit convention, with consistent conjugation on reversing the transition. Weak-field MCD separates the usual derivative, field-induced intensity, and thermal-population contributions.\cite{Stephens1970} Degenerate electronic subspaces are contracted through traces of the dipole and Zeeman matrices, preserving invariance to unitary changes of basis within each retained manifold. The required magnetic response of the electronic dipole is an input; it is not inferred from an electric transition moment alone.

Degenerate TPA uses a supplied symmetric electric response tensor; TPCD additionally retains both electric--magnetic and electric--quadrupole interference.\cite{Tinoco1975} The implementation concerns co-propagating photons and electronic tensors evaluated at an explicit photon frequency. A shared length gauge and tensor origin are required, but recording those conventions does not establish origin independence of approximate electronic response tensors. The present outputs are response strengths; absolute two-photon cross sections and resonant vibronic intermediate-state denominators are not supplied by this path. Synthetic covariance and TI/TD tests check the tensor contractions; the molecular pyrazine example also tests the constant quadratic-response vertex.

\subsection{Polarization and cross-section convention}
The tensor need not be symmetric or real near resonance. Isotropic rotational averages therefore retain its antisymmetric part as well as its symmetric traceless part. For 90-degree detection with linearly polarized incident light, the parallel and perpendicular analyzed components are reported separately. The vacuum photon-count differential cross section contains
\begin{equation}
 \frac{\omega_L\omega_s^3}{16\pi^2\epsilon_0^2c^4}
 \left\langle|\vect{e}_s^\dagger\vect{\alpha}\vect{e}_L|^2\right\rangle.
\end{equation}
Polarizabilities are converted from atomic to International System (SI) units before reporting square meters per steradian. Multiplication by $\omega_s/\omega_L$ gives the scattered-power convention. Positive and negative Raman shifts produce Stokes and anti-Stokes lines; retained thermal populations are not renormalized. Channels with nonpositive scattered-photon energy are excluded from the physical cross section. No solvent local-field correction is inferred from the linewidth.

\subsection{Cost, parallelism, and reproducibility}
The numerical kernels are implemented in C++17 and exposed through the
validated Python request layer. Reusable Gaussian workspaces serve the
factorizations, operator insertions and thermal resolvents; the latter use
the accretive rank-$N$ reduction described in Sec.~\ref{sec:thermal-resolvents}.
The implementation uses \href{https://www.netlib.org/lapack/}{LAPACK/BLAS} for dense kernels and \href{https://github.com/mreineck/pocketfft}{PocketFFT} for the
Hermitian transform. Exact-zero operator channels are removed structurally,
and numerical export retains 17 significant digits. Build, dependency and
restart details are recorded once in the standalone-distribution subsection
and are not repeated here.

\begin{figure}[!htb]
\centering
\includegraphics[width=.95\linewidth]{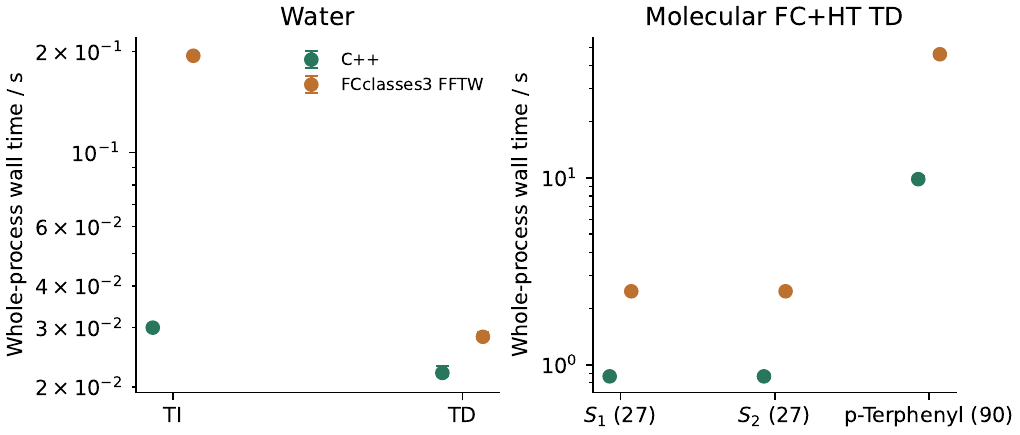}
\caption{Complete prepared-input execution times for the present implementation and an established harmonic solver. Points show medians and bars show ranges of three runs; conditions and profile errors are given in the text and timing tables.}
\label{fig:performance}
\end{figure}

Figure~\ref{fig:performance} compares complete prepared-input executions with
FCclasses3 built with optimization and \href{https://www.fftw.org/}{FFTW}. The three-mode water case and
the 27-mode optical models and 90-mode p-terphenyl field use the same physical inputs as the external
validation. The timing tables below report absolute wall times, exported
point counts and recovered intensity. Compilation and electronic acquisition
are excluded; model preparation and spectral export are included. Timings
are medians of three runs on a shared server and do not define a scaling law.

Complete prepared-input execution times for the general C++ kernels
and FCclasses3 with FFTW. Water uses 2501 exported points in both programs;
the 27- and 90-mode TD models use the same time step and 10000~fs window,
with 65536 exported spectral points per program. Profile residuals use the
common physical window; the FCclasses diagnostic export spans a wider
one-sided FFT interval. Points show
medians and bars the observed range of three runs, on a logarithmic time axis.
Different numerical profile errors are stated in the accompanying tables.

\subsection{Capabilities described for FCclasses3}
FCclasses3 is an established harmonic vibronic package with broad spectroscopy coverage.\cite{FCclasses2023} Table~\ref{tab:comparison} separates shared physical capabilities from the narrower scope and native integration of the present implementation. In particular, curvilinear coordinates, finite-temperature TI/TD, and the adiabatic/vertical families are common foundations. \SATI{} implements those capabilities in a shared ESALC/GF representation, allowing coherent reuse of molecular fields across harmonic and anharmonic models.

\begin{table*}[!htbp]
\caption{Functionality comparison with the published FCclasses3 description. Timing and recovery results are reported separately.}
\label{tab:comparison}
\begin{ruledtabular}
\begin{tabular}{p{0.25\textwidth}p{0.33\textwidth}p{0.33\textwidth}}
Feature & \SATI{} implementation & FCclasses3\\
\hline
Potential models & AH, AS, ASF, VH, VG, VGF and general local fields & Full adiabatic/vertical harmonic family and generalized vertical models\\
Nuclear representation & Native common ESALC-I coordinate representation; Wilson GF; explicit local-metric/Jacobian treatment & Cartesian and curvilinear valence/delocalized internal coordinates (DIC); reduced-space projectors\\
FC/HT and thermal spectra & Absorption/emission; independent TI and analytic TD; complex interference & TI/TD and temperature-dependent vibronic spectra\\
TI state reduction & Conditional Gaussian C2 priority tree; unrestricted excitation class within budgets; exact positive-strength omission audit & Established class-based transition selection/prescreening\\
Execution & General C++ state/amplitude kernels; reusable Gaussian workspaces; Hermitian FFT & Fortran implementation; concurrency requires version/build-specific assessment\\
Resonance Raman & Coherent Kramers--Heisenberg--Dirac (KHD) with FC/HT at both vertices; finite-intermediate TI and analytic-intermediate TD & TI/TD resonance Raman with mode selection and spectral analysis tools\\
Additional spectroscopies & ECD/CPL, weak-field MCD, degenerate TPA and TPCD from supplied electronic response tensors & Resonance Raman, chiral and two-photon spectroscopies\\
Anharmonic spectra & Two-surface VPT2; native curvilinear DVR and selected common-modal VCI & Quadratic nuclear models; projectors enable separate flexible-mode treatments\\
Evidence reported here & Analytic limits, coordinate/phase covariance, matched external harmonic executions, water VPT2/VCI controls, iodine model and molecular timings & Published molecular applications and broader spectroscopy demonstrations\\
\end{tabular}
\end{ruledtabular}
\end{table*}

The class strategy established in FCclasses and Gaussian organizes high-overtone C1/C2 reference calculations and higher excitation classes.\cite{Santoro2007,FCclasses2023} The present selector instead obtains its priorities from reduced Gaussian occupation marginals and a conditional-tree completion. Both approaches seek to avoid exhaustive high-class enumeration; a C2 guide does not mean that only two modes can be excited. The final external controls below report same-checkpoint water timings and finite cost--coverage curves for the executed $S_2$ selectors. Their budgets and maximum excitation classes differ, so equal-cost and equal-completeness conclusions are distinguished. Published FCclasses3 applications also identify large torsional rearrangement and anharmonicity as separate difficulties for p-terphenyl, including reduced-space and external torsional treatments.\cite{FCclasses2023} The current full-dimensional harmonic validation addresses numerical completeness of a specified model, not superiority over those anharmonic strategies.

Gaussian l718 and FCclasses3 supply harmonic FC/HT benchmarks; FCclasses3
projectors isolate a reduced harmonic subset for separate flexible-mode
treatment.\cite{FCclasses2023} Anharmonic comparisons require matched VCI/VPT2
or semiclassical models, including the extended thawed Gaussian approach with
HT terms.\cite{Patoz2018ETGA} Its phenyl-radical and benzene experimental tests
are published evidence; no execution benchmark against that method is performed
here. Timing comparisons require matched inputs, spectral accuracy and hardware.

\subsection{Relation to other implementations and anharmonic strategies}
ORCA ESD provides harmonic analytic path-integral spectra, with HT and optional
Duschinsky mixing, and extends its property framework to radiative and
nonradiative rates.\cite{DeSouza2018ESD,ORCA61ESD} FCclasses3 also covers
rate calculations and chiral spectra.\cite{FCclasses2023} Those observables
are broader than the molecular demonstrations in this article. SATI includes
ECD/CPL, MCD and two-photon operator contractions, validated here by synthetic
operator and covariance tests rather than molecular experimental examples.
IC/ISC rate applications are outside the scope of this spectroscopy paper.

The ezFCF component of ezSpectra is a harmonic progression reference; ezDyson
uses supplied Dyson orbitals for photoelectron cross sections and angular
observables.\cite{EzSpectra2022} This electronic continuum task is distinct
from nuclear FC/HT contraction. MolFC supplies harmonic FC factors including
Duschinsky mixing.\cite{MolFCSoftware} No execution or performance claim
against these programs is made here.

FCclasses3 permits flexible-mode projection and an optional replacement of
imaginary frequencies by real ones.\cite{FCclasses2023} Projection defines
a reduced nuclear space; absolute-frequency replacement changes the local
potential model. Neither operation supplies an inverted-oscillator correlation.
Changala \emph{et al.}\ instead derive stable, branch-cut-free correlations
including imaginary and unbound vibrations.\cite{Changala2022Imaginary}
SATI's unit-disk branch and its TI/TD validations in this article apply to
stable positive-frequency references. The periodic conditional DVR treatment
is a separate finite-amplitude nuclear model.

The hybrid internal-coordinate DVR--harmonic approach of Baiardi \emph{et al.}\
already includes path-dependent transverse normal modes.\cite{Baiardi2017HybridDVR}
The present conditional calculation is positioned as a joint variational
realization with a selected transverse basis and shared operator contractions.
The same-Hamiltonian tests, rather than this partition alone, support its
numerical solution. ViBra's quartic-field VSCF/selected-VCI framework uses
Epstein--Nesbet screening;\cite{ViBra2026} the occupation-limited common-modal
spaces used for the water validation are not presented as that algorithm.
Finite-temperature thawed Gaussian, double-well tests and Hagedorn SVL
fluorescence provide complementary trajectory-based references.\cite{Begusic2020Thermal,Begusic2022DoubleWell,Zhang2026Hagedorn}

\subsection{Standalone distribution and electronic-data interfaces}
SATI version 0.1.0rc8 is available from the public repository
\url{https://github.com/yogibubu/SATI-standalone}.\cite{SATISoftware2026}
The tagged release provides both a source archive and the offline Linux
x86\_64/CPython 3.11 distribution. Building from source first downloads standard
Python dependencies; installation from the complete wheelhouse is offline.
SATI is supplied independently of the parent software suite. Its source
package record pins the scientific revision and hashes the exported files; the
wheelhouse package record lists the build platform and dependency hashes. The
distribution includes the transitive package dependencies for coordinate
construction, Gaussian and ORCA input adapters, ZAFF, DVR, VPT2/VCI and
harmonic/general-operator spectra. Internal \texttt{matrix\_*} namespaces
retain their scientific owners and are included in the distribution.

Python 3.11 and compatible system libraries are prerequisites. The offline
installer verifies wheel hashes, installs into the user's chosen environment
without an index, and checks dependency consistency. Platform-specific wheels
are rebuilt for another Python application binary interface (ABI), operating system or processor. The standalone
harmonic executable and shared kernels are explicitly compiled from the
bundled C++17 sources with BLAS/LAPACK; no import triggers compilation.
Examples, source licenses and native/common-mode/variational regression checks
accompany the package. The distribution supplies 23 scientific components and 48 installation packages, including 24 third-party dependencies with recorded versions and checksums.
The archive identifier and full SHA256 are recorded in
\texttt{thermal\_release/standalone/distribution.json}. The rebuilt
executable and shared library reproduce the validated 27- and 90-mode
thermal correlations and spectra byte for byte; Section~\ref{sec:thermal-release}
reports the release and external checks.

On the reference Linux x86\_64/CPython 3.11 environment, a fresh installation installs offline
and passes the dependency check. Scientific imports resolve inside that
installed environment; ten current targeted-selection modules match their
exported source hashes byte for byte. Rebuilding the C++17 executable and
shared kernels is explicit. All 188 native/common-mode/variational regressions
pass (24.96~s reported by the test runner), followed by all 17 targeted-selection extension checks.
The two-state ESALC example retains maximum independent TI/TD correlation
error $1.89\times10^{-12}$.

A second, fully specified two-mode example exercises finite-temperature ECD
in two disjoint windows. Its 881 selected physical transitions recover
99.9264\% and 99.0042\% of the positive-bank reference; every signed
observable also passes the 2\% peak and 1\% integrated absolute profile
criteria. Repeating the command resumes the checkpoint with identical
state occupations, energies and strengths. The first run takes 1.85~s and
the replay 0.82~s in this installation check. The distribution supplies the
input specification, execution instructions, profiles, complete transition columns, FC/HT assignments, restart data and validation records.
These are installation and scientific-regression checks, separate from
molecular-model validation.

Gaussian formatted-checkpoint data supply geometries, Hessians, mode data and
available anharmonic fields. The ORCA adapter reads \texttt{.hess} force fields
and \texttt{.engrad} point results. The electronic-state interface also reads Gaussian formatted-checkpoint files and ORCA-to-Molden exports; the latter
requires the supplied \texttt{orca\_2mkl} conversion utility. These adapter capabilities
are distinguished from the Gaussian-based molecular force fields validated
in this article. Gaussian, ORCA and other electronic executables are not
redistributed. Continuum-operator coefficients can be ingested through the
general FC/HT interface; the electronic continuum solver is a separate project.

\FloatBarrier
\section{Additional coherent-optical and diatomic solver checks}
\subsection{Coherent two-vertex Raman amplitudes}
\subsection{Coherent KHD tensor}
Resonance Raman scattering requires two optical vertices and retains coherence over all intermediate vibronic states. One-photon intensities alone do not contain the relative phases needed for this sum. With $i$ and $f$ denoting initial and final vibrations of the same lower electronic surface, we evaluate the Kramers--Heisenberg--Dirac (KHD) tensor\cite{RamanTI2014,RamanTD2014}
\begin{align}
 \alpha^{fi}_{ab}(\widetilde\nu_L)=C_{\rm H}\sum_{e,m}
 \Bigg[&\frac{(D^e_{fm,a})^*D^e_{im,b}}
 {\Delta^e_m-E_i-\widetilde\nu_L-i\Gamma_e}\nonumber\\
 &+\frac{(D^e_{fm,b})^*D^e_{im,a}}
 {\Delta^e_m+\widetilde\nu_L-E_f+i\Gamma_e}\Bigg].
 \label{eq:raman}
\end{align}
Here $a$ and $b$ refer to scattered and incident polarizations, respectively; $\Delta^e_m$ includes the electronic origin and intermediate vibrational excitation; and all denominator energies are expressed as wavenumbers. The conversion factor $C_{\rm H}$ yields polarizability in atomic units. The positive amplitude HWHM $\Gamma_e$ represents lifetime/dephasing of the intermediate electronic surface and is distinct from a plotting linewidth for the final Raman band. The second, antiresonant time ordering is retained by default.

The complex dipoles $D^e$ include FC and HT contributions at both vertices. This retains mixed FC--HT terms and the product of two linear HT insertions. Different intermediate electronic surfaces are added at the amplitude level before squaring. They must share the same lower-state normal basis and phase convention, verified for native fields by a basis fingerprint; matching frequencies alone is insufficient.

\subsection{Independent TI and TD constructions}
The TI route uses the shared overlap recurrence to obtain complex amplitudes and explicitly sums retained intermediate states. For each external vibration, the missing dipole sum-rule strength $\delta S$ supplies a Cauchy bound on the omitted polarizability norm. For one time ordering and one electronic surface,
\begin{equation}
 \|\delta\vect{\alpha}^{fi}\|_{\rm F}
 \leq \frac{C_{\rm H}}{\Gamma_e}
 \sqrt{\delta S_f\,\delta S_i}.
\end{equation}
The bounds for the two time orderings and supplied electronic surfaces are added. This bound does not cover omitted electronic surfaces or unrequested final Raman transitions.

The TD route instead contracts the infinite intermediate harmonic manifold analytically in a two-sided Gaussian propagator. A compiled complex Hermite graph selects the external number states, and polynomial insertions retain both transition dipoles. This off-diagonal propagator differs from an absorption autocorrelation: it connects a specified lower-state initial vibration to a generally different final vibration. A cubic interpolation of the correlation is integrated against its exponential optical carrier analytically. Time blocks run concurrently within a shared estimated workspace budget. Lifetime damping supplies an absolute omitted-tail bound, while independent time-step refinement checks quadrature accuracy. External-state truncation remains explicit in both methods.

The same overlap machinery therefore connects one-photon spectra and a coherent two-vertex observable. Polarization averages, cross-section conventions and additional tensor observables are specified in the SI. The diatomic model below checks the resulting optical contraction.

\subsection{Iodine: a published-constant molecular model}
A compact molecular case uses the $X$ and $B$ surfaces of $^{127}$I$_2$, whose resonance Raman progression has been measured extensively.\cite{Kiefer1972,NISTIodine} The Huber--Herzberg/NIST compilation supplies $r_e=2.6663$ and $3.0247$~\AA, $\omega_e=214.502$ and $125.697$~cm$^{-1}$, and $T_e=15769.01$~cm$^{-1}$ for the two surfaces. The bond length is the common internal coordinate, its inverse reduced mass supplies $G$, and the harmonic curvatures are constructed from those frequencies. This is a spectroscopically parameterized molecular model, not a QC prediction. No iodine-specific solver term or fitted correction is introduced.

\begin{figure}[!htb]
\includegraphics[width=\columnwidth]{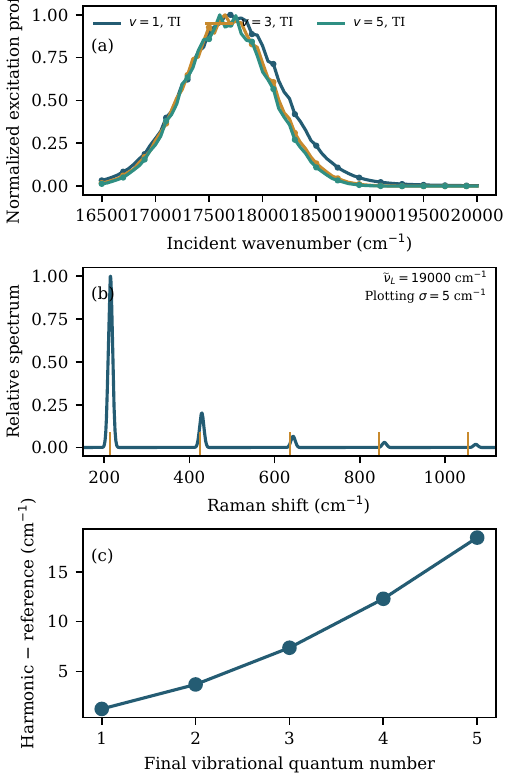}
\caption{Published-constant harmonic iodine model. (a) Excitation profiles for three final vibrations, individually normalized; lines are TI and open circles TD. (b) Raman spectrum at $\widetilde\nu_L=19000$~cm$^{-1}$, drawn with a separate Gaussian width $\sigma=5$~cm$^{-1}$. Short gold ticks indicate reference band positions only, with no experimental intensity implied. (c) Harmonic-minus-reference band-position differences.}
\label{fig:iodine}
\end{figure}

Figure~\ref{fig:iodine} compares excitation profiles and a harmonic Raman progression for this model. A constant unit effective transition dipole fixes an arbitrary intensity scale; it is not an independently computed spin--orbit transition moment. The illustrative amplitude width is $\Gamma=100$~cm$^{-1}$ and is not identified with a measured gas-phase predissociation lifetime. Consequently the profiles test the response of the specified harmonic model and are not a reproduction of experimental absolute intensities or measured excitation profiles.

Over 71 incident wavenumbers from 16500 to 20000~cm$^{-1}$, increasing the TI intermediate space from 100 to 140 quanta leaves the saved tensor unchanged. Independently evaluated 128-node Gauss--Hermite integrals reproduce the native TI tensor to $2.7\times10^{-14}$ relative to the largest reference tensor norm. The analytic-intermediate TD result differs by $3.0\times10^{-5}$ at a 0.5~fs step and $2.0\times10^{-6}$ at 0.25~fs, keeping the same 2048~fs window. This refinement separates time quadrature from intermediate-state completeness without additional electronic-structure calculations.

The physical harmonic error is much larger. Table~\ref{tab:iodine} compares harmonic shifts with reference shifts derived from the published lower-state Dunham constants, rather than digitized experimental sticks. The harmonic origin is 15724.6075~cm$^{-1}$, close to the tabulated 15724.57~cm$^{-1}$ because the two zero-point anharmonicities nearly cancel. That cancellation does not carry over to the Raman progression: the shift error grows from 1.23 to 18.44~cm$^{-1}$ between $v=1$ and 5. These differences are reported as model limitations and are not removed by changing the numerical convergence criteria.

A separate anharmonic check uses the measured I$_2$/CCl$_4$ Raman progression, keeping its environment distinct from the gas-phase model above.\cite{Kiefer1973Solution} Only the first two bands, 211.4 and 421.4~cm$^{-1}$, determine a Morse potential ($\omega_e=212.8$, $\omega_ex_e=0.7$~cm$^{-1}$). The ZAFF radial function and sine-basis nuclear solver then predict levels 3--15, excluded from calibration, with maximum spacing error 5.3~cm$^{-1}$; the corresponding harmonic model reaches 163~cm$^{-1}$. Basis and domain refinement change these spacings by less than $10^{-9}$~cm$^{-1}$. Because the Morse potential is calibrated on two solution bands, this check tests the nuclear solver and its anharmonic level predictions, rather than an electronic-structure model. It does not validate an ab initio solvent potential or Raman intensities. The full observations, exclusions and comparison are in the SI.

\begin{table}[!htbp]
\caption{Iodine Raman shifts (cm$^{-1}$). Reference values are derived from the compiled $X$-state constants as $\omega_e v-\omega_ex_e v(v+1)$; they are not newly measured line positions.}
\label{tab:iodine}
\begin{ruledtabular}
\begin{tabular}{rrrr}
$v$ & Harmonic & Reference & Difference\\
\hline
1 & 214.502 & 213.273 & 1.229\\
2 & 429.004 & 425.316 & 3.688\\
3 & 643.506 & 636.130 & 7.376\\
4 & 858.008 & 845.714 & 12.294\\
5 & 1072.510 & 1054.069 & 18.441\\
\end{tabular}
\end{ruledtabular}
\end{table}

\FloatBarrier
\section{Harmonic executable comparisons and performance}
\label{sec:external-performance}
The harmonic formulation shares its physical foundations with FCclasses3 and Gaussian's vibronic link l718.\cite{FCclasses2023} The standard FC/HT engines in these programs use harmonic nuclear models; FCclasses3 additionally provides projectors to remove flexible coordinates from the harmonic subset for separate anharmonic or mixed quantum--classical treatment.\cite{FCclasses2023} The present implementation connects the molecular-coordinate representation to reusable amplitudes and several nuclear approximations. Additional tests with 66-mode anthracene and 90-mode p-terphenyl fields verify selected TI expansions against independent TD correlations; their inputs, omission bounds and spectra are in the SI. They assess numerical recovery of specified harmonic models.

A same-checkpoint Cartesian AH/FC comparison with Gaussian l718 uses identical archived checkpoints. The native TI profile differs from the l718 TI result by $4.71\times10^{-4}$ of the external peak and $8.43\times10^{-4}$ of its integrated area, using the actual 298.15~K temperature and Gaussian HWHM 50~cm$^{-1}$. Absolute molar extinction is constructed without an area, peak or energy fit. This three-mode harmonic check uses the original B3LYP water checkpoints; it is separate from the rDSD/ZAFF/continuum water application. The SI records input hashes, units, retained strengths and the larger discrepancy against the default external TD execution. Water includes a computed continuum FC/HT operator, a direct internal-coordinate ZAFF/DVR calculation and a separate same-field local variational comparison, with an experimental comparison of water band positions and nuclear FC factors and separate electronic-model and nuclear-domain sensitivities. Published capabilities and input conventions of existing programs are compared in the SI. Execution benchmarks compare prepared physical inputs against the external programs, with hardware, numerical settings, build libraries and spectral recovery stated explicitly.

The l718 comparison also extends to the 27-mode $S_1$ absorption of
4-pyrone, using the same B3LYP/6-31+G(d) checkpoints, Cartesian AH fields,
complete FC+HT operator and 298.15~K temperature in both programs.
For 327 externally assigned transitions, the largest energy difference is
$5.7\times10^{-5}$~cm$^{-1}$ and the largest absolute dipole-strength
difference is $3.59\times10^{-4}$ of the strongest printed line.
The largest relative difference for the 326 resolved strengths is 0.224\%.
These line comparisons retain absolute strengths and require no energy or
intensity fit. The SI separates assignment agreement, progression coverage
and the independent l718 TD grid-index audit.

The available software landscape is broader than these two executable references. ORCA ESD (Excited State Dynamics) is the vibronic dynamics module of ORCA. It evaluates harmonic absorption and emission spectra by Fourier transformation of analytic time-correlation functions obtained with a path-integral formulation, including HT terms, optional Duschinsky rotation and internal coordinates. The full harmonic manifold enters the correlation without enumerating its individual transitions.\cite{DeSouza2018ESD,ORCA61ESD} The ezSpectra suite connects harmonic ezFCF progressions with ezDyson photoionization observables,\cite{EzSpectra2022} while MolFC computes harmonic FC factors with displacement and mode mixing.\cite{MolFCSoftware} Two additional 66-mode anthracene FC profiles execute ESD with the same Cartesian Hessians, gap and dipole at 0~K. The 512 and 1024~fs checks retain the actual ESD Gaussian width of 250~cm$^{-1}$; an endpoint-discretization audit is reported separately in the SI. Three subsequent interleaved repetitions per grid give native/ESD whole-process medians of 0.737/10.985~s at 512~fs and 1.347/15.685~s at 1024~fs: speed factors of 14.90 and 11.64 on the same server with one thread. Both timings include preparation and spectral export. The native export uses a finer 1~cm$^{-1}$ grid than the actual ESD FFT grid. Excluding ORCA front-end initialization and comparing only internal module timers still gives approximately nine- to eleven-fold shorter execution; the ESD internal timers are rounded to seconds. The SI preserves individual runs, hashes and the separate endpoint audit.

The executed ESD comparison also includes the complete 27-mode Cartesian
AH/FC+HT model of 4-pyrone $S_1$. The importer preserves the phases and ordering of
the stored mass-weighted normal vectors, and converts the dimensional
transition derivatives to dimensionless coordinates before transporting
them to the initial-state basis. FC, HT plus interference, and their sum
are compared independently at 0~K, $\sigma=100$~cm$^{-1}$ and 512~fs.
The largest unmodified total-profile difference is 1.34\% of the native
peak; the separately identified zero-endpoint contribution reduces the
residual to 0.076\%, with 0.071\% area difference and no fitted intensity
or energy adjustment. The FC+HT comparison figure in the SI retains both the raw
external profile and this discretization audit.

Table~\ref{tab:ESD-FCHT-timing} distinguishes the full prepared ORCA
workflow from direct execution of its spectral module. The reference
moment and all 27 derivatives are reused, excluding their electronic
acquisition from both timings. Nine interleaved direct-module repetitions
give a 6.26-fold advantage for the native executor, including its own
Gaussian preparation and the export of all three FC/HT components.
The native spectrum has a finer output grid, approximately 1~cm$^{-1}$
versus 65.15~cm$^{-1}$ for ESD. At 298.15~K, the same physical fields and FC/HT operator give a
3.09-fold direct-module advantage over nine additional interleaved runs.
The largest total-profile difference is 1.81\% before the endpoint audit
and 0.074\% after it; the audited integrated strengths differ by 0.068\%.
This comparison establishes a harmonic FC+HT gain at both temperatures
at the level of the spectral solver as well as the workflow.
An exact rank-$N$ reduction of the thermal Gaussian trace further
accelerates the correlation itself. Paired one-thread native measurements
at 298.15~K give correlation speed factors of 1.38 for the 27-mode
4-pyrone and 15.54 for 90-mode p-terphenyl; whole-process factors are
1.32 and 15.35. The external spectral-module timings use the released
solver; the SI gives the reduction algebra, conditioning guards,
preparation/output costs and numerical checks, preserving every archived
campaign with its build identity.

\begin{table*}[!htbp]
\caption{Matched 27-mode 4-pyrone $S_1$ FC+HT timings at one thread. Both programs use the same 512~fs window, 0.125~fs step, temperature and Gaussian broadening; medians include input preparation and spectral export.}
\label{tab:ESD-FCHT-timing}
\begin{ruledtabular}
\begin{tabular}{lrrrrr}
Executed comparison & $T$ (K) & Repeats & Native (s) & ESD (s) & Ratio\\
\hline
Prepared workflow & 0 & 3 & 0.182478 & 10.312455 & 56.51\\
Spectral module & 0 & 9 & 0.158490 & 0.991782 & 6.26\\
Spectral module & 298.15 & 9 & 0.345945 & 1.069277 & 3.09\\
\end{tabular}
\end{ruledtabular}
\end{table*}

Independent FCclasses3 3.0.4 executions also reproduce the complete 27-mode
harmonic $S_1$ and $S_2$ absorption models, including FC+HT. Frequencies,
the nonorthogonal mode map, displacement and the real transition operator
are transferred directly through its implicit-mode interface. The largest
FC+HT profile difference is $5.22\times10^{-4}$ of the external peak,
without separate area or energy adjustment. The SI reports the executed
inputs, absolute profiles, three-program water timings and recovered
intensities, and a separate cost--coverage comparison for $S_2$.

Table~\ref{tab:external-harmonic-timing} makes the harmonic performance
comparison explicit. For the 27-mode 4-pyrone and 90-mode p-terphenyl FC+HT TD profiles, both programs
include the complete harmonic manifold and use the same time step
and 10000~fs window (32768 intervals). FCclasses3 uses midpoint sampling;
the native grid includes the zero endpoint. The measured wall-time advantage over
FCclasses3 is 2.85 for both 4-pyrone states and 4.66 for p-terphenyl,
including preparation and spectral export. The independently measured $S_1$
and $S_2$ medians coincide only after rounding to the displayed millisecond;
both jobs have the same mode dimension, sampling and operator structure.
The unrounded medians are 0.866786045/2.471926679~s for $S_1$ and 0.866596396/2.472049448~s for $S_2$ (native/FCclasses3). Both programs export 65536 points;
profile residuals are assessed on their common physical spectral window. These are single-profile gains, with the three Cartesian
dipole components contracted together; they do not require amortizing
preparation over multiple independent operators. For water TI, recovery
exceeds 99.999999\% in both programs, and the corresponding ratio is 6.46.
This compares practically complete strength recovery, not identical
individual-line or broadened-profile errors. Water TD has different
sampling settings and is reported as a workflow comparison only.

Executed harmonic calculations from prepared physical inputs.
Wall times are medians of three processes on the same shared Linux server
(AMD EPYC 7543); ranges and input hashes are in the SI. Every time includes
startup, Gaussian preparation, spectral calculation and export, excluding
electronic calculations and checkpoint conversion. ``All'' denotes the
analytic infinite harmonic manifold in TD, not exact integration of a finite
spectral window. The 4-pyrone TI rows both pass a 54\% minimum recovery criterion, with
class C7 and a per-mode final occupation cap of 30 in both programs.
The 27- and 90-mode TD rows use the same time step and temporal window,
and 65536 exported points per program;
water grids and profile errors are validated separately in the text and SI.

\begin{table*}[!htbp]
\caption{Harmonic calculations from prepared physical inputs on the same AMD EPYC 7543 server. Times include preparation and export; TI recovery and TD spectral completeness are reported separately. Conditions, thread counts and unrounded repeats are given in the accompanying text.}
\label{tab:external-harmonic-timing}
\begin{ruledtabular}
\begin{tabular}{llcrrl}
System (modes), operator & Program & Route & Threads & Wall (s) & Strength recovery (\%)\\
\hline
Water (3), FC & SATI C++ & TI & 1 & 0.030 & $>99.999999$\\
 & FCclasses3 & TI & 1 & 0.194 & $>99.999999$\\
 & l718$^{b}$ & TI & 4 & 9.148 & 100.00 (printed)\\
Water (3), FC$^{a}$ & SATI C++ & TD & 1 & 0.022 & All\\
 & FCclasses3 & TD & 1 & 0.028 & All\\
 & l718$^{b}$ & TD & 4 & 0.798 & All\\
4-pyrone $S_2$ (27), FC+HT & SATI C++ & TI & 1 & 3.822 & 54.272\\
 & FCclasses3 & TI & 1 & 69.214 & 54.747\\
4-pyrone $S_1$ (27), FC+HT & SATI C++ & TD & 1 & 0.867 & All\\
 & FCclasses3 & TD & 1 & 2.472 & All\\
4-pyrone $S_2$ (27), FC+HT & SATI C++ & TD & 1 & 0.867 & All\\
 & FCclasses3 & TD & 1 & 2.472 & All\\
p-Terphenyl (90), FC+HT & SATI C++ & TD & 1 & 9.839 & All\\
 & FCclasses3 & TD & 1 & 45.845 & All\\
\end{tabular}
\end{ruledtabular}
\begin{minipage}{\textwidth}
\footnotesize
$^{a}$Water TD times are effectively comparable (0.022 versus 0.028~s).
Different sampling settings make these workflow observations, without a
TD speed-advantage claim.\par
$^{b}$l718 uses four threads rather than one and is included as a reference
for spectral agreement, not as a matched-resource speed benchmark.
\end{minipage}
\end{table*}

Both native and FCclasses3 builds use GNU 8.5.0 with \texttt{-O3} and
OpenBLAS/LAPACK; their Fourier transforms use PocketFFT and FFTW 3.3.11,
respectively. The one-thread limit includes the BLAS library. The four-thread l718 measurements use unchanged inputs. Each timed process produces one
profile from physical fields, without precomputed Gaussian parameters or
transition strengths. Water uses 298.15~K and the molecular FC+HT benchmarks 0~K, with Gaussian
HWHM 50~cm$^{-1}$. Thus the 27- and 90-mode TD comparison fixes analytic state
coverage and temporal resolution, rather than finite TI transition counts.

The long-window benchmark is supplemented by six nine-repeat, interleaved
campaigns using application-sized grids: 512 and 1024~fs, a 0.25~fs step
and Gaussian standard deviation 100~cm$^{-1}$, at 0~K. An additional
27-mode $S_1$ FC+HT campaign uses the same fields and grid at 298.15~K. Each program exports
4096 or 8192 spectral points, respectively. Table~\ref{tab:application-timing}
reports whole-process medians, including executable startup, preparation and export; no Python interpreter is started inside the timed command. A blocking process wait with a separate watchdog removes the up-to-50~ms polling latency of timeout-based waits. Native internal timers independently show the expected near-doubling of cost when the window doubles. The TD
speed ratios span 2.61--4.41 at 0~K on these shorter grids; the
298.15~K FC+HT control gives 1.15. The native half-step
checks change the profiles by at most $1.6\times10^{-15}$ of their peaks;
the larger interprogram profile residuals on the finite Fourier export
grids are recorded separately in the SI. The native TD endpoint rejection
correctly excludes a 512~fs window at the narrower original HWHM
50~cm$^{-1}$; the present 100~cm$^{-1}$ standard deviation makes both windows
pass its damping criterion.

\begin{table}[!htbp]
\caption{TD wall times on application-sized grids. One thread per process,
same server and physical fields as Table~\ref{tab:external-harmonic-timing}.
Nine interleaved repetitions per program and case use a blocking wait and the frozen native build archived with the manuscript. The window is in fs; Gaussian $\sigma=100$~cm$^{-1}$. All rows use
0~K except the marked final row at 298.15~K.}
\label{tab:application-timing}
\begin{ruledtabular}
\small\setlength{\tabcolsep}{3pt}
\begin{tabular}{lrrr}
System, window & Native (s) & FCclasses3 (s) & Ratio\\
\hline
$S_1$, 512 & 0.061818 & 0.163510 & 2.65\\
$S_1$, 1024 & 0.117366 & 0.317918 & 2.71\\
$S_2$, 512 & 0.062140 & 0.162163 & 2.61\\
$S_2$, 1024 & 0.116349 & 0.314998 & 2.71\\
p-Terphenyl, 512 & 0.657365 & 2.902187 & 4.41\\
p-Terphenyl, 1024 & 1.286177 & 5.585480 & 4.34\\
$S_1$, 512 (298 K) & 0.151038 & 0.174188 & 1.15\\
\end{tabular}
\end{ruledtabular}
\end{table}

The nearly equal costs of $S_1$ and $S_2$ follow from their equal dimensions
and operator structure: their native medians are 0.061818 and 0.062140~s
at 512~fs and 0.117366 and 0.116349~s at 1024~fs. Executable startup and
dispatch contribute only about 0.006--0.007~s. The released kernel is measured independently in these interleaved runs;
individual samples, quartiles, internal timers and executable hashes are
archived in the SI evidence. The server is shared, so the observed spread
is retained rather than inferred from rounded medians.

\begin{figure}[!htb]
\includegraphics[width=\columnwidth]{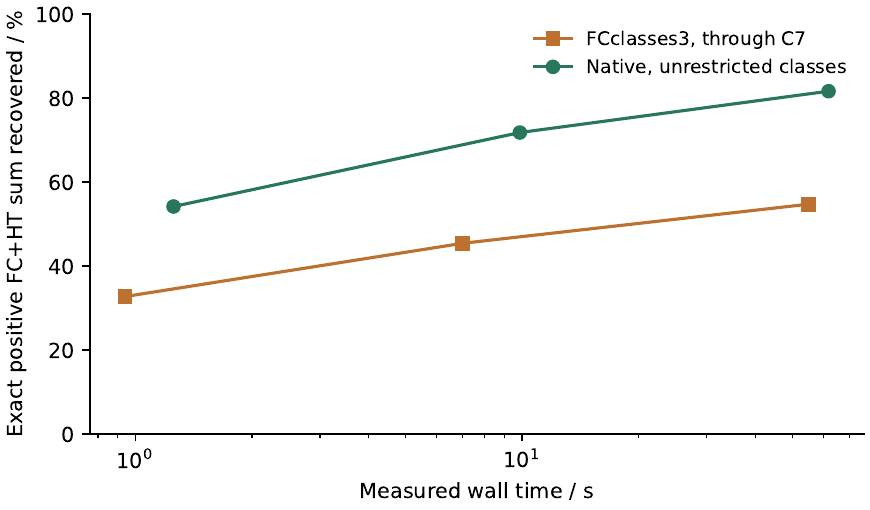}
\caption{TI time versus recovered positive FC+HT strength for harmonic $S_2$ absorption. Both programs use the same physical model and the stated class and occupation limits. Panel (b) resolves the low-coverage range. Settings and the distinction between incremental and complete-run times are specified in the text.}
\label{fig:s2-ti-cost-coverage}
\end{figure}

TI selection requires a separate cost--coverage assessment. A three-run
control requires both programs to recover at least 54\% of the same exact
$S_2$ FC+HT strength sum, with class C7 and a maximum final occupation of
30 per mode imposed in both programs. The native selector reaches
54.272\% in a median 3.822~s; FCclasses3 reaches 54.747\% in 69.214~s,
a measured factor of 18.11. Both use one thread and include setup,
assignments, a 3001-point Gaussian profile and export; the native time also
includes Python startup and a final checkpoint. FCclasses3 uses a budget
of $10^7$ integrals per class; the native selector evaluates 102400
transitions. This matches the class restriction and minimum recovered
strength, with the actual overshoot reported, rather than identical
transition sets. The unrestricted-class native control gave 22.50
under the same minimum-strength criterion; the C7 control shows that
removing the class restriction does not account for the observed advantage.
Figure~\ref{fig:s2-ti-cost-coverage} places this low-coverage ratio alongside
the available finite-budget curves. The largest plotted native calculation
recovers 81.656\%, while the largest FCclasses3 point recovers 54.747\%.
The separate 50-million-transition native expansion reaches 91.370\%.
No matched high-coverage FCclasses3 point is available, so the 18.11 ratio
does not establish its scaling towards 99\%. The SI supplies all budgets,
normalization conventions and per-process timings.

Measured time and recovered exact positive FC+HT strength for the
harmonic $S_2$ model. These archived finite-budget observations are separate
from the repeated 54\% control in Table~\ref{tab:external-harmonic-timing}.
Native costs are cumulative through resumed stages, including checkpoints
and retained-line export but excluding Python imports; FCclasses3 costs
include a complete process at each budget. Native execution uses one thread;
FCclasses3 retains a four-thread limit and serial main loop, with classes
through C7. This is a separate archived campaign from the repeated
one-thread FCclasses3 control in Table~\ref{tab:external-harmonic-timing}. Lines guide the eye between measured points; no time to 99\% or
high-coverage speed ratio is extrapolated.

For anharmonic nuclear models, relevant comparisons include same-field VPT2
and variational calculations, as well as semiclassical wavepacket approaches.
The on-the-fly extended thawed Gaussian method incorporates linear HT terms
and samples the potential locally along a classical trajectory. Patoz,
Begu\v{s}i\'{c} and Van\'{i}\v{c}ek demonstrated substantially improved agreement
with experimental phenyl-radical and benzene absorption spectra compared with
global harmonic models.\cite{Patoz2018ETGA} The propagated packet uses a time-dependent local quadratic potential,
providing a semiclassical approximation distinct from a variational
nuclear-state expansion. Later developments include finite-temperature extended
thawed Gaussians with HT terms,\cite{Begusic2020Thermal} tests on double-well
surfaces,\cite{Begusic2022DoubleWell} and local harmonic Hagedorn propagation
of non-Gaussian single-vibronic-level fluorescence states.\cite{Zhang2026Hagedorn}
The latter application uses the Condon approximation. These trajectory-based
models and the present finite-space VCI address anharmonicity with different
approximations; no performance ranking is inferred without matched executions.
Selected vibrational CI is also an active independent development: the ViBra
preprint combines VSCF modals with Epstein--Nesbet configuration screening
for quartic fields.\cite{ViBra2026} It is a methodological reference for
vibrational selection, rather than a two-surface vibronic benchmark executed here. Consequently, harmonic
FCclasses3/l718 comparisons and anharmonic VCI/VPT2 or thawed-Gaussian
comparisons test different levels of the present hierarchy. The executed
l718 and FCclasses3 checks support the specified harmonic models. The matched-step/window molecular TD and practically complete water TI comparisons
show a measured harmonic speed advantage in the tested cases. The common-threshold $S_2$
TI comparison additionally establishes a selector advantage at the tested
minimum recovery; the finite curves document the wider tradeoff. These results do not
establish a universal performance ranking across molecules and observables.
For the anharmonic applications, same-Hamiltonian accuracy and convergence
are the relevant controls; no equivalent FCclasses3 execution is timed.

The archived 90-mode molecular TI completeness calculation took approximately
217~s on an Apple arm64 host with two amplitude workers, excluding the full
transition-table export. This legacy observation is separate from the
resource-matched AMD EPYC executable benchmarks. The camphor directed
continuation took 19.83~s at one thread on the latter server, including
checkpoint restoration and reporting.

\FloatBarrier
\section{Spectral validation in optimized common modes}
\label{sec:commonmodespectralcheck}

\subsection{Matched harmonic water model}
The two descriptions use identical archived rDSD/jun-cc-pVTZ ESALC/GF fields,
the complete nonorthogonal coordinate map and the same complex continuum
transition operator. The operator bank consists of $1,q_1,q_2,q_3$ in the
initial dimensionless normal coordinates. Exact Hermitian channel Gram forms
retain FC, HT and signed interference separately at all five electron kinetic
energies. The common-mode calculation transports both quadratic potential
and kinetic matrices, displacements and electronic-property expansion points.
No intensity normalization, polar projection or frequency replacement is used.

The complete quadratic FC correlation agrees with the separate-state
Duschinsky result within $5.2\times10^{-12}$ at all 4097 time points,
spaced by 0.25~fs. Independent finite variational calculations use the same
oscillator products for both electronic states. Table~\ref{tab:commonwater}
compares their first 40 accepting levels with the exact harmonic spectrum.

\begin{table}[!htbp]
\centering
\caption{Harmonic common-mode VCI against the exact separate-state spectrum.
Intensity errors are maximum absolute errors divided by the corresponding
complete FCHT sum rule, including all four operator components and five
electron energies. Position errors cover the first 40 accepting levels.}
\label{tab:commonwater}
\begin{tabular}{rrrr}
\toprule
Total-quanta cap & Basis size & Position error (cm$^{-1}$) & Intensity error\\
\midrule
10 & 286 & 52.465 & $1.25\times10^{-5}$\\
14 & 680 & 0.7711 & $3.62\times10^{-11}$\\
18 & 1330 & 0.000445 & $1.68\times10^{-14}$\\
\bottomrule
\end{tabular}
\end{table}

The 40 retained lines recover 99.9953--99.9958\% of the FCHT strength.
Their reconstructed correlation differs from the complete independent TD
reference by at most $4.67\times10^{-5}$ of the sum rule, consistent with
the omitted lines. Reconstructing a correlation from finite VCI lines is
not an independent propagation test. The exact common-quadratic check
and the independently evaluated standard TI/TD comparison supply the
additional harmonic controls. Absolute energies include the zero-point
contribution once, through the Hamiltonian eigenvalues; the supplied scalar
gap is the electronic gap, rather than the already corrected band origin.

\subsection{Native anharmonic reference and molecular preparation}
The native common-mode reference evaluates the frozen water ZAFF functions
without a Taylor reexpansion and retains the complete curvilinear kinetic
energy. An affine common-coordinate box corresponds to a parallelepiped in
the original bond--angle coordinate representation. Boundary convergence must therefore be
established independently of basis convergence. In particular, an isotropic
common-mode box can include very short O--H distances at its corners and
make the variational problem stiff. The selected anharmonic extension is tested separately in
Sec.~\ref{sec:commonanharmonic}; perturbative dressing is outside those
variational tests.

For the planned 4-pyrone optical comparison, S$_1$ and S$_2$ geometries and
harmonic Hessians were computed with TD-B3LYP/6-31+G(d), targeting roots one
and two. Both final structures have 27 real frequencies. The lowest are
144.65 and 176.62~cm$^{-1}$, respectively. The S$_1$ structure is planar
within $3.6\times10^{-7}$~\AA\ for its heavy atoms. The initial S$_2$
stationary point had a $-144.56$~cm$^{-1}$ out-of-plane mode; following it
led to a nonplanar minimum with a maximum heavy-atom plane deviation of
0.0903~\AA. Geometry convergence and positive frequencies establish these
local nuclear references, not agreement of the two spectral formulations.
The S$_1$ checkpoint contains the target-state electric transition dipole
and its complete Cartesian derivative record. Phosphorescence additionally
requires a radiative operator induced by spin--orbit coupling; the scalar
coupling used in a nonradiative Golden Rule calculation does not supply it.

All numerical and electronic calculations were performed on the verified
remote compute host with preserved checkpoints. The general operator-bank
and affine-coordinate representation extensions pass 20 focused tests. Reproducible source,
input hashes and compact spectral evidence are archived in nano-MATRIX
commit \texttt{67b36d0a}, report
\texttt{common\_mode\_spectra\_20261027}, and benchmark
\texttt{vibronic\_4pyrone\_spectra}.

\FloatBarrier
\section{Anharmonic common-modal representation controls}
\label{sec:commonanharmonic}

For water, the common affine internal coordinate representation evaluates the unchanged accepted
ZAFF surfaces and the full coordinate-dependent curvilinear kinetic energy,
including the physical-volume connection. The expanded common-coordinate
intervals are $[-5.4,5.8]$, $[-4.2,6.0]$ and $[-4.8,4.8]$ in the stored
dimensionless coordinate representation. A full-product reference uses 24 sine functions and
52 Gaussian quadrature points per coordinate. Changing the basis/quadrature
from 20/44 to 24/52 changes the reported five-band position discrepancy
from 0.0102244 to 0.0102240~cm$^{-1}$. The relative FC+HT intensity
discrepancy against the separately converged bond--angle reference is below
$8.5\times10^{-6}$ at five electron energies.

Shared one-dimensional modals diagonalize harmonic box operators; selected
products of these functions are lifted to the primitive sine-DVR basis.
Application of the parent operator followed by the adjoint projection retains
every native potential and kinetic term in the declared finite space.
No PES is refitted or reexpanded. Both electronic surfaces use the same modals,
so nuclear overlaps and complex FC/HT banks contract directly.
Table~\ref{tab:common-anharmonic-water} compares this VCI contraction with
the identical full-product operator. Quadratic residuals are treated
variationally in this control; perturbative dressing requires a separate test.

\begin{table}[!htbp]
\centering
\caption{Selected common-modal water VCI versus its complete primitive-product
reference. Energy errors cover the first 12 accepting excitation energies;
intensity errors are maximum relative FC+HT errors on five assigned bands.}
\label{tab:common-anharmonic-water}
\begin{tabular}{rrrr}
\toprule
Modal excitation cap & Configurations & Energy error (cm$^{-1}$) & Intensity error\\
\midrule
8 & 165 & 135.224 & $4.22\times10^{-3}$\\
12 & 455 & 13.582 & $1.14\times10^{-4}$\\
16 & 969 & 1.131 & $3.94\times10^{-6}$\\
20 & 1771 & 0.1310 & $3.99\times10^{-7}$\\
23 & 2600 & 0.02015 & $6.29\times10^{-8}$\\
\bottomrule
\end{tabular}
\end{table}

\paragraph{Resolved operators and retained spectral profiles.}
A checkpoint-only audit reuses the 969-, 1771- and 2600-configuration
states and the identical primitive-product reference. It includes all 12
retained lines, at five electron kinetic energies. FC, HT and signed
interference are compared separately. No intensities are renormalized.
For the final 2600-configuration space, maximum line-strength errors,
divided by the retained positive FC+HT strength, are
$7.68\times10^{-9}$, $4.84\times10^{-10}$ and $2.74\times10^{-10}$
for FC, HT and interference, respectively. The corresponding FC+HT
error is $7.99\times10^{-9}$. The ground-to-ground energy error against
the complete common-product operator is $2.94\times10^{-8}$~cm$^{-1}$.

Gaussian profiles use $\sigma=50$~cm$^{-1}$ and a 1~cm$^{-1}$ grid over
$[-250,10000]$~cm$^{-1}$. Their integrated absolute differences, divided
by the retained positive FC+HT strength, are
$1.07\times10^{-7}$ (FC), $7.38\times10^{-9}$ (HT),
$3.02\times10^{-9}$ (interference) and $1.15\times10^{-7}$ (FC+HT)
against the complete common-product reference. The FC+HT differences
for caps 16, 20 and 23 decrease from $3.17\times10^{-5}$ to
$1.10\times10^{-6}$ and $1.15\times10^{-7}$.

Against the independent bond--angle calculation, the five assigned bands
agree within 0.010227~cm$^{-1}$ and $8.43\times10^{-6}$ relative FC+HT
strength. The 12-state FC+HT profile differs by at most
$6.65\times10^{-5}$ of its retained strength. The larger energy-ordered
manifold includes unassigned upper levels with differences up to
1.41~cm$^{-1}$ between coordinate domains. Their position accuracy is
therefore not assigned the tighter five-band validation. These tests
establish low-energy representation consistency and retained-profile
agreement, rather than completeness of an entire anharmonic spectrum.
Perturbative dressing is not used in these variational comparisons.

Anharmonic water FC+HT profiles at outgoing-electron energy
0.20~hartree, using 12 retained levels and Gaussian $\sigma=50$~cm$^{-1}$.
The bond--angle reference, complete common-product calculation and
2600-configuration common VCI are shown without intensity rescaling.
The lower panel displays profile differences divided by the maximum
bond--angle profile value. All profiles are referenced to their own
cation vibrational ground level; the separate absolute-origin check is
reported in the text. Upper unassigned levels are not given the tighter
five-band position validation.

\begin{figure}[!htb]
\centering
\includegraphics[width=.94\linewidth]{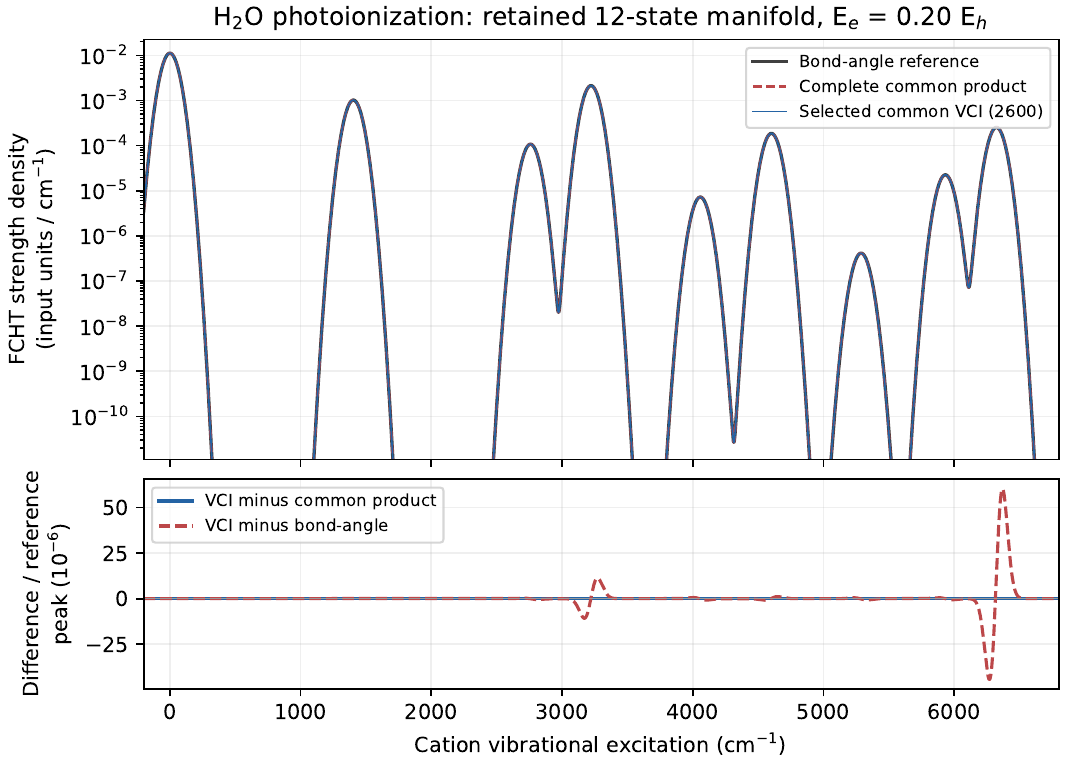}
\caption{Anharmonic water FC+HT profiles at electron energy 0.20~hartree. Selected common-mode states are compared with complete-product and bond--angle references, retaining the absolute FC, HT and signed interference contributions.}
\label{fig:si-water-common-profiles}
\end{figure}

Figure~\ref{fig:si-water-common-profiles} compares the selected common-mode and independent water spectral profiles.

Previous coupled anharmonic spectra must be distinguished from diagonal-only
models. Luis, Kirtman and Christiansen included Duschinsky rotations and
three-mode anharmonic couplings for furan.\cite{Luis2006CoupledFC}
Madsen and co-workers treated selected oligothiophene subspaces with coupled
two-mode PESs for both states in shared coordinates.\cite{Madsen2019Oligothiophene}
Their common-coordinate overlap formulation is not equivalent to discarding
quadratic mixing within that subspace. Molecular size alone therefore does
not support a priority claim; the number of coupled active modes and the
retained kinetic and property operators define the comparison.

Twenty-six focused tests on the compute server cover projected products,
complex modal embeddings, operator-bank contractions, affine curvilinear
transport and the two optional TI alternatives. No additional electronic point was
needed for these representation controls.

\section{4-pyrone optical models and numerical controls}
\label{sec:pyrone-optical-si}

All four optical models use B3LYP/6-31+G(d) fields and a single
curvilinear ESALC-I definition referenced to $S_0$. Each surface retains
its local Wilson metric and quadratic potential. Small residual
optimization gradients are relaxed by the local harmonic construction.
No unstable frequency is removed or reflected. The lowest frequencies
of the accepted $S_1$ and $S_2$ minima are 144.6487 and
176.6162~cm$^{-1}$, respectively. The first $S_2$ stationary point had
one imaginary frequency, $-144.5573$~cm$^{-1}$; its eigenvector has
out-of-plane Cartesian norm fraction 0.999999999997. A displacement
along this mode followed by optimization reaches the accepted nonplanar
minimum. The same electronic method and basis are used throughout.

Absorption properties are evaluated at the $S_0$ minimum by separate
target-state calculations; emission properties come from the appropriate
excited-state minimum. The property geometry and Hessian geometry are
therefore separately identified. Complete signed dipole derivatives are
transported before forming FC, HT and interference terms. These are
harmonic nuclear models with a linear transition operator; no anharmonic
optical spectrum or kinetic population model is implied.

The complete TD spectra use 4097 positive-time samples separated by
0.25~fs, a 1024~fs time window and Gaussian broadening with
$\sigma=50$~cm$^{-1}$. Independent refinement uses 8193 samples at
0.125~fs, preserving the time window and frequency resolution.
Table~\ref{tab:pyrone-optical-td} gives the largest change on the common
frequency grid, divided by the original peak density. The largest
refined-window tail ratio is $2.7\times10^{-22}$.

\begin{table}[!htbp]
\centering
\caption{Time-step refinement of complete 4-pyrone TD profiles.}
\label{tab:pyrone-optical-td}
\begin{tabular}{lr}
\toprule
Profile & Maximum density change / peak\\
\midrule
$S_0\to S_1$ absorption & $3.8\times10^{-16}$\\
$S_1\to S_0$ emission & $5.5\times10^{-16}$\\
$S_0\to S_2$ absorption & $5.2\times10^{-10}$\\
$S_2\to S_0$ emission & $3.1\times10^{-8}$\\
\bottomrule
\end{tabular}
\end{table}

Adaptive TI is controlled by the exact positive total dipole-strength
sum rule, without intensity renormalization or pruning based on a weak
parent amplitude. The $S_1$ absorption and emission expansions evaluate
143360 and 290816 transitions, respectively, to exceed 99\% completeness.
For $S_2$ absorption, expanding from 500000 to 5 million and 20 million
transitions increases completeness from 66.8779\% to 81.6559\% and
88.1086\%. The 20-million calculation resumes the stored selector,
overlap recurrence cache and retained transitions from the 5-million
checkpoint. Its additional TI stage takes 463.9~s on the Linux compute
server with eight amplitude workers, including checkpoint writes but
excluding subsequent compressed line-table export and correlation
comparison. Continuing the same checkpoint to 50 million transitions
raises completeness to 91.3698\%, with an omitted strength bound of
$3.39262\times10^{-4}$~a.u.$^2$. At 33 saved times the maximum
correlation difference is within that bound. This extension remains
below the 99\% target. These are stage timings for this workload.

At 33 saved times the finite 20-million-transition absorption correlation
differs from the independent complete TD correlation by at most
$4.67464\times10^{-4}$~a.u.$^2$, within the omitted positive strength
bound. This certifies consistency with the finite omission, rather than
99\% completeness. The initial $S_2$ emission TI expansion at 500000
transitions recovers 62.4177\%; the emission curve in the main text is
the complete TD result.

The harmonic common-mode comparison must retain both quadratic potential
and kinetic mixing and the same transition property. Exact common-mode
FC+HT correlations agree separately for the FC, HT and signed interference
contributions on all 4097 time samples. The largest component error divided
by total dipole strength is $1.68\times10^{-13}$ for absorption and
$9.92\times10^{-14}$ for emission. After the same Fourier transform and
broadening, the largest component density differences divided by the
complete spectral peak are $5.32\times10^{-14}$ and
$3.05\times10^{-14}$, respectively. Properties are transported to a common
frame centered between the two minima. Both states retain all quadratic
potential and kinetic mixing; no mixed coefficient is discarded.
The emission comparison uses the positive-photon-energy phase convention.
No finite occupation cutoff or finite-difference property is used.
Convergence of finite dressed-VCI wavefunctions remains a separate check.
The archived optical evidence stores
the TD arrays, source hashes, property expansion geometries, sum rules,
time-step refinements and TI omission bounds.

\subsection{Operator-aware two-mode transition ranking}
\label{sec:pyrone-block-selection}

A variant motivated by class-based prescreening was implemented and tested
against the existing conditional Gaussian selector. This is an internal
comparison of ranking strategies, not an execution benchmark against
FCclasses3 or the Dierksen--Grimme implementation. The variant evaluates
one- and two-mode number marginals of the complete linear-property
Gaussian state, including FC, HT and interference contributions. Analytic
spectator integration supplies reduced generating functions; padded
Cauchy/FFT coefficient extraction supplies the retained number blocks.
Exact operator-aware number covariances determine a maximum-correlation
tree. Its product of conditional blocks only ranks candidates; physical
line strengths are still evaluated with the shared overlap recurrence.
Neither strategy prunes a transition on the basis of a weak parent
amplitude. Both use the same exact positive total-strength sum rule,
without renormalization.

Table~\ref{tab:pyrone-block-selection} compares the two occupation-ranking strategies at fixed physical inputs.

\begin{table}[!htbp]
\centering
\caption{Matched $S_2$ absorption ranking comparison at fixed evaluated
transition counts. Completeness is relative to the exact FC+HT sum rule.}
\label{tab:pyrone-block-selection}
\begin{tabular}{rrr}
\toprule
Transitions & Existing guide (\%) & Operator-aware guide (\%)\\
\midrule
100000 & 54.2044 & 54.3677\\
500000 & 66.8779 & 67.0691\\
1000000 & 71.8156 & 72.0168\\
\bottomrule
\end{tabular}
\end{table}

For $S_2$, the gain at one million evaluated transitions is only
0.2012 percentage points. Total times through the three resumed stages
are 17.50 and 19.20~s for the existing and operator-aware guides,
respectively. For $S_1$, reaching 99\% requires 143360 transitions
with the existing guide and 157696 with the variant; the respective
completenesses are 99.00245\% and 99.00254\%, and times are
1.65 and 3.31~s. These single matched workload runs use the same stored
models and eight amplitude workers on the Linux compute server.
Times include selector preparation, recurrence evaluation, checkpoint
restoration where applicable and checkpoint writes; subsequent compressed
line export is excluded. They are not repeated timing benchmarks.
The variant therefore provides no substantial, systematic efficiency
improvement on these controls and is retained only as an optional
experimental strategy. The guide remains the default.
Thirty-four focused tests covering operator blocks, exact number moments,
line strengths, positive-norm operator-bank stopping and restart behavior
pass on the compute server.

\subsection{Exact prefix masses and loop-hafnian amplitudes}
The two alternatives use the identical stored 27-mode $S_2$ absorption model
at 0~K and the complete FC+HT operator. Neither modifies the production
checkpoints. For the prefix search, analytic spectator contraction leaves a
polynomial--Gaussian reduced density. Shared Hermite recurrences evaluate
its diagonal number coefficients. A prefix strength is an upper bound on
every descendant line strength, enabling best-first traversal without a weak
parent-amplitude test. This deterministic search is our adaptation of
conditional Gaussian-state sampling,\cite{QuesadaArrazola2020} rather than
a benchmark of the published sampler itself.

With a 45~s search budget, at most 1000 expanded nodes and a child cap of
12 quanta per mode, the search expands 468 nodes and makes 6085 mass
requests in 45.62~s. It emits one complete line, carrying 0.23554\% of the
total strength. The capped occupation tail is explicitly tracked;
retained lines, frontier and tail close the strength balance within
$1.13\times10^{-17}$~a.u.$^2$. Thus correctness does not translate into
an efficient selector for this broad progression.

The loop-hafnian experiment uses The Walrus 0.22.0, exact repeated
occupations and shared neighboring requests for the linear HT operator.
No weak matrix element is discarded. Table~\ref{tab:pyrone-hafnian}
compares it to the existing compiled, memoized recurrence on identical
candidate rows. Both kernels are warmed before measurement; each measured
recurrence run starts with a fresh cache. Three repetitions use one Numba
thread and one BLAS thread. The maximum absolute FC+HT line-strength
difference is $3.82\times10^{-21}$~a.u.$^2$. These are amplitude-bank
microbenchmarks, excluding transition selection and spectral transforms.
An initial oversubscribed diagnostic is excluded from the controlled timing
comparison and retained in the computational archive.

\begin{table}[!htbp]
\centering
\caption{Median amplitude-bank times on the same $S_2$ candidate rows.}
\label{tab:pyrone-hafnian}
\begin{tabular}{rrr}
\toprule
Lines & Shared recurrence (s) & Loop-hafnian (s)\\
\midrule
32 & 0.00354 & 0.00396\\
128 & 0.00384 & 0.01704\\
512 & 0.00547 & 0.07444\\
\bottomrule
\end{tabular}
\end{table}

At 512 lines the loop-hafnian implementation is approximately 14 times
slower. This result is specific to the tested implementation and low-total-
occupation batches, not a general complexity comparison. Together these
controls support retaining the existing selector and shared recurrence.

\subsection{Conditional anharmonic \texorpdfstring{$S_2$}{S2} Hamiltonian and low-band convergence}
\label{sec:pyrone-s2-conditional}

This additional model uses TDA-B3LYP/6-31+G(d), including the separately
computed TDA transition dipole and its Cartesian derivatives at the $S_0$
geometry. It is distinct from the TD-B3LYP harmonic models in the preceding
subsections. The acquisition is fixed at five analytic energy/gradient/
Hessian observations along the inversion path, at stored dimensionless
path parameters $t=1,-1,0,0.5,1.5$. All dense quadrature values are evaluations
of fitted analytic functions; basis refinement adds no electronic points.

Let $b\in\mathbb R^{26}$ denote the transverse coordinates. The potential is
\begin{equation}
 V(t,b)=V_{\rm inv}(t)-b_*(t)^{\mathsf T}H(t)b
                      +\tfrac12 b^{\mathsf T}H(t)b .
 \label{eq:pyrone-conditional-potential}
\end{equation}
Here the stored path parameter maps to the molecular inversion dihedral;
periodic functions are evaluated in that angle. $V_{\rm inv}$ is the
accepted periodic inversion fit. A periodic matrix-log interpolation keeps
$H$ positive, and a bounded periodic Shepard--Hermite response supplies
$b_*$. The linear force is $-H b_*$; no additional $b_*^{\mathsf T}Hb_*/2$
constant is added to the acquired scalar path potential. All 26 coordinates
and their mixed quadratic terms remain in the Hamiltonian.

The maximum excluded-inner-point transverse-Hessian component error of
the matrix-field fit is 4.8431~cm$^{-1}$ in the declared scaled coordinate representation.
The final mean-response fit imposes all five observed forces and mixed
inversion/transverse Hessians, with residuals below
$2.0\times10^{-13}$~cm$^{-1}$. Thus its agreement at these response nodes
is an interpolation constraint, not independent electronic validation.
At the minimum, reconstructing the complete harmonic problem with the
known inversion curvature gives a maximum frequency discrepancy of
0.000264~cm$^{-1}$. Unmeasured second response jets use a zero local
Hermite completion; they are not measured third- or fourth-order derivatives.

The kinetic operator retains the complete affine 27-coordinate/rotation
metric, its inversion dependence and mixed kinetic blocks, evaluated at
$b=0$. Its dependence on transverse displacement and the corresponding
transverse volume derivatives are omitted. Transverse self cubic and
quartic terms are also absent. The calculation is therefore a conditional
anharmonic model, rather than a complete 27-mode anharmonic force field.
The transverse zero-point contribution is generated by the Hamiltonian;
it is not added separately.

\paragraph{Basis hierarchy and retained couplings.}
The inversion coordinate uses a sine basis on $[-3.5,3.5]$, with 96
primitive functions and 256 Gaussian quadrature points. The zero-occupation
transverse Hamiltonian defines contracted inversion functions. The
transverse oscillator reference diagonalizes the full planar quadratic
potential/kinetic pair. Every one-quantum exterior intermediate required
by quadratic coordinate and momentum products is retained before
restriction to the variational space.

A full total-occupation-four space contains 27405 transverse configurations.
Subsequent spaces preserve all these configurations and add occupations
$n_i$ satisfying both a total-quanta ceiling and
\begin{equation}
 \sum_i w_i n_i\le B,\qquad
 w_i=\left\lceil\omega_i/(250~\mathrm{cm}^{-1})\right\rceil .
 \label{eq:pyrone-weighted-basis}
\end{equation}
In stored transverse-mode order, the weights are 1 for mode 1;
2 for modes 2--3; 3 for modes 4--10; 4 for modes 11--12;
5 for modes 13--15; 6 for modes 16--20; 7 for mode 21;
10 for mode 22; and 14 for modes 23--26.
The budgets $B=12$ and 16 are conservative nominal oscillator-energy
budgets of 3000 and 4000~cm$^{-1}$. Every added space is downward closed.
This basis selection removes no coordinate or Hamiltonian coupling.

Table~\ref{tab:pyrone-s2-basis-controls} summarizes independent basis, quadrature and nuclear-domain controls for the four low bands.

\begin{table}[!htbp]
\centering
\caption{Successive numerical controls for four low conditional-model
absorption bands. Position changes are absolute line changes, including the
origin. Strength changes are the largest relative FC+HT change on these
four lines. Overlap matching preserves state identity.}
\label{tab:pyrone-s2-basis-controls}
\begin{tabular}{lrr}
\toprule
Control & Position change (cm$^{-1}$) & Strength change (\%)\\
\midrule
Primitive/quadrature: 96/256 $\to$ 128/320 & 0.000080 & 0.000237\\
Contracted inversion: 24 $\to$ 32 & 0.010271 & 0.005003\\
Full occupation 3 $\to$ 4 & 5.886646 & 1.729971\\
Full 4 $\to$ weighted 5, $B=12$ & 0.054600 & 0.066487\\
Weighted 5 $\to$ 6, $B=12$ & 0.000850 & 0.000318\\
Weighted 6: $B=12\to16$ & 0.090638 & 0.054318\\
\bottomrule
\end{tabular}
\end{table}

Primitive/quadrature refinement uses 24 contracted functions and total
occupation three. The contraction control uses full occupation four; the
weighted controls use 24 contracted functions. The tested weighted spaces
contain 28409, 28914 and 43298 transverse configurations, respectively.
The final product dimension is 1039152. The initial full-three to full-four
change fails the 1~cm$^{-1}$/1\% criteria; the subsequent nested enlargements
pass both. The last matched-state overlaps exceed 0.9999865. These controls
support stability of the four low lines in the tested basis family; they
do not provide a rigorous error bound for every omitted occupation or for
the electronic potential.

\paragraph{Optical source and independent amplitude contraction.}
The initial state is the harmonic $S_0$ vacuum. Signed TDA dipoles and
linear HT derivatives are transported through a fixed planar ESALC tangent
embedding. This optical map is not a nonlinear, volume-corrected
all-coordinate transport of nuclear wavefunctions. The mixed inversion-
coordinate/transverse-Fock source is the analytic partial inverse of the
resident Gaussian generating function; all retained transverse words use
the shared TI recurrence. Projection onto the coupled eigenvectors gives
the line amplitudes directly, without separate intersurface normal-mode
overlap matrices.

An independent contraction first evaluates complete Gaussian TI
coefficients and then projects a one-dimensional harmonic-oscillator
expansion onto the inversion functions. Oscillator degree caps 32 and 64
give maximum component-strength discrepancies of $5.37\times10^{-14}$ and
$3.74\times10^{-16}$ relative to the four-line FC+HT sum, respectively.
This checks FC, HT and signed interference at fixed saved nuclear states;
it is not a second independent nuclear eigensolve. The four lines carry
1.6783\% of the full initial-state optical sum, while the full finite-source
projection carries 63.616\%. Neither quantity is renormalized. Low-line
basis convergence must therefore be distinguished from 99\% closure of a
complete anharmonic progression.

\paragraph{Solver cost and numerical optimization.}
All numerical work runs on the verified Linux compute server with four
BLAS threads and a 128~GB address-space limit. The general coupled-bath
owner compresses separable operator factors by singular-value decomposition,
with a conservative full-operator Frobenius error bound. For the final
space, 742 factors become 108; the bound is
$5.60\times10^{-6}$~cm$^{-1}$. No individual matrix element is screened and
the potential is unchanged. Lifted saved eigenvectors initialize a
preconditioned block LOBPCG solve. Residuals are then checked against the
original uncompressed Hamiltonian and are below
$9.41\times10^{-7}$~cm$^{-1}$ for the four final states.
The last assembly/solve takes 544.39~s and the separate optical evaluation
214.50~s. These are workload timings, not exclusive-host benchmarks.
The full-fourth-shell calculation gives an additional solver check:
checkpoint and generic starts agree within $2.08\times10^{-10}$~cm$^{-1}$;
their observed times are 372.87 and 1721.36~s. Checkpoint reuse, sparse
transverse algebra and bounded factor compression reduce repeated work
while retaining explicit residual and strength checks.

\FloatBarrier
\section{Representation cost, band character and model validation}
\label{sec:model-validation}

\subsection{Matched models and the scope of a computational advantage}
\label{sec:matched-cost}

The saved 27-mode harmonic $S_2$ absorption and emission models are each
compared at identical fields, displacements, transition dipoles and derivatives,
temperature, 4097 time samples and Fourier conventions. Both FC and HT and
their signed interference are retained. All residual quadratic position and
momentum terms are included in the exact common-Hamiltonian limit.
Table~\ref{tab:matched-cost} reports contractions from the existing
checkpoints, without electronic calculations or nuclear diagonalizations.
One warmup of each kernel precedes five measurements, with execution order
alternated on successive repeats. Four BLAS threads are used; the standard
routine also uses its four amplitude workers. These settings are specified
rather than treated as identical internal parallel algorithms.

\begin{table}[!htbp]
\centering
\caption{Matched harmonic correlation contraction times on the Linux compute
server, in seconds. Medians and observed ranges cover five warmed runs.
Common-frame preparation is separate; Fourier transforms and export are
excluded from both contraction timings.}
\label{tab:matched-cost}
\begin{tabular}{lrrr}
\toprule
Profile & Standard & Common & Frame preparation\\
\midrule
Absorption median & 0.9586 & 0.8315 & 0.8289\\
Absorption range & 0.9460--0.9636 & 0.8282--0.8345 & ---\\
Emission median & 0.9583 & 0.8318 & 0.6495\\
Emission range & 0.9561--0.9672 & 0.8293--0.8388 & ---\\
\bottomrule
\end{tabular}
\end{table}

The maximum component-correlation difference divided by total strength is
$1.7\times10^{-13}$; the maximum component-profile difference divided by
the complete peak is $5.4\times10^{-14}$. Thus the comparison holds accuracy
fixed. The contraction saving is approximately 13\% in these runs, while
frame preparation costs more than the saving for one contraction. The result
supports reuse when multiple contractions share the same frame; it is not
a repeated exclusive-host benchmark or a general asymptotic speedup claim.

For anharmonic water, the accepted native potentials, full curvilinear kinetic
operator, property expansion and coordinate domain are unchanged between the
complete common-product operator and its selected-modal projection. The
complete space has $24^3=13824$ products; the final selected space has 2600,
or 18.81\% of that dimension. The maximum first-12 excitation error is
0.020152~cm$^{-1}$ and the maximum relative error for five assigned FC+HT
strengths is $6.29\times10^{-8}$ against that complete-product reference.
The independent bond--angle calculation checks the representation at
0.010227~cm$^{-1}$ and $8.43\times10^{-6}$ relative strength on those bands.
The latter comparison includes distinct coordinate-domain boundaries and
is not an identical finite-basis timing comparison.

The complete-product calculation records 531.21 and 298.13~s for the neutral and cation solves, with 7.10 and 6.83~s of preparation. The selected 2600-configuration stage takes 30.91~s after preparation of the complete-product operator, including the two-state treatment, projected residual checks and optical contraction. Evaluation of spectra from the stored eigenstates takes 0.3645~s. These timings distinguish nuclear-state construction, selection and spectral contraction. The matched quantitative comparison is the 5.32-fold reduction of basis dimension at the stated spectral error.

\subsection{Character of the four low accepting states}
\label{sec:s2-character}

Let $C_{a\nu}^{(k)}$ expand normalized accepting state $k$ in contracted
inversion functions $\chi_a$ and transverse Fock words $\nu$. The functions
$\chi_a$ diagonalize the zero-occupation transverse Hamiltonian; their indices
are not harmonic inversion quantum numbers. The following diagnostics are
computed directly from the final saved eigenvectors:
\begin{align}
 p_{a\nu}^{(k)}&=|C_{a\nu}^{(k)}|^2, &
 \langle N_b\rangle_k&=\sum_{a\nu}p_{a\nu}^{(k)}\sum_i\nu_i,\\
 \rho^{(k)}_{aa'}&=\sum_\nu C_{a\nu}^{(k)}C_{a'\nu}^{(k)*}, &
 P_k&=\operatorname{Tr}[(\rho^{(k)})^2].
\end{align}
No normalization or orthogonalization is applied during analysis.
Populations and product participation depend on the modal basis. The purity
measures inversion/transverse correlation for this partition and is invariant
to separate unitary rotations within the retained factors. It is a diagnostic
of nuclear correlation, not a population of electronic states.

Table~\ref{tab:s2-character} assigns the dominant inversion and transverse character of the four accepting states.

\begin{table}[!htbp]
\centering
\caption{Checkpoint character of the four low accepting states.
$0$ denotes the transverse vacuum; $e_i$ is one quantum in transverse
reference mode $i$. Product weights are probabilities, not optical fractions.}
\label{tab:s2-character}
\begin{tabular}{rlrrr}
\toprule
$k$ & Leading product $(a;\nu)$ & Weight (\%) & $\langle N_b\rangle$ & $P_k$\\
\midrule
0 & $(0;0)$ & 95.42 & 0.05239 & 0.98183\\
1 & $(1;0)$ & 94.86 & 0.06760 & 0.95087\\
2 & $(0;e_1)$ & 95.44 & 1.05191 & 0.98140\\
3 & $(0;e_2)$ & 95.37 & 1.05041 & 0.98087\\
\bottomrule
\end{tabular}
\end{table}

Reference modes 1 and 2 have frequencies 204.216 and 381.113~cm$^{-1}$.
Cartesian displacements obtained from their kinetic/force normal modes and
the stored affine coordinate representation have mass-weighted out-of-plane fractions exceeding
0.999999999999 relative to the planar molecular frame. Their coupled
excitation intervals are 210.405 and 396.433~cm$^{-1}$. These differences
include the inversion-dependent field and kinetic couplings; they are not
isolated diagonal anharmonic frequency corrections.

Inversion reflection alone gives expectation values 0.98661, $-0.95848$,
0.98570 and 0.98475 for states 0--3. State 1 therefore has predominantly odd
inversion character, consistent with its leading $(1;0)$ product. Reflection
of the inversion coordinate alone is not asserted to be the full molecular
symmetry operation on every transverse coordinate, and these values are
not assignments of exact molecular irreducible representations.
The dominant Schmidt probabilities are 0.99085, 0.97492, 0.99063 and 0.99036.
Thus the first inversion excitation has the largest inversion/transverse
correlation within this low manifold, while each state remains dominated
by one product.

The lowest optical line has FC+HT strength divided by the incoherent sum
$S_{\rm FC}+S_{\rm HT}$ equal to 0.159315: destructive interference removes
84.07\% of that incoherent strength. The three upper lines are dominated
by HT and their interference is negligible at the reported scale. Wavefunction
character does not alone determine brightness; the signed property operator
must be contracted with the states. This analysis takes 0.472~s on the compute
server using the saved eigenvectors, with no new nuclear Hamiltonian solve or electronic point.

\subsection{Numerical sensitivities and unquantified model errors}
\label{sec:error-budget}

Table~\ref{tab:error-budget} separates measured numerical sensitivities from
physical-model assumptions for the $S_2$ low-band calculation. Entries are
not combined in quadrature: they are neither independent random variables
nor measured statistical standard deviations. A final basis change does
not bound all omitted configurations, and a fit residual at an acquired
node is not an independent validation error.

\begin{table}[!htbp]
\centering
\caption{Source-resolved assessment of the conditional $S_2$ calculation.
Position and strength changes refer only to four low bands.}
\label{tab:error-budget}
\begin{tabular}{p{.22\linewidth}p{.68\linewidth}}
\toprule
Source & Available control and scope\\
\midrule
Primitive/quadrature & Observed changes below $8.0\times10^{-5}$~cm$^{-1}$ and
$2.37\times10^{-4}$\% strength at the tested reference occupation space.\\
Inversion contraction & 24 to 32 functions: below 0.0103~cm$^{-1}$ and 0.00501\%
strength at full transverse occupation four.\\
Inversion domain & Expanding $[-3.5,3.5]$ to $[-4,4]$ at the final occupation
space changes positions by at most $6.56\times10^{-4}$~cm$^{-1}$ and strengths
by 0.00064\%; the potential and primitive/quadrature counts are fixed.\\
Transverse selection & Last nested enlargement: below 0.0907~cm$^{-1}$ and
0.0544\% strength; not a rigorous bound on every omitted word.\\
Operator compression and solve & Conservative compression bound
$5.60\times10^{-6}$~cm$^{-1}$; original-Hamiltonian eigenvector residuals
below $9.41\times10^{-7}$~cm$^{-1}$.\\
Amplitude contraction & Independent Gaussian-source contractions agree within
$3.74\times10^{-16}$ of the positive retained FC+HT sum.\\
Potential interpolation & Excluded-inner-point transverse-Hessian component
error 4.8431~cm$^{-1}$ in the scaled coordinate representation; final mean-response jets use all
observations. No propagated spectral error or unsampled-domain bound.\\
Conditional field and metric & Missing transverse self cubic/quartic terms and
transverse metric/volume dependence have no quantified spectral bound.\\
Optical embedding/operator & Fixed tangent transport. A five-point periodic
curvature sensitivity changes the cancelled lowest strength by a factor
2.06--2.14; upper strengths change by $<0.27$\%. No global nonlinear
property or nonlinear transport is validated.\\
Electronic surfaces & TDA-B3LYP conditional and TD-B3LYP harmonic models differ
in electronic approximation; subtraction does not isolate anharmonicity.\\
Electronic-state coupling & Separate adiabatic surfaces; no coupled-surface
nuclear calculation or validated nonadiabatic coupling field for $S_2$.\\
Spectral coverage & Four lines carry 1.6783\% of the full initial optical sum;
low-line stability is distinct from full progression completeness.\\
\bottomrule
\end{tabular}
\end{table}

\subsection{Operational model-selection and acceptance protocol}
\label{sec:model-selection-protocol}

\begin{enumerate}
\item Freeze the electronic method, geometries, coordinate representation, masses,
metric, potential and property-expansion reference. Preserve phases and units.
Check stationary-point stability or explicitly define any large-amplitude
coordinate; reflecting an imaginary frequency is not an accepted repair.
\item Establish the harmonic reference with independent TI and TD contractions.
Track the exact positive FC+HT sum, retain assignments and report any remaining
omission. The demonstrated 99\% harmonic stopping target applies to that
complete harmonic sum, not automatically to a later anharmonic window.
\item Use VPT2 for a local nonresonant target only after checking denominators,
wavefunction correction norms and observable stability. There is no universal
correction-norm threshold guaranteeing intensity accuracy. Qualify sensitive
weak bands against a variational calculation on the same local field; changing
the field at the same time cannot isolate the approximation error.
\item Use direct DVR/VCI when a periodic or large-amplitude motion, an active
resonance or an unstable perturbative intensity requires coupled states.
Define the finite electronic acquisition and analytic potential continuation
before refining the numerical basis. Validate that continuation separately.
\item For common modes, transform both surfaces, kinetic mixing and the full
transition property consistently. Verify the complete quadratic limit against
the standard representation, then compare selected anharmonic products with
their unchanged parent Hamiltonian. Treat resonant residual terms variationally;
accept perturbative dressing only after the target energies and amplitudes are
checked against enlargement of the variational space. Different states may
require different variational blocks. The present anharmonic controls use
variational residuals and do not establish a perturbative partition.
\item Refine numerical walls, primitive/quadrature resolution, contraction
and occupation/energy cutoffs separately where possible. Match states by
overlap and monitor absolute origins, intervals, FC, HT and signed interference.
For the present four-line window the adopted observed-change criteria are
1~cm$^{-1}$ and 1\% FC+HT strength. Require successive occupation enlargements;
retain original-operator residual checks and the independent amplitude test.
\item Stop within the declared spectral window when these controls pass.
Report coverage and the remaining model assumptions explicitly. Extend the
nuclear model or acquire an additional electronic dataset only if
an independent model-control discrepancy warrants it; numerical convergence
alone does not establish predictive accuracy or nonadiabatic validity.
\end{enumerate}

\FloatBarrier
\section{Experimental organic absorption and fluorescence comparisons}
\label{sec:organic-experiment-protocol}
The four experimental arrays are archived with the source URLs and SHA256
hashes in \texttt{organic\_experiment\_20261101}. They are the tabulated
arrays in the primary PhotochemCAD pages for anthracene (C02) and
p-terphenyl (B02), both measured in cyclohexane.\cite{PhotochemCAD2018}
The records identify R.-C.~A. Fuh, a Cary~3 absorption spectrometer and a
Spex FluoroMax emission spectrometer, with measurement dates 26 June 1995
and 13 June 1995 for anthracene and p-terphenyl, respectively.

Thermal spectra are evaluated from the two supplied minimum force
fields using standalone SATI on the compute server. The electronic
methods are PBE0/6-31G(d) for anthracene and PBE0/6-31G(d,p) for p-terphenyl.
The coordinate builder constructs the same curvilinear coordinate representation for both
surfaces, and the local quadratic relaxation retains residual gradients.
The calculation keeps all 66 or 90 modes. The optical-property expansion
is taken at the ground-state geometry for absorption and the excited-state
minimum for emission; each panel identifies whether complete HT derivatives
are included. The calculation summaries preserve every checkpoint hash and
operator convention.

Each native C++ execution uses one BLAS thread, 300~K, Gaussian standard
deviation 100~cm$^{-1}$ and a 1024~fs correlation window. Independent
0.25 and 0.125~fs grids contain 4097 and 8193 samples, respectively. The
exported wavenumber grid spans 14000--52000~cm$^{-1}$ in 2~cm$^{-1}$ steps.
Completed process outputs are reused when resuming the driver.

For absorption the comparison evaluates
$A(\lambda)\propto\widetilde\nu S(\widetilde\nu)$ with
$\widetilde\nu=10^7/\lambda$ for wavelength in nm.
For photon fluorescence the comparison evaluates
$F(\lambda)\propto\widetilde\nu^3 S(\widetilde\nu)
|d\widetilde\nu/d\lambda|$, with Jacobian
$10^7/\lambda^2$. Each curve is peak-normalized within the displayed
window. The raw and translated profiles are both retained in the numerical
archive. One rigid wavenumber translation per molecule aligns the lowest
resolved absorption band with the calculated origin for anthracene, and
the absorption-envelope maxima for p-terphenyl. This same translation is
used for emission. No vibrational-frequency scaling or separate fluorescence
alignment is applied. The archived comparison reports raw maxima, absorption
anchors, translations, prominent peaks and raw and translated profile errors.

Table~\ref{tab:organic-experiment-si} reports the shared absorption calibration and the resulting absorption/emission profile errors.

\begin{table}[!htbp]
\centering
\caption{Experimental organic spectra: wavelengths of the intensity maximum (nm) and RMS profile differences after peak normalization in the stated windows. Raw values retain the electronic energy origins. Translated values use one absorption-derived shift per molecule, shared with emission.}
\label{tab:organic-experiment-si}
\begin{tabular}{lrrrr}
\toprule
Molecule / spectrum & Measured & Raw & Translated & RMS\\
\midrule
anthracene / absorption & 356.25 & 381.74 & 377.00 & 0.127 \\
anthracene / emission & 396.50 & 405.65 & 377.50 & 0.183 \\
p-terphenyl / absorption & 276.25 & 292.77 & 276.25 & 0.030 \\
p-terphenyl / emission & 337.50 & 381.77 & 355.00 & 0.388 \\
\bottomrule
\end{tabular}
\end{table}

The translations are 1852.77 and 2043.10~cm$^{-1}$ for anthracene and
p-terphenyl, respectively. The absorption anchors are 376.25 and 276.25~nm.
The experimental anthracene absorption maxima are 323.25, 339.25, 356.25
and 376.25~nm; the archived peak lists retain the computed fine structure.
The experimental envelope Stokes shifts are 2849.49 and 6569.47~cm$^{-1}$;
the raw calculated values are 1544 and 7962~cm$^{-1}$, respectively.
The analysis evaluates maxima on the declared grids and keeps the full
experimental and calculated profiles in addition to these summary quantities.

\begin{figure}[!htb]
\centering
\includegraphics[width=.98\textwidth]{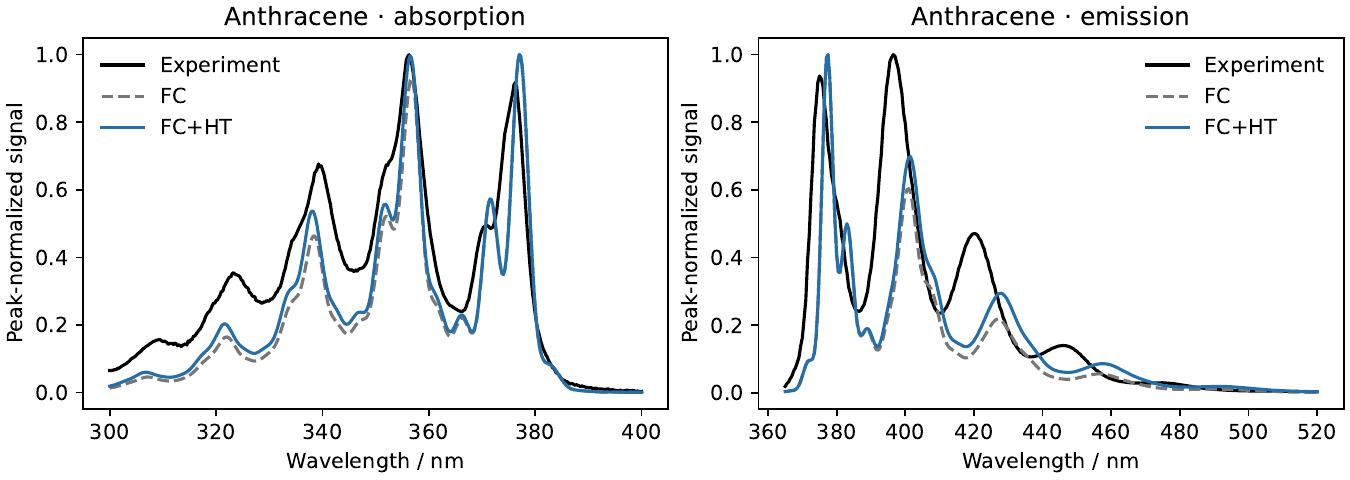}
\caption{Experimental anthracene absorption and fluorescence compared with FC and FC+HT calculations on the same nuclear Hamiltonians. All curves use the single absorption-origin translation and peak normalization. The HT derivatives are supplied by two analytic Gaussian calculations.}
\label{fig:anthracene-fcht-experiment-si}
\end{figure}

\begin{samepage}
The two Gaussian property calculations use the route
\begin{quote}\small\ttfamily
PBE1PBE/6-31G(d) TD=(Singlets,Root=1) Freq
\end{quote}
with 32 threads and 128~GB each, at the existing neutral anthracene geometries.
Both terminate normally. The property files preserve root~1 and transition
dipoles within $3.2\times10^{-5}$ relative to the supplied values, and contain
complete Cartesian derivative records. The same nuclear fields are retained
for the FC and FC+HT comparison in Fig.~\ref{fig:anthracene-fcht-experiment-si}.
\end{samepage}

\section{Spectral shape metrics and coordinate illustration}
\label{sec:shape-metrics}
The additional shape measures operate on the archived plotted profiles,
retaining their wavelength windows, photon-frequency factors and the
single absorption-derived energy translation. For nonnegative profiles
$I_a(\lambda)$ and $I_b(\lambda)$, define
\begin{align}
 P(\lambda)&=\frac{I_a(\lambda)}{\int I_a(\lambda)\,d\lambda},&
 Q(\lambda)&=\frac{I_b(\lambda)}{\int I_b(\lambda)\,d\lambda},\\
 \mathcal O&=\int\min(P,Q)\,d\lambda,&
 W_1&=\int|F_P-F_Q|\,d\lambda.
\end{align}
$\mathcal O$ lies between zero and one, and $W_1$ is in nm here. Both
normalization areas are stored. These are window- and coordinate-dependent
shape comparisons, with no new energy translation or fitted intensity
scale. Signed ECD/interference profiles are excluded from this probability-
density construction and retain their signed TD-referenced error measures.

The implementation integrates piecewise-linear densities. Overlap includes
every linear crossing exactly; the difference of cumulative areas is a
piecewise quadratic, integrated in absolute value after splitting at its
real roots. Four regressions check amplitude and axis-unit transformations,
translated triangular densities on nonuniform grids, exact interpolation
refinement, symmetry and rejection of signed/nonfinite/zero profiles.
All four pass on the compute server. The general function and tests,
input profiles, output areas and coordinate metadata are included in
\texttt{editorial\_focus\_20261102/}.

Table~\ref{tab:si-shape-metrics} compares shared-area and intensity-transport measures on the stated experimental windows.

\begin{table}[!htbp]
\centering
\caption{Shape comparisons with the archived molecular measurements.
Anthracene absorption uses 300--400~nm, emission 365--520~nm; p-terphenyl
absorption uses 240--320~nm and emission 300--410~nm. All anthracene profiles and
p-terphenyl harmonic absorption use 300~K; the matched p-terphenyl
fluorescence hierarchy uses 298.15~K. Wavelength-profile normalization
is used consistently within each comparison.}
\label{tab:si-shape-metrics}
\begin{tabular}{lrr}
\toprule
Profile/model & $\mathcal O$ & $W_1$ (nm)\\
\midrule
Anthracene absorption, FC & 0.85454 & 5.49921\\
Anthracene absorption, FC+HT & 0.87468 & 4.09518\\
Anthracene emission, FC & 0.74520 & 2.15697\\
Anthracene emission, FC+HT & 0.74376 & 4.29735\\
p-Terphenyl absorption, harmonic & 0.97661 & 0.36482\\
p-Terphenyl emission, harmonic & 0.70625 & 16.34695\\
p-Terphenyl emission, coupled DVR & 0.77293 & 12.61670\\
p-Terphenyl emission, bath response & 0.80631 & 10.68239\\
p-Terphenyl emission, displaced overlaps & 0.84336 & 8.64323\\
\bottomrule
\end{tabular}
\end{table}

The coordinate figure uses the original p-terphenyl coordinate representation built from its
archived reference geometry. Its bridge bonds are 1--7 and 10--13, each
with four proper-dihedral primitives. Four retained torsion-family rows
contain the plotted coefficients $\pm0.5$; all their nonzero terms are
shown. Their definitions and atom-index supports are stored verbatim, along
with the reference-file hash. The full coordinate representation retains both rigid bridge-
rotation directions. The plot visualizes the molecular support and family
identity of this block; spectral recovery remains the independent selector
comparison reported in the main text.

\section{Coordinate and spectral-model controls}
\label{sec:review-controls}
The numerical contractions use the compute server identified in the timing records. The archive
\texttt{review\_controls\_20261101} contains the exact JSON records, drivers,
prepared physical inputs, raw execution archives and numerical profiles.

\subsection{Identical-selector coordinate comparison}
The p-terphenyl comparison retains the supplied two minimum fields and the
same constant FC operator, independently of the FC+HT experimental profiles.
The temperature is 0~K. All coordinate representations use the same native C++ adaptive Gaussian-C2
proposal, exact positive-strength stopping, 4096-candidate batches, 100000
transitions, two million cached-integral budget, and a common requested
fraction 0.999999. The reported fractions stop at the transition budget;
none is renormalized. The per-mode occupation bound is 128 in each coordinate representation.
The TD profiles use 1024~fs, a 0.25~fs step and Gaussian
$\sigma=100$~cm$^{-1}$ on a common 14000--52000~cm$^{-1}$ output grid.

ESALC-I and the full-pool SVD reduction use the same nonlinear valence
primitive inventory and retain coordinate-curvature contributions when
transforming the Hessians. The Cartesian comparator separately projects
translations and rotations at each minimum after proper mass-weighted
Eckart alignment. Its map is the contraction of the two mass-weighted normal
frames, and its displacement is the Cartesian minimum separation projected
on the final frame. Residual gradients are relaxed within each harmonic
model. A second calculation with the initial Cartesian tangent held fixed
is retained as a separate diagnostic, not identified with the state-specific
Cartesian result.

Table~\ref{tab:si-coordinate-support} separates coordinate support and primitive-family composition from the intensity recovered at fixed transition budget.

\begin{table}[!htbp]\centering
\caption{Fixed-budget FC recovery and coordinate structure for p-terphenyl.
Atom support counts primitive coefficients larger than $10^{-8}$; family
counts use the same convention. These structural measures are independent
of intensity recovery.}
\label{tab:si-coordinate-support}
\begin{tabular}{lrrr}\toprule
Coordinate representation & Recovery (\%) & Mean atom support & Mean families\\\midrule
ESALC-I & 72.9447 & 9.2667 & 1\\
Delocalized valence & 73.4469 & 32 & 3\\
Cartesian Eckart & 0.0207182 & --- & ---\\\bottomrule
\end{tabular}
\end{table}
The delocalized TD profile differs from ESALC-I by 0.00788465 of its peak.
The Cartesian local model differs by 0.99987673 of that peak in the stated
spectral window. The latter comparison changes the finite-displacement
harmonic approximation; it is not an orthogonal rebasing or a comparison
against a measured spectrum. Coordinate covariance is tested separately
by rebasing a fixed nonlinear coordinate representation.

\subsection{Sensitivity of the experimental HT comparison}
For each anthracene spectrum, the finite analysis grid uses three edge
trims (0, 5 and 10~nm at both ends), three changes to the common
absorption-derived shift ($-100$, 0 and $+100$~cm$^{-1}$), and two
normalizations (unit peak or unit wavelength-integrated area), giving 18
variants. For each variant, FC and FC+HT use the identical experimental grid,
window, shift and normalization convention. HT decreases the RMS in all 18
variants for absorption and all 18 for emission. Within peak normalization,
the RMS gain ranges from 0.0166071 to 0.0227844 in absorption and from
0.00458459 to 0.0189084 in emission. The file
\texttt{metric-sensitivity.json} records every value. This finite-grid range
is not a statistical confidence interval and uses no experimental error model.

\subsection{Frequency-renormalized Gaussian water baseline}
The separate calculation retains the archived rectilinear normal axes,
minimum geometries, electronic gap, continuum amplitudes and physical
first derivatives. Gaussian anharmonic fundamentals replace the oscillator
frequencies: (1589.38936, 3650.90927, 3753.09516)~cm$^{-1}$ for water and
(1405.16157, 3223.56843, 3270.08622)~cm$^{-1}$ for its cation.
With unchanged dimensionful normal axes, the intersurface map is unchanged,
while $K_j$ is multiplied by $\sqrt{\nu'_j/\nu_j}$ on the final surface and
each initial dimensionless operator derivative by
$\sqrt{\nu_j/\nu'_j}$. This preserves the physical displacement and operator.
The original harmonic calculation is independently recovered to numerical
precision before evaluating the changed widths. No l718 calculation is
implied by this native baseline.

At 0.20~hartree electron energy, the 100 and 200 FC+HT strengths are
0.1149521602 and 0.0007667323. Their excitation energies are 1405.16157
and 2810.32314~cm$^{-1}$. The variational strengths on the original local
quartic fields are 0.0959981765 and 0.0006342903; strict wavefunction VPT2
gives 0.0993775569 and 0.0032579329. Thus the 200 intensity ratios are
1.2088 for the frequency-renormalized model and 5.1363 for strict VPT2.
These different approximations are kept separate in the main text.

\subsection{Application-sized TD timings}
The six comparisons use identical time grids, 0~K, Gaussian
$\sigma=100$~cm$^{-1}$ and equal spectral export counts in the two programs.
Their process order alternates over three repeats and their own timed
processes do not overlap. The server is shared. All six raw records retain
per-repeat times, binary hashes, physical-model hashes, exact strength sums,
profile errors and native half-step checks. Native closure errors are below
$10^{-12}$. At 512~fs, the largest interprogram profile residual is
0.009981 of the external peak for $S_1$, decreasing to 0.002124 at 1024~fs;
the corresponding 1024~fs residuals for $S_2$ and p-terphenyl are
0.00011670 and 0.00006940. These profile-export differences are stated
separately from the temporal half-step checks and the timing ratios.

\subsection{Molecular Gaussian comparison: 27-mode 4-pyrone}
The molecular control uses the B3LYP/6-31+G(d) $S_0$ and $S_1$
checkpoints, with every Cartesian vibrational mode retained. The frequency,
mode and property arrays are read from those same unmodified checkpoints
by both programs. A proper mass-weighted Eckart alignment is applied.
The comparison retains l718's final-state reference for the HT expansion
in absorption: the dipole and its derivatives are evaluated at the final
minimum, projected on its normal modes and re-expressed as the same linear
operator in initial dimensionless coordinates. All Cartesian derivative
columns are populated in the supplied records.

The electronic route is
\texttt{Freq=(FCHT,ReadFC,ReadFCHT) Geom=AllCheck}, with AH,
298.15~K, Gaussian HWHM 117.741002251547~cm$^{-1}$ and a relative
$[-1000,14000]$~cm$^{-1}$ window. Each job uses one thread and 128~GB.
The TI control uses \texttt{MaxC1=30}, \texttt{MaxC2=30}, and \texttt{MaxInt=1}. Gaussian interprets the last value in millions of
integrals.

The native common-recurrence engine independently evaluates the exact
initial/final quantum-number pairs printed by the completed $S_1$ TI job.
It checks 327 assignments using 982 cached Hermite integrals. The maximum
excitation-energy discrepancy is 0.000057~cm$^{-1}$. The maximum absolute
strength discrepancy divided by the largest printed strength is
0.000358943. A relative comparison excludes only the practically dark
0--0 line by requiring a printed strength above $10^{-6}$ of the largest;
for the remaining 326 lines its maximum is 0.00224081. The full assignment
list, unweighted native and external dipole strengths, and source hashes
are archived. This check is independent of which other transitions the
two selectors retain. Gaussian reports a final progression of 66.63\% for this
finite TI control. Its plotted finite progression is therefore compared
separately from the infinite-manifold native TD profile.

The two TD executions also include $S_2$. Their unshifted band origins
agree within $2.4\times10^{-5}$~cm$^{-1}$. The raw $S_1$ profile differs by
0.082727 of the external peak, with a native/external area ratio 0.999470;
the corresponding $S_2$ values are 0.069606 and 1.011002. The same native
profiles are unchanged within $6.2\times10^{-16}$ of their peaks on halving
the time step. No fit is applied to these raw comparisons.

A separate diagnostic inspects the installed l718 source and retains its
SHA256. In routine \texttt{TDOPDn}, the spectral reordering labels the
positive-frequency FFT index $k=i-1$ as $i+1/2$, and the negative branch
$k=i-1-N$ as $i-1/2-N$. The resulting axis offsets are $3\Delta\nu/2$
and $\Delta\nu/2$, respectively, with
$\Delta\nu=(cT)^{-1}=32.57461867$~cm$^{-1}$ for $T=1024$~fs.
Re-expressing those indices and their photon-frequency prefactor on the
corresponding Fourier axis uses zero fitted parameters. It reduces the
$S_1$ maximum profile discrepancy to 0.0027584; the $S_2$ discrepancy is
0.0520534. These source-index diagnostics are archived alongside the raw
spectra, not substituted for them. The $S_2$ TI execution automatically
terminates after C2 with 12.8\% progression and is not included among the
completed molecular TI validations.

\subsection{TI control with the same class restriction}
A second three-repeat $S_2$ FC+HT campaign imposes class C7 and a maximum
final occupation of 30 per mode in both selectors. The minimum recovery
is 54\% of the same exact positive strength sum, 0.003931097277643119.
One thread is used, including BLAS. Native times are 4.423630939,
2.568456972 and 3.822352337~s; FCclasses3 times are 83.131052471,
67.988756327 and 69.213648871~s. Their medians give a ratio of
18.10760568. Native recovery is 54.27163464\% with 102400 transitions;
FCclasses3 recovery is 54.74672319\%, using $10^7$ integrals per class.
Each process includes startup, selection, assignments, profile and export;
the native process includes its checkpoint. This campaign is distinct
from the unrestricted-class native control (22.50 ratio) and the archived
cost--coverage curves. The server is shared, and the per-repeat values are
retained in \texttt{S2-TI-C7.json} with binary and model hashes.

\subsection{Coupled torsions and emission-domain validation}
The finite p-terphenyl acquisition contains 44 completed analytic-force
calculations. The dataset contains 24 baseline points, two reference
checks, eight axial points, six mixed or near-gap $S_1$ points, and four
independent $S_1$ controls. Every calculation uses PBE0/6-31G(d,p),
with TD singlet root 1 for $S_1$, 32 threads and 128 GB. The bridge bonds
are 1--7 and 10--13 in the archived atom order. No electronic grid refinement
is used to converge the nuclear calculation.

The same-geometry reference checks give recomputed-minus-reference energies of
539.974 and 1722.272~cm$^{-1}$ for $S_0$ and $S_1$, respectively, and recomputed
torsional gradients of $(-82.242,82.182)$ and
$(-109.628,109.651)$~cm$^{-1}$ rad$^{-1}$. The torsional potentials
therefore use the coherent energy/gradient records, without imposing
the supplied Hessians. The latter supply the transverse harmonic bath
and the electronic gap, defining an explicitly composite nuclear model.

For $S_0$, a periodic Fourier field with axial order four and mixed order
one has 21 coefficients, three exact reference constraints and 18
identified free parameters. It fits 13 geometries and passes four
reserved controls with maximum energy and torque errors of 31.426~cm$^{-1}$
and 25.827~cm$^{-1}$ rad$^{-1}$. For $S_1$, the frozen local two-root
splitting prior plus periodic Hermite residual uses 23 training geometries,
kernel length 0.7 rad and regularization $10^{-8}$. Four newly acquired
controls remain excluded from fitting. The near-planar control has an
energy error of 8.807~cm$^{-1}$ and maximum torque error of
76.043~cm$^{-1}$ rad$^{-1}$. The maximum errors over all four controls
are 142.262~cm$^{-1}$ and 1234.589~cm$^{-1}$ rad$^{-1}$, so the
original global gates (50~cm$^{-1}$ and 250~cm$^{-1}$ rad$^{-1}$)
remain failed. The separately authorized validation concerns the
emission profile and its populated domain; it does not relabel this
global field test.

The explicit two-torsion Hamiltonian uses the Cartesian mass metric
of the actual rigid-fragment rotations. Translation is removed, and
the full internal/rotational metric supplies both its Schur-reduced
inverse kinetic tensor and its volume connection. A product sine basis
and oversampled Gaussian quadrature implement the resident curvilinear
weak-form solver. The two surfaces share absolute ring-plane angles;
the counterrotation from the $S_1$ to the $S_0$ reference angles is approximately $(28.4055,-28.4088)^\circ$.
Consequently their sine coefficients can be contracted on the same
physical interval, rather than overlapping two differently centered
coordinate boxes.

The two explicit rotor directions are removed from the common native
coordinate space. For a transverse embedding $S$, the bath uses
$F_b=S^TFS$ and $G_b=(S^TG^{-1}S)^{-1}$ on each surface. Its reference
shift compares the two geometries after rotating their terminal rings
to identical angles. This prevents a finite nonlinear rotor displacement
from being included a second time in the transverse bath shift.
All 88 transverse modes and their harmonic mode mixing are retained.
The rotor factor uses FC overlaps and the bath factor retains FC/HT;
the product correlation is broadened with a Gaussian standard deviation
of 100~cm$^{-1}$. The electronic gap and the previously determined
absorption calibration are retained, without fitting emission energies.

At 298.15~K, increasing the sine basis from 61 to 71 functions per torsion
changes the relative partition function from 29.905740 to 29.916335
(0.0354\%). The thermal coordinate probability at fitted potential values
above 2000~cm$^{-1}$ is $2.835\times10^{-5}$; that in the outermost
$10^\circ$ of the $\pm75^\circ$ box is $9.29\times10^{-8}$.
Smooth perturbations of $\pm500$~cm$^{-1}$, switched on between
1500 and 2500~cm$^{-1}$, change the first 24 levels by less than
$8\times10^{-6}$~cm$^{-1}$. These level and population checks are
supplemented by direct profile comparisons, including an expanded
$\pm85^\circ$ domain. All nuclear grids evaluate the unchanged analytic
fields, and every completed electronic record remains archived.

Table~\ref{tab:torsion-emission-validation} lists the basis and interpolation controls for the torsional emission model.

\begin{table}[!htbp]
\caption{Direct emission-profile validation. Differences refer to
unit-area dipole-strength densities on 14000--52000~cm$^{-1}$ at
298.15~K, with Gaussian standard deviation 100~cm$^{-1}$. The profile
column is the maximum absolute difference divided by the reference peak.
The fixed gates are 1\% in both profile metrics and 5~cm$^{-1}$ in
centroid. Every row passes.}
\label{tab:torsion-emission-validation}
\begin{ruledtabular}
\begin{tabular}{lrrr}
Control & $L^1$ difference & Profile (\%) & Centroid (cm$^{-1}$) \\
\hline
61 versus 71 & $8.86\times10^{-6}$ & 0.001073 & 0.02791 \\
High region $-500$ & $1.57\times10^{-4}$ & 0.01549 & 0.51909 \\
High region $+500$ & $5.38\times10^{-5}$ & 0.005265 & $-0.17827$ \\
Box $85^\circ$, basis 81 & $5.75\times10^{-7}$ & 0.0000971 & $-0.00175$ \\
Bins 0.5 versus 1 cm$^{-1}$ & $2.90\times10^{-7}$ & 0.0000656 & $-0.00061$ \\
\end{tabular}
\end{ruledtabular}
\end{table}

The reference uses $\pm75^\circ$ and 71 functions per rotor; the enlarged
box uses $\pm85^\circ$ and 81 functions. Between $2.0\times10^3$ and
$2.7\times10^3$ thermally weighted initial states suffice in the reported
profile controls. Each trace discards less than $10^{-9}$ initial
population and retains the torsional FC sum to within $10^{-9}$.
The initial-state density diagnostic uses the stricter $10^{-11}$ trace
cutoff. The independent thermal and profile truncations are recorded
separately. All control profiles share the identical bath correlation,
energy origin, broadening and optical operator. The experimental
comparison retains the original absorption-derived translation
$+2043.095$~cm$^{-1}$; no new fluorescence shift is fitted.

\subsection{Matched-temperature optical and bath-response refinements}
The refined harmonic, two-torsion and bath-response spectra all use
298.15~K, the fixed $+2043.095$~cm$^{-1}$ absorption translation and
Gaussian standard deviation 100~cm$^{-1}$. A 512~fs correlation window,
0.25~fs time step and fourfold FFT padding are shared by these calculations.
The harmonic reference is recalculated at this temperature, rather than
using a profile evaluated at 300~K. All potential and optical data come
from the completed acquisition; no electronic point or keyword is added.

For the active/bath FC+HT contraction, write
$A_{fi}=\langle f|i\rangle$ and
$M^a_{fi}=\langle f|\mu_a(q)|i\rangle$. Five active traces contain
$|A|^2$, $\sum_a|M^a|^2$ and the three real $M^aA^*$ products.
A four-channel Gaussian operator bank contains the bath identity and
three linear optical insertions. Its five Hermitian forms give the
scalar trace, the summed bath-HT trace and three signed identity--HT
cross traces. Their contraction retains active--bath optical interference,
using one C++ Gaussian preparation for all forms. Active coordinate
operators use exact sine-basis position matrices. When torsional
property derivatives vanish, the correlation recovers the factorized
FC-rotor/FC+HT-bath calculation within $2.5\times10^{-16}$ relative error.
The full linear optical operator changes the zero-time strength from
16.29808548 to 16.53371885~a.u.\ and preserves the finite-space trace.

For the potential response, the existing Cartesian gradients are aligned
to their actual rigid-rotor geometry and projected onto co-rotated
transverse normal displacements. A periodic force increment $g_j(q)$
is anchored to zero at the reference and fitted with eight identified
coefficients per bath mode. Its basis contains single-rotor sine/cosine
functions and their two-rotor products; only the existing training
geometries enter the fit. The fitting weights are
$[1+(V/1500~\mathrm{cm}^{-1})^2]^{-1}$. There are 13 training and four
reserved geometries for $S_0$, and 23 training and four reserved for
$S_1$. Across all 88 components, training/reserved force RMS errors are
13.90/42.31~cm$^{-1}$ for $S_0$ and 68.94/85.32~cm$^{-1}$ for $S_1$,
per dimensionless normal displacement. The near-planar $S_1$ reserved
control has 15.95~cm$^{-1}$ RMS. These force-fit residuals characterize
the added response model; they are separate from spectral basis error.

For dimensionless bath coordinates $z_j$, completing
$\omega_j z_j^2/2+g_j(q)z_j$ gives the displacement
$d_j(q)=-g_j(q)/\omega_j$ and the potential contribution
$-g_j(q)^2/(2\omega_j)$. The added thermal diagonal displacement term is
\begin{equation}
 V_{\rm BH}(q)=\tfrac14\sum_{j,ab}
 G_{ab}(q)\,\partial_a d_j(q)\,\partial_b d_j(q)
 \coth\!\left[\frac{\omega_j}{2k_BT}\right].
\end{equation}
It is evaluated with the same rigid-rotor kinetic tensor and periodic force
functions. This is a leading adiabatic energy/diagonal-kinetic response;
the optical bath trace retains its reference geometry and mode map.
A complete coupled torsion--bath VCI is a different nuclear model.

Forces are ranked independently for each surface by the initial-density
mean of $g_j^2/(2\omega_j)$. The complete mean response magnitudes are
190.95~cm$^{-1}$ on $S_0$ and 58.11~cm$^{-1}$ on $S_1$ in the emission
region. The three leading frequencies are 823.0, 215.7 and 1219.8~cm$^{-1}$
for $S_0$, and 659.7, 727.9 and 200.6~cm$^{-1}$ for $S_1$. Three forces
retain 75.8\% and 72.3\% of these response magnitudes. A cumulative 95\%
selection (9 and 15 forces) passes the profile gate but changes its centroid
by 11.59~cm$^{-1}$, exceeding the fixed 5~cm$^{-1}$ criterion. The next
99\% selection retains 16 and 24 forces and passes every profile gate.
Selection uses calculated response and numerical profile differences;
experimental intensities do not enter it.

Table~\ref{tab:si-torsion-response} resolves the fluorescence changes associated with coupled torsions and bath response.

\begin{table}[!htbp]
\caption{Matched-temperature fluorescence and response controls.
The shoulder/main area ratio integrates 318--328 and 332--345~nm,
respectively. Every curve uses the same fixed absorption calibration.}
\label{tab:si-torsion-response}
\begin{ruledtabular}
\begin{tabular}{lrrr}
Model & Maximum (nm) & RMS & Shoulder/main \\
\hline
Experiment & 337.5 & --- & 0.632 \\
Harmonic FC+HT & 355.0 & 0.38657 & 0.393 \\
Two DVR, rotor FC & 343.5 & 0.28921 & 0.418 \\
Two DVR, full FC+HT & 343.5 & 0.28967 & 0.417 \\
Three bath forces/state & 341.5 & 0.26576 & 0.443 \\
Six bath forces/state & 341.0 & 0.26021 & 0.449 \\
Selected 16/24 forces & 341.0 & 0.25626 & 0.454 \\
Complete 88 forces/state & 341.0 & 0.25591 & 0.454 \\
\end{tabular}
\end{ruledtabular}
\end{table}

The selected response differs from the 88-force reference by
$7.01\times10^{-4}$ in unit-area $L^1$ distance, 0.08316\% of the reference
peak and $-2.221$~cm$^{-1}$ in centroid. Increasing the complete-response
basis from 51 to 61 functions per rotor gives
$6.20\times10^{-5}$, 0.00789\% and 0.1954~cm$^{-1}$, respectively.
The full FC+HT calculation with no bath response changes by
$8.92\times10^{-6}$ and 0.00108\% between 61 and 71 functions.
Finite-window/normalization/translation controls give an RMS increase
of 0.00043--0.00052 upon adding torsional HT; this example does not
show an HT improvement. The adiabatic bath response is the appreciable
additional nuclear effect. Minima searches from 25 seeds per surface
retain four $S_0$ and two $S_1$ interior minima in the accepted
$\pm75^\circ$ coordinate representation, with positive curvature eigenvalues, for both the
complete and selected response models. No additional interior minimum
is found in this domain.

The three scientific services (finite active/Gaussian-bath operator traces, periodic adiabatic bath response and rigid-rotor geometry/metric) are included in the standalone SATI distribution. A fresh offline installation imports all three without the parent checkout; the package and clean-installation records accompany the numerical evidence.

\subsection{Explicit oscillator checks and translated Gaussian optical fibers}
The selected response is checked independently by promoting its strongest
force below 300~cm$^{-1}$, then that mode and the leading force contribution,
to explicit oscillators. For the selected set $E$, the contracted operator is
\begin{equation}
 H^{\rm exp}_s=H_{a,s}+V^{\rm ad}_{s,\overline E}
       +\sum_{j\in E}\{\omega_{sj}N_j+g_{sj}(q)Q_j\}.
\end{equation}
The selected modes' own completed-square and Born--Huang contributions are
removed before their explicit force is added; bath zero-point energies are
subtracted consistently. The remaining selected forces retain their
adiabatic response. Saved 61-function-per-rotor eigenvectors provide the
active contraction, enlarged through 96, 160, 256, 384 and 512 states.
Total oscillator occupations are enlarged through 6, 10 and 14 quanta.
The adiabatic reference includes the oscillator-occupation-dependent factor
$(2n_j+1)$ multiplying each mode's diagonal Born--Huang term, before the
combined levels are ordered. Thus extra oscillator bands are compared with
extra bands, rather than being mistaken for additional torsional states.

Table~\ref{tab:si-explicit-bath} compares the adiabatic response with explicit variational bath excitations.

\begin{table}[!htbp]
\caption{Zero-bath-temperature variational checks with two explicit force
couplings. The maximum difference concerns the first 24 excitation energies.
The final contraction uses 512 active states and 14 total oscillator quanta.}
\label{tab:si-explicit-bath}
\begin{ruledtabular}
\begin{tabular}{lrrr}
Surface & Frequencies (cm$^{-1}$) & $\delta E_0$ (cm$^{-1}$) & Max. difference (cm$^{-1}$)\\
\hline
$S_0$ & 215.7, 823.0 & $-0.00523$ & 0.29232\\
$S_1$ & 200.6, 659.7 & $-0.04470$ & 0.25387\\
\end{tabular}
\end{ruledtabular}
\end{table}
The single slow-mode checks give maximum differences of 0.29060 and
0.23202~cm$^{-1}$, respectively. Between 384 and 512 active states the
largest change in these excitation energies is 0.00017~cm$^{-1}$;
increasing the oscillator cutoff from 10 to 14 at 256 active states also
passes the fixed 1~cm$^{-1}$ gate. Final eigenvector residual norms are below
$2\times10^{-8}$~cm$^{-1}$. These are low-energy checks of the specified
linear-force Hamiltonian at zero bath temperature; the optical comparison
below retains the common 298.15~K population model.

For the optical extension, each harmonic fiber is displaced by
$d_s(q)=-g_s(q)/\omega_s$, using the same independently selected 16/$S_0$
and 24/$S_1$ forces as its active Hamiltonian. Its curvatures and reference
Duschinsky map are retained. Main-text Eq.~\eqref{M-eq:translated-bath-shift} gives
the resulting Gaussian shift. The electric operator includes its initial
fiber displacement, so its constant part acquires $D_i d_i(q)$ as well as
the explicit torsional contribution. Both linear HT vertices and their
signed interference are included.

Write the bath transition amplitude as
$O_{mn}(q)\simeq\sum_\alpha L_\alpha(q)O_{mn}(q_\alpha)$, where the
$L_\alpha$ are products of Chebyshev--Lobatto cardinal functions.
Interpolation acts on amplitudes, preserving their cross terms, rather
than on intensities. For real active eigenvectors the correlation is
\begin{equation}
 C(t)=\sum_{\alpha\beta}A_{\alpha\beta}(t)K_{\alpha\beta}(t),
\end{equation}
where $A_{\alpha\beta}$ contains the signed products
$\langle f|L_\alpha|i\rangle\langle f|L_\beta|i\rangle$, thermal active
populations and phases, while $K_{\alpha\beta}$ traces the two translated
bath amplitudes over the harmonic manifold. Unequal-source Bargmann
contraction evaluates $K$ with one shared Gaussian factorization per time;
no bath number-state cutoff is introduced. The analytic reference scalar
trace supplies the established determinant branch. Active energy cross
histograms are batched, and the three Cartesian optical components share
the same source solves.

Table~\ref{tab:si-displaced-convergence} reports independent interpolation and nuclear-basis controls for the displaced-bath profile.

\begin{table}[!htbp]
\caption{Independent convergence controls for the displaced-fiber profile.
Differences are measured between unit-area spectra on the common photon
wavenumber grid. Peak differences are percentages of the reference peak.}
\label{tab:si-displaced-convergence}
\begin{ruledtabular}
\begin{tabular}{lrrr}
Control & $L^1$ & Peak (\%) & Centroid (cm$^{-1}$)\\
\hline
13 to 17 nodes/rotor, basis 61 & $2.36\times10^{-5}$ & 0.00286 & 0.0743\\
51 to 61 sine functions/rotor & $6.87\times10^{-5}$ & 0.00807 & 0.2088\\
1 to 0.5~cm$^{-1}$ energy bin & $1.62\times10^{-7}$ & 0.000056 & $-0.00029$\\
\end{tabular}
\end{ruledtabular}
\end{table}
The final 61-function calculation retains 1665 initial torsional states,
with discarded population below $10^{-7}$. Its zero-time strength is
16.51960871~a.u.; the relative difference from direct quadrature of the
local thermal optical sum rule is $5.35\times10^{-6}$. At zero displacement,
the 88-mode kernel recovers the preceding reference correlation within
$1.75\times10^{-15}$ relative error. A separate one-dimensional check uses
220 coordinate quadrature nodes, 60 initial and 100 final oscillator states,
three unequal pairs of translations, unequal frequencies and all three
linear optical components. The largest absolute difference is
$4.34\times10^{-14}$ across the five sampled times.

The final translated-fiber fluorescence maximum is 340.5~nm, its
peak-normalized experimental RMS is 0.20109 and its shoulder/main area
ratio is 0.48795. The temperature, broadening and absorption-derived
translation are unchanged. No additional electronic points are used.
The general translated-kernel, signed cross-trace and explicit-bath services
are included in the standalone scientific distribution.

\subsection{Interleaved application-window timing audit}
Nine repetitions of each prepared physical input were interleaved across
systems and across native/FCclasses3 execution order. The executables,
thread limits, time grids and spectral parameters were unchanged. Each
measurement covers spawn through successful exit with complete profile
export; Python orchestration is outside the timed command. POSIX
\texttt{subprocess.run(timeout=...)} uses a polling wait, which can add
up to about 50~ms to short jobs. The revised driver instead uses blocking
\texttt{wait()} and a separate watchdog; direct kernel timers provide an
independent check of the measured scaling. The previous polling-based
short-window medians are superseded by this audit.

Table~\ref{tab:si-window-timings} resolves the execution cost on the physical time windows used for molecular spectra.

\begin{table}[!htbp]
\caption{Application-window timing audit in seconds. Parentheses give
first and third quartiles of nine processes, not confidence intervals.
The executable startup/dispatch contribution is the median of the
process-minus-internal native timer, about 0.006--0.007~s in all cases.}
\label{tab:si-window-timings}
\begin{ruledtabular}
\begin{tabular}{lrrr}
Case & Native & FCclasses3 & Native internal\\
\hline
$S_1$, 512~fs & 0.06302 (0.06244, 0.06941) & 0.16445 (0.16322, 0.18668) & 0.05599\\
$S_2$, 512~fs & 0.06337 (0.06315, 0.06399) & 0.16286 (0.16228, 0.18174) & 0.05644\\
p-Terphenyl, 512~fs & 0.65472 (0.65379, 0.68507) & 3.09330 (2.92618, 3.23481) & 0.64789\\
$S_1$, 1024~fs & 0.12884 (0.11764, 0.12945) & 0.35483 (0.33819, 0.37387) & 0.12266\\
$S_2$, 1024~fs & 0.12886 (0.11767, 0.13263) & 0.34101 (0.31737, 0.35080) & 0.12259\\
p-Terphenyl, 1024~fs & 1.43250 (1.34218, 1.49644) & 6.38078 (6.04217, 6.40055) & 1.42667\\
\end{tabular}
\end{ruledtabular}
\end{table}
The raw 108 process records include host load, input/executable checksums,
individual sample values and internal native timings in
\texttt{numerical\_evidence/revision\_checks\_20261102/}.
The service is general: \texttt{matrix\_core.process\_timing.time\_process}
accepts any executable argument vector and output file without a shell.
The same code commit adds a molecular-agnostic ORCA Hessian exporter and
a product-state VSCF diagnostic for arbitrary finite tensor-product
Hamiltonians. Eight independent regression tests pass on the compute server, including
negative curvature retention, complex Hamiltonians and timeout handling.

\subsection{Product-state torsional ablation}
The complete saved 51-by-51 sine-basis eigensystems reconstruct the same
active Hamiltonians without new potential evaluations or electronic points.
Alternating modal minimization produces two self-consistent one-dimensional
anharmonic problems. Two independent starts from the Schmidt factors of the
coupled ground state are compared. Frozen modal excitations then supply a
product spectrum with the ground-state double-counting correction removed.
This is a correlation ablation within the same two-coordinate Hilbert space,
not a one-dimensional DVR plus a harmonic replacement of the other rotor.
The product ground energies exceed the coupled values by 2.65092 and
1.59142~cm$^{-1}$ for $S_0$ and $S_1$, respectively. Stationarity residuals
are below $4.7\times10^{-9}$~cm$^{-1}$.

The optical calculation retains the same 16/$S_0$ and 24/$S_1$ force sets,
translated Gaussian kernels, full linear FC/HT operator, temperature,
broadening and absorption-derived shift. The product calculation retains
1378 initial states; its relative zero-time sum-rule residual is
$5.85\times10^{-5}$. Its maximum/RMS/shoulder ratio is
341.5~nm/0.219893/0.465183, compared with
340.5~nm/0.201045/0.488000 for the equally resolved coupled calculation.
These differences exceed the reported interpolation and basis residuals.
The general product solver accepts arbitrary finite Hermitian tensor
Hamiltonians; it does not embed molecular or torsional coefficients.

\subsection{Executed ORCA ESD Cartesian FC audit}
ORCA 6.1.1 executes both anthracene inputs with the specified
\texttt{NOITER} keyword, imported Cartesian Hessians and supplied adiabatic
gap/electric dipole. No SCF iterations or new force acquisitions determine
the spectral model. The actual Gaussian width printed by ESD is
250~cm$^{-1}$: its inhomogeneous width is distinct from the Lorentzian
\texttt{LINEW} parameter. Native controls use this same Gaussian width,
0~K, 512/1024~fs and a 0.125~fs sampling step. ESD's actual exported
spacings are approximately 65.1492/32.5746~cm$^{-1}$ despite the requested
5~cm$^{-1}$ resolution; comparison uses the exported energy points.

The maxima coincide on both common grids. The peak-normalized raw-profile
difference is 1.51\%; original ESD intensities are retained unchanged.
An independent endpoint diagnostic computes, without fitting a background,
\begin{equation}
 \delta\epsilon(\widetilde\nu)
 =P\,\widetilde\nu\,C(0)\,c\,\Delta t,
\end{equation}
where $P$ is the physical molar-extinction prefactor, $C(0)=0.767721637$
is the FC sum rule, $c$ is in cm/fs and $\Delta t=0.125$~fs.
This is the extra zero-time rectangle contribution relative to a
trapezoidal one-sided transform. Its wavenumber coefficient is
0.00202337224. Subtracting this analytically specified contribution for
the audit gives integrated native/ESD agreement to 0.0081\% and maximum
absolute differences of 0.1105/0.1113\% of the native peak, without energy
or multiplicative intensity adjustment. This numerical agreement supports
the endpoint interpretation; it is not evidence from ORCA source code.
Both unmodified external profiles and the diagnostic are archived.

Three subsequent interleaved repetitions per physical case use cold work
directories, the same one-thread limit and a blocking process wait. Input,
executable and Hessian-dependency hashes are checked before checkpoint
reuse. The CPU is the same AMD EPYC 7543 server used above. No electronic
model data are generated in these \texttt{NOITER} comparisons.

Table~\ref{tab:si-anthracene-timings} compares prepared-input and internal-module execution times for the same anthracene FC model.

\begin{table}[!htbp]
\caption{Executed 66-mode anthracene Cartesian AH/FC timing at 0~K,
Gaussian $\sigma=250$~cm$^{-1}$ and 0.125~fs step. Whole-process times are
medians of three cold executions; internal ESD timers are rounded to seconds.
Both programs prepare their own harmonic correlation from physical fields.}
\label{tab:si-anthracene-timings}
\begin{ruledtabular}
\begin{tabular}{lrrrrr}
Window (fs) & Native (s) & ESD (s) & Ratio & Native internal (s) & ESD internal (s)\\
\hline
512 & 0.737248 & 10.984527 & 14.90 & 0.730745 & 8\\
1024 & 1.347214 & 15.685114 & 11.64 & 1.338754 & 12\\
\end{tabular}
\end{ruledtabular}
\end{table}

The three native wall times are (0.737248, 0.893264, 0.713526)~s at
512~fs and (1.347214, 1.360555, 1.219306)~s at 1024~fs. The corresponding
ESD times are (11.076329, 10.094535, 10.984527)~s and
(16.742782, 15.485936, 15.685114)~s. Native processes export 20914/20979
points at approximately 1~cm$^{-1}$; ESD exports 322/645 points on its
coarser actual FFT grid. Thus the native advantage does not come from a
smaller exported grid. All native executions recover the analytic FC
strength to within $1.6\times10^{-13}$ relative error. The approximate
spectral-module ratios, 10.95 and 8.96, separately show that ORCA front-end
initialization does not explain the observed difference. These timings support the harmonic FC benchmark. The separately executed
FC+HT comparison and direct-module timings are given in Sec.~\ref{sec:three-controls}.

\FloatBarrier
\section{Matched FC+HT ESD comparison and finite electronic controls}
\label{sec:three-controls}

\subsection{Phase-preserving FC/HT import}
The 27-mode 4-pyrone $S_1$ calculation uses B3LYP/6-31+G(d), the
previously computed Cartesian Hessians, full TD-DFT singlet root 1,
Duschinsky mixing and 0~K. Its 54 central displacements acquire the
27 transition-dipole derivatives. Subsequent spectral comparisons reuse
these derivatives; no new force-field point is added.
The comparison has a 512~fs window, 4096 intervals of 0.125~fs and
Gaussian $\sigma=100$~cm$^{-1}$. The separately printed homogeneous
width of 50~cm$^{-1}$ is inactive for the selected Gaussian line shape.
The actual ESD FFT grid has approximately 65.1492~cm$^{-1}$ spacing.
The explicit original \texttt{DELE}, 28976.024634909365~cm$^{-1}$,
is transferred without the correlation summary's two-decimal rounding.
The native exported grid is approximately 1~cm$^{-1}$; residuals are
computed at the external grid points, with no energy or intensity fit.

The files contain both the eigenvectors of the mass-weighted Hessians
and the complex transition-dipole derivatives. Retaining their stored
phases and ordering is essential. The stored derivative is dimensional:
if $D_f=\partial\boldsymbol\mu/\partial\boldsymbol Q_f$ uses atomic
mass-weighted coordinates, its dimensionless counterpart is
\begin{equation}
 H_f=D_f\,\operatorname{diag}(\omega_f^{-1/2}),\qquad
 \omega_f\text{ in atomic units}.
\end{equation}
The dimensionless displacement step used for acquiring the derivatives
must not be confused with this storage convention. An independent audit
reconstructs all 27 derivatives from the 54 printed plus/minus dipoles
with $\Delta q=0.01$. Its maximum difference is
$4.22\times10^{-8}$ atomic units, or $1.10\times10^{-6}$ relative to
the largest derivative, consistent with the dipoles' printed precision.

With $\boldsymbol q_f=A\boldsymbol q_i+\boldsymbol k$, where
$A=\omega_f^{1/2}J\omega_i^{-1/2}$, the imported operator is
\begin{equation}
 \boldsymbol\mu(\boldsymbol q_i)
 =\boldsymbol\mu_{\rm ref}+H_f\boldsymbol k+H_fA\boldsymbol q_i.
\end{equation}
This transfers the operator used by ESD into the native initial-state
basis. It does not replace that operator by the separate Gaussian
optical-property calculation. Three Hermitian observable forms share the
same native Gaussian: FC, HT including FC--HT interference, and total.
Their zero-time strengths are $3.2089848613\times10^{-6}$,
$1.2064141273\times10^{-3}$ and $1.2096231122\times10^{-3}$ atomic
units, respectively.

Table~\ref{tab:si-pyrone-fcht-errors} compares the absolute FC+HT components and their endpoint-discretization contributions.

\begin{table}[!htbp]
\centering
\caption{Absolute 4-pyrone FC+HT comparison. Differences are normalized
by the corresponding native component's peak. The endpoint audit uses
the analytic rectangle contribution given in the preceding FC audit;
the unmodified ESD profiles are retained separately.}
\label{tab:si-pyrone-fcht-errors}
\begin{tabular}{lrrr}
\toprule
Component & Raw maximum difference (\%) & Audit difference (\%) & Audit area ratio\\
\midrule
FC & 0.8524 & 0.0475 & 1.000466\\
HT + interference & 1.3392 & 0.0755 & 1.000712\\
Total & 1.3414 & 0.0755 & 1.000711\\
\bottomrule
\end{tabular}
\end{table}

\begin{figure}[!htb]
\centering
\includegraphics[width=.92\textwidth]{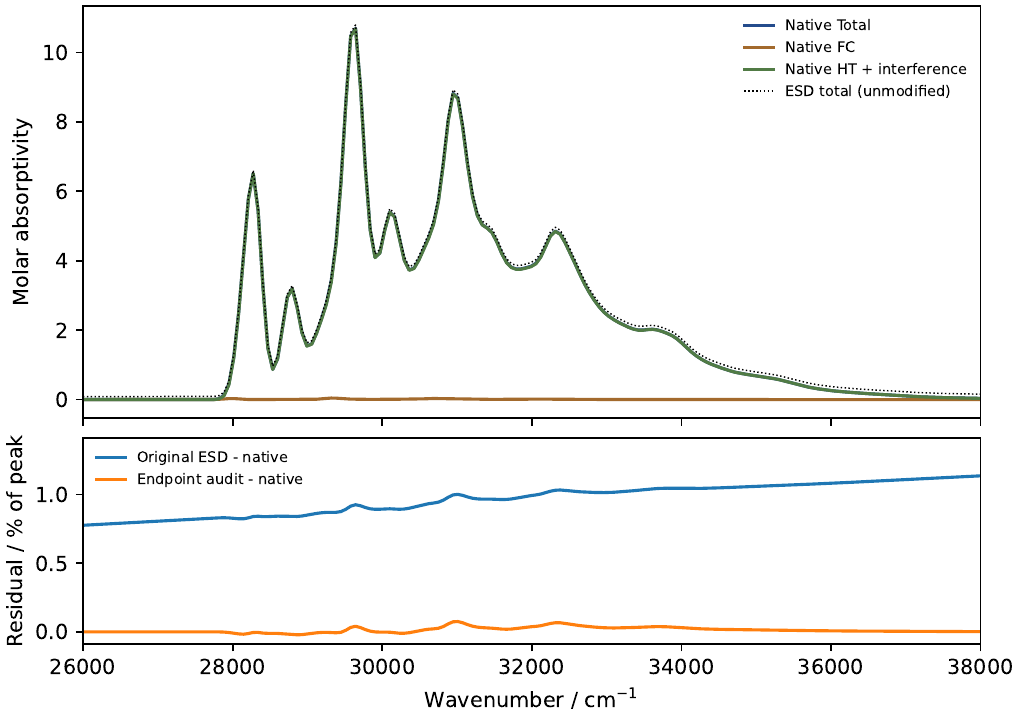}
\caption{Matched 27-mode 4-pyrone $S_1$ FC+HT spectra at 0~K.
Top: native FC, HT including interference and total molar absorptivities;
the dotted curve is the unmodified ESD total. Bottom: residuals before
and after the separately specified zero-endpoint audit. No multiplicative
normalization or energy shift is fitted.}
\label{fig:ESD-FCHT}
\end{figure}

Figure~\ref{fig:ESD-FCHT} compares the complete FC+HT profiles and the separately evaluated endpoint-discretization correction.

\subsection{Prepared workflow and directly executed spectral module}
The prepared input reuses the moment
\[
 \boldsymbol\mu_{\rm ref}=(7.51413,7.22112,-178.833)\times10^{-5}
 \quad\text{atomic units},
\]
with \texttt{TDIP}, together with \texttt{NOITER} and the stored Hessian
transition derivatives. This excludes the derivative acquisition from
both timing protocols. The whole ORCA process still includes startup
and the reference TD dispatch associated with its route. The direct
\texttt{orca\_esd} protocol reads the captured ESD/main inputs, the same
Hessians, and their restart dependencies; it excludes those preceding
steps. All measured direct-module runs complete without electronic
mode-displacement calculations.

Both protocols use the AMD EPYC 7543 server, a one-thread OpenMP/BLAS
limit and separate fresh output directories. Native Gaussian preparation
remains inside the timed executable, and the three physical components
are exported in one shared calculation. Input, executable and dependency
hashes are checked before resuming an existing timing campaign.
Native and external runs alternate order between repetitions. Whole
prepared-workflow medians are 0.182478048 and 10.312454842~s for three
repetitions, a factor 56.513. Direct spectral-module medians over nine
repetitions are 0.184837308 and 1.608866516~s, a factor 8.7042.
The module's output also passes the independent FC/HT comparison:
the largest endpoint-audited total residual is 0.0755\%.
Every unrounded measurement, recorded server load and source identity
is archived in \texttt{three\_controls\_20261102/timings/}.

\subsection{One explicit DVR torsion versus two}
This ablation differs from the product-state VSCF test above. Exactly
one of the two torsions is represented by DVR; the other returns to the
harmonic bath. The complementary bath consequently has 89 modes,
rather than 88, and the total remains 90. Both one-rotor choices retain
the complete linear optical operator, 298.15~K, Gaussian
$\sigma=100$~cm$^{-1}$ and the fixed absorption-derived calibration.
The one-coordinate and two-coordinate baselines both omit the subsequent
adiabatic/displaced-bath response, isolating the number of explicit rotors.

For a column-orthonormal common-coordinate bath basis $B$, the restricted force
matrix and induced inverse metric are
\begin{equation}
 F_b=B^TFB,\qquad G_b=(B^TG^{-1}B)^{-1}.
\end{equation}
The latter is generally different from $B^TGB$. The common coordinate representation retains
the excluded rotor's harmonic displacement, while the explicit rotor
uses its fitted periodic potential and curvilinear kinetic operator.

Table~\ref{tab:si-rotor-ablation} isolates the spectral contribution of the second explicit torsion.

\begin{table}[!htbp]
\centering
\caption{Matched p-terphenyl fluorescence ablation. RMS uses
peak-normalized profiles over 300--410~nm. The shoulder/main ratio uses
the integrated 318--328 and 332--345~nm windows. No emission calibration
is adjusted between models.}
\label{tab:si-rotor-ablation}
\begin{tabular}{lrrr}
\toprule
Model & Maximum (nm) & RMS & Shoulder/main\\
\midrule
Experiment & 337.5 & --- & 0.632\\
90 harmonic modes & 355.0 & 0.3866 & 0.3930\\
One DVR rotor A + 89 harmonic & 351.0 & 0.3266 & 0.4098\\
One DVR rotor B + 89 harmonic & 351.0 & 0.3260 & 0.4113\\
Two DVR rotors + 88 harmonic & 343.5 & 0.2897 & 0.4169\\
Two DVR rotors + responding bath & 340.5 & 0.2010 & 0.4879\\
\bottomrule
\end{tabular}
\end{table}

The one-rotor basis is enlarged through 61, 81 and 101 sine functions.
Between 81 and 101, the two choices give normalized-profile $L^1$
changes $1.97\times10^{-8}$ and $2.93\times10^{-8}$, maximum
peak-relative differences $2.14\times10^{-8}$ and $3.17\times10^{-8}$,
and centroid changes below $1.01\times10^{-4}$~cm$^{-1}$.
The discarded initial thermal populations are below $8.4\times10^{-10}$.
Thus the difference from the two-coordinate spectrum is resolved well
beyond the numerical basis error.

\subsection{Finite CAM-B3LYP torsional controls and electronic identity}
The finite sensitivity acquisition contains eight existing geometries
per electronic state: six fit geometries and two reserved controls.
CAM-B3LYP/6-31G(d,p) is used for $S_0$, with TD-CAM-B3LYP singlet root 1
for $S_1$. The PBE0 and CAM calculations have identical atomic geometries
and Gaussian AO bases. All jobs complete normally with 32 threads and
128~GB; a maximum of two jobs is active at a time.

The initial difference-field test uses the same fixed periodic wavevectors
$(2,0)$, $(4,0)$, $(0,2)$, $(0,4)$, $(2,2)$ and $(2,-2)$ on each surface. There are
13 coefficients, 18 training energy/gradient observations and a fixed
0.15~rad derivative weight. Both design matrices have rank 13 and
condition number 57.09. Each model is serialized and hashed before
opening its reserved-point fields. For $S_0$, maximum training errors
are 0.25~cm$^{-1}$ and 8.51~cm$^{-1}$/rad; reserved-point errors are
10.04~cm$^{-1}$ and 42.55~cm$^{-1}$/rad. For $S_1$, the corresponding
reserved errors are 540.33~cm$^{-1}$ and 3076.82~cm$^{-1}$/rad.
These results concern the finite difference-field representation.

An independent state-character audit reuses the Gaussian checkpoint parser
and the general transition-fingerprint overlap routine. The common AO metric is recovered from the complete restricted
MO set and checked against the other functional's orbitals. The maximum
orthonormality residual is below $2\times10^{-6}$. Transition-density
fingerprints are transported with the resulting MO cross-overlap, so
orbital signs and mixing are retained. Printed TD configurations remain
thresholded diagnostics rather than exact many-electron overlaps.
In every sampled geometry the PBE0 $S_1$ matches CAM $S_1$, with
negligible overlap with CAM $S_2$ and $S_3$.

Table~\ref{tab:si-cam-state-character} follows the bright electronic state across the eight torsional geometries.

\begin{table}[!htbp]
\centering
\caption{CAM-B3LYP $S_1$ character at the eight existing torsional
geometries. Angles are rotations relative to the existing $S_1$ reference,
not absolute ring-plane dihedrals. The last two rows are reserved controls.
Overlap is the AO-metric transition-density fingerprint overlap with PBE0
at the identical geometry. All eight states are dominated by HOMO--LUMO.}
\label{tab:si-cam-state-character}
\begin{tabular}{rrrrr}
\toprule
$(\Delta\phi_1,\Delta\phi_2)$ (deg) & $E_{01}$ (eV) & $E_{12}$ (eV) & $f_{01}$ & Overlap\\
\midrule
$(0,0)$ & 3.7925 & 0.8408 & 1.1793 & 0.98923\\
$(30,0)$ & 4.0025 & 0.7579 & 1.1432 & 0.99961\\
$(0,30)$ & 3.8593 & 0.8211 & 1.1714 & 0.98924\\
$(60,0)$ & 4.2994 & 0.5517 & 0.9841 & 0.99970\\
$(0,60)$ & 4.1408 & 0.6686 & 1.0839 & 0.99966\\
$(60,60)$ & 4.8363 & 0.3101 & 0.9555 & 0.98864\\
$(30,30)$ & 4.0808 & 0.7432 & 1.1395 & 0.98922\\
$(90,30)$ & 4.4891 & 0.4593 & 0.9139 & 0.99974\\
\bottomrule
\end{tabular}
\end{table}

The direct CAM surface is also fitted diagnostically with the same
13-parameter Fourier representation and the same six training geometries;
the controls never enter the objective. Its maximum reserved energy
error is 80.81~cm$^{-1}$, versus 540.33~cm$^{-1}$ for the difference fit.
However, the unconstrained least-squares fit changes the reference
forces. Retaining the reference energy and both gradients exactly instead
gives reserved errors of 26.93 and 163.85~cm$^{-1}$, with maximum gradient
errors 173.85 and 667.51~cm$^{-1}$/rad, respectively. The latter control
is 6134~cm$^{-1}$ above the CAM reference. This representation diagnostic
was performed after the initial difference-field validation; the training
set and reserved controls are retained and reported for both fits.
The fluorescence calculations use the emission-validated PBE0 field.
The finite controls identify field representation as the source of the
fitting discrepancy, with electronic character retained throughout the sample.

The interpolation error can be isolated algebraically. At the same six
training geometries, the same weights and the same linear Fourier design,
least squares obeys
\begin{equation}
 \mathcal F(E_{\rm CAM}-E_{\rm PBE0})
 =\mathcal F(E_{\rm CAM})-\mathcal F(E_{\rm PBE0}).
\end{equation}
The coefficient identity holds to $1.63\times10^{-17}$ Hartree. At the
high control, the direct CAM energy error is +80.81~cm$^{-1}$, whereas
the reduced-basis PBE0 error is +621.14~cm$^{-1}$; their difference is
the observed $-540.33$~cm$^{-1}$ error. The large difference-field
error therefore inherits the sparse Fourier representation error of
PBE0. The full emission-validated PBE0 potential uses its original,
more complete representation.

\subsection{Camphor: fixed-field functional sensitivity}
Two additional property/Hessian calculations at the B3LYP ground-state
geometry use TD-PBE0-D3(BJ) and TD-CAM-B3LYP-D3(BJ), both with def2-TZVP.
The E1/M1 moments and their derivatives are transported into the same
B3LYP initial-state modes. Ground-state SCF energies, rather than the
excited-state \texttt{Total Energy} field, define each vertical excitation.
The vertical-gap change shifts the existing adiabatic gap, keeping the
B3LYP nuclear fields and relaxation fixed. The spectra use 298.15~K,
0.25~fs, 4097 time points and $\sigma=100$~cm$^{-1}$. No energy shift or
intensity factor is fitted to experiment.

Table~\ref{tab:si-camphor-functional} compares camphor band strengths at fixed nuclear fields with three electronic functionals.

\begin{table}[!htbp]
\centering
\caption{Camphor fixed-nuclear-field FC+HT controls. $D$ is in atomic
units; $R$ is reported in $e^2a_0\hbar/m_e$. The band ratio uses the
experimental electric-dipole denominator convention, $4|R|/D$.
The maxima concern the frequency-dependent dissymmetry.}
\label{tab:si-camphor-functional}
\begin{tabular}{lrrrrr}
\toprule
Property functional & $\Delta E_{\rm vert}$ (cm$^{-1}$) & $D$ & $|R|$ & $4|R|/D$ & Maximum (nm)\\
\midrule
B3LYP & 0.00 & 0.004930 & 0.009828 & 0.05819 & 285.00\\
PBE0 & +149.92 & 0.004685 & 0.009059 & 0.05645 & 285.25\\
CAM-B3LYP & +680.15 & 0.003765 & 0.006616 & 0.05130 & 279.00\\
Experiment & --- & 0.005150 & 0.009580 & 0.05400 & $\sim303$\\
\bottomrule
\end{tabular}
\end{table}

\begin{figure}[!htb]
\centering
\includegraphics[width=.92\textwidth]{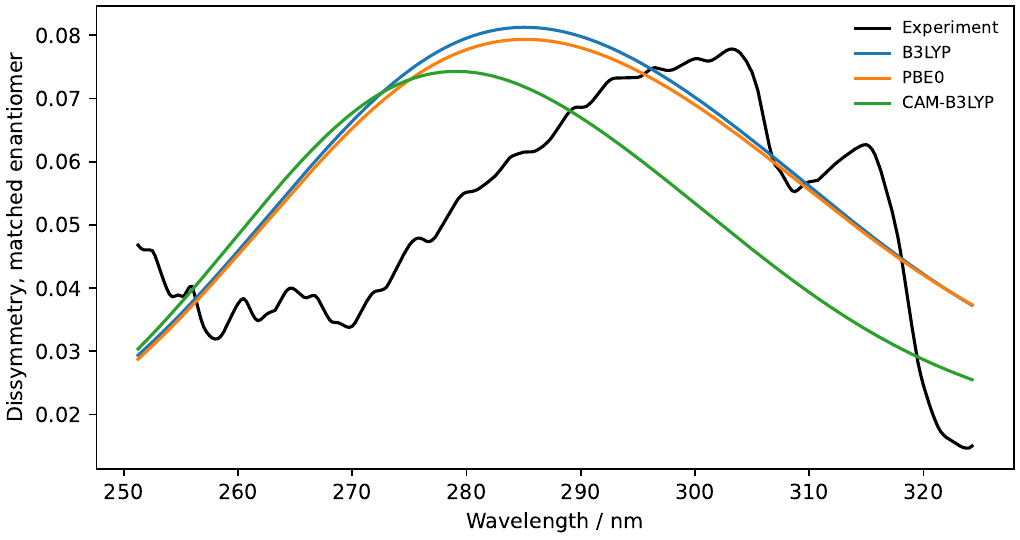}
\caption{Unshifted camphor dissymmetry with three property functionals
on fixed B3LYP nuclear fields. The respective vertical-gap changes are
included while the nuclear relaxation is kept fixed. Experimental
points and enantiomer matching are the same as in the main ECD figure.
The comparison separates operator/gap sensitivity from changes of the
nuclear Hamiltonians.}
\label{fig:camphor-functional}
\end{figure}

Figure~\ref{fig:camphor-functional} resolves how the electronic functional changes camphor absorption, ECD and dissymmetry.

\subsection{Reproducibility and general implementation}
The initial controls use source settings; the general
follow-up and paired sensitivity analysis use \texttt{d3fca339} and
\texttt{98512e0b}. General
interfaces own phase-preserving ESD import, Hermitian FC/HT observable
forms, harmonic subspace restriction with the induced metric, periodic
energy/force fitting with optional exact reference constraints, and
restartable interleaved process timing. Molecular coefficients and
finite acquisition plans occur in the explicit benchmark data.
Seven regression tests pass on the compute server, including complex operator phases,
malformed matrix rejection, exact periodic E/G recovery and exact reference
energy/gradient retention. Numerical data, source identities, unrounded
measurements and frozen model coefficients are archived under
\texttt{numerical\_evidence/three\_controls\_20261102/}.

\subsection{Direct periodic CAM field and paired spectral sensitivity}
The direct $S_1$ refinement uses the same six training geometries and no
additional electronic acquisition. Its generic Fourier E/G fitter minimizes
\begin{equation}
 \sum_a\left[(V_a-E_a)^2+w_g^2\lVert\nabla V_a-\boldsymbol g_a\rVert^2\right]
 +\lambda\sum_{\boldsymbol k}|\boldsymbol k|^4
 (c_{\boldsymbol k}^2+s_{\boldsymbol k}^2),
\end{equation}
subject to exact reference E/G constraints, with $w_g=0.15$ rad.
The Laplacian penalty allows regularized underdetermined expansions.
Candidate maximum Fourier orders are 1--4, smoothing parameters span
$10^{-8}$ to 1, and inversion-even and unrestricted fields are considered.
All candidate choices use grouped deletion of one training geometry,
keeping its energy and both gradients together; the exact reference stays
in each training fold. A 41-by-41 periodic landscape screen rejects
fields falling below $-50$~cm$^{-1}$. The selected field has Fourier order
two, inversion symmetry about the planar absolute torsions and
$\lambda=10^{-8}$. Its selection score is 7.895 in normalized training
deletion units, so it is not a claim of a globally predictive six-point
force field. Neither previously examined control enters selection, and
neither is relabelled as a fresh blind test.

The energy residuals at the two controls are +0.720 and +5.599~cm$^{-1}$.
The corresponding gradient residuals are $(-312.1,248.1)$ and
$(-1109.9,-564.5)$~cm$^{-1}$/rad; the global 250~cm$^{-1}$/rad gate is
therefore not passed. The reference energy and gradient are retained
algebraically. Periodic-grid local minima, followed by analytic-gradient
minimization, give two symmetry-related wells at absolute angles
approximately $(8.05,-8.06)$ and $(-8.05,8.06)$ degrees. Both are
0.148~cm$^{-1}$ below the reference and have positive Hessian eigenvalues
1198 and 3728~cm$^{-1}$/rad$^2$; no additional grid-seeded minimum is found.

The sensitivity calculation combines this direct $S_1$ field with the
previously validated $S_0$ CAM correction. Its comparison member retains
both original PBE0 torsional fields. Both use the same 88-mode harmonic
bath, the same complete linear FC/HT operator, 298.15~K,
$\sigma=100$~cm$^{-1}$ and the unchanged absorption calibration of
2043.095~cm$^{-1}$. Neither member adds adiabatic or displaced-bath response;
this isolates the field change from those later refinements. The reference
CAM energy increments imply an electronic-gap change of
+1924.678~cm$^{-1}$, which is applied separately.

Table~\ref{tab:si-torsion-field-gap} separates torsional-surface and electronic-gap effects on fluorescence.

\begin{table}[!htbp]
\centering
\caption{Paired fixed-bath torsional-field sensitivity. Intensities are
peak-normalized over the same 300--410~nm experimental window. The gap
column denotes a change relative to the PBE0 reference.}
\label{tab:si-torsion-field-gap}
\begin{tabular}{lrrrr}
\toprule
Model & Gap change (cm$^{-1}$) & Maximum (nm) & RMS & Shoulder/main\\
\midrule
PBE0 & 0 & 343.5 & 0.2896 & 0.4169\\
CAM shape & 0 & 344.0 & 0.3018 & 0.4021\\
CAM shape and gap & 1924.68 & 323.0 & 0.2196 & 0.8169\\
\bottomrule
\end{tabular}
\end{table}

\begin{figure}[!htb]
\centering
\includegraphics[width=.94\textwidth]{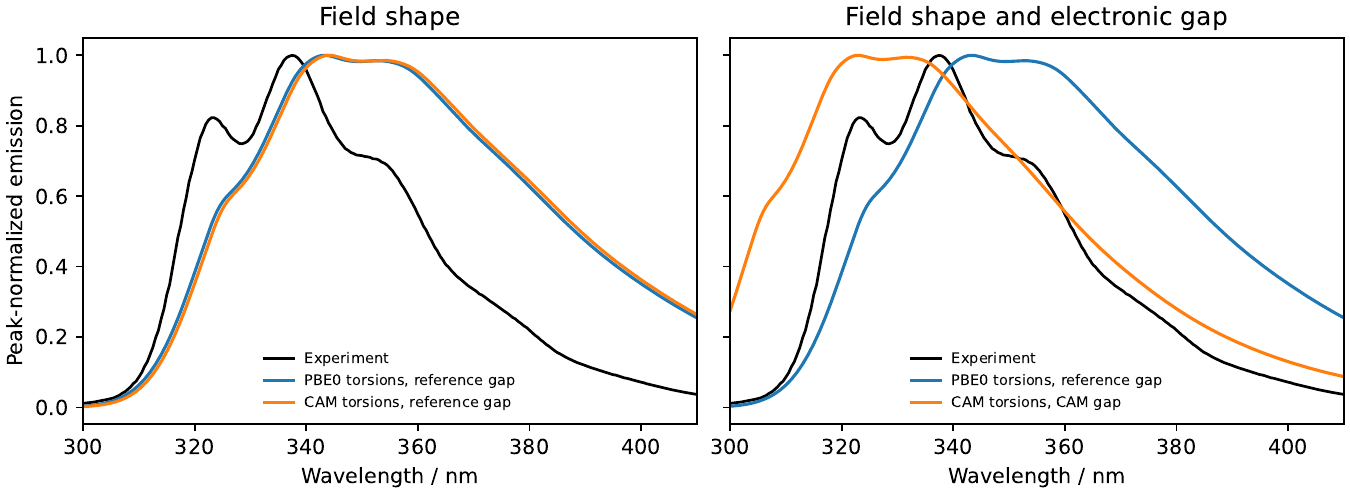}
\caption{Matched torsional-field sensitivity with a fixed harmonic bath.
Left: field-shape change at fixed gap. Right: the additional CAM electronic
gap change with the same absorption calibration. Experimental curves and
peak normalization are identical in both panels. The responding-bath final
PBE0 model is not replaced by this diagnostic.}
\label{fig:torsional-functional-sensitivity}
\end{figure}

For the 41-to-51-function refinement, peak differences are 0.0467\%
(CAM) and 0.0481\% (PBE0); the normalized $L^1$ differences are
$3.72\times10^{-4}$ and $3.77\times10^{-4}$ and centroid changes
are 1.17 and 1.19~cm$^{-1}$. The discarded initial thermal population
is below $10^{-9}$ in each case. These are basis controls for the
specified paired Hamiltonians, separate from the force-field checks.
Figure~\ref{fig:torsional-functional-sensitivity} compares the torsional-surface and electronic-gap contributions to the p-terphenyl fluorescence response.

\subsection{Room-temperature FC+HT ESD validation}
The prepared 27-mode input changes only the temperature from 0 to
298.15~K. The complete FC/HT operator, fields, gap, 512~fs window,
4096 time intervals and Gaussian width are unchanged. No new electronic
displacement is evaluated. Nine interleaved direct-module repetitions
give median native and ESD times of 0.510949438 and 1.200982039~s,
a ratio of 2.35049, at one thread on the same EPYC server in that archived
campaign. Section~\ref{sec:thermal-release} renews the direct comparison
with the released thermal solver.
Every repeated spectrum is byte-identical within each program.

Table~\ref{tab:si-thermal-esd-errors} gives the corresponding absolute spectral comparison at 298.15~K.

\begin{table}[!htbp]
\centering
\caption{Absolute ESD comparison at 298.15~K; conventions match the
zero-temperature table. The endpoint audit remains separately labelled.}
\label{tab:si-thermal-esd-errors}
\begin{tabular}{lrrr}
\toprule
Component & Raw difference (\%) & Audit difference (\%) & Audit area ratio\\
\midrule
FC & 1.1163 & 0.0507 & 1.000462\\
HT + interference & 1.8124 & 0.0739 & 1.000680\\
Total & 1.8132 & 0.0739 & 1.000680\\
\bottomrule
\end{tabular}
\end{table}
The total zero-time strength is $2.1953710613\times10^{-3}$ atomic units.
Both spectra peak at 29577.754~cm$^{-1}$ on the common external grid.
All unrounded timings, input identities, raw spectra, the endpoint audit,
frozen fits and general source are retained in the
\texttt{followup/} subdirectory of the closing-control dataset.

\begin{figure}[!htb]
\centering
\includegraphics[width=.92\textwidth]{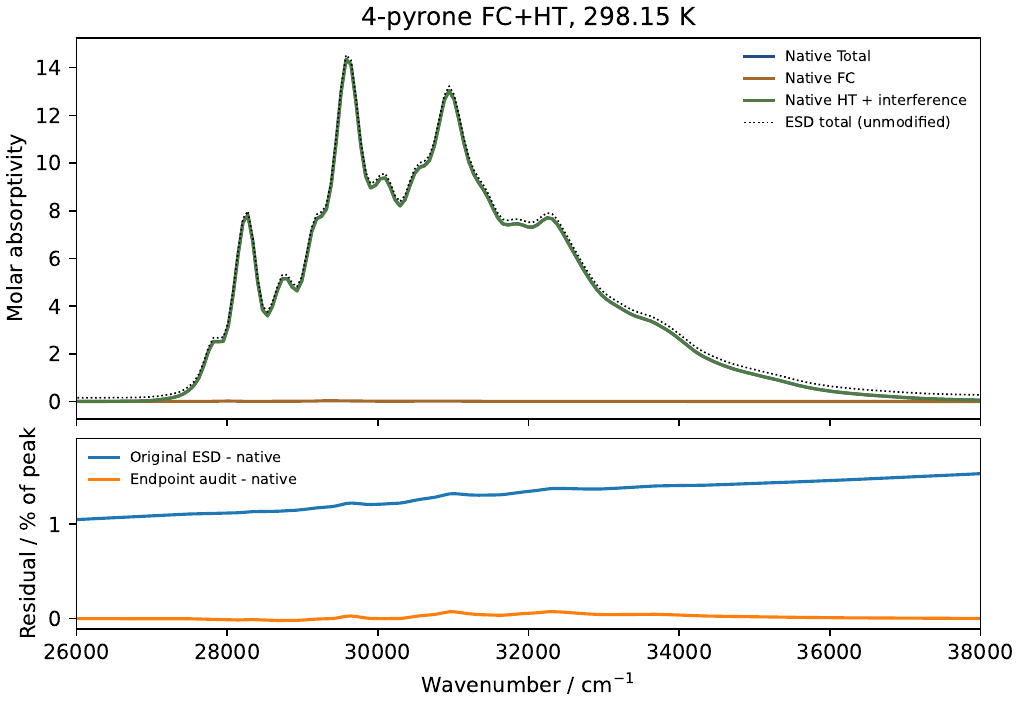}
\caption{Complete FC+HT ESD comparison at 298.15~K. Component curves,
raw external spectrum and separately labelled endpoint audit use the
same conventions as the zero-temperature comparison. No intensity or
energy adjustment is fitted.}
\label{fig:ESD-FCHT-room-temperature}
\end{figure}

Figure~\ref{fig:ESD-FCHT-room-temperature} shows the corresponding complete FC+HT comparison at room temperature.

\subsection{Training-only E/G weight selection and representation spread}
\label{sec:weighted-periodic-controls}
The lower-order direct field's small errors at the two control energies
are accompanied by maximum training residuals of 252.4~cm$^{-1}$ in
energy and 1118.1~cm$^{-1}$/rad in gradients. A second finite analysis
therefore applies the training E/G gates as well as the landscape check.
It considers Fourier orders 2--4, both inversion-even and unrestricted
fields, gradient weights 0.03, 0.1, 0.3 and 1 rad, and Laplacian penalties
$10^{-8}$, $10^{-4}$ and $10^{-2}$: 72 candidates in total. Linear
inequalities constrain the potential above $-50$~cm$^{-1}$ on a
25-by-25 periodic interpolation grid. These are evaluations of the fitted
function, not additional electronic acquisitions. The reference energy and
both gradients remain exact.

Eligibility requires maximum training errors below 50~cm$^{-1}$ and
250~cm$^{-1}$/rad. Grouped deletion of the non-reference training
geometries selects the minimum worst scaled E/G RMS score; fields within
1.25 times this score are retained. The complete training-selected set is
frozen before the previously examined control residuals are computed.
A 61-by-61 periodic landscape check followed by analytic-gradient
stationary-point refinement retains six fields with the required two
positive-curvature wells. The three spectral representatives are the
best training-deletion score among these fields, the smallest lowest
well curvature, and the largest sum of well curvatures. These criteria
select fields 67, 60 and 69 without using the control residuals.

Table~\ref{tab:si-periodic-fit-checks} reports energy and gradient residuals for the three periodic-field representations.

\begin{table}[!htbp]
\centering
\caption{Finite E/G fit checks for the three spectral representatives.
All fields have Fourier order four and inversion symmetry. Energy errors
are in cm$^{-1}$ and gradient errors in cm$^{-1}$/rad. ``Training'' and
``control'' denote maximum absolute residuals within their respective sets.}
\label{tab:si-periodic-fit-checks}
\begin{tabular}{rrrrrrr}
\toprule
Field & $w_g$ (rad) & $\lambda$ & Training E & Training G & Control E & Control G\\
\midrule
67 & 0.3 & $10^{-4}$ & 28.875 & 113.987 & 463.322 & 1223.538\\
60 & 0.03 & $10^{-8}$ & 0.00583 & 5.267 & 729.467 & 1735.553\\
69 & 1 & $10^{-8}$ & 0.02006 & 0.01667 & 767.928 & 1718.170\\
\bottomrule
\end{tabular}
\end{table}
The training forces improve, but none of these fields passes the
external gates. Consequently no improved CAM production field is
accepted from this finite analysis. In particular, exact interpolation of
the training forces does not establish the quality of the interpolated
field between them.

The spectral comparison keeps the existing $S_0$ correction, 88-mode
harmonic bath, linear FC/HT operator, electronic gap, absorption calibration,
298.15~K and Gaussian width unchanged. All three maxima remain at
344.0~nm. Their pointwise peak-normalized range is at most 1.8443\%,
with RMS deviations from experiment of 0.2984--0.3056 and shoulder/main
ratios of 0.3980--0.4070. The lowest $S_1$ torsional splitting ranges
from 10.9649 to 11.4401~cm$^{-1}$, compared with 10.2365~cm$^{-1}$ for
the lower-order direct reference. The 41-to-51-function profile changes
are at most 0.05244\% of the peak and 1.283~cm$^{-1}$ in centroid.
The representation spread therefore exceeds the numerical basis error.
This finite representation set is a sensitivity analysis, not a
statistical confidence interval or a force-field validation.

\begin{figure}[!htb]
\centering
\includegraphics[width=.92\textwidth]{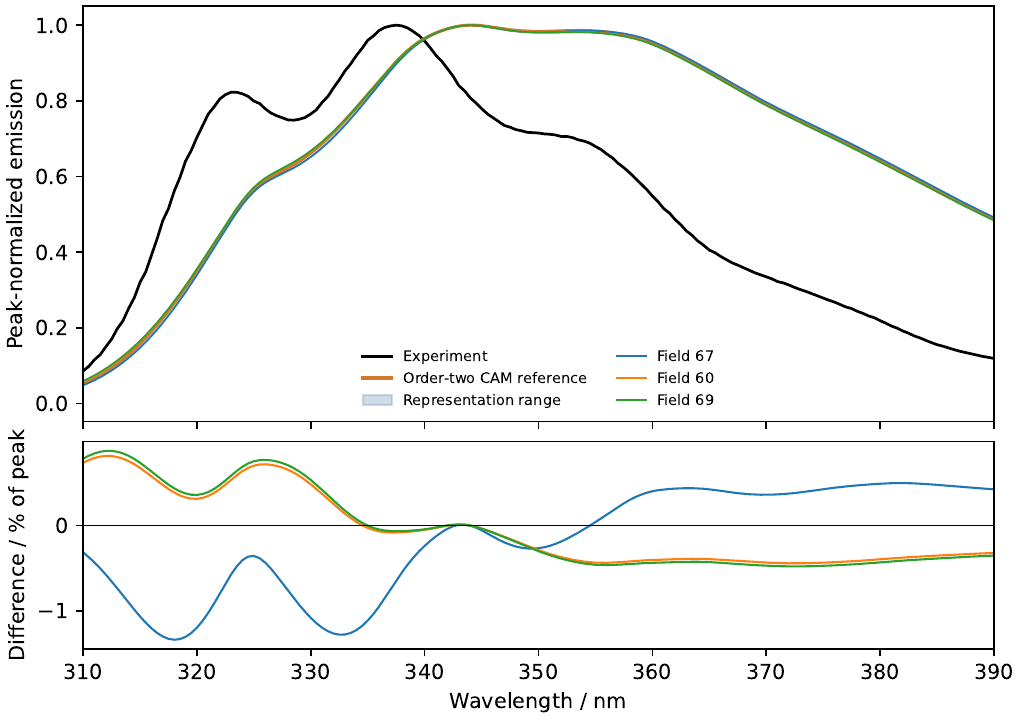}
\caption{Torsional representation sensitivity at fixed gap, harmonic bath
and FC/HT operator. Top: peak-normalized profiles; shading spans the three
training/landscape-selected fields. Bottom: differences from the order-two
CAM reference on the same wavelength grid. Every field uses 51 sine
functions per rotor and the same absorption calibration.}
\label{fig:torsional-representation-envelope}
\end{figure}

Figure~\ref{fig:torsional-representation-envelope} shows the spread of spectra obtained from the tested periodic potential representations.

\FloatBarrier
\subsection{Stage-resolved general C++ TD execution}
\label{sec:TD-stage-costs}
The general Hermitian operator-bank executor records input, Gaussian
preparation, operator preparation, correlation, FFT/interpolation and
numeric export separately. A small correlation buffer separates numeric
output from Gaussian evaluation. The same implementation applies to
arbitrary Hermitian forms and signed operator channels, at zero or finite
temperature. No molecular coefficient or operator-specific shortcut enters
the timing path.

Nine interleaved repetitions per executable and temperature verify the
same prepared physical inputs. Correlations agree byte for byte with
the frozen reference executor, and spectra agree numerically exactly;
each executor also reproduces its own outputs in every repetition.
The stages sum to the recorded internal execution time for each process.
The table gives medians of individual stages; their sum need not equal the
median of the whole process. Process startup and final metadata output
remain in the wall time.

Table~\ref{tab:si-correlation-stages} resolves preparation, thermal correlation and output costs.

\begin{table}[!htbp]
\centering
\caption{One-thread C++ TD stage medians in seconds on the EPYC server
for the 27-mode, three-component FC/HT bank, 512~fs and 4096 intervals.
The physical grids and operators match the ESD comparison.}
\label{tab:si-correlation-stages}
\begin{tabular}{lrr}
\toprule
Stage & 0 K & 298.15 K\\
\midrule
Input & 0.000320 & 0.000336\\
Gaussian preparation & 0.000404 & 0.000537\\
Operator preparation & 0.000134 & 0.000201\\
Correlation & 0.102731 & 0.426756\\
FFT and interpolation & 0.006015 & 0.006240\\
Numeric export & 0.047171 & 0.051069\\
Remaining internal work & 0.001694 & 0.001667\\
Whole process & 0.166185 & 0.492552\\
\bottomrule
\end{tabular}
\end{table}

At 298.15~K, correlation accounts for 87.7\% of the sum of median
internal stages, numeric export for 10.5\%, and FFT/interpolation for
1.28\%. Gaussian and operator preparation together account for about
0.15\%. The thermal Gaussian trace is therefore the principal target for
further acceleration on this case. All paired measurements, stage values,
build settings and representative raw outputs are archived in \texttt{weighted\_and\_stages/}. Thirteen focused regression tests pass, including complex signed operator banks in TI/TD and absorption/emission.

\subsection{Exact reduction of the thermal Gaussian correlation}
\label{sec:thermal-resolvents}
The following optimization changes the analytic Gaussian contraction,
with all populations, mode mixing and operator insertions retained.
It reuses the same Gaussian workspace and supports both electric-dipole
correlations and arbitrary Hermitian operator banks. Archived external campaigns retain their build identities; this subsection
supports the native reduction on identical inputs. Renewed external timings
with the released solver follow in Section~\ref{sec:thermal-release}. No FFT algorithm is changed.

Let $N$ be the number of molecular modes and $M$ the real symmetric
Hermite quadratic matrix. The preparation gives
\begin{align}
 A&=\Omega_i+J^T\Omega_fJ,\qquad
 L=[\sqrt{2\Omega_i},J^T\sqrt{2\Omega_f}],\\
 M&=L^TA^{-1}L-I=W^TW-I,\qquad WW^T=2I,
\end{align}
where $W=\operatorname{chol}(A)^{-1}L$. Thus $M+I$ has rank $N$,
although the finite-temperature trace generally contains $2N$ active
Hermite variables. For diagonal thermal phase factors $Z$ and
$E=Z^{1/2}$, set $P=ZM$ and $S=EME$. The required contractions obey
\begin{align}
 (I-P^2)^{-1}(I+P)Zr &= E(I-S)^{-1}Er,\\
 (I-P^2)^{-1}Z &= \tfrac12 E[(I-S)^{-1}+(I+S)^{-1}]E.
\end{align}
The two Gaussian means coincide. Factoring the complex symmetric
resolvents $I-S$ and $I+S$ therefore removes the explicit matrix square.
The implementation uses pivoted LAPACK LDL$^T$ factors and symmetric
solves, rather than an additional dense matrix product and general LU.

Writing $U=WE$ and $\sigma=\pm1$ gives
\begin{align}
 I+\sigma S&=D_\sigma+\sigma U^TU,\qquad D_\sigma=I-\sigma Z,\\
 C_\sigma&=I+\sigma U D_\sigma^{-1}U^T.
\end{align}
Woodbury solves and the determinant lemma reduce each factorization to
$N$ dimensions. Weighted $W$ columns avoid forming $E^{-1}$, including
for very small thermal populations. If a diagonal denominator has modulus
below $10^{-3}$, the full symmetric resolvents are reconstructed instead.
This switches the numerical solver without altering a population or a mode.
Small systems and zero-temperature reduced traces keep their existing route.

For the stable positive-frequency domain, $\lVert S\rVert<1$ and
$I\pm S$ have positive-definite real part. Their symmetric LDL$^T$
Schur blocks retain this property. Scalar pivots have positive real part;
a two-dimensional block has eigenvalues in the right half-plane, so its
determinant remains off the negative real axis. Summing the principal
block logarithms therefore gives the analytic, unwrapped determinant
logarithm. The sum is not folded back into $(-\pi,\pi)$. The reduced
factors are also accretive: with the initial/final columns of $W$ split,
\begin{equation}
 \operatorname{Re}C_\sigma
 =W_i\operatorname{diag}\!\left[
 \frac{1-|z_i|^2}{2|1-\sigma z_i|^2}\right]W_i^T>0.
\end{equation}
The zero-population limit has coefficient $1/2$. Numerical pivot blocks
are checked; a failed positivity/factorization check uses the established
LU/spectral route. This local certificate also covers models outside the
sufficient global LU-phase bound, avoiding repeated eigenvalue calculations.

Thermal-correlation refinement on the compute server/AMD EPYC 7543, one
thread, identical prepared physical fields and optical operators at
298.15 K. Whole processes include input, preparation and output. The
27-mode bank exports FC, HT/interference and total; the 90-mode bank
exports the complete FC+HT observable. Both use 512~fs and Gaussian
$\sigma=100$~cm$^{-1}$. Time steps are 0.125 and 0.25~fs, respectively.
$t_C$ and $t_P$ denote correlation and complete-process times in seconds;
ref and red denote the reference and reduced solvers.

Table~\ref{tab:si-thermal-correlation} compares thermal-correlation and complete-process timings at fixed physical inputs.

\begin{table}[!htbp]
\centering
\caption{Thermal-correlation timings on the AMD EPYC 7543 server with one thread, identical physical inputs and shared preparation. Medians cover interleaved runs; separate columns report correlation and complete-process costs. Operator components and exported spectra retain their original numerical conventions.}
\label{tab:si-thermal-correlation}
\begin{tabular}{lrrrrrr}
\toprule
System & Runs & $t_C^{\rm ref}$ & $t_C^{\rm red}$ & $t_P^{\rm ref}$ & $t_P^{\rm red}$ & Ratio\\
\midrule
4-pyrone & 9 & 0.404977 & 0.293375 & 0.467471 & 0.355116 & 1.32\\
p-Terphenyl & 3 & 29.928869 & 1.926462 & 29.951272 & 1.950893 & 15.35\\
\bottomrule
\end{tabular}
\end{table}
The corresponding correlation speed factors are 1.3804 and 15.5357.
The 90-mode reference has global log-determinant bound 20.317, exceeding
$\pi$; the accretive route evaluates all 2049 points without a contraction
spectrum. Of these points, 1945 use reduced resolvents and 104 use the
full symmetric solve near diagonal poles. The 27-mode thermal case uses
4023 reduced and 74 full-symmetric evaluations. The zero-temperature
27-mode outputs remain byte-identical with the retained route.

Maximum correlation differences, normalized by the reference peak, are
$2.03\times10^{-14}$ and $7.10\times10^{-14}$; spectral differences
are $9.44\times10^{-16}$ and $1.01\times10^{-14}$. Thirty-six focused
regressions pass, covering nonorthogonal maps, complex/signed operators,
absorption and emission, temperatures through 30000~K, phase-pole grids,
TI compatibility and the shared-library preparation interface. Build settings, physical inputs, unrounded process records and representative correlations/profiles are recorded in \texttt{thermal\_resolvents/}. These comparisons reuse the supplied electronic data and apply no spectral fit.

\subsection{Released thermal solver: external profiles and full execution costs}
\label{sec:thermal-release}
The standalone thermal solver supplies the thermal reduction above.
All comparisons run on the verified AMD EPYC 7543 server with one
thread, including BLAS, from immutable physical inputs. Each of nine
interleaved repetitions measures the complete executable process: input,
Gaussian/operator preparation, correlation, transform and spectral export.
There is no Python interpreter in the timed command and no new electronic
acquisition. The server is shared; individual samples, quartiles and host
load remain in \texttt{thermal\_release/}.

The main-text application table contains six FC+HT molecular comparisons
with FCclasses3 at 0~K. Native profiles are byte-identical to their archived
profiles. An additional 27-mode $S_1$ FC+HT check changes only the existing
temperature parameter to 298.15~K. Nine interleaved runs give native/FCclasses3
medians 0.151038/0.174188~s (factor 1.1533). The native spectrum agrees
with the preceding kernel within $1.58\times10^{-15}$ of the output-column
peak. Its absolute external peak residual is 0.54245\%, integrated absolute
residual 0.14961\%, and area ratio 1.00001049. A separate finite-temperature
FC check on water at 298.15~K gives
0.025844/0.030803~s (native/FCclasses3), retained as a workflow comparison.
Its two native output columns change by at most $1.22\times10^{-16}$ of
their respective peaks. Absolute interprogram residuals on each actual
finite Fourier/export grid are retained without fitted shifts or intensity
normalization:

Table~\ref{tab:si-distributed-profile-check} checks absolute spectra produced by the distributed implementation against archived reference profiles.

\begin{table}[!htbp]
\centering
\caption{Released native versus archived absolute FCclasses3 profiles.
Maximum residuals are percentages of the external peak on the common
physical window. These finite-grid residuals are unchanged by the release;
they are separate from native time-step refinement and complete-manifold
coverage.}
\label{tab:si-distributed-profile-check}
\begin{tabular}{lrr}
\toprule
System & Window (fs) & Maximum residual (\%)\\
\midrule
4-pyrone $S_1$ & 512 & 0.9981\\
4-pyrone $S_1$ & 1024 & 0.2124\\
4-pyrone $S_2$ & 512 & 0.0502\\
4-pyrone $S_2$ & 1024 & 0.0117\\
p-Terphenyl & 512 & 0.0255\\
p-Terphenyl & 1024 & 0.0069\\
4-pyrone $S_1$, 298 K & 512 & 0.5424\\
Water & 1024 & 0.3439\\
\bottomrule
\end{tabular}
\end{table}

The acceptance check retains each archived external residual within
$2\times10^{-10}$ of the external peak, in addition to comparing the
native profile directly with the archived native output. Thus its threshold
is a numerical regression criterion, rather than a claim that every export
grid has the same discretization error. The general auditor takes explicit
file paths, energy conversions and thresholds; four unit tests verify
unit-aware cancellation tails and rejection of changed spectra, shifted
axes, external mismatches and nonfinite inputs.

The direct ESD FC+HT module is also rerun nine times per temperature with
unchanged fields, dipole, all derivatives and captured restart data.
Main-text medians are 0.158490/0.991782~s at 0~K (factor 6.2577) and
0.345945/1.069277~s at 298.15~K (factor 3.0909). These are measured external
ratios, distinct from the paired native algorithm speedups above. The
room-temperature total-profile residual remains 1.8132\% before and
0.073945\% after the separately labelled analytic zero-endpoint audit;
the audited integrated ratio is 1.00067991. FC, HT/interference and total
are all checked independently; no amplitude or energy is fitted.

The offline distribution is built from committed sources in an isolated directory.
It supplies 23 bundled components and 48 wheels; all 24 third-party wheel
versions and hashes match the prior distribution. A fresh isolated
Linux x86\_64/CPython 3.11 environment installs using only these wheels,
passes the dependency check and resolves scientific imports within the
installed environment. All 188 shipped regressions and 17 targeted-selection
extension checks pass, as do the ESALC example and checkpoint replay with
identical state-to-state columns. The native executable and shared library
are rebuilt from the installed sources. Three further installed-kernel
repetitions reproduce correlation and profile files byte for byte for
0~K and thermal 27-/90-mode cases.

For the installed 27-mode thermal bank, median correlation, transform and
export costs are 0.274795, 0.005849 and 0.045270~s. For 90 modes they are
1.740646, 0.000188 and 0.006178~s. Input, Gaussian and operator preparation
sum to 0.001100 and 0.009174~s, respectively; complete-process medians are
0.337160 and 1.761759~s. Stage medians are reported separately rather than
summed as a synthetic measured process. These three-repeat installation
probes are distinct from the nine-repeat external timing campaigns and
from the paired algorithm validation. The source, wheelhouse,
build and archive hashes, full timing records, actual physical inputs and
representative spectra accompany the evidence packet. Python, a compatible
system BLAS/LAPACK and the documented explicit C++ build remain platform
prerequisites; electronic programs are not redistributed.

\FloatBarrier
\section{Camphor: adiabatic E1/M1 spectra}
\label{sec:camphor-ecd-si}
The (1$S$,4$S$) starting geometry is taken from Table III of the
published camphor supporting data.\cite{Kerber2025ECD} It is reoptimized
at B3LYP/def2-TZVP with D3(BJ), and both electronic-state harmonic fields
are calculated with analytic Hessians. The three electronic routes
are S$_0$ \texttt{B3LYP/def2TZVP EmpiricalDispersion=GD3BJ Opt Freq},
S$_1$ \texttt{B3LYP/def2TZVP EmpiricalDispersion=GD3BJ
TD=(Singlets,Root=1) Opt Freq}, and the S$_1$ property calculation at the
converged S$_0$ geometry with the latter route without \texttt{Opt}.
Each electronic job uses 32 threads and 128 GB on the compute server.

The S$_1$ minimum is obtained with internal-coordinate optimization and AO transition-density overlap tracking across geometries. Charge model 5 (CM5) populations accompany the electronic calculations, which use
\texttt{B3LYP/def2TZVP EmpiricalDispersion=GD3BJ
TD=(Singlets,NStates=6,Root=1) Pop=CM5 Guess=Read Force}.
The final point replaces \texttt{Force} by \texttt{Freq=Analytic}.
An exact final Hessian verifies the minimum. All 75 frequencies are
positive; the lowest Cartesian frequency is 118.2179~cm$^{-1}$.
The final projected maximum and RMS gradients are
$2.206\times10^{-4}$ and $3.594\times10^{-5}$ hartree/bohr.
The normal modes used in spectroscopy are recomputed through Wilson GF
in the common curvilinear ESALC-I coordinate representation, including the residual-gradient
curvature correction and local quadratic relaxation. No imaginary mode
is reflected or discarded.

\subsection{Magnetic convention and operator transport}
The 16-component Gaussian ETran state records provide the length-gauge
electric moment in columns 2--4 and the real magnetic-generator record
in columns 8--10 (one-based indexing). The Cartesian derivative block
contains the same ordered components for the target state. In the
$\langle i|\hat O|f\rangle$ convention of the general operator bank,
the electric-equivalent magnetic moment is
\begin{equation}
 \vect m/c=\frac{i\alpha}{2}\vect M_{\rm G},\qquad
 R=\operatorname{Im}[\vect\mu\cdot(\vect m/c)^*]
   =-\frac{\alpha}{2}\vect\mu\cdot\vect M_{\rm G}.
\end{equation}
The same factor multiplies its Cartesian derivatives. Proper
mass-weighted Kabsch rotations align the property frame; both vector
indices of each derivative transform. The common Wilson right inverse
then maps both E1 and M1 derivatives into the same initial normal
coordinates. Their constant terms are transported to the relaxed initial
minimum by the same linear expansion.

An independent conversion to the units printed by Gaussian gives
$R/(10^{-40}\,\mathrm{erg\,esu\,cm/G})
=-235.7217\,\vect\mu\cdot\vect M_{\rm G}$.
This reproduces the printed length-gauge rotatory strengths of all three
available singlets within their four-decimal printing precision. The
target-state printed value is $-4.5222$ in these cgs units. This check
fixes the sign and factor of two before any vibronic contraction.
Electric, magnetic and rotatory strengths are three Hermitian quadratic
forms of the same six-channel complex operator bank. The ECD is retained
as a signed spectrum, and $g=4R/(D+M)$ includes the small M1 contribution
to the unpolarized strength. The often-used $4R/D$ convention is recorded
separately in the evidence. Published camphor comparisons must also
account for enantiomer choice and the 2026 correction to the reference
hotFCHT input.\cite{Kerber2025ECD,Kerber2026Correction}

\subsection{Spectral controls and reproducibility}
Both TD grids span 1024~fs at Gaussian $\sigma=100$~cm$^{-1}$, with
steps of 0.25 and 0.125~fs. The 0~K TI correlation audit uses 41 times
up to 100~fs and positive-norm omission bounds. All spectral refinements
reuse the same electronic data. Model arrays, compact spectra, source
hashes and validations are archived in
\texttt{numerical\_evidence/camphor\_ecd\_20261101/}.
Camphor band strengths without fitted energy or intensity changes.
Calculated $R$ is reported as a magnitude for comparison between enantiomers.
$D$ is in $10^{-3}e^2a_0^2$, and $|R|/\alpha$ in
$10^{-3}e^2a_0\hbar/m_e$. The last column follows the conventional
electric-dipole denominator used in the reference. Experimental values
are those of Pulm \textit{et al.}\cite{Pulm1997Camphor} as numerically
reported in the source of Fig.~1 by Kerber \textit{et al.}\cite{Kerber2025ECD}

Table~\ref{tab:si-camphor-strengths} compares integrated electric, rotatory and dissymmetry values with the enantiomer-matched measurements.

\begin{table}[!htbp]
\centering
\caption{Camphor integrated band strengths with absolute intensity and energy scales. The experimental sign is matched to the calculated enantiomer. Electric strength, rotatory strength and dissymmetry use the conventions defined in the text; discrepancies refer to the stated gas-phase measurements.}
\label{tab:si-camphor-strengths}
\begin{tabular}{lrrrr}
\toprule
Model & $T$ (K) & $10^3D$ & $10^3|R|/\alpha$ & $4|R|/D$\\
\midrule
FC & 0 & 2.18199 & 9.55937 & 0.127880\\
FC+HT & 0 & 4.33441 & 9.71602 & 0.065431\\
FC & 298.15 & 2.18199 & 9.55937 & 0.127880\\
FC+HT & 298.15 & 4.93012 & 9.82792 & 0.058187\\
Experiment & room temperature & 5.15 & 9.58 & 0.054\\
\bottomrule
\end{tabular}
\end{table}

The 0.25-to-0.125~fs step refinement changes the electric and signed
rotatory-strength profiles by less than $2\times10^{-15}$ of their
respective peaks on the common grid in 25000--45000~cm$^{-1}$.
Full FFT-grid integration reproduces all three analytic operator sum
rules to $1.3\times10^{-13}$ relative error. The positivity bound
$|g|\leq2$ is satisfied wherever the numerical ratio is resolved.
The finite TI control evaluates 500000 transitions and 835411 cached
integrals, retaining 14.6936\% of the positive E1+M1 norm. Its unbinned
correlation differences remain within the corresponding omitted-norm
bounds. Extending the operator-aware selector from two to five million
transitions retains 28.8242\% of that norm. The continuous profiles in the main text are the complete
analytic TD profiles, without truncation to this selected line list.

The general operator-bank TI implementation avoids sorting repeated
initial occupation vectors when the initial population is entirely in
the vibrational ground state. In a fixed 500000-transition camphor control,
this reduces the measured execution from 30.32 to 12.82~s while leaving
all line energies, strengths and occupations bitwise identical.
A finite-temperature coupled three-mode control retains the original
distinct-initial-state calculation and is also bitwise identical.
The continuation from two to five million camphor transitions takes
47.09~s on one Oracle thread, including checkpoint recovery and writing.
These timings describe the recorded controls and exclude subsequent
compression of the line archive. The code and validation record are
included in the numerical evidence.

\paragraph{Experimental curve extraction.}
The burgundy experimental $g$ trace in Fig.~6 of Kerber
\textit{et al.}\cite{Kerber2025ECD} reproduces the gas-phase camphor
measurements of Pulm \textit{et al.}\cite{Pulm1997Camphor} The arXiv
source archive contains it as \texttt{camphor\_lorentz0.08eV.pdf}.
The original vector path supplies 156 vertices from 251.232 to
324.365~nm. Axis tick-line positions fix a linear calibration in
wavelength and $g$; no visual pixel tracing, smoothing or fitted
curve is used. The experimental enantiomer has the opposite ECD sign
to the calculated (1$S$,4$S$) species, as explicitly stated in the
source caption, so the experimental sign is reversed for the main
figure. The gray absorption and ECD traces in that source figure are
theoretical; they are excluded from the experimental dataset.

For the profile metric, the measured vector trace is interpolated on a
uniform 0.25~nm grid over 251.25--324.25~nm. Using $4R/D$, the FC and
FC+HT RMS differences are 0.077715 and 0.017051, respectively. No
energy alignment or amplitude fit is applied. The FC+HT $|g|$ maximum
occurs at 285.0~nm, compared with 303.31~nm on the experimental trace.
These wavelengths characterize the $g$ profile, not the maximum of
the unpolarized absorption band. The archive records the vector-PDF
checksum, calibration anchors, extracted values, enantiomer convention
and comparison script. The precision of the line vertices is not
assigned as an experimental uncertainty.

\subsection{General window-directed TI selection}
\label{si:targeted-ti}
The goal is an assigned state-to-state TI transition list whose completeness
and broadened profile are checked against TD within a specified window.
The option is shared by the ordinary FC/HT and Hermitian operator-bank
workflows, including ECD and CPL. It accepts one or
more disjoint intervals $W_r=[l_r,u_r]$, an area-recovery target $\eta$, the
line width and explicit transition/integral budgets. The molecular model,
physical amplitudes and thermal populations are unchanged. The default
unrestricted calculation retains its original selection order.

For amplitudes $A_{if}^{a}$, the positive bank strength of a transition is
$n_{if}=p_i\sum_a|A_{if}^{a}|^2$. Its contribution to a broadened window is
$n_{if}B_r(\nu_{if})$, where, for Gaussian width $\sigma$,
\begin{equation}
 B_r(\nu)=\Phi\!\left(\frac{u_r-\nu}{\sigma}\right)
             -\Phi\!\left(\frac{l_r-\nu}{\sigma}\right).
\end{equation}
Here $\Phi$ is the standard normal cumulative distribution. The pure Lorentz
case uses the corresponding arctangent difference. Transitions outside a window
retain their contribution through the line tails. Neither physical line
strengths nor broadened profiles are renormalized.

A same-model analytic TD calculation supplies $n_{\mathrm{TD}}(W_r)$.
An additional identity quadratic form obtains this positive reference even
when every requested observable is signed. The selected transition sum
must satisfy
\begin{equation}
 \frac{\sum_{if\in\mathcal S} n_{if}B_r(\nu_{if})}
 {n_{\mathrm{TD}}(W_r)}\geq\eta\qquad\text{for every }r.
\end{equation}
This is a numerical reference-based recovery criterion for the requested
windows. The exact global positive-norm omission bound remains separately
reported. A cancelling rotatory-strength integral is never used as a
completeness criterion. For every requested observable, the output also
compares broadened TI and TD profiles on 65 points per window by default,
reporting relative peak and integrated absolute errors. Both are stopping
criteria, with defaults of 2\% and 1\%, respectively, for every observable
and window. The peak error is the maximum absolute TI--TD difference
divided by the TD peak magnitude; the integrated error uses trapezoidal
integration of the absolute difference divided by the integral of the
absolute TD profile. A configurable absolute tolerance and a roundoff
floor of $256\epsilon$ times the operator norm and TD bank peak handle
vanishing channels without dividing by a cancelling signed area. The TD values and
window integrals are evaluated from its periodic Fourier polynomial,
avoiding linear-interpolation errors between FFT nodes. The TD time step,
frequency aliasing and temporal tail still require their own validation.

Each requested window has a distinct ranking built with the existing C2
occupation tree. Unfinished windows are served in round-robin order;
windows satisfying all goals pause. A single shared Hermite cache supplies
all rankings. Packed occupation vectors identify physical transitions,
with full vector comparisons after hashing, so a transition proposed by
several trees is evaluated and counted only once. The budget counts unique
physical transitions, separately from recurrence dependencies. The Gaussian
operator source is exponentially tilted by
$\exp[-\beta(E_f^{\mathrm{vib}}-E_i^{\mathrm{vib}})/2]$ at amplitude level,
with energies in cm$^{-1}$. Thus $\beta$ has units of cm. A deterministic
initial bias follows from the TD bank mean and variance and the mean within
each requested window. Its magnitude is capped at three inverse window
widths and reduced until the tilted Gaussian is normalizable. Positive
per-channel ranking weights may balance operators with different strength
scales. Both the tilt and these weights affect ranking only; the physical
recurrence uses the original source. No joint-intensity factorization or
weak-parent pruning is assumed.

Checkpoints preserve all occupation heaps, deduplication keys and the
shared physical line data. Restoration recomputes the window sums from stored physical
lines. Window goals may therefore change while retaining the same
model, domain and occupation ordering. A changed ranking bias is rejected
by the checkpoint identity. Each retained line keeps both vibrational-state labels. The same controls
are available through the ESALC and optical requests and Python interfaces.
Validation covers FC/HT, ECD/CPL, both temperatures and transition
directions, multiple windows and restart consistency; default outputs
remain bitwise unchanged.

The automatic assignment report retains full initial and final occupation
vectors and readable mode labels. For each window it reports signed
observable and positive-bank totals over all selected transitions,
separating line centers inside the interval from lower and upper external
tails. The displayed assignments are ranked by their broadened window
contributions (default: 50 per observable and bank); this display limit
does not truncate any total or the complete transition columns.
FC, HT and interference components are evaluated for the listed assignments
using the same exact recurrence. Export to JSON is atomic.

Seventeen extension checks pass. These include native and Python paths,
restart consistency, optical labels, component sums, empty budgets and
multiword occupation packing for 150 coordinates. A two-window finite-domain
control yields 1396 unique transitions and removes 40 duplicate candidates.
An area recovery above 99\% with a signed profile error of 2.54\% correctly
fails the combined criterion; enlarging the domain closes both tests.

The camphor continuation reuses the four-million-line checkpoint
($\beta=0.001$~cm), adding 417,792 transitions. At 4,417,792 transitions,
the positive-bank recovery in 30000--33000~cm$^{-1}$ is 99.2758\%,
with 25.0392\% global bank recovery. The E1, M1 and signed ECD relative
peak errors are 2.0000\%, 1.7860\% and 1.5607\%; their integrated absolute
errors are 0.7260\%, 0.7114\% and 0.6186\%, respectively. All stopping
tests pass using unrounded values. The incremental continuation takes
19.83~s on the compute server with one thread, including restoration, report generation
and checkpoint writing; this is not the cost of generating the full list
from scratch. A zero-new-transition replay verifies the exported report.

\paragraph{Assignment stability and window-edge validation.}
The final controls use 129 profile points per window. A zero-new-transition
replay of the 4,417,792-line checkpoint first reproduces its accepted
profile (the small L1 change reflects the denser comparison grid).
The molecular inputs, physical property expansion, $\beta=0.001$~cm,
0~K populations and $\sigma=100$~cm$^{-1}$ remain fixed.
The final continuation budget is twelve million unique transitions and
32 million cached integrals. The original checkpoint is preserved.
With a 99.5\% bank target, 1\% peak and 0.5\% L1 tolerances, the
30000--33000~cm$^{-1}$ calculation closes at 8,689,408 transitions:
99.6551\% recovery, 0.7673\% ECD peak error and 0.2928\% ECD
L1 error. A preliminary eight-million-line control gives 1.0918\% electric peak
error and therefore does not pass the tighter stopping test. Continuing
that checkpoint takes a further 55.13~s,
including restoration, profiles, assignments and checkpoint writing.

The wider interval retains the 99.5\% bank goal and uses the original
2\% peak/1\% L1 tolerances; the central interval is subsequently checked
against the tighter 1\%/0.5\% tolerances. Widening to
29750--33250~cm$^{-1}$ closes at 11,393,024
transitions with 99.5000\% recovery, 1.1334\% ECD peak error and
0.4131\% ECD L1 error in that wider interval. The wider continuation
takes 101.01~s. A zero-new-transition replay
in the original interval gives 99.7470\% recovery, 0.5700\% ECD
peak error and 0.2149\% ECD L1 error. Its profile differs from the
tighter central calculation by at most 0.1973\% of the TD peak.
These are incremental timings on the compute server with one BLAS/OpenMP thread,
not independent from-scratch benchmark costs.

The 20 leading positive-bank assignments and, separately, the 20 leading
absolute ECD window contributions have identical occupation sets before
and after each control. Comparing the full initial/final occupation
vectors also confirms unchanged transition energies and physical FC, HT
and interference strengths (absolute checks of $10^{-8}$~cm$^{-1}$
and $10^{-14}$ in the supplied strength units). The archived reports
retain ranking order and signed contributions for independent inspection.
The comparison concerns the same original interval after the enlarged-window
continuation, so an edge-weight change is not mistaken for a change of
physical assignment.

Table~\ref{tab:camphor-targeted-assignments} illustrates the resulting
state-to-state assignments. Mode labels follow the ordered final harmonic
modes; every initial state is the vibrational ground state. The full
machine-readable report also contains window contributions and FC/HT
components.
\begin{table}[!htbp]
\centering
\caption{Six leading positive-bank assignments in the targeted camphor
window, sorted by transition energy. Signed rotatory strengths use the electric-equivalent atomic-unit
convention of the operator bank.}
\label{tab:camphor-targeted-assignments}
\begin{tabular}{llrr}
\hline
Label & Final occupations & $\nu$ (cm$^{-1}$) & $R$ ($10^{-9}$ a.u.)\\
\hline
A & $v_{1}^{3}\,v_{2}^{2}\,v_{9}^{1}$ & 31469.96 & $-1.1261$\\
B & $v_{1}^{4}\,v_{2}^{2}\,v_{9}^{1}$ & 31588.07 & $-1.3073$\\
C & $v_{1}^{4}\,v_{2}^{3}\,v_{9}^{1}$ & 31757.52 & $-1.2698$\\
D & $v_{1}^{3}\,v_{2}^{2}\,v_{9}^{1}\,v_{15}^{1}$ & 31983.21 & $-1.2868$\\
E & $v_{1}^{4}\,v_{2}^{2}\,v_{9}^{1}\,v_{15}^{1}$ & 32101.32 & $-1.5034$\\
F & $v_{1}^{4}\,v_{2}^{3}\,v_{9}^{1}\,v_{15}^{1}$ & 32270.77 & $-1.4464$\\
\hline
\end{tabular}
\end{table}

\paragraph{Physical content of the assigned modes.}
Table~\ref{tab:camphor-assigned-mode-character} identifies the four final-state
modes appearing in Table~\ref{tab:camphor-targeted-assignments}. The final
ESALC-GF field and its mass-metric Cartesian tangent map are reconstructed
from the same Gaussian data and frozen coordinate representation as the production spectrum.
All 75 frequencies reproduce the archived ordering with zero observed
difference; the normalized mass-weighted Cartesian modes are orthonormal
to $5.6\times10^{-14}$. These are the transported ESALC-GF frequencies
used by TI/TD. The archived modal vectors, geometry and animated XYZ
sequences preserve their one-based atom numbering and arbitrary modal phase.

\begin{table}[!htbp]
\centering
\caption{Physical interpretation of the final-state modes used in the
camphor state-to-state assignments. Labels identify their collective motion.}
\label{tab:camphor-assigned-mode-character}
\begin{tabular}{lrp{0.67\linewidth}}
\hline
Mode & $\omega$ (cm$^{-1}$) & Character\\
\hline
$v_1$ & 118.109 & Carbonyl/cage wag with gem-methyl torsion.\\
$v_2$ & 169.447 & Coupled torsion of the two geminal methyl groups.\\
$v_9$ & 297.993 & Cage/methylene deformation with methyl mixing.\\
$v_{15}$ & 513.254 & Carbonyl bend/pyramidalization and cage-angle motion.\\
\hline
\end{tabular}
\end{table}

For a unit mass-weighted Cartesian mode $e$, the atomic kinetic
participation is $P_a=\sum_\alpha |e_{a\alpha}|^2$.
Carbonyl O1 carries 38.18\% of this norm in $v_1$; the C6H$_2$ group
(C6, H13, H15) carries 32.90\% in $v_9$; and carbonyl C4 carries
20.67\% in $v_{15}$. Localized methyl-rotation vectors are constructed
by rotating each methyl's hydrogen atoms about its C--C axis and
normalizing in the same mass metric. The two geminal axes C3--C9 and
C3--C11 have summed squared overlaps with $v_2$ of 68.97\%, compared
with 15.90\% in $v_1$. Normalized analytic primitive-gradient projections
identify O1--C4--C2 bending and the out-of-plane coordinate
$U(4,2,7,1)$ among the leading contributions to $v_{15}$.
Atomic kinetic fractions, rigid-rotor overlaps and primitive-direction
projections are distinct geometric measures; the latter are not summed
as a potential-energy distribution. The archived report retains their
individual values, source FCHK/coordinate representation hashes and the molecular bond graph.

\FloatBarrier
\section{Electronic continuity, operator sensitivity and external controls}
\label{sec:final-external-controls}

\subsection{Electronic identity along the frozen $S_2$ chord}
The five existing TDA-B3LYP/6-31+G(d) observations are reused without
additional electronic points. Each logged manifold is selected by matching
all three excitation energies to its formatted checkpoint, at the precision
printed in the log. Selecting the last manifold indiscriminately would mix
the requested point with displaced frequency subcalculations. The electronic-state interface imports
the AO basis and orbitals, restores the intended Cartesian frame, and computes
cross-geometry AO overlaps. For printed single-excitation coefficient
matrices $X^L_p,X^R_q$, the normalized fingerprint is
\begin{equation}
 O_{pq}=\frac{\operatorname{Tr}[(X^L_p)^\dagger S_{LR}^{\rm MO}
                X^R_q(S_{LR}^{\rm MO})^\dagger]}
 {\|X^L_p\|_F\|X^R_q\|_F},\qquad
 S_{LR}^{\rm MO}=C_L^\dagger S_{LR}^{\rm AO}C_R.
\end{equation}
Thus orbital phases and mixing are transported rather than inferred from
unchanged orbital numbers. The printed configurations are thresholded;
these fingerprints are not complete many-electron wavefunction overlaps.

Table~\ref{tab:si-pyrone-state-tracking} follows electronic-state identity along the inversion path.

\begin{table}[!htbp]
\centering
\caption{Adjacent electronic fingerprints for root 2. The margin is its
absolute overlap minus the largest overlap with a different printed root.}
\label{tab:si-pyrone-state-tracking}
\begin{tabular}{lrr}
\toprule
Chord interval $t_L\to t_R$ & $|O_{22}|$ & Margin\\
\midrule
$-1\to0$ & 0.92972 & 0.80817\\
$0\to0.5$ & 0.98086 & 0.91777\\
$0.5\to1$ & 0.98322 & 0.91708\\
$1\to1.5$ & 0.98586 & 0.91600\\
\bottomrule
\end{tabular}
\end{table}

Root 2 remains the best match in every comparison. Its nearest electronic
gap ranges from 0.45329~eV at the planar point to 0.09595~eV at the outer
point. This supports the identity of the sampled adiabatic surface; it
does not bound derivative couplings or exclude unsampled avoided crossings.
The single-surface spectrum therefore retains its stated nonadiabatic
limitation, particularly beyond the acquired chord. Analytic transition
derivatives are complete at all five points. Phase-matched finite dipole
secants agree with the corresponding averaged analytic path derivatives
in sign and trend; the stored differences also resolve finite-path
nonlinearity. This is not an infinitesimal derivative-step convergence
test. These path properties are distinct from the separate $S_0$-geometry
linear property expansion used for the optical calculation.

\subsection{Conditioning of the destructive FC/HT interference}
The final saved FC and HT vector amplitudes are contracted without changing
nuclear states. A simultaneous complex phase applied to both components
changes total strengths by at most $2.3\times10^{-16}$ relatively. Changing
the HT sign alone instead changes the physical operator and is retained
only as a negative control. For the lowest line,
$S_{\rm FCHT}/(S_{\rm FC}+S_{\rm HT})=0.159315$, as already reported.
Scaling every HT amplitude by $1\pm0.01$ changes its intensity by
$-2.385\%$ and $+2.413\%$, respectively; a $\pm5\%$ scale variation gives
$-11.64\%$ and $+12.36\%$. The ordering here is increasing and decreasing
HT scale. The other three, HT-dominated lines show the expected approximately
$\pm2\%$ change for a 1\% amplitude variation. The logarithmic first-line
condition number with respect to the HT scale is $-2.3991$.
These are deterministic conditioning probes, not electronic-structure
uncertainty estimates. Together with the independent Gaussian-source and
basis controls, they distinguish numerical phase consistency from sensitivity
to the chosen property model.

\subsection{Same-checkpoint Gaussian l718 comparison}
\label{sec:l718-execution}
The two Gaussian executions use the same original
B3LYP/6-31+G(d) H$_2$O and H$_2$O$^+$ checkpoints as the native Cartesian
AH/FC comparison. They use a unit electronic transition moment, not the
continuum photoionization operator of the main water application. Both
executions terminate normally. Although their additional input requests
zero temperature, their simulation summaries report 298.15~K; the native
comparison therefore uses 298.15~K. The l718 TD output supplies separate
zero-temperature and finite-temperature columns. Only the latter enters
this comparison. Input routes, checkpoint hashes and original output logs
are archived with the nano-MATRIX evidence.

Gaussian broadening uses HWHM 50~cm$^{-1}$.
The requested window extends from $-500$ to $12000$~cm$^{-1}$ relative to
the origin. The native calculation enumerates 33915
transitions with initial/final total-occupation limits 4/16 and all three
excitation classes. Its retained population and FC strength differ from
unity by less than $4\times10^{-15}$; neither is renormalized. Its origin
is 101071.61226~cm$^{-1}$. The three nonzero printed l718 line spacings,
1483.1719, 3284.1761 and 4767.3479~cm$^{-1}$, agree with the native values
within $5.1\times10^{-5}$~cm$^{-1}$, consistent with their output precision.

Molar extinction is obtained independently from
$f=(2/3)(\widetilde\nu/E_h^{\rm cm})S$ and
\begin{equation}
 \int\epsilon(\widetilde\nu)\,d\widetilde\nu
 =\frac{N_A e^2}{40\epsilon_0m_ec^2\ln 10}\,f,
\end{equation}
with $\epsilon$ in dm$^3$~mol$^{-1}$~cm$^{-1}$ and wavenumber in
cm$^{-1}$. The integral coefficient is derived from SI constants.
There is no fitted intensity, area or origin factor.

Table~\ref{tab:si-gaussian-profile-errors} compares absolute profiles at the same physical temperature.

\begin{table}[!htbp]
\centering
\caption{Same-checkpoint profile differences at the actual reported
temperature. Peak errors use each external profile's maximum; integrated
errors are absolute differences divided by its area.}
\label{tab:si-gaussian-profile-errors}
\begin{tabular}{lrrr}
\toprule
External reference & Peak error & Integrated error & Native/external area\\
\midrule
l718 TI & $4.7084\times10^{-4}$ & $8.4284\times10^{-4}$ & 1.000025\\
l718 TD, defaults & $7.4632\times10^{-3}$ & $8.0390\times10^{-3}$ & 0.999472\\
\bottomrule
\end{tabular}
\end{table}

Each native profile is evaluated on the corresponding external output grid;
TI and TD have different grid spacings. Interpolating external TD onto the
TI grid gives an external TI/TD peak difference of $7.2046\times10^{-3}$.
Thus the default external TD execution is not treated as an independently
converged reference at the native TI precision. No additional
TD control keywords are introduced. This is an accuracy comparison, not a
program-to-program performance benchmark. The additional executable and
timing controls below address these separate questions.

\begin{figure}[!htb]
\centering
\includegraphics[width=.9\linewidth]{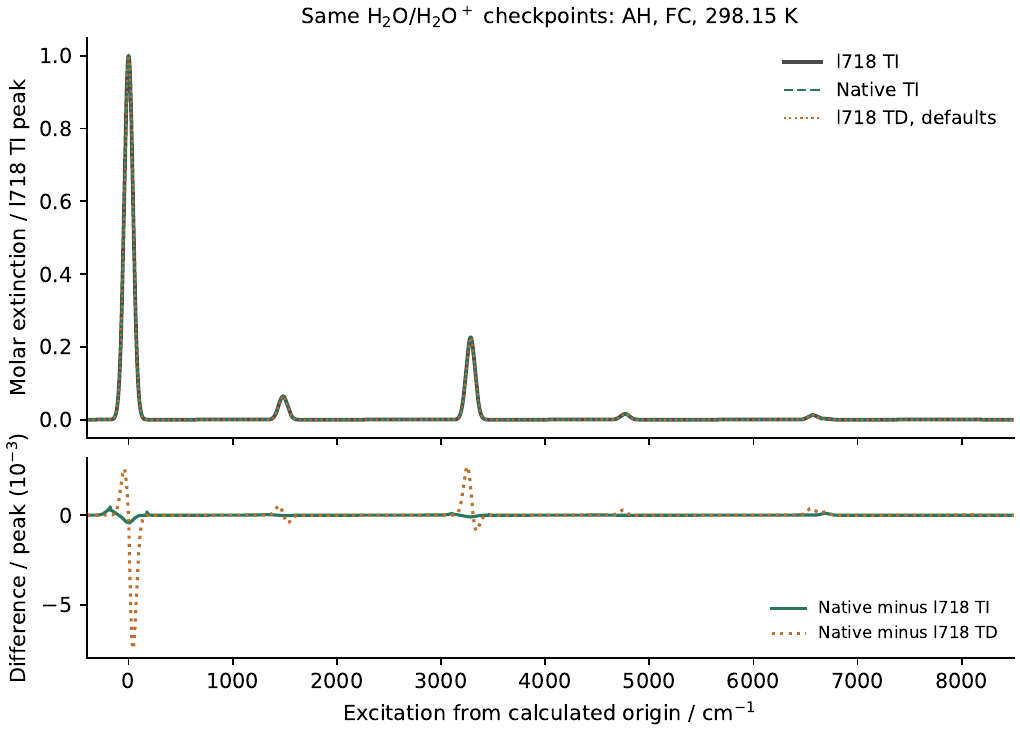}
\caption{Same-checkpoint AH/FC profiles. A single common display divisor,
the l718 TI peak, is applied after the absolute comparison; profiles are
not individually normalized. Residuals retain the larger default external
TD discrepancy.}
\label{fig:si-gaussian-profiles}
\end{figure}

Figure~\ref{fig:si-gaussian-profiles} displays the assigned Gaussian reference and the independent harmonic profiles.

\subsection{An experimental progression excluded from calibration}
Table~I of Kiefer and Bernstein reports the I$_2$/CCl$_4$ Raman shifts
used here.\cite{Kiefer1973Solution} The two lowest bands alone set
$\omega_e=212.8$ and $\omega_ex_e=0.7$~cm$^{-1}$ in the Morse form
$V=D[1-\exp(-a(R-R_0))]^2$, with
$D=\omega_e^2/(4\omega_ex_e)=16172.8$~cm$^{-1}$.
The resident ZAFF Morse function and sine-basis weak kinetic operator
give the levels, with the $^{127}$I$_2$ reduced mass and $J=0$.
The other 13 measured bands, $v=3$--15, are excluded from calibration.
Their maximum error is 5.3~cm$^{-1}$, compared with 163~cm$^{-1}$ for
the harmonic model retaining the same curvature.

The 192/384 and 256/512 basis/quadrature pairs use radial walls
$[3.5,8.5]$~bohr; the final 320/640 pair expands them to
$[3.2,9.0]$~bohr. The last change is $4.2\times10^{-11}$~cm$^{-1}$,
and the numerical spacings agree with the exact Morse expression within
$2.9\times10^{-11}$~cm$^{-1}$. The reference distance merely locates
this nonrotating single-surface well. This empirical progression check
does not establish an ab initio solvent potential, optical intensities or
a polyatomic coupled-surface validation. No gas-phase intensities from
the separate iodine example are assigned to this solution experiment.
The transcribed observations and unrounded predictions are archived.

\begin{figure}[!htb]
\centering
\includegraphics[width=.94\linewidth]{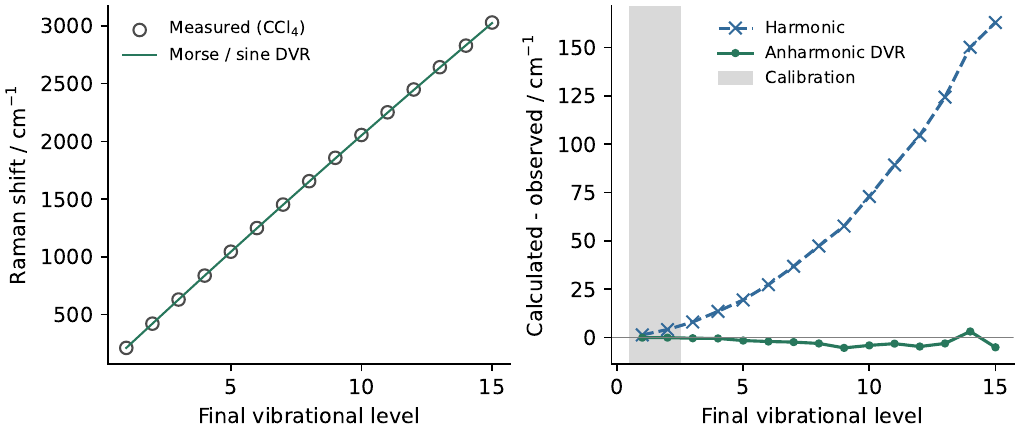}
\caption{I$_2$/CCl$_4$ experimental progression and the two-band-calibrated
anharmonic calculation. The shaded region contains the only calibration
bands. Higher overtones test extrapolation of this empirical potential;
symbols show tabulated positions, without invented experimental error bars.}
\label{fig:si-iodine-holdout}
\end{figure}

Figure~\ref{fig:si-iodine-holdout} distinguishes the two calibration bands from the higher iodine overtones used for testing.

\subsection{Final inversion-domain and optical-curvature controls}
The final 43298-word, 24-contracted-function calculation is repeated with
inversion walls $[-4,4]$, instead of $[-3.5,3.5]$, using the same periodic
potential, 96 primitive functions and 256 quadrature points. The reference
states provide a warm start through the physical sine-function overlap on
the intersection of the two intervals. Coefficient vectors at different
walls are not identified by index. The largest original-operator residual
is $8.87\times10^{-7}$~cm$^{-1}$. The four absolute positions change by
0.000656, 0.000388, 0.000634 and 0.000634~cm$^{-1}$; their relative FC+HT
strength changes are $6.36$, $5.27$, $3.14$ and $3.23\times10^{-6}$,
respectively. An independent 64-degree oscillator-source contraction
agrees with the direct coordinate/Fock contraction within
$3.0\times10^{-16}$ of the retained positive strength. This is a domain
control at fixed primitive/quadrature counts, complementary to the resolution controls; it does not support the potential outside the acquired
electronic domain.

The five existing dipoles and analytic path derivatives also permit a
finite optical-curvature sensitivity. AO-transported root-2 transition
fingerprints anchor the path to the separate $S_0$ property reference
with overlap $-0.96655$; the common phase is then fixed before contracting
amplitudes. Write the path as $\mathbf x(t)=\mathbf x_0+t\mathbf d$ and
its physical inversion dihedral as $\phi(t)$, with $\phi(0)=0$. Subtract
$\boldsymbol\mu(0)+t\boldsymbol\mu'(0)$ from the observed dipoles and
subtract $\boldsymbol\mu'(0)$ from their derivatives. The resident
multi-component constrained-SVD property fitter fits these residuals to
the periodic angular features
\begin{equation}
 f_p(\phi)=\frac{[2\sin(\phi/2)]^p[\cos(\phi/2)]^{p\bmod2}}{s^p},
 \qquad p=2,\ldots,d,\quad d=3,4,
\end{equation}
where $s$ is the largest sampled absolute angle. Value and derivative
rows use relative weights 1 and 0.5 in the path coordinate. Both added
responses have zero value and first derivative at the planar anchor;
the existing linear optical operator is retained and the fitted curvature
is added coherently. The scalar mixed Gaussian overlap, obtained from the
same recurrence as FC/HT, carries this additional coordinate-dependent
operator without division by a dipole component. Its independent FC-channel
check has maximum absolute error $4.34\times10^{-19}$.

The $t=0.5$ point, including its slope, is excluded in the diagnostic fit.
For orders 3 and 4 the held-out dipole-vector errors are
$5.62\times10^{-4}$ and $1.38\times10^{-4}$~a.u.; slope errors are
$1.91\times10^{-3}$ and $3.01\times10^{-4}$~a.u. per unit $t$.
The final fits use all five points. Their lowest-line strengths are
$4.2230\times10^{-6}$ and $4.3778\times10^{-6}$~a.u.$^2$, compared with
$2.0487\times10^{-6}$ for the linear operator. Relative changes of the
three upper lines are, respectively, $(0.2675,0,-0.0062)$\% and
$(-0.2504,0,0.0061)$\%. The nuclear states and all potential parameters
remain unchanged. This large response of the cancelled weak line is a
model sensitivity, not a confidence interval or a replacement validated
spectrum. The continuation beyond $t\in[-1,1.5]$ and nonlinear transverse
properties remain unvalidated; the two orders do not bound the exact
operator. The main-text table therefore retains the explicitly named
linear FC/HT model.

\subsection{Executed FCclasses3 harmonic comparisons}
FCclasses3 3.0.4 is compiled on the Linux compute server with GNU Fortran
8.5.0 and OpenBLAS. The official download endpoint was unavailable during
acquisition; the archived provenance identifies the public source mirror,
archive SHA-256 and installed executable SHA-256. No third-party source
modification is made. Serial compilation resolves the Fortran-module build
ordering, with user-local BLAS/LAPACK library links.

The implicit-mode interface receives the identical 27-mode $S_1$ and $S_2$
AH models and their real transition operators. If the native mass-weighted
map is $\mathbf Q_f=J_n\mathbf Q_i+\mathbf k_n$, FCclasses receives
$J_F=J_n^{-1}$ and $\mathbf K_F=-J_n^{-1}\mathbf k_n$, in mass-weighted
atomic units. Dimensionless initial-mode derivatives are converted by
$\partial\mu/\partial Q_{i,k}=g_k\sqrt{\omega_{i,k}/C}$, where
$C=219474.6313705$~cm$^{-1}$/hartree. The frequencies, displacement and
nonorthogonal matrix are not refitted or orthogonalized. Both programs use
0~K and Gaussian HWHM 50~cm$^{-1}$, with 32768 positive-time midpoint
samples in FCclasses and the corresponding spacing plus the zero endpoint
in the native TD implementation. Profiles are compared as absolute molar
extinction coefficients on the FCclasses grid, without peak, area or
energy adjustment.

Table~\ref{tab:si-fcclasses-profile-errors} reports absolute profile residuals against the established harmonic reference.

\begin{table}[!htbp]
\centering
\caption{Absolute harmonic profiles against executed FCclasses3 TD.
Errors are normalized by the external peak or integrated area.}
\label{tab:si-fcclasses-profile-errors}
\begin{tabular}{lrrr}
\toprule
Model & Peak error & Integrated error & Native/external area\\
\midrule
$S_2$ FC & $2.3852\times10^{-4}$ & $3.5663\times10^{-5}$ & 1.0000096\\
$S_2$ FC+HT & $7.9172\times10^{-5}$ & $1.7802\times10^{-5}$ & 1.0000100\\
$S_1$ FC+HT & $5.2164\times10^{-4}$ & $1.8571\times10^{-4}$ & 1.0000091\\
\bottomrule
\end{tabular}
\end{table}
The diagnostic FCclasses \texttt{corr.dat} precedes the final spectral
normalization. It is archived but is not identified with a normalized
native autocorrelation. Its raw values are not used as a convergence
criterion. FCclasses' exported spectrum already includes its map-determinant
factor; applying it again would spoil the absolute comparison.

\begin{figure}[!htb]
\centering
\includegraphics[width=.97\linewidth]{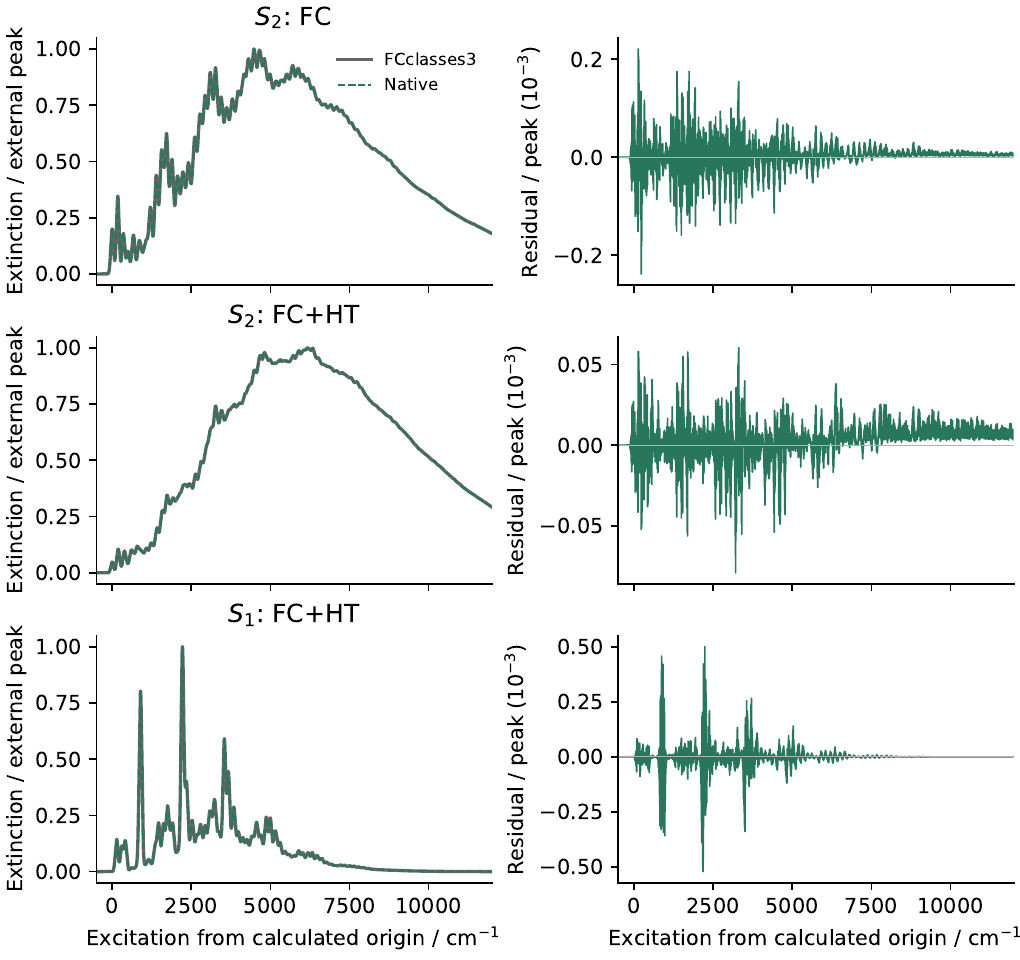}
\caption{Independent 27-mode harmonic calculations with the same supplied
mode map and transition operator. A common external-peak divisor per panel
is used only for display. The right-hand panels show absolute residuals on that
same scale, magnified by $10^3$; curves are not separately normalized.}
\label{fig:si-fcclasses-profiles}
\end{figure}

Figure~\ref{fig:si-fcclasses-profiles} compares complete 27-mode profiles and their residuals on identical physical inputs.

\subsection{Timing and recovered intensity: three programs}
The original water checkpoints permit a common Cartesian AH/FC model in
all three programs, using a unit transition dipole, 298.15~K and Gaussian
HWHM 50~cm$^{-1}$. The original l718 inputs are reused unchanged.
FCclasses uses \texttt{BOLZTHR=1e-12}, retaining the small hot-state
populations, and \texttt{CLASSIC=NO}, excluding unrelated classical
simulations. Its default Boltzmann threshold 0.8 would retain only the
initial vacuum here. Prepared-input runs are executed three times on the
same shared server. The C++ and FFTW-enabled FCclasses measurements use
one thread; the archived l718 measurements retain their four-thread
inputs. Both C++ and FCclasses export 2501 points across the same interval;
C++ uses a 5~cm$^{-1}$ grid and FCclasses reports photon energies to its
printed precision. The l718 output grid remains as specified. Wall times include program startup, input and spectrum
export, but exclude installation, electronic calculations and checkpoint
conversion. They are not exclusive-host timings. Output grids and TI
screening settings remain explicit in the archived inputs.

Table~\ref{tab:si-water-execution} reports water execution times together with recovered strength.

\begin{table}[!htbp]
\centering
\caption{Water execution costs and strength recovery for prepared physical inputs. Wall times are medians, with observed ranges in parentheses.
TD includes the infinite harmonic manifold rather than selecting a finite
list of transitions. The last column separately measures the exported
profile error relative to the dense native TI reference.}
\label{tab:si-water-execution}
\begin{tabular}{llrrr}
\toprule
Program & Route & Wall seconds & TI recovery (\%) & Peak error (\%)\\
\midrule
FCclasses3 & TI & 0.194 (0.194--0.197) & $\simeq100$ & 0.0206\\
C++ & TI & 0.030 (0.030--0.031) & $\simeq100$ & $<10^{-10}$\\
l718 & TI & 9.148 (7.954--10.465) & 100.00 printed & 0.0471\\
FCclasses3 & TD & 0.028 (0.028--0.029) & infinite manifold & 0.3505\\
C++ & TD & 0.022 (0.022--0.023) & infinite manifold & 0.0358\\
l718 & TD & 0.798 (0.796--0.868) & infinite manifold & 0.7463\\
\bottomrule
\end{tabular}
\end{table}
The native TI retains 33915 transitions and recovers
$0.9999999999999967$ of its exact positive strength sum. The FCclasses
thermally weighted stick zero-order moment is 0.9999999999974113.
The l718 progression is rounded to 100.00\%, so its printed precision is
not equated with machine-precision closure. A finite output window also
differs from full TI recovery: the native TI retains 99.9928\% after
convolution within the exported $[-500,12000]$~cm$^{-1}$ interval relative
to the origin. TD has complete analytic state coverage, while its finite
grid still controls the numerical profile error shown above. These runs
therefore do not constitute a performance comparison at identical profile
accuracy. Gaussian preparation is included in each measured C++ process;
no transition strengths or Gaussian recurrence parameters are supplied
as precomputed input.

The standalone C++ backend has dynamic mode dimensions and exact
occupation keys, with declared state, transition and cache budgets.
It accepts the physical frequencies, full nonorthogonal mode map,
displacement and complex linear position/momentum operators, and
reconstructs the same Gaussian internally. LAPACK supplies dense
factorizations and PocketFFT supplies the Fourier transform. The latter
is distributed with its license and pinned source checksum. Short
occupation keys use inline storage as an allocation optimization;
larger systems follow the same recurrence with dynamic storage.
The regression tests cover TI assignments and FC/HT/interference
components for 1, 3, 7 and 27 modes, thermal complex TD correlations,
signed tensor observables, emission, Voigt profiles and the cache-budget
failure boundary. TD samples the damped water correlation at 4097 times
and pads the converged window by eight for the required Fourier spacing;
its final relative damped amplitude is $2.1\times10^{-15}$. The
independent 32769-sample control reproduces the profile within
$8.1\times10^{-15}$ of the peak. TI retains 33915 transitions.

A general Hermitian operator bank shares the Gaussian factorization and
exact recurrence across all electronic tensor channels. Signed bilinear
forms retain electric/magnetic and multipole interference without clipping
or strength renormalization. This representation supplies the TI and TD
kernels for absorption, emission, ECD/CPL, weak-field MCD and TPA/TPCD.
Gaussian, Lorentz and Voigt profiles are evaluated in C++; the Voigt
kernel uses the MIT-licensed Faddeeva implementation by Steven G. Johnson,
with its source and license included in the software distribution.
TI calls that request individual transitions do not construct an unused
broadened spectrum.

Table~\ref{tab:si-molecular-td-execution} compares complete harmonic TD executions at matched time steps and windows.

\begin{table}[!htbp]
\centering
\caption{Matched-step/window molecular FC+HT TD executions. All use one
thread, 32768 intervals over 10000~fs and 65536 exported spectral points per
program. FCclasses uses midpoint sampling and native TD includes the zero
endpoint. Both times include preparation and spectral export. Parentheses
show the observed range of three runs; peak errors refer to absolute
profiles on the common physical comparison window.}
\label{tab:si-molecular-td-execution}
\begin{tabular}{lrrr}
\toprule
Model & C++ seconds & FCclasses3 seconds & Peak error (\%)\\
\midrule
$S_1$, 27 modes & 0.867 (0.867--0.867) & 2.472 (2.472--2.472) & 0.05213\\
$S_2$, 27 modes & 0.867 (0.867--0.867) & 2.472 (2.472--2.472) & 0.007956\\
p-Terphenyl, 90 modes & 9.839 (9.789--10.238) & 45.845 (45.751--46.049) & 0.003653\\
\bottomrule
\end{tabular}
\end{table}
The median ratios are 2.85 for each 27-mode input and 4.66 for p-terphenyl.
Integrated native strengths close the exact analytic sums within
$2\times10^{-13}$ relative error. The native half-step controls change the
profiles by at most $2.2\times10^{-15}$ of their peaks. The integrated absolute
native/FCclasses profile differences are 0.01867\%, 0.001592\% and 0.001463\%
of the external area, respectively. These are full harmonic-manifold TD
calculations, not intensity-selected TI lists.

The p-terphenyl benchmark reuses the supplied 90-mode generalized-vertical
field and its FC+HT operator. Its effective quadratic pair is passed through
the same general IMPLICIT adapter, without additional electronic points or a new
adiabatic optimization. The time step and window are identical in the two
programs. FCclasses writes a full one-sided FFT diagnostic interval extending
beyond the native Nyquist interval. Native output retains the declared physical
window from $-1000$ to $14000$~cm$^{-1}$ relative to the model origin,
but matches the total exported point count; absolute profile residuals use
the common window without endpoint extrapolation or fitted normalization.
Thus temporal resolution and spectral-output counts are matched, while the
exported frequency ranges are recorded separately.

Each timed process produces one FC+HT vector profile with its three Cartesian
components. The gains do not require reuse across multiple independent
operator banks. Alternating program order and preserving all per-process
records gives the ranges above; the server is shared, without exclusive-host
or general scaling claims. Executable hashes, inputs, profiles, refinement
controls and raw outputs are frozen in
\texttt{fcclasses\_complete\_20261030}.

The TD logarithm has an explicit branch certificate. For the thermally
weighted real symmetric Gaussian matrix $C=DMD$, the singular values $s_j$
with $s_j<1$ give the bound
\begin{equation}
 \left|\log\det[I-(ECEC)]\right|
 \le -\sum_j\log(1-s_j^2),
 \qquad |E_{jj}|=1.
\end{equation}
It follows by expanding the analytic logarithm and applying Schatten
H\"older to each trace. If the bound is strictly below $\pi$, the principal
LU determinant is on the required analytic branch at every sampled time.
The archived comparison builds evaluate the spectral logarithm when
this sufficient bound fails. The further thermal evaluator in
Sec.~\ref{sec:thermal-resolvents} obtains the same analytic branch from
accretive symmetric pivot blocks. The bounds are 0.380 and
2.437 for the $S_1$ and $S_2$ models, and 0.185 for p-terphenyl. This certificate avoids phase unwrapping
and preserves the complete harmonic manifold.

The C++ Raman TI kernel sums intermediate-state amplitudes coherently. Its
TD counterpart performs an analytic partial Gaussian trace, retains the
external-state Hermite coefficients and integrates the carrier exactly over
cubic interpolants. Multiple electronic surfaces and the resonant and
antiresonant terms are combined before rotational averaging. Adaptive optical
TI uses native C1/C2 marginal recurrences, a conditional priority heap and a
collision-checked multiword Hermite cache. The established checkpoint schema
retains pending states, accumulated strengths and line assignments. Regression
checks include continuation across numerical interfaces, resource-budget
failures, signed optical forms and order-by-order VPT2 contractions.

\subsection{TI comparison at a common minimum recovery}
A separate three-run control requires at least 54\% of the exact positive
$S_2$ FC+HT sum, $0.003931097277643119$~a.u.$^2$, in both programs.
One thread is used for selection, recurrence and BLAS. Each process starts
from physical input and exports assignments plus a 3001-point Gaussian
profile. Native Python imports, preparation, uncompressed line export and
the final restart checkpoint are included in its wall time. Program order
is alternated between repetitions, and no measured process overlaps another
process of this benchmark campaign.

Table~\ref{tab:si-matched-ti-execution} compares TI costs at the common minimum intensity threshold, retaining each measured overshoot.

\begin{table}[!htbp]
\centering
\caption{Common-minimum TI recovery, with actual overshoots retained.
Wall times are medians of three fresh processes; parentheses give observed
ranges on the shared server. The two programs need not select identical
transitions.}
\label{tab:si-matched-ti-execution}
\begin{tabular}{lrr}
\toprule
Program & Wall seconds & Exact strength recovered (\%)\\
\midrule
SATI C++ & 2.621 (2.571--2.721) & 54.063\\
FCclasses3 & 58.975 (58.337--71.902) & 54.747\\
\bottomrule
\end{tabular}
\end{table}

The median ratio is 22.50. Native selection uses 4096-transition batches
and stops at 98304 evaluated transitions. FCclasses3 retains its tested
C7 selector with \texttt{C1MAX=C2MAX=30} and
\texttt{MAXINT=10000000} per class. Its stick moment is converted with the
same map determinant and divided by the same exact operator sum as above;
its estimated full-spectrum normalization is not substituted for that sum.
This establishes a speed advantage for these settings at a common minimum
recovery, not an optimum over all FCclasses3 controls or a measured time to
99\%. Per-process records, source scripts, physical inputs and unrounded
fractions are frozen in \texttt{fcclasses\_complete\_20261030}.

\subsection{A finite 27-mode TI cost--coverage curve}
The harder $S_2$ FC+HT model tests whether short execution times also
recover substantial intensity. FCclasses' executed selector treats
classes through C7; its \texttt{MAXINT} is a per-class prescreening budget,
not the total number of transitions. The native selector permits all
classes within its transition budget. Both receive the same model. The C++ run uses one thread; the archived
FCclasses curve retains its four-thread resource limit and serial main loop. Native rows resume a saved checkpoint; the
reported cost is the cumulative wall time through that budget, including
checkpoint and retained-line export. FCclasses rows are complete separate
processes. Native line columns are disk backed and the reported export uses
lossless compression level 6. These are single shared-server observations,
with distinct algorithmic costs.

Table~\ref{tab:si-finite-ti-coverage} gives the measured finite-budget TI cost and intensity-recovery curve.

\begin{table}[!htbp]
\centering
\caption{$S_2$ finite TI cost and recovered positive FC+HT strength.
Recovery uses the same exact total $0.003931097277643119$~a.u.$^2$.
The native transition budget and FCclasses per-class budget are distinct.}
\label{tab:si-finite-ti-coverage}
\begin{tabular}{lrrr}
\toprule
Program & Budget & Wall seconds & Recovery (\%)\\
\midrule
FCclasses3 & $10^5$/class & 0.939 & 32.760\\
FCclasses3 & $10^6$/class & 6.995 & 45.434\\
FCclasses3 & $10^7$/class & 54.862 & 54.747\\
C++ & $10^5$ transitions & 1.254 & 54.204\\
C++ & $10^6$ transitions & 9.855 & 71.816\\
C++ & $5\times10^6$ transitions & 61.909 & 81.656\\
\bottomrule
\end{tabular}
\end{table}
FCclasses evaluates 0.643, 4.464 and 34.577 million integrals in these
three runs. The faster individual recurrence is therefore distinct from
the efficiency of the occupation selector. Its own printed full-spectrum
recoveries are 33.0, 45.8 and 55.2\%; the table instead divides the
map-normalized stick zero-order moment by the identical exact operator
sum used for the native calculation. This distinction matters for the
nonorthogonal coordinate map and the program's estimated total intensity.
The previously archived 50-million-transition native result reaches
91.370\%, with contributions through C14; it is not rerun or labeled
99\% converged. These finite curves show the actual cost--coverage
tradeoff for this input and do not supply an equal-completeness speedup
or a general scaling law.

The corresponding finite-budget cost--coverage curves are shown in the main text.

\subsection{Batch execution, block storage and compatible-request reuse}
A single C++ batch performs conditional selection, packed recurrence,
complex dipole amplitudes, thermal populations, transition energies and
class reductions, preserving candidate order and positive-strength stopping.
Checkpoints stream blocks into standard \href{https://numpy.org/}{NumPy} archives with integrity checks
and atomic replacement. Disk-backed output appends fields per batch and
returns read-only memory maps. Finite TI also returns columns directly,
avoiding a Python dictionary for each transition.

The integral workspace persists across batches. A transient index combines an
8-bit hash fingerprint and a node identifier in one 32-bit word when the cache
capacity fits 24 node bits; larger capacities use 64-bit entries. Matching
fingerprints always require a full packed-key comparison. The checkpoint keys,
node identifiers and hash function remain unchanged. Lowering masks, square
roots, iterative stacks and index buffers are reused, with explicit rebuilds
when their backing storage changes.

An optional degree-level path discovers the same dependency graph and cache
insertion order as the direct recurrence. It retains the linear-first,
ascending-axis edge order for each node, then contracts nodes by increasing
total occupation. Each dependency lowers that occupation by one or two.
Nodes within a level are therefore independent. A persistent executor accepts
1--32 workers and parallelizes only levels with at least 256 nodes. Discovery
and cache insertion stay serial. Per-batch graph buffers are reused rather
than retained for the entire transition set.

Automatic execution tiles are fixed by the selected numerical route: 4096
transitions for native direct contraction, 8192 for degree-level contraction,
and 256 for the independent Python reference. They are capped by the declared
state and transition budgets and remain explicitly overridable. These choices
change execution granularity, not the proposal or the recurrence. The positive
strength stopping certificate is evaluated at batch boundaries; fixed-budget
checks compare identical ordered transitions, whereas target-based stopping
may retain an additional final batch. Scalar and class reductions can differ
at floating-point rounding precision when the grouping changes.

Asynchronous checkpoints use one writer and no request backlog. Snapshots are
taken after a completed batch: append-only tree/cache prefixes and line blocks
retain frozen extents, while mutable totals, components and pending states are
copied. Heap arrays are exported on the calculation thread. Generic callers
without the append-only ownership contract receive fully copied snapshots.
Hashing, archive writing, file synchronization and atomic replacement execute
in the background. Busy periodic requests are skipped; the final save waits
for all required writes before returning or releasing file-backed blocks.
Writer failures propagate and preserve the last complete archive. The archive
schema and physical-identity checks are unchanged. Background write durations
overlap the recurrence and are not additive critical-path costs; a final-only
save cannot hide its own completion time.

Table~\ref{tab:ti-export-cost} separates the output cost from the numerical
work. Each row retains the same five-million-transition frontier and the
same 81.6559\% of the exact strength sum; all formats are lossless. The
cumulative calculation, checkpoint and disk-backed columns cost is 34.21~s
across the three resumed stages. Compression is a storage choice, and its
cost must be included when comparing full workflow times.
\begin{table}[!htbp]
\centering
\caption{Cumulative TI workflow costs through five million transitions.
The calculation is identical in each row. Archive size refers to the final
five-million-transition line table; export time includes all three stages.}
\label{tab:ti-export-cost}
\begin{tabular}{lrrr}
\toprule
Compression level & Export seconds & Total seconds & Final archive (MB)\\
\midrule
6 & 27.70 & 61.91 & 294\\
1 & 11.50 & 45.71 & 304\\
0 (stored) & 1.15 & 35.36 & 590\\
\bottomrule
\end{tabular}
\end{table}
A field-by-field check against the archived transition table gives identical
occupations, ordering and FC overlaps. The maximum strength difference is
$1.7\times10^{-21}$~a.u.$^2$ and the maximum position difference is
$2.9\times10^{-11}$~cm$^{-1}$. An archived v1 checkpoint continues from
one million to 1000001 transitions with its source file preserved.

The preparation context retains immutable Gaussian geometry and finite
overlap indices/values in a least-recently-used cache limited to 128~MiB and
32 entries. Exact keys include physical arrays, temperature, origin and
binary identity; overlap reuse also requires the same occupations and active
insertions. TD keys include the full operator, time grid and direction,
with broadening applied afterward. Changed inputs invalidate reuse, and
shared-context calls serialize. The size limit applies to retained cache data.

For 27-mode requests with the same finite TI basis and changed operator
coefficients, subsequent column-output calls take 11--13~ms, compared with
20--22~ms without reuse. For new widths on the same 32769-point TD grid,
subsequent calls take 6--8~ms, compared with approximately 700~ms without
reuse. Initial context fills cost 54~ms and 717~ms, respectively, and are
not counted as cache hits. Strength differences remain below
$2.6\times10^{-21}$~a.u.$^2$ and TD profiles are unchanged in this check.
These repeated-request costs are distinct from the cold-process external
benchmarks. All 223 regression tests pass on the compute server, including
restart across recurrence modes, deliberate hash-fingerprint collisions,
32/64-bit index storage, parallel ordered sums, immutable asynchronous
snapshots, injected write failures, disk-backed storage and invalidation checks.

\subsection{Batch-size and asynchronous-restart controls}
Table~\ref{tab:ti-batch-size} tests execution tiles without changing the
27-mode input, candidate ordering or finite transition budget. Each entry is
a single 100000-to-1000000 advance, including checkpoint completion,
disk-backed line materialization and stored lossless export; Python imports
are excluded. Direct and single-worker level contractions use CPU 49; the
four-worker executor uses CPUs 49, 51, 104 and 120. The host is nonexclusive.
The sweep does not establish a universal optimal tile. In particular, larger
serial tiles do not uniformly reduce elapsed time, and I/O contention is
visible in the wall/CPU differences. Tile size remains an explicit control.

\begin{table}[!htbp]
\centering
\caption{TI tile sweep. Entries are wall seconds, with process CPU seconds
in parentheses. Process CPU includes all recurrence threads. A dash denotes
an unexecuted setting. All cases retain 71.8156\% of the exact strength sum
at one million transitions.}
\label{tab:ti-batch-size}
\begin{tabular}{rrrr}
\toprule
Tile & Direct, 1 worker & Levels, 1 worker & Levels, 4 workers\\
\midrule
1024 & 7.25 (6.93) & 7.37 (7.03) & 5.59 (5.23)\\
4096 & 7.14 (6.82) & 7.93 (7.10) & 5.92 (5.65)\\
8192 & -- & 8.48 (6.82) & 5.46 (5.20)\\
16384 & -- & 10.06 (6.39) & 5.55 (5.24)\\
\bottomrule
\end{tabular}
\end{table}

Table~\ref{tab:ti-async-checkpoint} isolates restart scheduling at a fixed tile
of 4096. The one-million run saves every 128 batches and at completion. The
one-to-five-million advance resumes identical copies of a one-million
checkpoint, saves every 256 batches and at completion, with time gating set
to zero. Both writer modes complete the same number of saves and skip none.
The final completion barrier and lossless line export are included. These
are single shared-server observations: asynchronous writing is slightly
slower in the smaller control and modestly faster in the larger one.
\begin{table}[!htbp]
\centering
\caption{Matched periodic-checkpoint controls. CPU seconds are process sums,
including the background writer. No pending write is excluded from wall time.}
\label{tab:ti-async-checkpoint}
\begin{tabular}{llrrr}
\toprule
Advance & Writer & Wall (s) & CPU (s) & Saves\\
\midrule
$0\to10^6$ & sync & 8.84 & 8.39 & 2\\
$0\to10^6$ & async & 9.23 & 8.92 & 2\\
$10^6\to5\times10^6$ & sync & 45.47 & 40.44 & 4\\
$10^6\to5\times10^6$ & async & 43.70 & 34.33 & 4\\
\bottomrule
\end{tabular}
\end{table}

All line columns agree byte for byte across the tested tiles and writer
modes. At a fixed tile, all 27 numerical checkpoint fields agree byte for
byte between synchronous and asynchronous execution. Across tile sizes,
accumulated reductions differ by at most $1.6\times10^{-17}$~a.u.$^2$;
class summaries are compared by class label. Cache keys, values and ordered
line data remain exact. The five-million strength recovery is unchanged at
81.6559\%, so these controls support execution and restart rather than
completeness of the molecular spectrum.

\subsection{Frozen-data reproduction}
\texttt{python3 scripts/reproduce\_paper.py} verifies the SHA-256 package record,
checks the remote host identity, and regenerates every external figure on
the compute server in an isolated directory. It then compiles the article
and SI locally using the frozen \href{https://www.tug.org/texlive/}{TeX} table sources and regenerated figures.
Selected unrounded numerical tables are also exported as TSV files.
The replay records Python, NumPy, \href{https://scipy.org/}{SciPy}, \href{https://matplotlib.org/}{Matplotlib} and TeX versions,
source and evidence revisions, and input/output hashes. It launches no
electronic calculation, potential fit, nuclear diagonalization or sampling.
This is publication replay from frozen evidence, not a rerun of the
electronic and nuclear acquisition campaigns. PDF timestamps and renderer
versions can change binary hashes without changing the numerical inputs.

\FloatBarrier
\section{Continuum implementation: separating electrostatic representations}
\label{sec:continuum-electrostatic-validation}
Carbon monoxide (CO) provides a continuum-operator comparison between the independent implementation used here and the established \href{https://doi.org/10.1016/j.cpc.2023.109038}{TIRESIA} program. The same imported Gaussian electronic
density, nuclear geometry, central radial B-splines, angular channels and
integration grid were used. The central spaces contain 1430, 1320 and 1210
functions for $\Sigma$, $\Pi$ and $\Delta$, respectively. The comparison uses
the same Bloch boundary term in both Hamiltonians.

A comparison with separately constructed electrostatic potentials gives maximum matrix-element differences of
$6.38\times10^{-5}$, $5.22\times10^{-5}$ and $3.70\times10^{-5}$~hartree.
These arise from different representations of the Hartree field: the independent continuum solver
evaluates the Coulomb potential of the Gaussian AO density analytically,
whereas TIRESIA solves Poisson's equation in a finite multicentre spline
space. The LB94 exchange--correlation contribution was verified separately;
its small difference due to rounded legacy constants does not explain the
electrostatic residual.

The distinction was tested directly by importing the already computed
TIRESIA Hartree plus exchange--correlation multipoles, before addition of
the nuclear potential. With these identical electronic fields, the independent continuum assembly agrees with the reference central Hamiltonians to
$1.14\times10^{-12}$, $1.37\times10^{-12}$ and
$1.14\times10^{-12}$~hartree for $\Sigma$, $\Pi$ and $\Delta$.
The first five central-only $\Sigma$ eigenvalues and first $\Pi$ eigenvalue
then agree within $4.6\times10^{-6}$~cm$^{-1}$; with separately constructed fields,
the second $\Sigma$ root differed by 35.57~cm$^{-1}$. These diagnostics test the restricted central spaces.

Two further controls check the analytic Hartree evaluation independently.
Increasing the polar quadrature from 40 to 80 nodes in each of six panels
changes Hamiltonian entries by at most $4.58\times10^{-11}$~hartree.
Increasing radial quadrature from 12 to 24 and then 48 nodes per interval
changes entries by $1.38\times10^{-8}$ and $4.53\times10^{-11}$~hartree,
respectively, while leaving the original reference discrepancy unchanged.
Furthermore, integration of $\rho(\mathbf r)V_H(\mathbf r)/2$ on independent
atom-centred DFT grids converges to the Coulomb self-energy obtained from
four-centre Gaussian integrals, 76.2021397770~hartree: errors on successively
larger grids are $3.63\times10^{-6}$, $-6.92\times10^{-8}$ and
$-8.25\times10^{-10}$~hartree.

Accordingly, the analytic Gaussian Hartree field is retained in production.
The reference-field substitution is confined to verification: it establishes
assembly equivalence at matched fields without forcing the production
potential to reproduce the finite Poisson approximation. This central-block
control complements the shared-operator continuum-solver comparison;
it does not by itself establish independent full multicentre spectral
convergence. Raw numerical results and reproduction scripts are retained with
the numerical electrostatic-validation record. The controls reuse the stored electronic calculations.

\subsection{Molecular tails and phase-consistent exterior matching}
The molecular continuum uses a separate inner spline box and exterior
propagation region. Complete channel-potential matrices are sampled once,
interpolated cubically with no extrapolation, and passed to the coupled
outgoing-wave propagator. Direct evaluations at intermediate radii check the
interpolation. At a fixed local field, moving the inner boundary from 20 to
25~bohr changes the complex amplitudes by less than $4\times10^{-9}$ in relative
norm in the water control. This separates inner-box convergence from the
choice of the exterior potential.

The raw LB94 potential of a finite Gaussian density requires an explicit
asymptotic prescription: its Gaussian density limit is $-2/r$, whereas an
exponential density gives $-1/r$. An optional quintic switch joins the XC
contribution to $-1/r$ over a declared interval, with continuous first and
second derivatives. Nuclear and Hartree multipoles are retained separately.
The interval is a parameter of the local model. In the present water controls,
changing it from 8--12 to 10--15~bohr changes summed strengths by up to 4\%;
this physical-model sensitivity is distinct from propagation convergence.
The dilute-density gradient magnitude is evaluated with a scaled Euclidean
norm to prevent underflow of otherwise finite Cartesian components.

The leading molecular exterior is represented as
\begin{equation}
 \mathbf V(r)=-\frac{Q}{r}\mathbf I+\frac{\mathbf D}{r^2},\qquad
 \operatorname{diag}[l(l+1)]+2\mathbf D
   =\mathbf U\operatorname{diag}(\lambda_a)\mathbf U^{\mathsf T}.
\end{equation}
Here $\mathbf D$ is the angular matrix of minus the molecular dipole dotted
into the radial unit vector; the dipole is evaluated with the same AO density
and origin as the local field. Each $\lambda_a$ defines an independent outgoing
Coulomb equation, integrated from its asymptotic expansion. Negative
$\lambda_a$, including values below $-1/4$, are retained directly in this real
radial equation. After transformation back, the original spherical-channel
Coulomb phases are restored. Thus the changing eigenvector gauge of
$\mathbf U$ does not enter the geometry-dependent photoionization amplitudes.
Higher multipoles are included through exterior propagation of the full field.
Independent Whittaker-function evaluations and invariance of coupled amplitudes
and scattering under a change of box radius verify this boundary construction.

For water, increasing the exterior reference radius from 60 to 120~bohr changes
complex amplitudes by at most 0.030\% and summed strengths by less than
0.0007\% with the dipole-aware reference. With a pure Coulomb reference, the
amplitude difference is about 32 times larger. The five-energy controls use
36 spherical channels and a 3168-function inner box; they take approximately
11--13~s on an Apple M3 Ultra workstation with four BLAS threads. Increasing
$l_{\max}$ from 5 to 6 changes summed strengths by at most 0.21\% over the five
energies. These are central-continuum convergence controls rather than timings
for an entire FCHT spectrum.

An equilibrium rDSD/jun-cc-pVTZ single point with the current correlated
density supplies the electronic field independently of the anharmonic force
calculation. Its imported AO density has 9.99999999996 electrons and agrees
with the producer's dipole, including the symmetry-forbidden transverse
components. The complete signed NR2 Dyson orbital is imported in the AO basis
of its own checkpoint; its squared norm agrees with the full pole strength
within $5.7\times10^{-10}$. Neither density nor Dyson orbital is rescaled.
The following controls validate geometry derivatives and their harmonic nuclear
FC/HT contraction separately from these central-field tests.

\subsection{Normal-coordinate amplitude derivatives and harmonic FC/HT}
The continuum operator is expanded in all three dimensionless initial-state
normal coordinates,
\begin{equation}
 d_{\alpha c}(\boldsymbol q;E)
 =C_{\alpha c}(E)+\sum_i H_{\alpha c,i}(E)q_i,
\end{equation}
where $\alpha$ denotes the Cartesian dipole component and $c$ the outgoing
angular channel. NR2/jun-cc-pVTZ Dyson orbitals and current
rDSD/jun-cc-pVTZ densities are evaluated at $q_i=\pm0.01,\pm0.02$;
the equilibrium point is reused. The 24 displaced electronic single points
terminate normally, with 128~GB and 32 threads allocated to each sequential
job. No displaced geometry is optimized. The electronic templates and all
formatted checkpoints are archived with the validation data.

Each checkpoint is rigidly aligned to its intended displaced geometry.
The corresponding AO rotation is applied to density and Dyson tensors, and
cross-geometry AO overlaps follow the signed Dyson orbital. All normalized
overlaps exceed 0.99999995, without rescaling pole strengths. The spline trial
space is fixed throughout the stencil; moving nuclear radii introduce only
integration cuts. Both real and imaginary amplitudes enter a fourth-order
central first derivative. Comparing the $h=0.02$ and $h/2$ central derivatives
gives a relative derivative-vector difference below $1.62\times10^{-5}$ over the three
modes and five electron energies. This diagnoses step sensitivity, rather than
a statistical uncertainty. The 13 continuum configurations at five energies
were evaluated in 46.1~s on the M3 Ultra workstation with four workers and four
threads per worker; electronic jobs and parent-process imports are excluded,
while worker startup is included. Per-point
content hashes allow completed amplitudes to be reused after file transfer.

The electronic-channel sum is reduced exactly to $n+1$ nuclear operators
$\boldsymbol O=(1,q_1,\ldots,q_n)$:
\begin{equation}
 W_{ab}(E)=\sum_{\alpha c}B_{\alpha c,a}(E)
 B_{\alpha c,b}^{*}(E),\qquad
 B_{\alpha c}=(C_{\alpha c},H_{\alpha c,1},\ldots,H_{\alpha c,n}).
\end{equation}
The convention for a nuclear transition is
$\sum_{ab}O_aW_{ab}O_b^*$. Separate constant, linear--linear and cross blocks
retain FC, HT and their interference. For water, 108 electronic components
therefore require only four nuclear operators, with no rank truncation.
Automatically generated valence ESALC coordinates and state-specific Wilson metrics
supply the nuclear model to the existing TI and TD implementations. The
Gaussian-normal to ESALC-normal displacement transport closes within
$4.2\times10^{-10}$; its full metric-dependent mode map is retained.

At zero temperature, TI includes 680 final states with total occupation at
most 14. It captures the FCHT sum rule to better than $10^{-14}$ relatively.
The independently evaluated TD correlation uses 4097 time points separated by
0.25~fs and Gaussian broadening with standard deviation 50~cm$^{-1}$. Across
all five electron energies (0.05, 0.10, 0.20, 0.50 and 1.00~hartree), the
largest TI/TD correlation difference is $3.4\times10^{-12}$ relative to the
FCHT sum rule, over every time point. Fourier-transformed TD densities agree
with independently broadened TI lines within $6.7\times10^{-14}$ of the peak
on all 8192 spectral points. The combined nuclear calculation and these
checks take approximately 0.30~s after imports and compiled-kernel caching;
this is a single small-molecule timing, not a scaling benchmark.

\begin{figure}[!htb]
\centering
\includegraphics[width=0.84\linewidth]{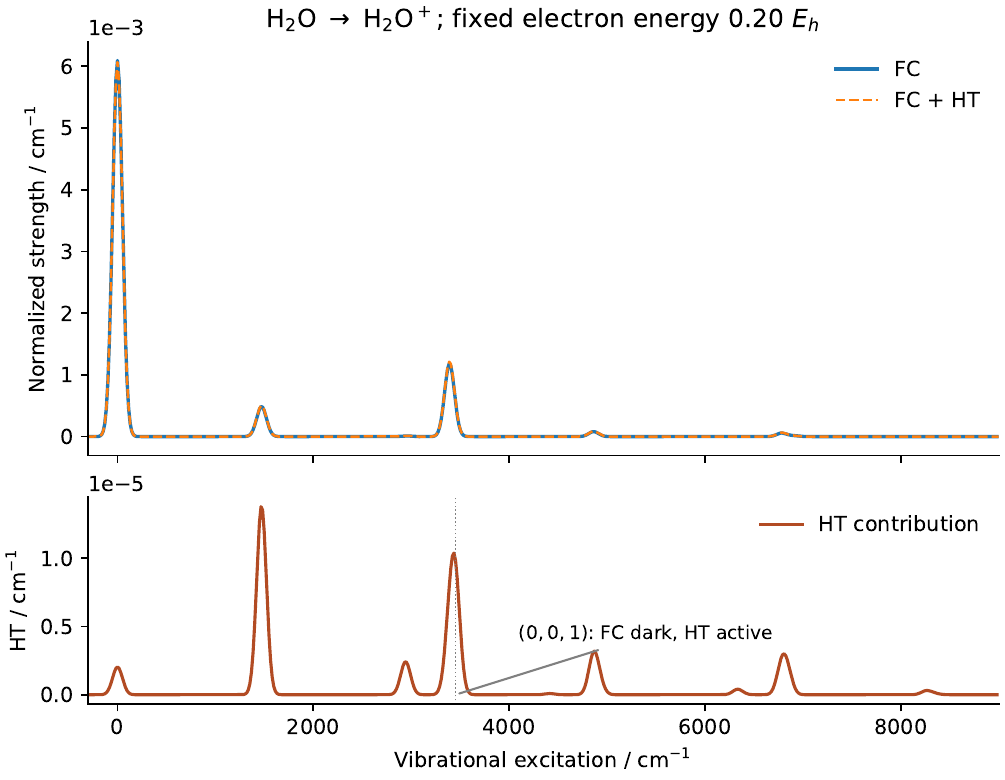}
\caption{Harmonic nuclear FC and linear-HT dipole-strength distributions for
water ionization at fixed outgoing-electron energy 0.20~hartree. Both panels
use $l_{\max}=5$, the same integrated FCHT normalization and Gaussian standard deviation
50~cm$^{-1}$. The lower panel resolves the HT contribution, including the
FC-dark antisymmetric fundamental. These are conditional electronic-energy
slices, without photon, spin or orientation prefactors.}
\label{fig:water-continuum-fcht}
\end{figure}

For the $l_{\max}=5$ stencil, the integrated HT/FC ratio decreases from 0.00771 at 0.05~hartree to 0.00293
at 1.00~hartree. At 0.20~hartree it is 0.00488. The antisymmetric fundamental
$(0,0,1)$ is FC-dark but HT-active, with summed dipole strength
$1.85\times10^{-3}$ in the continuum normalization used here, although the
first derivative of its pole strength vanishes by symmetry. Thus pole-strength
derivatives alone cannot represent the HT operator. Figure~\ref{fig:water-continuum-fcht}
shows the resulting conditional vibrational distribution.

Separate seven-configuration angular controls reuse the same electronic
points at $q_i=\pm0.02$. Increasing $l_{\max}$ from 5 to 6 changes the total
FCHT strength by less than 0.20\%, but the antisymmetric HT fundamental by up
to 4.92\%. Increasing it from 6 to 7 reduces these changes to 0.098\% and
2.39\%, respectively. The total integrated HT contribution changes by up to
1.78\% in the latter comparison. Thus convergence of the dominant integrated
strength does not certify equal relative accuracy for a weak HT line. At
$l_{\max}=7$ and 0.20~hartree, HT/FC is 0.00506 and the antisymmetric
fundamental strength is $1.96\times10^{-3}$. All angular controls retain the
same excellent independent TI/TD agreement. The seven-configuration runs take
81 and 177~s for $l_{\max}=6$ and 7, respectively, on the shared workstation;
these timings are recorded separately from the five-point baseline.

Changing the explicit XC-tail switching interval from 8--12 to 10--15~bohr
changes the integrated FCHT strength by up to 4.1\%, much more than the
finite-difference or TI/TD numerical residuals. Both versions are archived.
The present validation establishes a harmonic nuclear FC/linear-HT
calculation at fixed electron kinetic energy. Absolute cross sections,
fixed-photon-energy spectra and anharmonic nuclear dressing require their
corresponding additional inputs or calculations; they are not inferred from
the numerical TI/TD agreement.

\subsection{Accepted internal-coordinate ZAFF SPF potentials and direct DVR}
\label{sec:water-internal-dvr}
The molecular anharmonic calculation uses $q=(R_1,R_2,\theta)$, with distances
in angstrom and angle in radians. Atom ordering is O,H,H. The fixed embedding
is $\mathbf{x}_{\rm O}=0$, $\mathbf{x}_{\rm H1}=(0,0,R_1)$ and
$\mathbf{x}_{\rm H2}=(R_2\sin\theta,0,R_2\cos\theta)$. Both electronic
surfaces share this physical coordinate representation. The rDSD/jun-cc-pVTZ reference values are
$R_1=R_2=0.96103895$~\AA, $\theta=104.4709^\circ$ for the neutral and
$R_1=R_2=1.00077570$~\AA, $\theta=109.4663^\circ$ for the cation; precise
values and isotope masses are retained in the numerical archive.

\paragraph{Finite information and functional form.}
Per surface, the frozen data budget contains one reference geometry,
19 finite-energy fitting geometries, six previously computed finite E/G/H
records, and four excluded energy checks. The original acquisition comprised
23 single points per surface, reduced by hydrogen-exchange symmetry. The
additional six observations are recovered from the already completed
anharmonic Gaussian logs, not calculated anew. Printed Cartesian displacements,
paired D2Numr markers, forces, Hessians and nuclear repulsions establish their
identity and frame. A common recovered amplitude is approximately
0.01889726123~bohr, with maximum nuclear-repulsion mismatches below
$2.2\times10^{-9}$~hartree. The antisymmetric $\pm$ pair is related by
hydrogen exchange. No third- or fourth-order electronic derivative is a fit
target, and no dense electronic grid is acquired for DVR convergence.
The four excluded points were used in model assessment and are
excluded fitting checks, not a newly blinded validation set.

Define the resident SPF radial variable and native angular variable as
\begin{equation}
 y_a=1-\frac{R_{a0}}{R_a},\qquad
 x_\theta=\frac{\cos\theta_0-\cos\theta}{\sin\theta_0}.
\end{equation}
The potential consists of pure radial SPF series, a native angular series,
and connected mixed native products,
\begin{equation}
 V(q)=\sum_{a=1}^2\sum_n c_{an}y_a^n+
 S_\theta(q)V_\theta(x_\theta)+
 \sum_t C_t S_t(q)\prod_{i\in t}u_i^{p_{ti}},
 \label{eq:water-spf}
\end{equation}
where $u_a=R_{a0}y_a$ for a bond and $u_\theta=x_\theta$.
Support factors are the existing native functions
$S_t=\exp[-\sum_{a\in\mathcal B_t}(\Delta R_a/R_{a0})^4]$,
with $\mathcal B_t$ the union of defining bonds. Pure radial SPF terms have
no extra damping envelope. The angular reference quadratic is a cosine
function, and reference off-diagonal force constants supply mixed quadratic
products. The reference gradient and full Hessian are fixed, including the
nonzero-gradient curvature terms required by the coordinate change.
Higher pure powers 3--6 and connected mixed degrees 3--5 are fitted. These
powers refer to the native variables, not to a global Cartesian Taylor
polynomial. Exact H-exchange ties leave 29 identified coefficient groups.
The mixed-degree hierarchy 4,5,6 selects the lowest degree passing calibration;
degree 4 fails for both states, and degree 5 passes. The four excluded energies
are not used in this selection.

The general native multi-anchor E/G/H fitter supplies the weighted linear
solve. Cartesian derivatives are transformed using both the coordinate
Jacobian and its second derivative. Conditioning scales are 0.1~\AA,
0.1~\AA\ and $10^\circ$. Energy tolerance is 5~cm$^{-1}$; derivative
weights propagate declared coordinate uncertainty $10^{-8}$~\AA, force
rounding $5\times10^{-10}$~Eh/bohr and Hessian rounding
$5\times10^{-7}$~Eh/bohr$^2$, with safety factor five. These weights describe
source precision and fitting priorities; they are not statistical confidence
intervals for the PES.

Radial shape regularization adds the first eligible positive even power from
8,10,12,14,16,18,20. Eligibility requires one radial minimum, no maximum,
positive leading compression behavior, and protected-energy changes within
the remaining calibration headroom. Order 14 is selected for both surfaces.
The additions preserve reference derivatives through fourth order and change
fitting energies by at most 1.2174 and 0.1079~cm$^{-1}$ for neutral and
cation. Actual analytic E/G/H are checked again after regularization.
Maximum fitting-energy errors are 3.570 and 4.634~cm$^{-1}$; maximum excluded
errors are 5.0213 and 3.4871~cm$^{-1}$. The neutral exceeds the nominal
5~cm$^{-1}$ target by 0.0213~cm$^{-1}$. The fields are accepted at this
stated accuracy for the present finite molecular calculation.

For each state, 27 bounded minimizations in the central domain and 27 in the
sampled box converge to one interior minimum, with no additional minimum
found. The sampled box has $\Delta R\in[-0.30,0.50]$~\AA\ and
$\Delta\theta\in[-45,55]^\circ$. These finite checks do not prove global
uniqueness. The SPF limits are finite, but fragment energies and physical
long-range dispersion coefficients are not independently determined by this
near-molecular data set.

\paragraph{Complete kinetic operator and direct transition integrals.}
The full six-dimensional vibration/rotation metric is formed from the
Cartesian embedding and isotope masses and inverted before selecting its
$J=0$ vibrational block. Thus the kinetic matrix $G^{ij}(q)$ includes all
mixed terms. With $\rho\propto R_1^2R_2^2\sin\theta$ and
$\widetilde\psi=\sqrt\rho\,\psi$, the flat-measure weak form is
\begin{equation}
 \langle\widetilde\phi|T|\widetilde\psi\rangle
 =\frac12\int dq\,
 \left[(\partial_i-\tfrac12\partial_i\log\rho)\widetilde\phi\right]^*
 G^{ij}
 \left[(\partial_j-\tfrac12\partial_j\log\rho)\widetilde\psi\right].
\end{equation}
The orthogonal sine DVR evaluates this form on an oversampled tensor Gaussian
quadrature. Values and derivatives share the same quadrature
orthonormalization. Tensor contractions apply the operator without a dense
Hamiltonian. A square first-derivative collocation matrix is avoided because
it can introduce null modes for odd basis sizes.

The continuum derivatives already calculated in initial dimensionless normal
coordinates are transported by
$J_{ai}=\partial q_a/\partial Q_i$ at the neutral reference:
$\mu(q)=\mu_0+\sum_i h_i[J^{-1}(q-q_0)]_i$.
The same signed complex channel amplitudes, derivatives, and Hermitian
FC/HT/interference forms are retained. Both nuclear wavefunctions are
contracted directly in their common flat measure; no second volume factor,
harmonic projection cutoff, or grid Duschinsky rotation is inserted.
This specifies a linear internal-coordinate continuation of the operator,
not a claim that all nonlinear property derivatives have been computed.

\paragraph{Numerical closure.}
Dirichlet domains are expanded independently of basis and quadrature:
\begin{center}
\begin{tabular}{lcc}
\hline
Domain & Both bond intervals (\AA) & Angle interval (degrees)\\
\hline
Central & $[0.75,1.40]$ & $[75,150]$\\
Wide & $[0.705,1.455]$ & $[65,159]$\\
Extended & $[0.65,1.55]$ & $[50,175]$\\
Outer & $[0.60,1.60]$ & $[40,178]$\\
\hline
\end{tabular}
\end{center}
The last two domains deliberately test the analytic continuation outside
parts of the sampled molecular box. They use only Eq.~\eqref{eq:water-spf};
no additional electronic points or automatic electronic refinement are introduced.
Agreement of their localized bands is a numerical-domain test of the fitted
model, not independent evidence for extrapolated PES accuracy.

The basis sequence is 12,16,20,24 on the central domain, 16,20,24 on the wide
domain, 20,24,28 on the extended domain, and 24,28 on the outer domain.
The usual quadrature is $2N+4$ points per coordinate; an independent check
uses 60 instead of 52 points at fixed $N=24$ in the extended domain.
Twelve eigenpairs per surface are saved and resumed. The adopted result is
the outer-domain $N=28$, 60-point quadrature calculation.

For the five assigned bands, basis enlargement from 24 to 28 in the outer
domain changes spacings by at most $1.21\times10^{-6}$~cm$^{-1}$ and FC/HT
strengths by at most $2.0\times10^{-9}$ relatively. Fixed-basis quadrature
refinement changes these quantities by at most $5.3\times10^{-11}$~cm$^{-1}$
and $1.9\times10^{-13}$. Enlarging the extended to the outer domain changes
spacings by at most 0.01506~cm$^{-1}$ and strengths by at most
$1.987\times10^{-5}$ relatively over the five electron energies.
The selected eigenpair residuals are below $1.1\times10^{-6}$~cm$^{-1}$;
quadrature norm errors are below $1.7\times10^{-14}$. Probabilities in the
outermost tenth of any coordinate interval are below $5.3\times10^{-5}$
for the five assigned cationic states and smaller for the ground state.

The twelve final states retain 99.0748--99.1912\% of the linear-operator sum
rule. Their highest levels are less localized: the largest final-domain
spacing change among all twelve is 3.94~cm$^{-1}$, compared with
0.01506~cm$^{-1}$ for the five bands in the main table. Therefore the
retained sum-rule fraction is reported separately from band-specific
convergence, and a complete anharmonic high-energy spectrum is not claimed.

The adopted cationic fundamentals are 1403.42, 3221.34 and
3268.89~cm$^{-1}$; the bending overtone is 2758.59~cm$^{-1}$.
The main table and figure compare this internal-coordinate ZAFF Hamiltonian
with the ESALC/GF harmonic model. The following rectilinear-quartic section
instead compares VPT2 and local variational wavefunctions on identical
polynomial fields. Its weak-band discrepancy remains a test of that
perturbation approximation and is not transferred to the different ZAFF
Hamiltonian as an accuracy estimate.

\paragraph{Implementation and reproducibility.}
The general ZAFF potential routines supply SPF values and analytic derivatives. The native batch energy path avoids constructing Cartesian
gradients at dense DVR nodes. The nuclear solver evaluates the complete metric and
matrix-free kinetic contractions and exports quadrature wavefunctions for
direct property integrals. The continuum transition-operator set combines the
unchanged complex continuum channels. The molecular driver only specifies
sources, domains, finite basis controls and assignments. Thirty-one focused
tests pass, including analytic model derivatives, batch/scalar equality,
triatomic metric identities, independent weak-form matrix checks, quadrature
norms, fit regression and cached Gaussian data integrity.

On the Apple workstation (Apple M3 Ultra), with one numerical thread per process,
preparation plus twelve-state solutions on both surfaces and all five-energy
FC/HT contractions take 24.4~s at $N=24$ in the extended domain and 46.9~s at
$N=28$. A separate 52-to-60-point quadrature check at fixed $N=24$ takes
38.3~s. These are observed single-run wall times after imports, not repeated
hardware benchmarks or comparisons with other programs. Checkpoints, fitted
coefficients, source hashes and all numerical controls are archived; all
electronic points are reused.

\subsection{Provenance of the external water comparison}
The experimental check uses Reutt \emph{et al.}'s spacings as tabulated
in Table~II of Engin \emph{et al.}, and the Brundle--Turner FC factors
in Table~2a of Botter and Rosenstock.\cite{Reutt1986Water,Engin2019Water,BotterRosenstock1969}
The original tables order symmetric stretch before bend; occupations
are permuted to the bend-first convention used in this paper.
Observed energies 174.4, 397.4 and 344.0~meV correspond to our bands
100, 010 and 200. The conversion is 8.065544005~cm$^{-1}$/meV.
Reported energy error limits are 0.4--0.5~meV. Experimental FC error
limits retain their original probable-error interpretation.
No FC intensity is supplied for our 200 bending overtone: the 200
intensity in the original stretch-first table instead refers to the
symmetric-stretch overtone and is not reassigned to our bending band.

Calculated FC factors are the squared constant-operator integrals from
the already saved internal-DVR contractions. They are not obtained by
normalizing the five displayed FC/HT strengths. The external comparison
therefore tests the nuclear Condon sector, separately from the specified
energy-dependent continuum calculation. Relative factors in the main
figure are divided by each data set's origin FC factor; no three-band
area or PES parameter is fitted. Machine-readable observations, units,
source identifiers and calculated differences are preserved in
\texttt{numerical\_evidence/h2o\_ionization/experimental\_comparison\_20261027/}.
The local-quartic matched-Hamiltonian VPT2 control is reported in
the main text as well as detailed in the following subsection.

\subsection{Two-surface anharmonic FC/HT and a finite-domain variational control}
The continuum amplitudes also dress the archived rDSD/jun-cc-pVTZ cubic and
quartic fields of both water species. Each field contains all six
symmetry-allowed cubic and nine quartic derivatives of the three-mode model.
The printed harmonic/anharmonic index correspondence is applied explicitly.
Here the nuclear Hamiltonians use the original rectilinear Gaussian normal
coordinates and constant diagonal kinetic metrics. Their matching harmonic
reference is recomputed in that same representation. The curvilinear harmonic
model is a separate coordinate-model control;
its frequencies or metric are not silently combined with these rectilinear
higher derivatives. The extension reuses the stored electronic calculations.

The signed continuum operator is expanded in the original initial Gaussian
modal gauge, using the $l_{\max}=7$ amplitudes and derivatives. The
two-surface SAM contraction acts on both independently corrected nuclear
wavefunctions. Four polynomial operators are evaluated in two blocks of the
native three-component implementation, then combined with the same Hermitian
channel forms as above. Strengths are strictly truncated at mechanical order
two. Squares of summed corrected amplitudes are retained separately and are
not substituted for the perturbative observable. Twelve explicit targets
include the ground band, fundamentals, overtones, combinations and the third
and fourth bending excitations. Independent 16-point-per-axis Gauss--Hermite
integration checks their FC amplitude orders within $5.2\times10^{-16}$.
The complete selected-target FC/HT calculation, matching harmonic reference
and this quadrature check take 0.62~s on the Apple workstation after imports.

Both unchanged quartic potentials have negative leading rays: the minimum
sampled contractions of the fourth-derivative tensors are approximately
$-512$ and $-500$~cm$^{-1}$ for neutral and cation, respectively. Thus a global
bound-state interpretation of these polynomials is inappropriate. A local
variational comparison instead uses the existing sine-spectral kinetic
operator with Dirichlet boundaries at $q_i=\pm L$. The analytic quartic is
sampled directly on that nuclear grid; no electronic points, coefficient fits
or asymptotic completion are introduced. The domain/grid sequence is
$(L,N)=(3.5,35),(4,39),(4.5,43),(4.5,51),(5,55)$ on each surface, with 12
saved eigenstates per calculation.

Saved eigenvectors are projected onto harmonic occupations by tensor Hermite
transforms. Their raw projection coefficients and norms are retained without
normalization or orthogonalization. The full two-surface overlap matrix is
reused for every FC and linear-HT operator. Coordinate ladder action extends
the initial occupation basis before constructing the overlap, retaining the
contribution beyond its top shell. Harmonic projection cutoffs of 12, 14 and
16 provide an independent check. At $L=5$ the neutral ground-state norm loss
is $3.7\times10^{-10}$; the largest loss among the five assigned cationic
states is $1.3\times10^{-5}$. These are diagnostics of a finite projection,
not global wavefunction error bounds.

For the five assigned bands, increasing the harmonic projection cutoff from
14 to 16 changes FCHT strengths by less than 0.00040\% over all five electron
energies. At $L=4.5$, increasing $N$ from 43 to 51 changes the five spacings by
less than $4.6\times10^{-5}$~cm$^{-1}$. The final domain expansion from
$L=4.5$ to 5 changes fundamental spacings by at most 0.052~cm$^{-1}$ and their
FCHT strengths by less than 0.007\%. The bending overtone remains more
sensitive: its spacing changes by 1.63~cm$^{-1}$ and its intensity by up to
1.47\% across the five electron energies (1.21\% at 0.20~hartree). Its outer
box probability at $L=5$ is $9.5\times10^{-4}$. These observed domain changes
are reported explicitly rather than interpreted as certified global errors.

Table~\ref{tab:water-anharmonic-continuum} resolves the assigned water bands into their nuclear-model positions and conditional FC+HT strengths.

\begin{table}[!htbp]
\centering
\caption{Conditional FC/HT dipole strengths for five assigned water-ionization
bands at outgoing-electron energy 0.20~hartree. All columns use the same
continuum operator and matching rectilinear nuclear models. The variational
column uses $L=5$, $N=55$ and harmonic projection cutoff 16; it is a local
finite-domain result for the unchanged quartic.}
\label{tab:water-anharmonic-continuum}
\begin{tabular}{lrrr}
\hline
Final occupation & Harmonic & Strict VPT2 & Local variational\\
\hline
000 & 1.434910 & 1.446318 & 1.448219\\
100 & 0.117384 & 0.099378 & 0.095998\\
010 & 0.288322 & 0.266244 & 0.275825\\
001 & 0.001959 & 0.001789 & 0.001810\\
200 & 0.000974 & 0.003258 & 0.000634\\
\hline
\end{tabular}
\end{table}

The local variational fundamental spacings are 1379.70, 3267.47 and
3307.97~cm$^{-1}$, compared with potential-only VPT2 values of 1393.14,
3223.57 and 3258.06~cm$^{-1}$. Gaussian's total anharmonic frequencies include
additional rovibrational kinetic contributions and are not identified with
either potential-only column. The important discrepancy concerns the weak
bending overtone: strict VPT2 gives about 5.1 times the local variational
strength, whereas the latter changes by only 1.2\% in the last domain check
at this electron energy. Its cationic first- and second-order correction
norms are 0.279 and 0.120. Thus moderate correction norms and machine-precision
contraction checks do not ensure accurate relative intensities for weak
transitions. The three-mode comparison separates this perturbative limitation
from numerical overlap errors and from the local-field tail sensitivity.

\begin{figure}[!htb]
\centering
\includegraphics[width=0.84\linewidth]{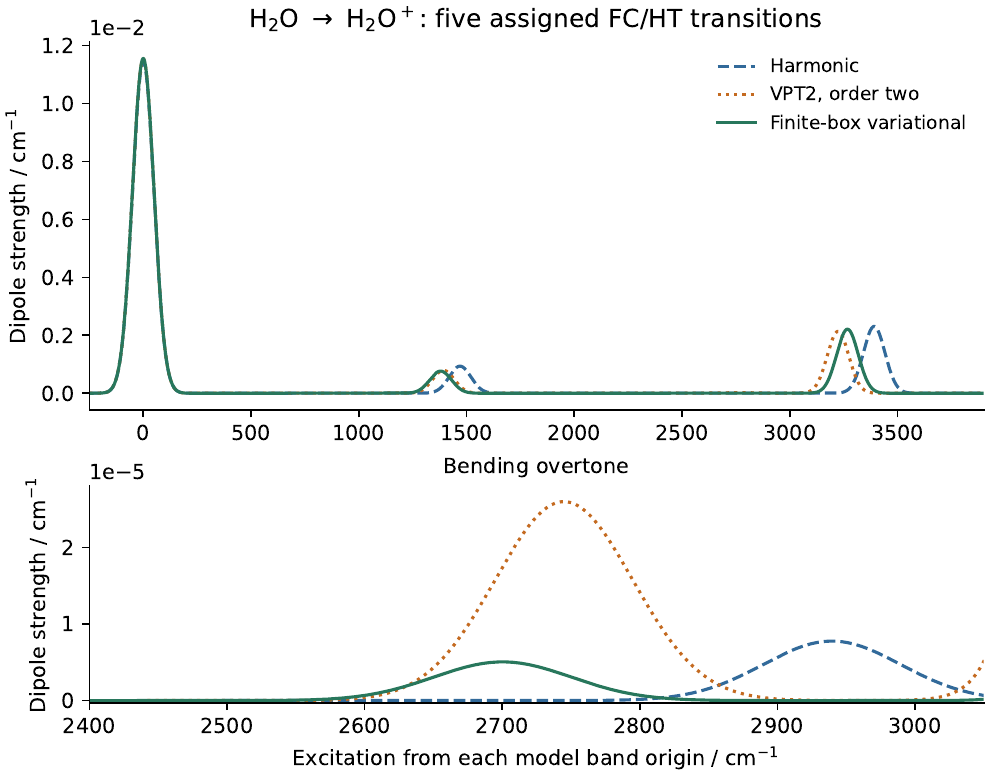}
\caption{The five assigned FC/HT transitions of
Table~\ref{tab:water-anharmonic-continuum}, with Gaussian standard deviation
50~cm$^{-1}$ and each model's own band origin as zero. Absolute conditional
dipole strengths are retained, without area normalization. The lower panel
resolves the bending overtone and its large perturbative error. These partial
profiles are neither complete anharmonic spectra nor experimental comparisons.}
\label{fig:water-anharmonic-continuum}
\end{figure}

Figure~\ref{fig:water-anharmonic-continuum} compares the retained harmonic and anharmonic water photoionization bands.

\section{Pyrazine two-photon vibronic interface test}
\label{sec:pyrazine-tpa-si}

The geometry is taken from the 2025 library of chemical building blocks (LCB25), using its Pisa composite scheme (PCS2) entry, and was canonicalized to $D_{2h}$ by ORACLE before conversion to \href{https://www.turbomole.org/}{TURBOMOLE} coordinates.\cite{Lazzari2025LCB25} The electronic response is recomputed with CAM-B3LYP/aug-cc-pVTZ, which provides diffuse functions for the Rydberg orbital. The state-specific geometry and Hessian are obtained with Gaussian for the TDDFT root identified as $n\rightarrow3s$; the direct quadratic response tensor is evaluated with TURBOMOLE at the same $D_{2h}$ reference geometry. At the planar reference, the equal-frequency tensor is
\begin{equation}
 S=\operatorname{diag}(-2.3300466588,-21.6527668910,10.5759893099)\ \mathrm{a.u.},
\end{equation}
and the corresponding CAM-B3LYP root is 54\,110.4~cm$^{-1}$. The state assignment is supported by the dominant $21\rightarrow24$ configuration in the diffuse orbital space. The experimental notation is $n\rightarrow3s$, $^1A_g\leftarrow{}^1A_g$ and the measured origin is 50\,835~cm$^{-1}$.\cite{Dion1995Rydberg}

For degenerate photons, the electronic vertex is the symmetric rank-two amplitude $S_{ab}=S_{ba}$. The rotationally averaged response is written in terms of
\begin{equation}
 d_F=\frac1{30}\sum_{ab}S_{aa}S_{bb},\qquad
 d_G=\frac1{30}\sum_{ab}S_{ab}S_{ab},\qquad
 d_H=\frac1{30}\sum_{ab}S_{ab}S_{ba},
\end{equation}
with polarization factors
\begin{equation}
 \delta_{\parallel}=2d_F+2d_G+2d_H,\qquad
 \delta_{\perp}=-d_F+4d_G-d_H,\qquad
 \delta_{\mathrm{circ}}=-2d_F+3d_G+3d_H.
\end{equation}
These expressions are the isotropic orientation average used by the TURBOMOLE response output. In the present nuclear contraction the tensor is a constant FC vertex: for a final vibrational state $\vect{n}$,
\begin{equation}
 A_{ab}(\vect{n})=S_{ab}\,\langle\vect{n}_f|\vect{0}_i\rangle,
\end{equation}
The propagated profiles are the parallel channel and the two circular helicities; the perpendicular linear strength is retained as the electronic invariant above. No $\partial S_{ab}/\partial Q_j$ terms are included; therefore this test is FC-only and does not test two-photon HT derivatives.

The reported profile uses the full adiabatic harmonic Gaussian CAM-B3LYP/aug-cc-pVTZ field: two optimized geometries, both analytic Hessians, the mass-weighted displacement converted to dimensionless oscillator coordinates, and the orthogonalized Duschinsky map. The map orthogonality residual is $8.6\times10^{-15}$. The FC calculation starts from the vibrational ground state; TD uses a 0.25~fs step, 8192 Fourier points and Gaussian width 80~cm$^{-1}$. The electronic invariants are $\delta_{\parallel}=90.1326$, $\delta_{\perp}=52.6209$ and $\delta_{\mathrm{circ}}=105.2417$. Their integrated values follow directly from completeness for a constant vertex. Numerical comparison therefore uses the selected TI intensities and the broadened TD profile. For the constant tensor, each polarization is the corresponding electronic invariant multiplied by the same FC probability. The final adaptive TI selects 1024 physical transitions and recovers 0.9920415325 of the unit FC sum, with omitted fraction 0.0079584675. The total-occupation and per-mode caps are both 64, with all 24 modes available and no class restriction; selected states visit classes through six. The search evaluates 1024 candidates and 1672 cached integrals. The calculation took 2.05 s on the compute server with 32 worker threads. The line list is stored in columns with explicit initial and final occupation numbers. No intensity normalization is applied. With the same 80~cm$^{-1}$ Gaussian broadening, the maximum TI--TD profile difference is 0.22213\% of the TD peak and the integrated absolute residual is 0.0079584675. The Gaussian peak bound from the omitted FC probability is $3.97\times10^{-5}$~cm, compared with an actual maximum residual of $1.60\times10^{-6}$~cm on the unit-strength scale. The selected line list, profiles and diagnostics are archived together.

\begin{figure}[!htb]
\centering
\includegraphics[width=0.95\linewidth]{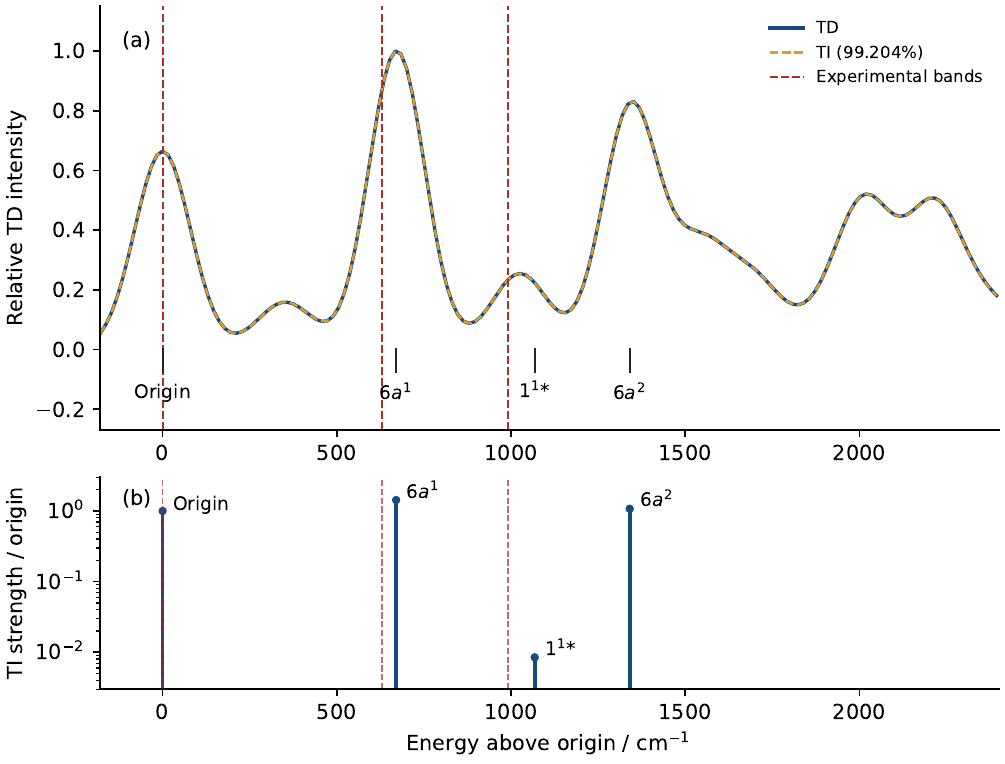}
\caption{Pyrazine FC two-photon spectrum from the full Gaussian CAM-B3LYP/aug-cc-pVTZ field. (a) TD profile aligned to the experimental origin, with measured bands indicated. (b) Wilson assignments and TI strengths relative to the origin; the asterisk marks the proposed $1^1$ assignment.}
\label{fig:pyrazine-tpa-si}
\end{figure}

The numerical record also retains a native TDA tight-binding $n\to\pi^\ast$ field for provenance. The physical Rydberg comparison and Figure~\ref{fig:pyrazine-tpa-si} use only the complete diffuse-basis Gaussian field.

The cold-molecule two-photon/one-photon ionization experiment reports the $n\rightarrow3s$, $^1A_g\leftarrow{}^1A_g$ origin at 50\,835~cm$^{-1}$ and assigned bands at 51\,464 and 51\,827~cm$^{-1}$. The full Gaussian origin is 53\,347.3~cm$^{-1}$, 2512.3~cm$^{-1}$ (0.312 eV) above the experimental origin. TI assigns $6a^1$ ($a_g$, program index 6) at 670.3377~cm$^{-1}$ and $6a^2$ at 1340.6754~cm$^{-1}$. The experimental $6a^1$ spacing is 629~cm$^{-1}$. The proposed $1^1$ ring-breathing fundamental ($a_g$, program index 12) at 1068.0389~cm$^{-1}$ is the candidate counterpart of the experimental $1^1$ band at 992~cm$^{-1}$. The corresponding strengths relative to the origin are 1.4280 and 0.00848 for $6a^1$ and the proposed $1^1$, and 1.0741 for $6a^2$. The proposed $1^1$ strength is only 0.59\% of $6a^1$, making the observed ring-breathing band weak in the constant-FC model. Tensor derivatives $\partial S_{ab}/\partial Q_j$ enter through the existing HT interface and add the coordinate-dependent two-photon intensity-borrowing terms needed to test this discrepancy. These assignments distinguish individual lines from maxima of the broadened spectrum. The profile is shifted to the experimental origin for comparison.

\FloatBarrier
\begingroup\small
\bibliographystyle{unsrt}
\bibliography{references,validation_references}
\endgroup